\documentclass[lettersize,journal]{IEEEtran}
\usepackage{amsmath,amsfonts}
\usepackage{algorithmic}
\usepackage{array}
\usepackage[caption=false,font=normalsize,labelfont=sf,textfont=sf]{subfig}
\usepackage{graphicx}
\usepackage{caption}
\graphicspath{{./figure/}}
\usepackage{soul}
\usepackage{color}
\usepackage{enumerate}
\usepackage{multirow}
\usepackage{amsmath}

\usepackage{amssymb}
\usepackage[linesnumbered,ruled]{algorithm2e}
\usepackage{booktabs}
\usepackage{array}
\usepackage{subfig}
\usepackage{url}
\usepackage{tabularx}
\usepackage{listings}
\usepackage{bm}
\usepackage{enumitem}
\usepackage{pifont}
\newcommand{\matr}[1]{\bm{#1}}
\newcommand{\vect}[1]{\bm{#1}}
\usepackage{xcolor}
\usepackage{textcomp}
\usepackage{makecell}
\usepackage{mathrsfs}
\def\BibTeX{{\rm B\kern-.05em{\sc i\kern-.025em b}\kern-.08em
    T\kern-.1667em\lower.7ex\hbox{E}\kern-.125emX}}
\usepackage{balance}

\begin{document}
\title{A Systematic Literature Survey of Sparse Matrix-Vector Multiplication}

\author{Jianhua Gao, Bingjie Liu, Weixing Ji, Hua Huang
\thanks{
Weixing Ji is the corresponding author.}
\thanks{Jianhua Gao, Weixing Ji, and Hua Huang are with the School of Artificial Intelligence, Beijing Normal University, Beijing 100875, China (e-mail: gaojh@bnu.edu.cn, jwx@bnu.edu.cn, huahuang@bnu.edu.cn).}
\thanks{Bingjie Liu is with the School of Computer Science and Technology, Beijing Institute of Technology, Beijing 100081, China (e-mail: liubj@bit.edu.cn).}
}


\maketitle

\begin{abstract}
  Sparse matrix-vector multiplication (SpMV) is a crucial computing kernel with widespread applications in iterative algorithms. Over the past decades, research on SpMV optimization has made remarkable strides, giving rise to various optimization contributions. However, the comprehensive and systematic literature survey that introduces, analyzes, discusses, and summarizes the advancements of SpMV in recent years is currently lacking. Aiming to fill this gap, this paper compares existing techniques and analyzes their strengths and weaknesses. We begin by highlighting two representative applications of SpMV, then conduct an in-depth overview of the important techniques that optimize SpMV on modern architectures, which we specifically classify as classic, auto-tuning, machine learning, and mixed-precision-based optimization. We also elaborate on the hardware-based architectures, including CPU, GPU, FPGA, processing in Memory, heterogeneous, and distributed platforms. We present a comprehensive experimental evaluation that compares the performance of state-of-the-art SpMV implementations. Based on our findings, we identify several challenges and point out future research directions. This survey is intended to provide researchers with a comprehensive understanding of SpMV optimization on modern architectures and provide guidance for future work. 
\end{abstract}

\begin{IEEEkeywords}
Sparse matrix-vector multiplication, sparse matrix, parallel computing, parallel architecture
\end{IEEEkeywords}

\section{Introduction}
Sparse matrix-vector multiplication (SpMV) is one of the essential and dominant computing kernels in many fields, such as scientific and engineering computing, data mining, and graph analysis. Equation \ref{equ:linearSystem} gives its formal representation, where $\matr{A}$ is a sparse matrix of $m\times n$, $\vect{x}$ and $\vect{y}$ are two dense vectors of $n\times 1$ and $m\times 1$, respectively. 

\begin{equation}\label{equ:linearSystem}
    \vect{y}=\matr{A}\vect{x}
\end{equation}

SpMV optimizations have garnered considerable attention from researchers since the 1970s. The challenge of SpMV optimization stems from the non-uniform distribution of non-zero elements (abbreviated as non-zeros) in sparse matrices, various hardware architectures, and diverse applications. Over the years, extensive research has been conducted on SpMV optimization, resulting in a number of valuable contributions and discussions. However, this abundance of research makes it challenging for researchers who are interested in SpMV optimization to gain a concise, systematic, and comprehensive understanding. 

Currently, several SpMV survey papers have been published, including works by Kulkarni et al. \cite{kulkarni2014survey}, Grossman et al. \cite{grossman2016survey}, Filippone et al. \cite{DBLP:journals/toms/FilipponeCBF17}, Wang et al. \cite{wang2020research}, and Li et al. \cite{10.1145/3604606}. The first one \cite{kulkarni2014survey} focused on performance modeling and optimization techniques for SpMV on GPUs, and their coverage was limited to the year of 2014. The second one 
 \cite{grossman2016survey} examined SpMV performance on only six large sparse matrices and provided limited experimental details. The third one \cite{DBLP:journals/toms/FilipponeCBF17} provided the most comprehensive survey of SpMV optimization known to date, particularly encompassing a wide range of new compression formats. However, their discussion is only limited to GPGPU. In \cite{wang2020research}, limited SpMV researches are discussed. Xiao et al. \cite{10.1145/3604606} presented a systematic literature survey on accelerating parallel sparse linear algebra operations, providing a general survey and summary of SpMV. In comparison, we provide a more extensive, systematic, and targeted survey of SpMV. This encompasses emerging trends in machine learning and mixed-precision algorithms, alongside optimizations tailored for CPU, GPU, FPGA, heterogeneous, and distributed architectures.

The primary objective of this survey paper is to address the challenge of SpMV optimization and provide a clear and systematic overview. Moreover, this article seeks to inspire experts in the field by introducing recent optimizations and encouraging new contributions. It answers the following research questions (RQ):

\textbf{RQ1}: What are the applications of SpMV, and how are they formulated in practice?

\textbf{RQ2}: What are the latest advancements in SpMV research?

\textbf{RQ3}: How do the state-of-the-art SpMV implementations perform?

\textbf{RQ4}: What challenges and potential directions for future research can be inferred from the current research efforts?

\begin{figure*}[!htbp]
	\centering
	\includegraphics[width=0.75\textwidth]{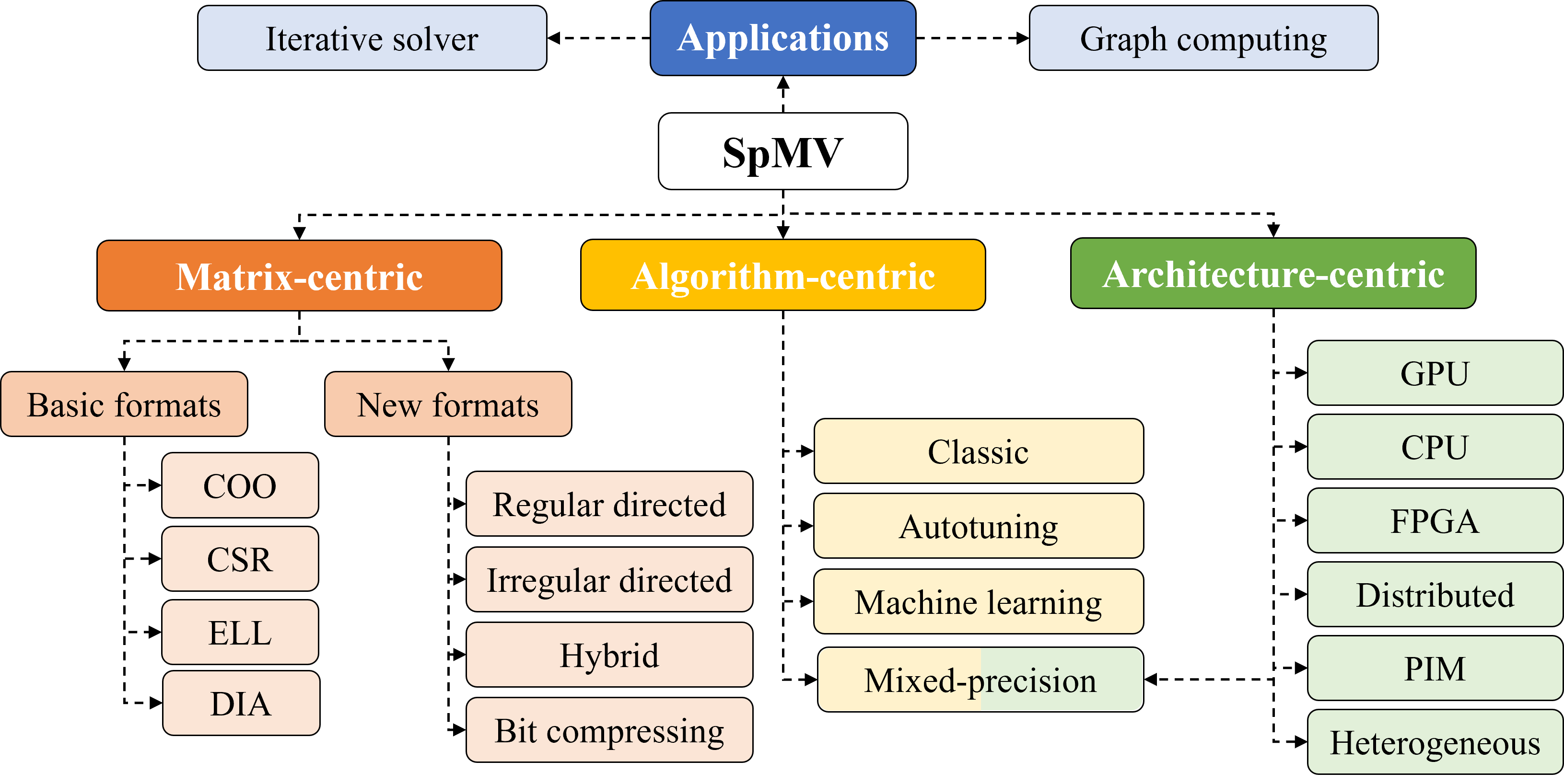} 
	\caption{Classification framework.}
	\label{fig:overview}
\end{figure*}

In relation to RQ1, we provide two typical applications of SpMV, including numerical iterative solvers and the PageRank algorithm. As for RQ2, we offer a comprehensive and systematic introduction of SpMV optimization efforts from various perspectives. RQ3 entails an extensive performance evaluation, comparing several prominent SpMV implementations on two different hardware platforms. Lastly, for RQ4, we engage in a comprehensive discussion and summary of key challenges and prospective research directions.

Figure \ref{fig:overview} presents the core framework of SpMV optimization introduced in this survey. We categorize the introduction of SpMV into four aspects: applications of SpMV, sparse matrices-centric, SpMV algorithms-centric, and architectures-centric optimizations over the years.

1) \textbf{Applications}: The application of SpMV is extensive and includes scientific and engineering computing, graph computing, and linear programming, among others. We illustrate the utilization of SpMV through two prominent examples: iterative solvers commonly found in scientific and engineering computing, and the well-known PageRank algorithm in graph computing.

2) \textbf{Matrix-centric}: The significant sparsity of sparse matrices often necessitates the adoption of compression storage. Researchers have proposed various sparse compression formats to effectively handle different distributions of non-zeros and capitalize on the unique features offered by different hardware architectures. Besides four early proposed basic formats including COO, CSR, ELL, and DIA, many new compression formats based on regular slicing and blocking, irregular partitioning, bit compressing, and hybrid compressing are proposed. 

3) \textbf{Algorithm-centric}: The performance of SpMV relies not only on compression formats but also on the design of the corresponding SpMV algorithm. In addition to classical optimization strategies for SpMV, researchers have proposed automatic tuning techniques or utilized machine learning methods to select the optimal compression format, parameter configuration, or SpMV algorithm for a given sparse matrix and hardware platform. Also, mixed-precision-based optimization is discussed. 

4) \textbf{Architecture-centric}: The performance of SpMV is greatly impacted by hardware architecture, making architecture-centric research a prominent and widely pursued direction. We elaborate on the SpMV optimization based on different hardware architectures, including CPU, GPU, FPGA, processing in memory (PIM), heterogeneous, and distributed platforms. 

The remainder of this article is structured as follows: Section \ref{sec:applications} provides an overview of typical applications of SpMV. Section \ref{sec:sparseFormats} summarizes existing sparse compression formats. In Section \ref{sec:csr-format}, the classic SpMV optimizations are discussed. Section \ref{sec:auto-tuning} presents researches that explore auto-tuning techniques for SpMV optimization, while Section \ref{sec:ML} delves into machine learning-based approaches. Section \ref{sec:hardware} introduces hardware architecture-oriented optimization. Furthermore, Section \ref{sec:mixed-precision} details how mixed-precision techniques are employed to accelerate SpMV and SpMV-centered iterative solvers. Subsequently, Section \ref{sec:performance} presents a comprehensive performance evaluation. Section \ref{sec:challenge} discusses the challenges and future research directions in SpMV. Finally, our conclusions are provided in Section \ref{sec:conclusion}.

\section{Typical Applications of SpMV}\label{sec:applications}
SpMV plays a crucial role as one of the most important components in various applications. In this section, we provide detailed insights into two typical applications to exemplify the significance of SpMV: iterative solver of sparse linear systems and graph computing.

\subsection{Iterative Solver of Sparse Linear System}
\begin{figure*}[!htbp]
	\centering
	\includegraphics[width=0.85\textwidth]{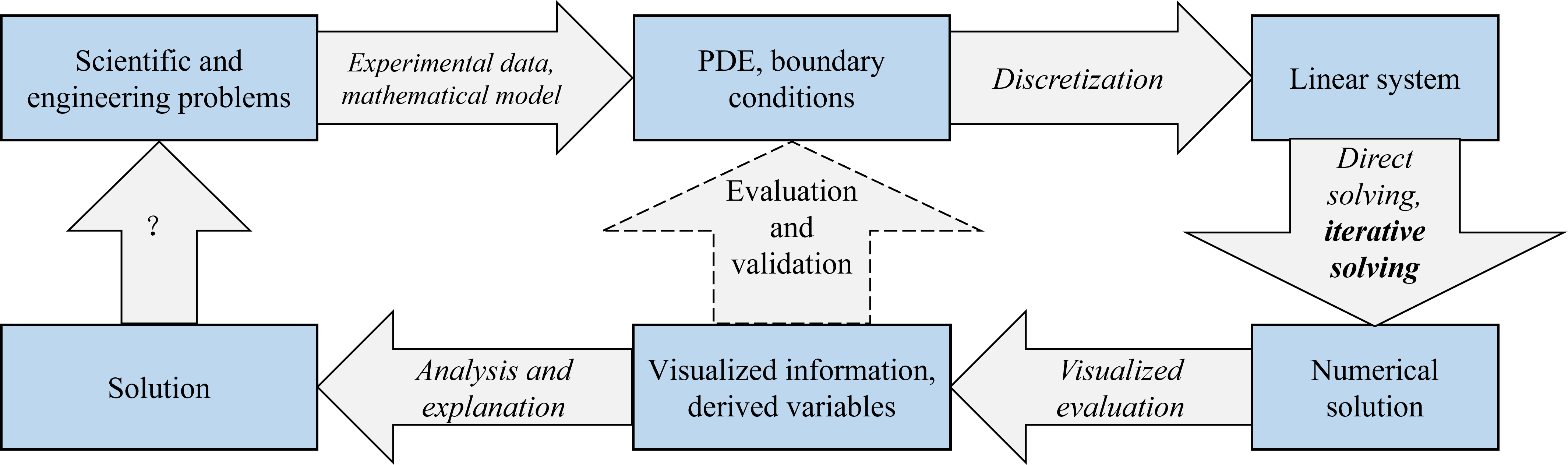} 
	\caption{Workflow of numerical solving for scientific and engineering computational problems.}
	\label{fig:numericalSimulation}
\end{figure*}

Figure \ref{fig:numericalSimulation} illustrates the general workflow of numerical solving for scientific and engineering computational problems. Initially, a real-world problem undergoes conversion into a mathematical model with specific conditional constraints, often expressed as a partial differential equation (PDE) and accompanied by boundary conditions derived from experimental data or observations. Subsequently, the mathematical model is discretized using the finite difference method, finite element method, or finite volume method, resulting in a sparse system of linear equations. In the numerical solving stage, direct and iterative solving methods are commonly employed. Following this, post-processing of the numerical solving may involve error calculation or solution visualization to assess the feasibility of the mathematical model. With further analysis and interpretation, scientists or engineers can find the problem's solution.

The solution of linear systems is crucial and has a significant impact on overall workflow efficiency. Direct and iterative methods are two primary and effective approaches for solving linear systems. Direct methods, such as Gaussian elimination, LU decomposition, and Cholesky decomposition, involve matrix decomposition, forward and backward substitution, and solution verification. Since computational complexity and memory footprints of direct methods are dependent on the size of the linear system, they are particularly suitable for small-scale linear systems \cite{DBLP:conf/sc/FuZWLLYZGLZ0023}. Differently, the Krylov-based conjugate gradient method (CG) and the generalized minimal residual method (GMRES), take an iterative approach to converge towards an accurate solution. They excel in large-scale linear systems, as their computation and memory overhead is primarily affected by the number of non-zeros (NNZ) of the input sparse coefficient matrix. 

For a large sparse linear system of $\matr{A}\vect{x}=\vect{b}$, where $\matr{A}\in\mathbb{R}^{n\times n}$ is a nonsingular sparse matrix and $\vect{x},\vect{b}\in \mathbb{R}^n$, Krylov methods find an approximate solution $\vect{x}_i$ in an order-$i$ Krylov subspace $\mathcal{K}_i(\matr{A},\matr{r}_0)\equiv\textup{span}\{\vect{r}_0,\matr{A}\vect{r}_0,...,\matr{A}^{i-1}\vect{r}_0\}$. The process starts with an initial guess $\vect{x}_0$ and an initial residual $\vect{r}_0\equiv\vect{b}-\matr{A}\vect{x}_0$.

\textbf{PCG}: preconditioned conjugate gradient (PCG) is a widely used variant of CG, and it significantly decreases the number of CG iterations by employing an appropriate preconditioner. Algorithm \ref{alg:PCG} outlines the iterative process of PCG, where $\matr{M}$ denotes the preconditioner matrix, and the output vector $\vect{\hat{x}}$ represents the approximate solution that satisfies the residual threshold. Notably, the SpMV calculation is used in lines 1, 3, and 6 of the algorithm. 

\begin{algorithm}
    \KwIn{coefficient matrix $\matr{A}$, right-hand-side vector $\vect{b}$, preconditioner $\matr{M}$, initial guess $\vect{x}_0$, tolerance threshold $\epsilon$, and maximum iterations $MAX$}
    \KwOut{approximate solution $\hat{\vect{x}}$}
    $\vect{r}_0=\vect{b}-\matr{A}\vect{x}_0$; $\vect{z}_0=\matr{M}^{-1}\vect{r}_0$; $\vect{p}_0=\vect{z}_0$; \tcc*[f]{\textbf{SpMV}} \\
    \For{\textup{(}$i=1$; $i\le MAX\textup{ or }\Vert \vect{r}_i\Vert<\Vert\vect{b}\Vert\epsilon$; $i++$\textup{)}}
    {   
        $\alpha_{i-1}=\frac{\vect{r}_{i-1}^T\vect{z}_{i-1}}{\vect{p}^T_{i-1}\matr{A}\vect{p}_{i-1}}$; \tcc*[f]{\textbf{SpMV}} \\
        $\vect{x}_{i}=\vect{x}_{i-1}+\alpha_{i-1}\vect{p}_{i-1}$;\\
        $\vect{r}_{i}=\vect{r}_{i-1}-\alpha_{i-1}\matr{A}\vect{p}_{i-1}$; \\ 
        $\vect{z}_{i}=\matr{M}^{-1}\vect{r}_{i}$; \tcc*[f]{\textbf{SpMV}} \\
        $\beta_i=\frac{\vect{r}_i^T\vect{z}_i}{\vect{r}_{i-1}^T\vect{z}^{i-1}}$; \\
        $\vect{p}_{i}=\vect{z}_{i}+\beta_{i}\vect{p}_{i-1}$; \\
    }
    $\hat{\vect{x}}=\vect{x}_{i-1}$; \\
    \caption{PCG algorithm}
    \label{alg:PCG}
\end{algorithm}

\textbf{GMRES}: The PCG algorithm is suitable for linear systems with symmetric positive definite coefficient matrices, while GMRES \cite{saad1986gmres} is designed for solving indefinite nonsymmetric linear systems of equations. GMRES consists of two steps. The first step, known as the Arnoldi process, calculates an orthonormal basis of $\mathcal{K}_i(\matr{A},\matr{r}_0)$ using the modified Gram-Schmidt (MGS) method. The second step solves a linear least squares problem with an upper Hessenberg coefficient matrix using Given rotations to minimize the residual vector. Algorithm \ref{alg:GMRES} presents the restart GMRES with left preconditioning \cite{DBLP:journals/jcam/SuzukiFI23}. Notably, SpMV operations are used in lines 2, 4, and 15 of the algorithm.

\begin{algorithm}
        \KwIn{coefficient matrix $\matr{A}$, initial guess $\vect{x}_0$, right-hand side vector $\vect{b}$, tolerance factor $\epsilon$, precondition matrix $\matr{M}$}
        \KwOut{approximate solution $\hat{\vect{x}}$}
        $\vect{z}_0=\vect{b}-\matr{A}\vec{x}_0$; \\
        $\vect{r}_0=\matr{M}^{-1}\vect{z}_0$;
        $\beta = \Vert \vect{r}_0 \Vert$; 
        $\vect{v}_1=\vect{r}_0/ \beta$;\tcc*[f]{\textbf{SpMV}} \\
        \For(\tcc*[f]{Arnoldi process}){$j=1,2,...,m$}
        {
            $\vect{w}_j=\matr{M}^{-1}\matr{A}\vec{v}_j$; \tcc*[f]{\textbf{SpMV}} \\
            \For(\tcc*[f]{Modified Gram-Schmidt (MGS) process}){$i=1,2,...,j$}
            {
                $\matr{H}_{ij}=(\vect{w}_j,\vect{v}_i)$;
                $\vect{w}_j=\vect{w}_j-\matr{H}_{ij}\vect{v}_i$; \\
            }
            $\matr{H}_{j+1,j}=\Vert \vect{w}_j\Vert$;
            $\vect{v}_{j+1}=\vect{w}_j/\matr{H}_{j+1,j}$\;   
            \If{$\matr{H}_{j+1,j}==0$\textup{ or }$<Stopping$\textup{ }$criteria$\textup{ }$satisfied>$}
            {
                $m=j$;
                break; \\
            }
        }
        $\matr{V}_m=[\vect{v}_1,...,\vect{v}_m]$; \\
        find $\vect{y}^m$ that minimizes    
        $\Vert \beta\vect{e}_1-\matr{H}_m\vect{y}\Vert$; \\ 
        $\vect{x}_m=\vect{x}_0+\matr{V}_m\vect{y}^m$; \tcc*[f]{\textbf{SpMV}}\\
        \eIf{$\Vert \vect{b}-\matr{A}\vect{x}_m\Vert\le \epsilon\Vert\vect{b}\Vert$}{
            $\hat{\vect{x}}=\vect{x}_m$; 
            break;\\
        }{
            $\vect{x}_0=\vect{x}_m$\tcc*[c]{Restart} 
            \textbf{Goto} 1; \\
        }
        \caption{Restarted GMRES algorithm with the left preconditioning.}
        \label{alg:GMRES}
        \end{algorithm}

\begin{figure*}[!h]
	\centering
	\includegraphics[width=0.9\textwidth]{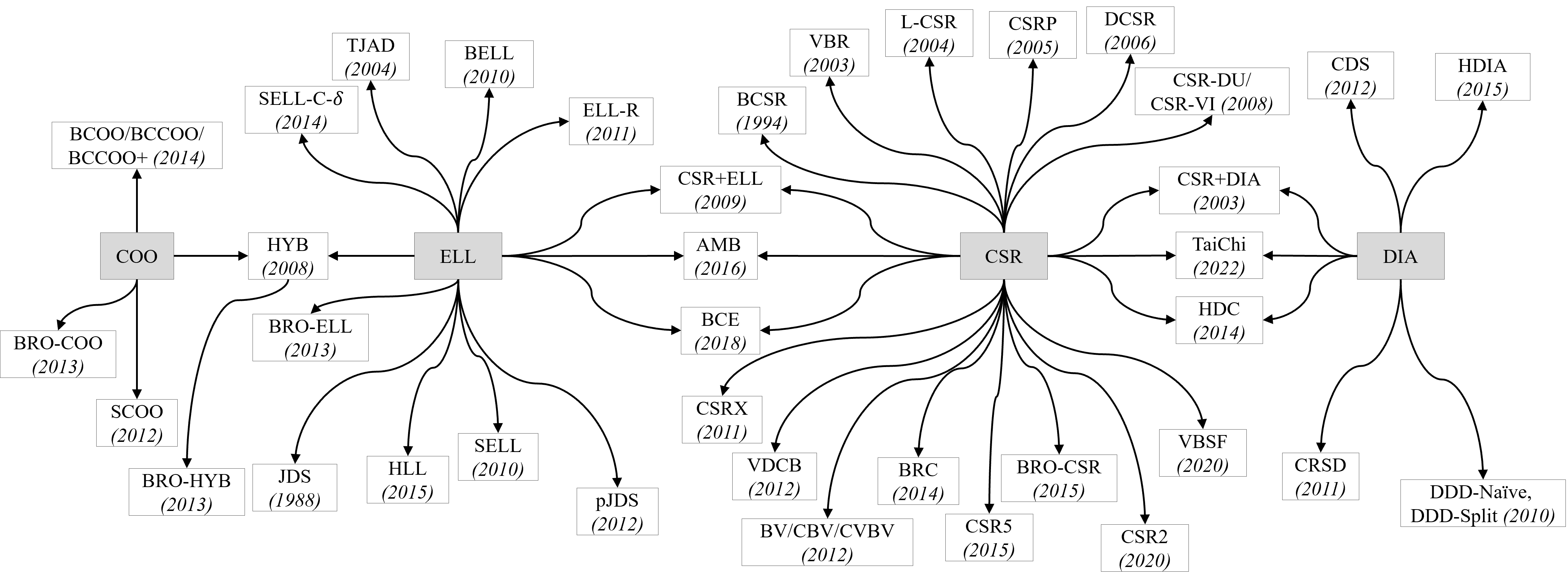} 
	\caption{Sparse compression formats and their derivative relationship.}
	\label{fig:format-year}
\end{figure*}

\subsection{Graph Computing}
    SpMV is also an important operator in graph applications, particularly in graph-based social networks, web searching, and web link mining \cite{DBLP:journals/csur/ShiZZJHLH18}. Sparse matrices are frequently used as adjacency matrices for graphs. For example, in a social network, the vertices represent social participants, and the edges represent various relationships (such as likes or friendships) between participants. Large-scale social network graphs are typically represented as sparse matrices, allowing linear algebra computations to be used as an alternative to graph calculations. As the number of participants in a social network increases, the corresponding sparse matrix exhibits a higher degree of sparsity. This is because, in general, an individual's friends constitute only a small fraction of the entire network. SpMV plays a crucial role in various algorithms such as the Restart Random Walk (RWR) \cite{DBLP:journals/kais/TongFP08}\cite{DBLP:conf/sc/AshariSEPS14}, the PageRank \cite{DBLP:conf/icpp/WuWSYWX10}, and the Hyperlink-Induced Topic Search (HITS) algorithm \cite{DBLP:journals/jacm/Kleinberg99}\cite{DBLP:journals/cn/BrinP12}. These algorithms are commonly used for ranking web pages and calculating the closeness between two nodes in a graph.

Taking the PageRank algorithm as an example, Algorithm \ref{alg:pagerank} outlines a power method of the algorithm. In this algorithm, $n$ represents the total number of web pages to be ranked. The importance value of each page $i$ is represented by the $i$-th element of a column vector $\vect{x}$ with $n$ dimension. PageRank algorithm solves $\vect{x}$ using the iterative process $\vect{x}_{i+1}=q\matr{A}\vect{x}_i$, where $q$ represents the dominant eigenvalue of matrix $\matr{A}$, and matrix $\matr{A}$ has dimensions $n\times n$. If page $i$ is linked to page $j$, the corresponding element $\matr{A}_{ij}$ is set to $1/O_j$, where $O_j$ represents the number of out linkages from page $j$. Conversely, if there is no linkage, $\matr{A}_{ij}$ is set to $0$. Besides, an escape vector $\vect{e}$, representing the possibility of one page randomly jumping to another page, is introduced to increase numerical stability \cite{DBLP:conf/icpp/WuWSYWX10}. It is worth noting that $\matr{A}$ is typically a sparse matrix, and the performance of the PageRank algorithm is primarily dominated by the SpMV operation in line 2.

\begin{algorithm}
    \KwIn{$\matr{A}$, $q$, $\epsilon$}
    \KwOut{$\hat{\vect{x}}$}
    \While{\textup{(}$\left\|\vect{x}_{i}-\vect{x}_{i-1}\right\|_1>\epsilon$\textup{)}} 
    {
        $\vect{y}=q\matr{A}\vect{x}_i$ \tcc*[r]{\textbf{SpMV}}
        $d=\left\|\vect{x}_i\right\|_1 - \left\|\vect{y}\right\|_1$\;
        $\vect{x}_{i+1}=\vect{y}+d\vect{e}$\;
        $i=i+1$\;
    }
    $\hat{\vect{x}}=\vect{x}_i$; \\
    \caption{Power method of the PageRank algorithm}
    \label{alg:pagerank}
\end{algorithm}

\section{Sparse Compression Formats}\label{sec:sparseFormats}

Sparse matrices are typically stored in compressed formats to minimize storage, transmission, and memory access overhead due to their massive zero elements. Currently, there are several compression formats available, which we can categorize into two groups. The first group is referred to as basic compression formats, consisting of four straightforward and early-proposed formats. The second group is referred to as new compression formats, which are typically derived from the basic formats. Figure \ref{fig:format-year} illustrates the derivative relationship between these basic and new compression formats. Table \ref{tab:sparse-formats} provides a summary of the existing sparse compression formats.

\begin{table*}[!htbp]
    \centering
    \caption{Classification of sparse compression formats.}\label{tab:sparse-formats}
        \begin{tabular}{m{3.5cm}m{12cm}}
        \toprule
            \multicolumn{1}{c}{\textbf{Formats}} & \multicolumn{1}{c}{\textbf{Contributions}} \\ \midrule
            Basic formats & COO,CSR,ELL,DIA \cite{paolini1989data}\cite{DBLP:journals/pc/Peters91}\cite{saad1994sparskit}\cite{BellG09}\cite{Sengupta_Harris_Zhang_Owens_2007}\cite{Garland_2008}
            \\ \midrule
            Regular slicing & 
            Sliced COO\cite{DBLP:journals/procedia/DangS12},
            Sliced CSR\cite{DBLP:conf/iccS/DAzevedoFM05},
            Sliced ELL\cite{DBLP:conf/hipeac/MonakovLA10}\cite{barbieri2015three}\cite{pJDS2012}\cite{kreutzer2014unified},
            Sliced DIA\cite{DBLP:conf/hpcc/YuanZSW10}\cite{DBLP:conf/europar/SunZWLZL11}\cite{DBLP:conf/icpp/SunZWZYR11}\cite{barbieri2015three}
            \\ \midrule
            Regular blocking & 
            Blocked CSR\cite{saad1994sparskit}\cite{DBLP:conf/sc/PinarH99}\cite{DBLP:journals/tjs/LiXCLYLGGX20}\cite{vuduc2003automatic}\cite{AshariSES14_BRC}\cite{DBLP:conf/reconfig/Jain-MendonS12},
            Blocked COO\cite{Yan2014},
            Blocked DIA\cite{DBLP:conf/asplos/GodwinHS12},
            Blocked ELL\cite{DBLP:conf/ppopp/ChoiSV10}
             \\ \midrule
            Irregular representing & CSR5\cite{CSR5}, CSR2\cite{DBLP:conf/ccgrid/BianHDLW20}, CSRX\cite{DBLP:conf/ppopp/KourtisKGK11}, Cocktail\cite{cocktail2012} 
            \\ \midrule
            Hybrid combining & COO+ELL\cite{bell2008efficient}\cite{AnztCYDFNTTW20},
            CSR+ELL\cite{DBLP:conf/hipc/MaringantiAP09}\cite{DBLP:journals/jcss/YangLL18}\cite{2016_AMB},
            CSR+DIA\cite{bolz2003sparse}\cite{DBLP:journals/ijhpca/YangLLSW14}\cite{DBLP:journals/tpds/GaoJTWS22},
            COO+CSR\cite{DBLP:conf/cgo/TangZLLHLG15},
            More than two formats\cite{DBLP:conf/samos/MonakovA09}\cite{TAN2020521}\cite{NiuLDJ0T21_TileSpMV}
            \\ \midrule
            Bit/Byte compressing & Delta Coding\cite{DBLP:conf/ics/WillcockL06}\cite{DBLP:conf/cf/KourtisGK08}\cite{TangTRWCKGTW13}\cite{Tang_TPDS_2015},
            Bitmask\cite{DBLP:conf/fccm/KesturDC12}
            \\ \midrule
            Other variants & L-CSR\cite{DBLP:journals/ijhpca/Mellor-CrummeyG04}, ELL-R\cite{DBLP:journals/concurrency/VazquezFG11},
            JDS and its variants\cite{DBLP:journals/ijhsc/AndersonS89}\cite{DBLP:journals/pc/Melhem88}\cite{DBLP:journals/ipl/MontagneE04}
            \\ 
            \bottomrule
        \end{tabular}
\end{table*}

\subsection{Basic Compression Formats}

\begin{figure*}[!htbp]
	\centering
	\includegraphics[width=0.7\textwidth]{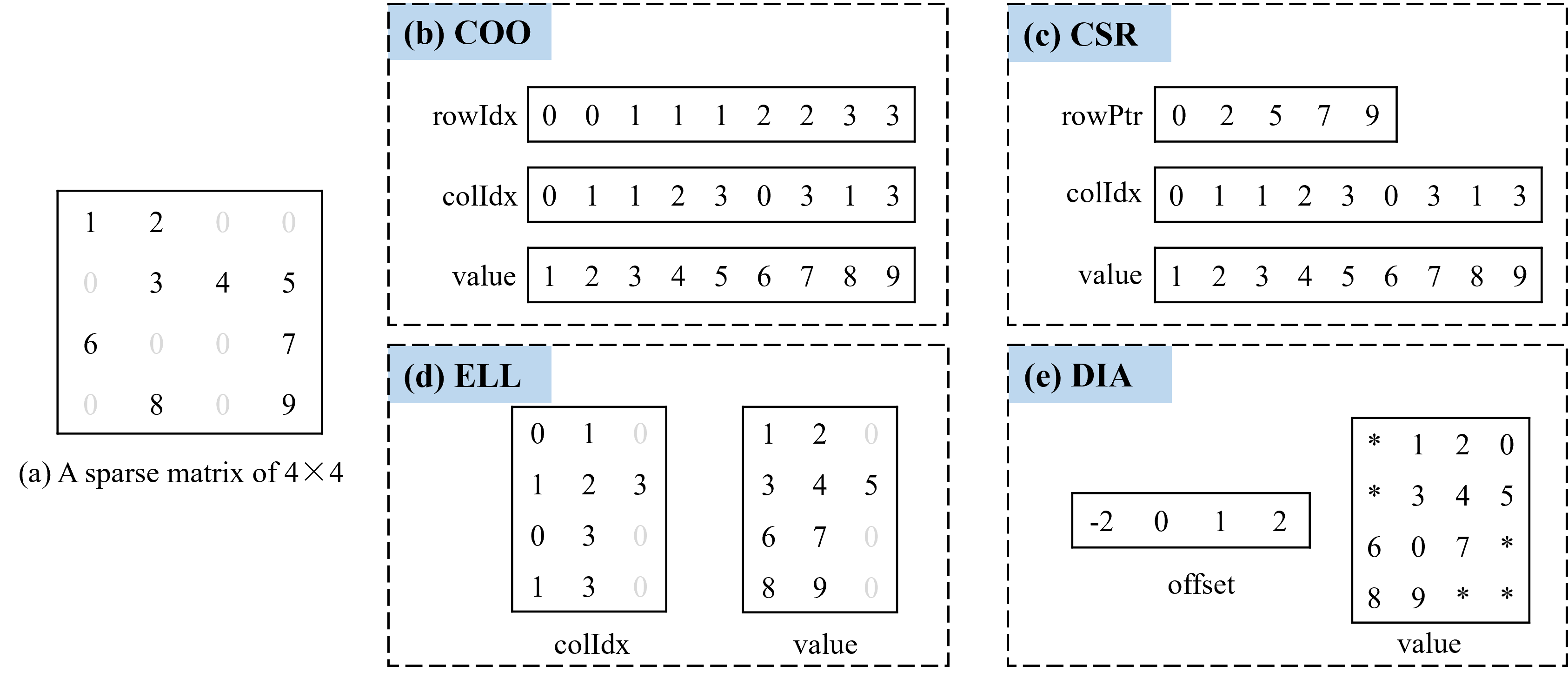} 
	\caption{Examples of four basic compression formats.}
	\label{fig:4basicFromats}
\end{figure*}

These basic sparse compression formats were proposed around 1990s \cite{paolini1989data}\cite{DBLP:journals/pc/Peters91}\cite{saad1994sparskit}. Their examples are illustrated in Figure \ref{fig:4basicFromats}, with Figure \ref{fig:4basicFromats}(a) depicting a sparse matrix of size 4$\times$4.

\textbf{COO}. The COO format comprises three arrays: $\vect{rowIdx}$, $\vect{colIdx}$, and $\vect{value}$, representing row indices, column indices, and values of non-zeros, respectively. These arrays contain the same number of elements. Figure \ref{fig:4basicFromats}(b) illustrates the COO encoding of a sparse matrix example. For a sparse matrix of size $m\times n$ with $nnz$ non-zeros, assuming each row index and column index require $S_{idx}$ storage size, and each non-zero element requires $S_{val}$ storage size, the storage size $S_{COO}$ of a sparse matrix compressed in the COO format can be calculated using Equation  (\ref{equ:S_COO}).

\begin{equation}\label{equ:S_COO}
    S_{COO}=(2\times S_{idx}+S_{val})\times nnz
\end{equation}

Although the COO format and its corresponding SpMV are straightforward and easy to understand, they do have certain limitations. Firstly, if a sparse matrix contains numerous non-zeros in certain rows or columns, storing the same row index for each non-zero element in these rows or columns leads to significant redundant storage and memory access overhead. Secondly, COO-based SpMV algorithms typically use a task division strategy with an equal NNZ. This requires atomic operations to merge local results, which significantly impacts SpMV performance. Subsequent segment reduction methods \cite{BellG09,Sengupta_Harris_Zhang_Owens_2007}\cite{Garland_2008} eliminate the need for atomic operations.

\textbf{CSR}: The CSR format optimizes the COO format by compressing the row indices array while keeping the other two arrays unchanged. Instead of storing the row index for each non-zero element, CSR uses a row pointer array $\vect{rowPtr}$ to store the global position of the first non-zero element in each row within $\vect{colIdx}$ and $\vect{value}$ arrays. Figure \ref{fig:4basicFromats}(c) illustrates the CSR encoding of the sparse matrix example. The storage size $S_{CSR}$ of the CSR format is given by Equation (\ref{equ:S_CSR}).

\begin{equation}\label{equ:S_CSR}
    S_{CSR}=(m+1)\times S_{idx}+nnz\times(S_{val}+S_{idx})
\end{equation}

The CSR format takes less storage than the COO format for most sparse matrices and is less sensitive to the data distribution compared with the later-introduced ELL and DIA formats. The CSR format is widely used in practical applications. 
However, for irregular sparse matrices with a significant variation in the NNZ per row, CSR-based SpMV suffers from load imbalance when using a row-wise task partitioning method that assigns each thread to process an equal number of rows.

\textbf{ELL}: The ELLPACK format (ELL) consists of two arrays of the same size: $\vect{colIdx}$ and $\vect{value}$, which store the column indices and values of non-zeros, respectively. Figure \ref{fig:4basicFromats}(d) illustrates the ELL format, where all non-zeros per row of a sparse matrix are shifted to the leftmost positions. The two-dimensional array structure naturally retains the row indices of non-zeros, eliminating the need for additional space to store it. Additionally, Figure \ref{fig:4basicFromats}(d) shows that the width of the two arrays in the ELL format is determined by the longest row, which contains the maximum NNZ ($max$). Padding zeros is necessary for rows with fewer non-zeros than $max$. The storage size $S_{ELL}$ of the ELL format is calculated as Equation (\ref{equ:S_ELL}).

\begin{equation}\label{equ:S_ELL}
S_{ELL} = (S_{idx}+S_{val})\times 2\times m\times max
\end{equation}

It is evident that the storage size of the ELL format is sensitive to $max$. If there are numerous rows with a much smaller NNZ than $max$, zeros padding results in a significant increase in storage space and memory access overhead.

\textbf{DIA}: The DIA format arranges non-zeros in a diagonal-wise pattern and consists of two arrays: $\vect{offset}$, which stores the offset of each non-zero diagonal, and $\vect{value}$, which stores the corresponding non-zeros. Figure \ref{fig:4basicFromats}(e) illustrates an example of the DIA format. In a sparse matrix, the main diagonal typically has an offset of 0, while the diagonals in the lower triangular region have negative offsets and those in the upper triangular region have positive offsets. For a sparse matrix of size $m\times m$, with $N_{dias}$ diagonals that have at least one non-zero element, the storage size $S_{DIA}$ of the DIA format is calculated as Equation (\ref{equ:S_DIA}).

\begin{equation}\label{equ:S_DIA}
S_{DIA} = N_{dia}\times m\times S_{val}+N_{dia}\times S_{idx}
\end{equation}

\begin{figure*}[!htbp]
	\centering
	\includegraphics[width=\textwidth]{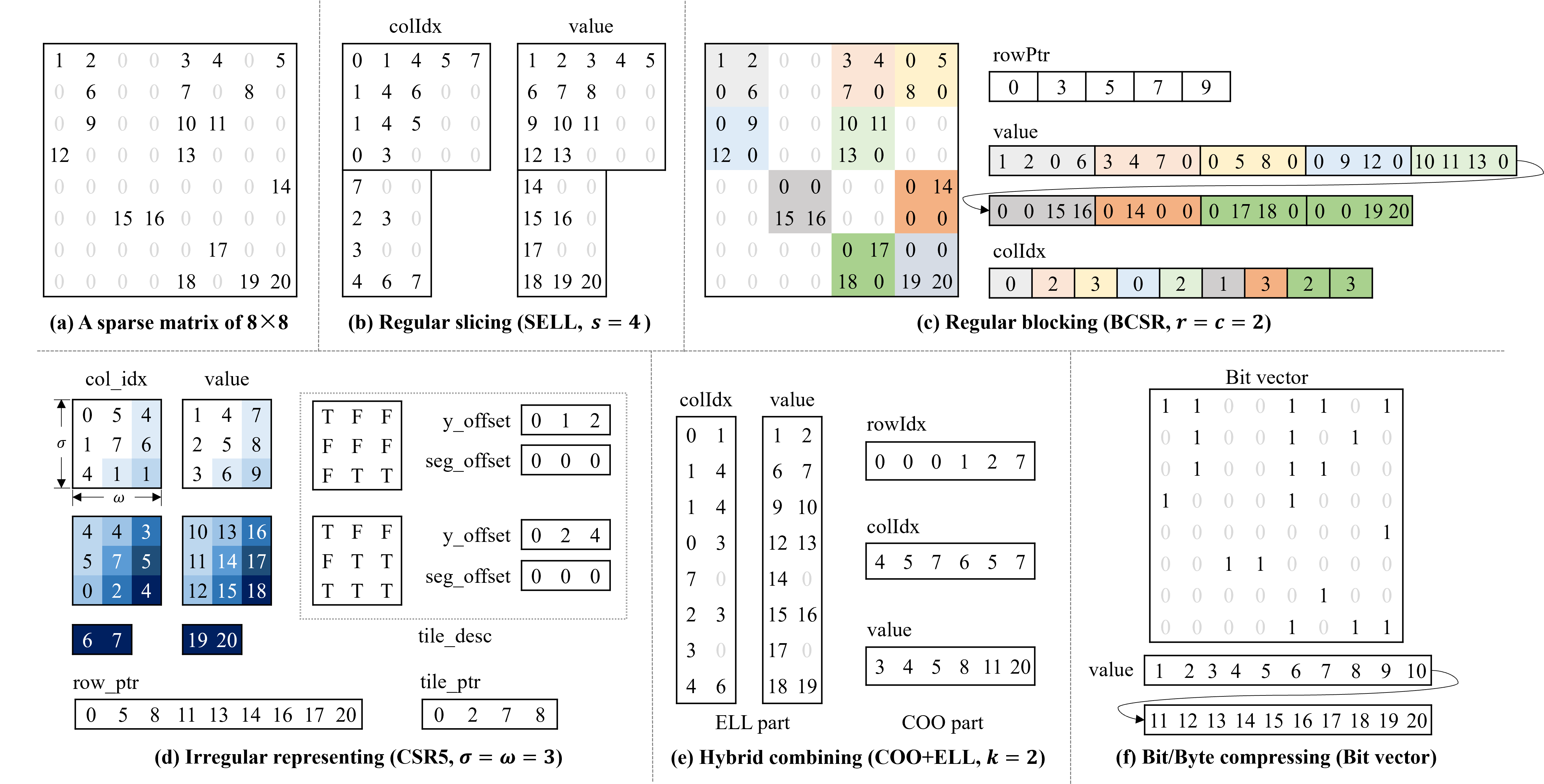} 
	\caption{Examples of new compression formats.}
	\label{fig:newFormats}
\end{figure*}

The DIA format is well-suited for sparse matrices where non-zeros are primarily concentrated near the main diagonal. However, if a sparse matrix has non-zeros scattered across numerous diagonals, the padding of numerous zeros significantly increases the storage overhead of the DIA format. Due to the varying lengths of these diagonals, bounds checking on the column indices is necessary for the DIA-based SpMV. It is worth noting that zero padding not only increases the storage overhead of the DIA format but also results in a considerable amount of unnecessary memory access and computation.

\subsection{New Compression Formats}
The new compression formats can be categorized into six classes: regular slicing, regular blocking, irregular representing, bit/byte compressing, hybrid combining, and other variants. The first two classes are derived from basic formats by one-dimensional (regular slicing) or two-dimensional (regular blocking) partitioning in a sparse matrix. They allow for better utilization of local distribution and load balancing through different slicing and blocking sizes. However, they may still encounter load imbalances when processing highly irregular sparse matrices. Therefore, irregular representing is proposed to tackle the issue, which partitions a sparse matrix by NNZ, instead of regular 1D or 2D partitioning. Basic compression formats are suitable for unified non-zeros distributions. However, in some cases, a sparse matrix may exhibit a hybrid distribution. Therefore, the hybrid format is proposed to address this. It uses two or more formats (including basic and new formats) to encode different areas of a sparse matrix. Besides, several bit/byte-based compression schemes that leverage the two-dimensional structure of the matrix and the locality distribution characteristics of non-zeros are introduced. Figure \ref{fig:newFormats} provides examples of these new formats, with Figure \ref{fig:newFormats}(a) showing an 8$\times$8 sparse matrix.


\subsubsection{Regular Slicing}
Regular slicing partitions the matrix in one dimension. In Figure \ref{fig:newFormats}(b), we take sliced ELL (SELL) format for example. The sparse matrix is split into two slices, with each slice containing four rows encoded using the ELL format. It is evident that the second slice has a smaller $max$ value compared with the first one. This sliced partitioning and separate storage effectively reduces zero padding for the ELL format.

\textbf{Sliced COO}: Dang et al. \cite{DBLP:journals/procedia/DangS12} introduce the sliced COO format (SCOO), which partitions a sparse matrix into multiple slices based on rows. Within each slice, the non-zeros are arranged according to the size of their column indices.

\textbf{Sliced CSR}: The sliced CSR was originally introduced and referred to as CSRP (CSR with Permutation) in \cite{DBLP:conf/iccS/DAzevedoFM05}. This approach performs a virtual sorting of the sparse matrix rows and subsequently groups rows with similar non-zeros into slices.

\textbf{Sliced ELL}: As the example illustrated in Figure \ref{fig:newFormats}(b), SELL \cite{DBLP:conf/hipeac/MonakovLA10}, also referred to as Hacked ELL (HLL) in \cite{barbieri2015three}, stores each slice in the ELL format, resulting in a significant reduction in memory usage compared with the classical ELL format due to reduced zero filling. Kreutzer et al. \cite{pJDS2012} also propose a similar format called padded jagged diagonal storage (pJDS). In pJDS, rows are first sorted in descending order of NNZ per row. Then, consecutive rows are grouped into slices and padded to match the length of the longest row in each slice. Another format similar to SELL is SELL-$C$-$\sigma$ \cite{kreutzer2014unified}, which sorts consecutive $\sigma$ rows in descending order of non-zeros and grouping them into multiple chunks.

\textbf{Sliced DIA}: Yuan et al. \cite{DBLP:conf/hpcc/YuanZSW10} introduce two enhanced formats, namely DDD-Naive (direct dense diagonal) and DDD-Split, based on the classical DIA format. DDD-Naive utilizes a one-dimensional array, $\vect{Val}$, to store each diagonal of the sparse matrix, with the space allocated for each diagonal determined by its position and the matrix size. Instead of storing the offset of each diagonal from the main diagonal, DDD-Naive stores the global position of the first element of each diagonal in $\vect{Val}$. DDD-Split partitions the sparse matrix into multiple row groups, ensuring that each row within the same group has the same non-zeros. Consequently, DDD-Split only requires storing the non-zeros in the first row of each row group, effectively reducing the redundant storage of non-zeros. In addition, Sun et al. \cite{DBLP:conf/europar/SunZWLZL11,DBLP:conf/icpp/SunZWZYR11} propose the CRSD (compressed row segment with diagonal pattern) format. CRSD groups the sparse matrix by row slices and diagonal distribution patterns, capitalizing on the local data distributions to improve efficiency. Hacked DIA (HDIA) presented in \cite{barbieri2015three} splits the original sparse matrix into equally sized ``hacks" (row groups) and stores each hack in DIA format.

\subsubsection{Regular Blocking}
Regular blocking partitions a sparse matrix into two-dimensional (2D) blocks and utilizes a basic format to store these non-zero blocks. In Figure \ref{fig:newFormats}(c), we take blocked CSR (BCSR) format for example to illustrate the concept of regular blocking.

\textbf{Blocked CSR}: The Blocked CSR (BCSR) format divides a sparse matrix into dense 2D blocks of size $r\times c$ and stores each block, containing at least one non-zero element, in CSR format \cite{saad1994sparskit}. Similar to CSR, BCSR has three arrays, but the original sparse matrix of size $m\times n$ is considered as a sparse matrix of size $\lceil \frac{m}{r}\rceil \times \lceil \frac{n}{c}\rceil$. In each stored block, the positions without non-zeros are filled with zero elements. BCSR requires less memory for the row pointer and column index arrays compared with CSR, but the explicit filling of zeros in $\vect{value}$ array incurs extra memory and computation overhead. In the example presented in Figure \ref{fig:newFormats}(c), let $r=c=2$, the size of $\vect{rowPtr}$ and $\vect{colIdx}$ are reduced by approximately half compared with the typical CSR format, but $\vect{value}$ contains additional 16 zeros. The choice of block size in BCSR is crucial for performance, and the optimal block size varies for different sparse matrices. Pinar et al. \cite{DBLP:conf/sc/PinarH99} introduce a variation of BCSR where the blocks are one-dimensional and can vary in size. They rearrange the input sparse matrix to enlarge dense blocks and treat the reordering problem as the traveling salesperson problem (TSP). The Variable Blocked-$\sigma$-SIMD (VBSF) format, proposed by Liu et al. \cite{DBLP:journals/tjs/LiXCLYLGGX20}, is another BCSR-style format with variable block sizes. In VBSF, $\sigma$ represents the number of operands that can be processed by a single vector instruction. Compared with BCSR, VBSF avoids excessive zero filling and the need to select an optimal block size. To address the issue of excessive zero filling in BCSR, Vuduc et al. \cite{vuduc2003automatic} introduce the Variable BCSR (VBR) format, while Ashari et al. \cite{AshariSES14_BRC} propose the Block format BRC (Blocked Row Column). In the BRC format, the rows of a sparse matrix are first sorted in descending order of their non-zeros. Then, all non-zeros are shifted horizontally to the leftmost side, and the reordered and rearranged matrix is stored using the BCSR-style storage. In \cite{DBLP:conf/reconfig/Jain-MendonS12}, a variable dual compressed block (VDCB) format is proposed. It divides a sparse matrix into multiple variable-sized blocks, and each block is accompanied by auxiliary descriptions comprising three components: Header (containing block size, row and column start, and total NNZ), Bitmap (indicating the presence of a non-zero element with a value of one), and non-zeros.

\textbf{Blocked COO}: Yan et al. \cite{Yan2014} present the blocked COO (BCOO) format, which partitions a sparse matrix into fixed-size blocks and adopts a format similar to COO to store the row and column indices and values of each non-zero block. Similar to BCSR, BCOO also incorporates zero filling. Additionally, they propose the Blocked Compressed COO (BCCOO) format, which utilizes binary bits to represent the row index information of a block. BCCOO achieves a compression ratio of 32 times for the row index array compared with BCOO. To enhance access locality during vector multiplication, they introduce the extended format BCCOO+. In BCCOO+, the sparse matrix is vertically split into slices, which are then reassembled from top to bottom. The rearranged matrix is stored using the BCCOO format, while the encoding information of the column indices remains unchanged.

\textbf{Blocked DIA}: Godwin et al. \cite{DBLP:conf/asplos/GodwinHS12} introduce the Column Diagonal Storage (CDS) format, a blocking storage format based on the DIA format. It is designed to enhance the performance of stencil calculations on sparse matrices that typically exhibit a diagonal structure. In CDS, the sparse matrix is initially partitioned into multiple 2D blocks of equal size, which are then stored using the DIA format.

\textbf{Blocked ELL}: Choi et al. \cite{DBLP:conf/ppopp/ChoiSV10} propose a blocked ELL storage format. Similar to BCSR, they begin by dividing a sparse matrix into dense blocks of size $r\times c$. They then sort the blocked rows in descending order based on the number of blocks per row. Next, they group consecutive blocked rows into sub-matrices and partition the sorted results into multiple sub-matrices. Finally, each sub-matrix is stored using the ELL format or the $r\times c$ blocked ELL format.

\subsubsection{Irregular Compressing}
Irregular compressing overcomes the limitations imposed by the rows, columns, or 2D structure of a matrix, making it particularly advantageous for sparse matrices with irregular distributions of non-zeros. One notable compression format in this regard is CSR5, introduced by Liu et al. \cite{CSR5}. An example of CSR5 is illustrated in Figure \ref{fig:newFormats}(d). CSR5 partitions all nonzeros into equally-sized 2D tiles, as depicted in the figure, while introducing the computational and storage cost of auxiliary arrays ($tile\_desc$). Another format, CSR2 \cite{DBLP:conf/ccgrid/BianHDLW20}, also utilizes a similar 2D-tile storage approach but requires explicit zero padding. Kourtis et al. \cite{DBLP:conf/ppopp/KourtisKGK11} propose the CSRX format, which identifies various structures within a sparse matrix and stores them using different storage techniques. These structures can include horizontal, vertical, diagonal, and two-dimensional patterns. Furthermore, Su et al. \cite{cocktail2012} introduce the Cocktail format, which employs different compression formats (supporting nine different formats) to encode different regions of a sparse matrix, leveraging the unique advantages of each format.

\subsubsection{Bit/Byte Compressing}
Bit/Byte compressing mainly utilizes the techniques of delta coding, run-length encoding, and bitmask to compress index arrays of traditional sparse formats.

\textbf{Delta Coding}: The DCSR format, proposed by Willcock et al. \cite{DBLP:conf/ics/WillcockL06}, is a byte-oriented encoding scheme designed to efficiently represent the differences between the column indices of adjacent non-zeros within a single row. This format achieves compactness by minimizing the number of bytes required to encode these differences and also combines non-zeros in consecutive positions into a single entity for further compression. Kourtis et al. \cite{DBLP:conf/cf/KourtisGK08} propose two compression techniques for the index and value arrays in CSR format: CSR Delta Unit (CSR-DU) and CSR Value Indexed (CSR-VI). Utilizing a similar approach to DCSR, CSR-DU replaces the $\vect{rowPtr}$ and $\vect{colIdx}$ arrays using a byte array. On the other hand, CSR-VI stores unique non-zero values, at the expense of additional index information required to identify the corresponding rows and columns. Tang et al. \cite{TangTRWCKGTW13,Tang_TPDS_2015} present a family of bit-representation-optimized (BRO) compression formats, including BRO-CSR, BRO-ELL, BRO-COO, BRO-HYB. 

\textbf{Bitmask}: Kestur et al. \cite{DBLP:conf/fccm/KesturDC12} propose three bit-based sparse formats: bit-vector (BV), compressed bit-vector (CBV), and compressed variable-length bit-vector (CVBV). BV format encodes a sparse matrix using a dense bitmask, where binary ``0" indicates a zero element, while binary ``1" indicates a non-zero element. The CBV uses a fixed-width run-length encoding to compress redundant binary ``0"s or ``1"s in BV representation. Differently, CVBV uses a variable-length data field to store the run-length encoding.

\subsubsection{Hybrid Encoding}
Hybrid encoding usually matches different compression formats to varying distributions within a sparse matrix, efficiently mitigating excessive memory footprints and zeros paddings in ELL, SELL, BCSR, and so on.

\textbf{COO+ELL}: Bell et al. \cite{bell2008efficient} introduce the HYB format, which combines COO and ELL formats. In HYB, the first $k$ non-zeros of each row are encoded using the ELL format, while the remaining non-zeros exceeding $k$ are encoded using the COO format. When the NNZ in a row is less than $k$, HYB requires zero padding. The choice of the parameter $k$ significantly impacts the storage size of the HYB format and the performance of HYB-based SpMV \cite{AnztCYDFNTTW20}. Figure \ref{fig:newFormats}(e) illustrates an example of HYB encoding, where $k$ is set to 2.

\textbf{CSR+ELL}: Maringanti et al. \cite{DBLP:conf/hipc/MaringantiAP09} propose a hybrid storage format called CSR+ELL, which reorders a sparse matrix by the NNZ per row and partitions it into two parts. The first part comprises rows with NNZ greater than 32, compressed using the CSR format, while the second part consists of the remaining rows, stored in the ELL format. Yang et al. \cite{DBLP:journals/jcss/YangLL18} propose BCE format, a hybrid and blocked storage of CSR and ELL. They reorder the sparse matrix and combine rows including similar NNZ into blocks. Each block is stored in ELL format, while the starting and ending positions are stored in CSR pattern. Adaptive multi-level blocking (AMB), proposed by Nagasaka et al. \cite{2016_AMB}, has three levels of division and blocking. In the first level, a sparse matrix is divided into column partitions, and each partition is then sliced into row segments in the second level. Finally, in the third level, each segment is divided into blocks by column indices. It can be seen as a hybrid format of CSR and ELL.

\textbf{CSR+DIA}: Bolz et al. \cite{bolz2003sparse} introduce the hybrid DIA+CSR format, which stores the non-zeros of the main diagonal using the DIA format and the remaining non-zeros using the CSR format. In contrast, HDC \cite{DBLP:journals/ijhpca/YangLLSW14}, also a combination of CSR and DIA, categorizes each diagonal of a sparse matrix as either the DIA part or the CSR part based on a predefined threshold. TaiChi format proposed by Gao et al. \cite{DBLP:journals/tpds/GaoJTWS22} also takes a combination of DIA+CSR, which stores dense diagonals using the DIA format and sparse diagonals using the CSR format.

\textbf{COO+CSR}: Tang et al. \cite{DBLP:conf/cgo/TangZLLHLG15} design a vector format, named Vectorized Hybrid COO+CSR (VHCC). They first divide a sparse matrix into multiple vertical panels and then partition each vertical panel into multiple 2D blocks.

\textbf{More than two formats}: Monakov et al. \cite{DBLP:conf/samos/MonakovA09} propose a hybrid blocked format that combines BCSR, BCOO, and ELL formats. It divides a sparse matrix into multiple strips, with each strip containing multiple blocked rows. The blocks within each strip are encoded using the BCOO format, and their column indices and offsets from the top row of the strip are stored in a single word. Additionally, an array similar to the CSR's $\vect{rowPtr}$ is used to store the index of the first block in each strip. To minimize the impact of zero filling, blocks with a filling ratio greater than a given threshold $f$ are stored in ELL format. This approach avoids storing significant numbers of zeros caused by block storage. Tan et al. \cite{TAN2020521} propose a matrix partitioning method based on mathematical morphology theory, which partitions a matrix into diagonal, rectangle, and triangle areas and stores these areas using different compression formats. In a similar vein, Liu et al. \cite{NiuLDJ0T21_TileSpMV} propose a blocked compression format that partitions a sparse matrix into multiple 16$\times$16 blocks. For each block, the optimal encoding method is selected from seven existing formats. This blocked storage leverages the local distribution of sparse matrices to enhance the performance of SpMV operations.

\subsubsection{Other variants}
\textbf{L-CSR}: Paper \cite{DBLP:journals/ijhpca/Mellor-CrummeyG04} introduces a length-grouped CSR (L-CSR) format. Given that sparse matrices often contain a considerable number of rows with identical NNZ, the study initially performs a virtual grouping of the matrix rows based on their NNZ. Subsequently, three arrays are introduced to replace the $\vect{rowptr}$ in CSR. These arrays include $\vect{length}$ array that stores the length of rows within a group, $\vect{lenstart}$ array that stores the starting and ending indices of non-zeros within the group, and $\vect{rowindex}$ array that stores the original row indices for each row within the group.

\textbf{ELL-R}: V{\'{a}}zquez et al. \cite{DBLP:journals/concurrency/VazquezFG11} tackle the efficiency concerns associated with the ELL format, which can be significantly impacted by the maximum NNZ per row of a sparse matrix. To mitigate this issue, they propose the ELLPACK-R (ELL-R) format, which incorporates an additional array that records the NNZ for each row. This modification aims to reduce the computation of invalid zeros, thereby improving the overall efficiency of SpMV.

\begin{figure*}[!htbp]
	\centering
	\includegraphics[width=0.8\textwidth]{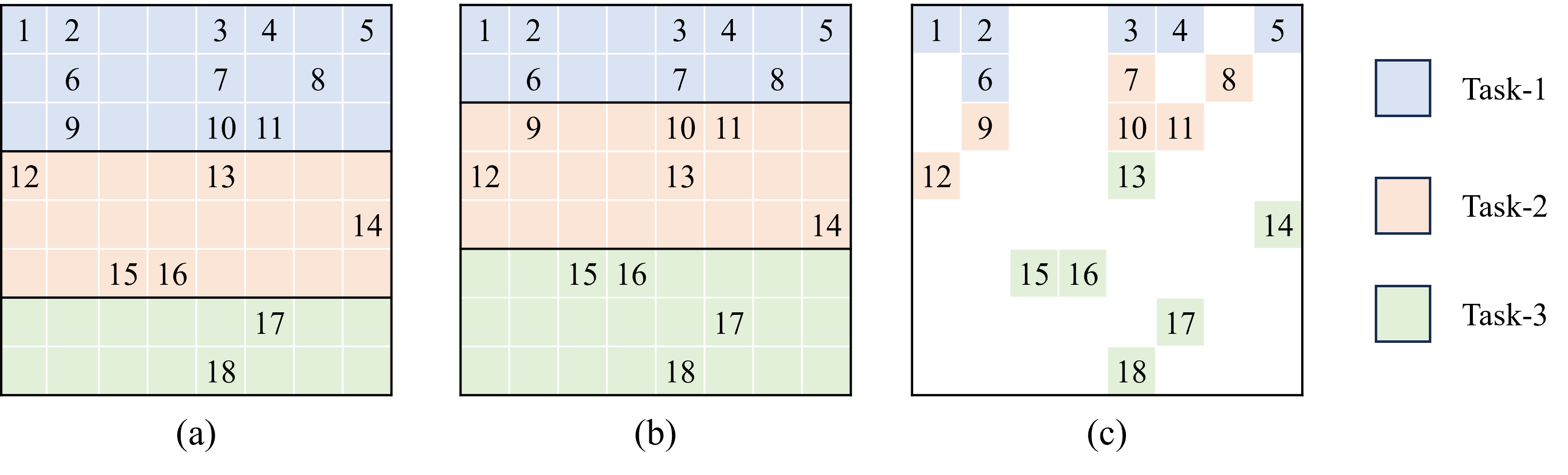} 
	\caption{Examples of different work partitioning approaches in CSR-based SpMV. (a) Fixed rows-based partitioning: dividing work by matrix rows. (b) Relaxed NNZ-based partitioning: dividing work by NNZ without breaking apart matrix rows. (c) Strict NNZ-based partitioning: strictly dividing work by NNZ, even breaking apart matrix rows.}
	\label{fig:CSR-spmv}
\end{figure*}

\textbf{JDS and its variants}: Jagged Diagonal Storage (JDS) \cite{DBLP:journals/ijhsc/AndersonS89,DBLP:journals/pc/Melhem88} is a variant of the ELL format by sorting the matrix rows in descending order of their degrees. Later on, Montagne et al. \cite{DBLP:journals/ipl/MontagneE04} drew inspiration from the CSC format and introduced the transpose JAD (TJAD) format.

\section{Classical Algorithm}\label{sec:csr-format}
Given that CSR is the most widely used compression format, there has been extensive research focused on optimizing CSR-based SpMV. As shown in Table \ref{tab:CSR format-based}, these studies can be categorized into three groups based on their approach to work partitioning:

(1) \textbf{Fixed rows-based partitioning}. This category of methods partitions a sparse matrix based on rows, aiming to assign an equal number of rows to each working thread. However, it is important to note that an equal number of matrix rows does not necessarily correspond to the same workload. Figure \ref{fig:CSR-spmv}(a) illustrates an example of this type of method, where each work partition has three rows, except for the third partition. In this example, the first partition contains 11 non-zeros, while the second partition only has 5 non-zeros. This demonstrates that an equal number of matrix rows does not always result in an equal workload distribution.

(2) \textbf{Relaxed NNZ-based partitioning}. This category of methods combines multiple matrix rows into a single block, aiming to achieve a balanced workload by ensuring a similar NNZ in each block. Each working thread is then assigned to process one block. Figure \ref{fig:CSR-spmv}(b) illustrates an example of this algorithm, where the goal is to include about 6 non-zeros in each block. However, as observed from the figure, this method still faces the challenge of workload imbalance. Furthermore, if a sparse matrix contains very long rows with tens of thousands or even millions of non-zeros, these methods can only achieve limited performance improvements.

(3) \textbf{Strict NNZ-based partitioning}. This category of methods adopts a more aggressive approach to work partitioning in a sparse matrix, prioritizing the inclusion of the same NNZ in each work partition, regardless of the constraints imposed by matrix rows. Figure \ref{fig:CSR-spmv}(c) illustrates an example where each work partition has 6 non-zeros. However, it is important to note that one matrix row may be involved in two or more work partitions.

\begin{table}[!htbp]
    \centering
    \caption{Classification of CSR format-based.}\label{tab:CSR format-based}
        \begin{tabular}{m{2.5cm}m{5.5cm}}
        \toprule
            \textbf{Classification} & \multicolumn{1}{c}{\textbf{Contribution}} \\ \midrule
            Fixed rows-based & \cite{bell2008efficient}
            \cite{baskaran2008optimizing}
            \cite{BellG09}
            \cite{DBLP:conf/ccgrid/ZeinR10}
            \cite{DBLP:journals/concurrency/ZeinR12}
            \cite{Cusp}
            \cite{DBLP:conf/iccsa/MukunokiT13}
            \cite{giles2012efficient}
            \cite{YoshizawaT12}
            \cite{CMRS_Koza_2014}
            \cite{DBLP:conf/icpp/WuWSYWX10}
            \cite{AnztCYDFNTTW20}
            \cite{anzt2020ginkgo} 
            \cite{DBLP:conf/sc/FlegarA17} 
            \\ \midrule
            Relaxed NNZ-based & \cite{DBLP:conf/hpcc/LiuZSQ09}
            \cite{Feng_Jin_Zheng_Hu_Zeng_Shao_2011}
            \cite{greathouse2014efficient}
            \cite{CSR_adaptive_improved}
            \cite{DBLP:conf/icpads/GaoJLSWS21}\cite{LiuS15}\cite{liu2018lightspmv}
            \\  \midrule
            Strict NNZ-based &  \cite{CSR5}
            \cite{DBLP:conf/ccgrid/BianHDLW20}\cite{MergedSpMV_SC16}\cite{DBLP:conf/europar/FlegarQ17}\cite{SteinbergerZS17}
            \\
            \bottomrule
        \end{tabular}
\end{table}

CSR-Scalar \cite{bell2008efficient} (or CSR-Naive \cite{baskaran2008optimizing}) and CSR-Vector \cite{baskaran2008optimizing}\cite{bell2008efficient}\cite{BellG09} are representative algorithms that utilize fixed rows-based work partitioning. These algorithms assign an equal number of threads to each matrix row. CSR-Scalar assigns one thread per row, while CSR-Vector assigns a group of threads per row. However, both methods exhibit limited performance when handling extremely irregular sparse matrices that contain a small number of long rows. Considering that different sparse matrices exhibit preferences for distinct thread configurations or SpMV algorithms, several studies have proposed empirically-based decision rules to improve performance. Zein et al. \cite{DBLP:conf/ccgrid/ZeinR10}\cite{DBLP:journals/concurrency/ZeinR12} select CSR-Scalar or CSR-Vector algorithm, along with specific options within each method, based on the average NNZ per row ($avg\_nnz$), the number of matrix rows and columns. CSR-Vector implementation in the CUSP library \cite{Cusp} employs $avg\_nnz$ to determine the number of threads allocated per row ($TpR$), and the same thread configuration is used by Mukunoki et al. \cite{DBLP:conf/iccsa/MukunokiT13}. Differently, Reguly et al. \cite{giles2012efficient} configure $TpR$ by the square root of $avg\_nnz$. The detailed formulas for these two rules are provided in Equation \ref{equ:TpR_mean} and \ref{equ:TpR_SqMean}. Yoshizawa et al. \cite{YoshizawaT12} set $TpR$ based on the maximum NNZ per row ($maxrl$), as shown in Equation \ref{equ:T-maxrl}. Koza et al. \cite{CMRS_Koza_2014} also utilize the CSR-Vector algorithm but first fold adjacent multi-rows into a stripe to increase the row density. Wu et al. \cite{DBLP:conf/icpp/WuWSYWX10} first sort matrix rows in ascending order by NNZ per row, and then handle tiny rows (with NNZ per row no more than 6) using CSR-Scalar method, small rows (with NNZ per row more than 6 and no more than 96) using CSR-Vector with $TpR$ of 16, and normal rows (with NNZ per row more than 96) using CSR-Vector with $TpR$ of 64. The Ginkgo library \cite{AnztCYDFNTTW20}\cite{anzt2020ginkgo} provides an automatic selection strategy for CSR-based SpMV algorithms, which chooses between CSR-Scalar and CSR-I based on the number of matrix rows and the maximum NNZ per row. The selection strategy is outlined in Algorithm \ref{alg:csr-spmv-selection}. Based on the observation of experimental results, Flegar et al. \cite{DBLP:conf/sc/FlegarA17} determine the ratio of the number of allocated threads to the number of physical cores according to the NNZ of a sparse matrix.

\begin{equation}\label{equ:TpR_mean}
TpR\_mean=\min(2^{\lceil\log_2^{\lfloor\frac{nnz}{m}\rfloor}\rceil},32),
    \end{equation}
\begin{equation}\label{equ:TpR_SqMean}
	TpR\_SqMean=\min(2^{\lceil\log_2^{\lfloor\sqrt{\frac{nnz}{m}}\rfloor}\rceil},32).
\end{equation}
\begin{equation}\label{equ:T-maxrl}
        T^* =\left\{
        \begin{aligned}
            &16 & (maxrl \ge 32) \\
            &2^{\lceil log_2{maxrl}\rceil-2} & (maxrl < 32)
        \end{aligned}\right .
\end{equation}

\begin{algorithm}
\caption{The selection strategy of CSR-based SpMV algorithms in Ginkgo library \cite{AnztCYDFNTTW20}}
	\label{alg:csr-spmv-selection}
	\eIf{$m>10^6$}
	{
		call CSR-I\;
	}{
        \eIf{$max>64$}
        {
            call CSR-I\;
        }{
            call CSR-Scalar\;
        }
    }	
    \end{algorithm}

Having the same number of rows does not necessarily mean having the same workload. The NNZ accurately indicates the workload. Work \cite{DBLP:conf/hpcc/LiuZSQ09} compares this approach with three scheduling methods provided in OpenMP and concludes that the relaxed NNZ-based method outperforms the others. In \cite{Feng_Jin_Zheng_Hu_Zeng_Shao_2011}, matrix rows are reordered in descending order by NNZ, and consecutive rows are grouped into segments to ensure each segment has a similar NNZ. The CSR-Adaptive method proposed by Greathouse et al. \cite{greathouse2014efficient}\cite{CSR_adaptive_improved} is a notable example of the relaxed NNZ-based methods. It combines the traditional CSR-Vector and CSR-Stream proposed in their work. The CSR-Stream fixes the NNZ assigned to each thread block and then partitions the sparse matrix by rows to ensure similar NNZ in each partition. Liu et al. \cite{LiuS15}\cite{liu2018lightspmv} dynamically distribute matrix rows over vectors to tackle the limitations of CSR-Vector. Gao et al. \cite{DBLP:conf/icpads/GaoJLSWS21} propose an adaptive multirow folding method referred to as AMF-CSR to address the load imbalance in fixed multirow folding (CMRS) \cite{CMRS_Koza_2014}. It folds adaptive multi-rows into one row and then utilizes a relaxed NNZ-based blocking method. However, these methods, which preserve the integrity of matrix rows, still face performance bottlenecks caused by extremely long rows.

To achieve a balanced workload among working threads, a more aggressive approach is strict NNZ-based work partitioning. CSR5, proposed by Liu et al. \cite{CSR5}, is a typical representative of this method. They divide all non-zeros into two-dimensional sub-blocks of equal size, with each parallel unit processing one or more sub-blocks, ensuring strict load balance. Another approach, CSR2 proposed by Bian et al. \cite{DBLP:conf/ccgrid/BianHDLW20}, also guarantees strict workload balance but introducing a small increase in storage overhead due to significant zeros padding. Merged-based SpMV, proposed by Merrill et al. \cite{MergedSpMV_SC16} represents CSR-based SpMV as a logical merge of two lists: row pointer and column index arrays. By allocating equal-sized portions of the merge to each processing element, the algorithm ensures strict workload balance. Additionally, CSR-I, proposed by Flegar et al. \cite{DBLP:conf/europar/FlegarQ17}, Hola-SpMV proposed by Steinberger et al. \cite{SteinbergerZS17} also employ strict NNZ-based work partitioning strategy.

\section{Auto-Tuning Based Algorithm}\label{sec:auto-tuning}
\begin{figure*}
    \centering
    \includegraphics[width=0.85\textwidth]{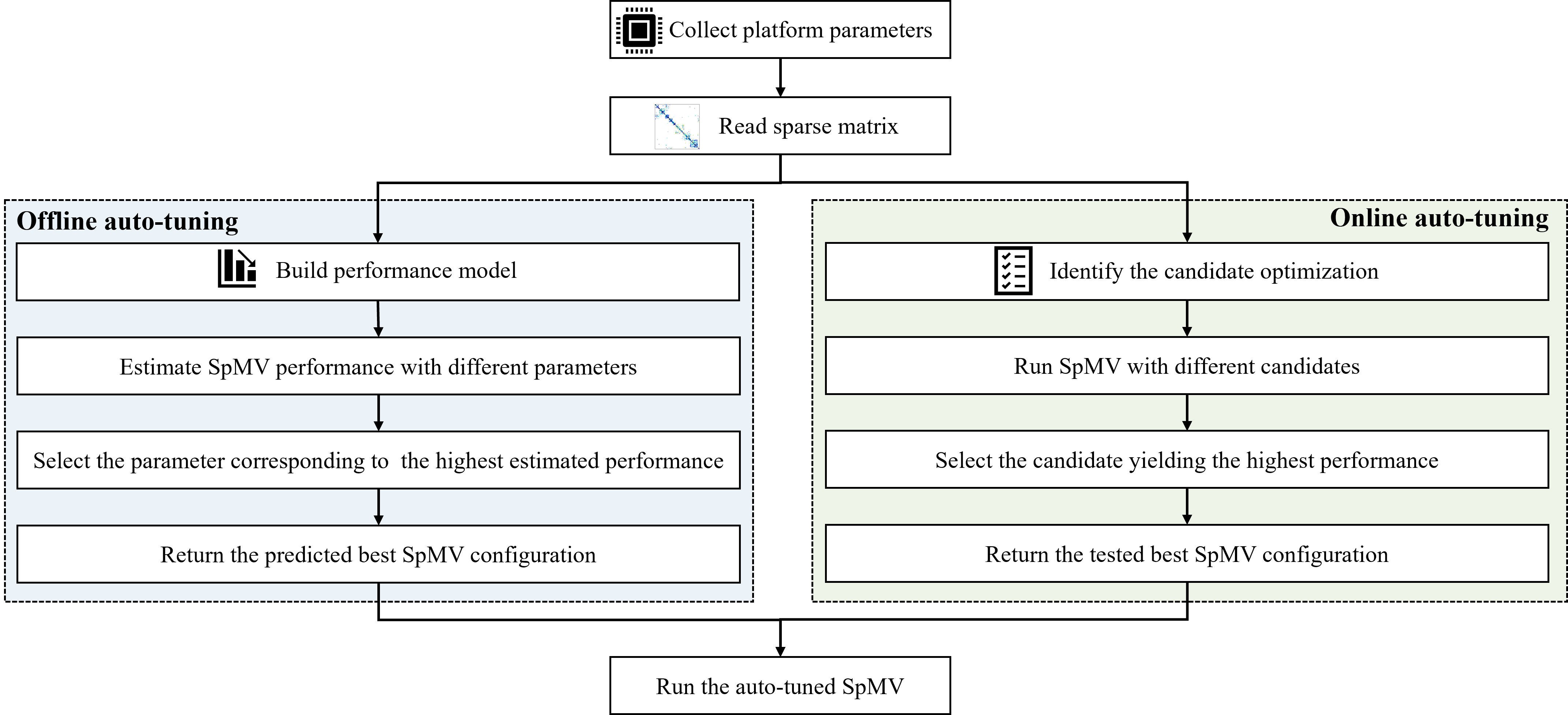}
    \caption{General workflow of offline and online auto-tuning.}
    \label{fig:auto-tuning}
\end{figure*}

Researchers have put forth numerous storage formats and SpMV algorithms to tackle the challenges of SpMV. However, it is important to note that different compression formats or SpMV algorithms are better suited for sparse matrices with specific non-zero distributions. At present, there is no single format, parameter setting, or SpMV algorithm that can consistently outperform others across all matrices. As a result, researchers propose various auto-tuning techniques to find the optimal configuration for a given matrix. As summarized in Table \ref{tab:auto-tuning}, we classify these techniques into two categories: offline and online auto-tuning. Their workflows are illustrated in Figure \ref{fig:auto-tuning}.

\begin{table}[!ht]
    \centering
    \caption{Classification of auto-tuning based optimization.}\label{tab:Classification of Auto-Tuning Based Optimization}
        \begin{tabular}{m{2.5cm}m{5cm}}
        \toprule
            \textbf{Classification} & \multicolumn{1}{c}{\textbf{Contribution}} \\ \midrule
            Offline auto-tuning & \cite{DBLP:conf/iccS/ImY01}
            \cite{DBLP:journals/ijhpca/ImYV04}
            \cite{DBLP:conf/sc/VuducDYKNL02}
            \cite{vuduc2003automatic}
            \cite{vuduc2005oski}
            \cite{DBLP:journals/aaecc/NishtalaVDY07} 
            \cite{zhang2009automatic}
            \cite{DBLP:conf/ppopp/ChoiSV10}
            \cite{DBLP:journals/tpds/GuoWC14}
            \cite{LiKenli-TPDS2016}
            \cite{VazquezFG12}
            \cite{DBLP:conf/IEEEcit/VazquezOFG10}
            \cite{DBLP:journals/chinaf/LiHZZ15}
            \cite{DBLP:conf/hpcc/ChenXXY19}
            \\ 
            \midrule
            Online auto-tuning & \cite{guo2010auto}
            \cite{giles2012efficient}
            \cite{DBLP:conf/supercomputer/Abu-SufahK13}
            \cite{DBLP:conf/hpcc/Abu-SufahK12}
            \cite{DBLP:conf/europar/SunZWLZL11}
            \cite{Yan2014}
             \\             
            \bottomrule
        \end{tabular}
        \label{tab:auto-tuning}
\end{table}

\subsection{Offline Auto-Tuning}
Im et al. \cite{DBLP:conf/iccS/ImY01}\cite{DBLP:journals/ijhpca/ImYV04}\cite{DBLP:conf/sc/VuducDYKNL02}\cite{vuduc2003automatic}\cite{vuduc2005oski} develop a performance model to predict SpMV performance for different block sizes, aiming to determine the optimal block size in the blocked format-based SpMV. The performance model is represented as the ratio of $flop$ to $fill$, where $flop$ denotes the performance of a dense matrix in $r\times c$ sparse blocked format, and $fill$ estimates the overhead of zeros filling for $r\times c$ blocking. Later, they \cite{DBLP:journals/aaecc/NishtalaVDY07} extend this model by taking the TLB misses into account to determine when to apply the cache-blocking optimization technology. Zhang et al. \cite{zhang2009automatic} also utilize a similar $flop/fill$ based approach to determine the block size in BCSR-based SpMV. Choi et al. \cite{DBLP:conf/ppopp/ChoiSV10} introduce a performance model and consider additional parameters such as block size, sub-matrix size, and the number of thread blocks. Guo et al. \cite{DBLP:journals/tpds/GuoWC14} design performance models specific to different compression formats-based SpMV kernels (including COO, CSR, ELL, and HYB) to predict their execution time for a given sparse matrix. They develop a dynamic programming-based auto-selection algorithm to choose the optimal configuration, including whether to partition the matrix, the optimal compression format, and the predicted execution time. For the same group of SpMV algorithms, Li et al. \cite{LiKenli-TPDS2016} utilize a probability mass function (PMF) to model their performance. They combine the model with GPU hardware parameters to predict the performance of each SpMV algorithm for a given sparse matrix. V{\'{a}}zquez et al. \cite{VazquezFG12} construct a performance model based on memory access overhead to predict two parameters in ELLR-T \cite{DBLP:conf/IEEEcit/VazquezOFG10}: the block size of CUDA and the number of threads assigned to process each row. Li et al. \cite{DBLP:journals/chinaf/LiHZZ15} develop a performance model for SpMV on multi-core clusters to configure the best threads, processes, and communication patterns. Chen et al. \cite{DBLP:conf/hpcc/ChenXXY19} present an auto-tuning framework to determine the optimal threshold in HYB format, based on the execution time analysis of HYB-based SpMV.

\subsection{Online Auto-Tuning}
Guo et al. \cite{guo2010auto} determine the optimal thread configuration for CSR-Vector by launching the SpMV kernel with different thread configurations (16 or 32 threads) during the initial iterations of the iterative algorithm. Similarly, the dynamic rule proposed by Reguly et al. \cite{giles2012efficient} determines the optimal $TpR$ by executing the SpMV kernel multiple times with different thread configurations ($TpR=2,4,8,16,32$). Kunkel et al. \cite{DBLP:conf/supercomputer/Abu-SufahK13} use heuristic exhaustive searching to determine the optimal compression format and related parameters. Their search space for compression formats or SpMV algorithms includes CSR (vector kernel), COO, DIA, ELL, ELL-R, BELL, and their previously proposed blocked TJAD (BTJAD) \cite{DBLP:conf/hpcc/Abu-SufahK12}. Additionally, parameters such as kernel block size, number of launched threads, threads assigned to each matrix block, matrix block size, and number of registers are considered. Sun et al. \cite{DBLP:conf/europar/SunZWLZL11} present an auto-tuning framework for diagonal pattern-dominated applications, which searches the optimal parameter and generates corresponding code for each application-specific diagonal pattern. Yan et al \cite{Yan2014} propose an auto-tuning framework to select the compression format from BCCOO and BCCOO+, compute strategy, and associated parameters.

\section{Machine-Learning Based Algorithm}\label{sec:ML}
\begin{figure}[!htbp]
    \centering
    \includegraphics[width=1\linewidth]{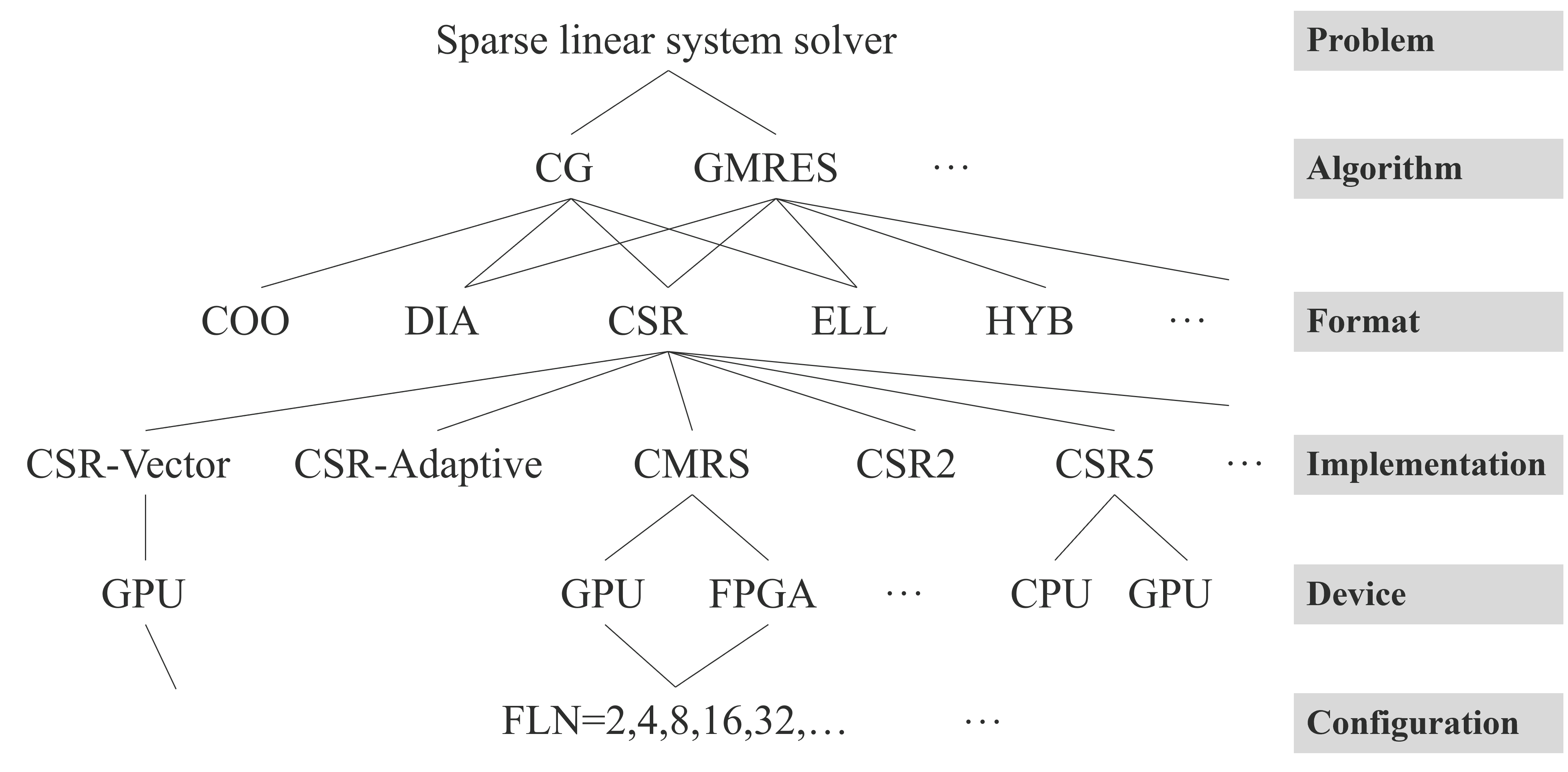}
    \caption{The hierarchical design choices for sparse linear system solvers}
    \label{fig:solver}
\end{figure}

Over an extended period of accumulation, a highly substantial body of research outcomes on SpMV has currently been established. Hence, multiple pathways exist in the design and implementation of a sparse linear solver. As shown in Figure \ref{fig:solver}, from a top-down perspective, aside from direct solvers, various iterative solving algorithms (such as CG and GMRES) are currently known, each with distinct applicability conditions and requirements (e.g., sparse matrix symmetry and positive definiteness). Regarding SpMV alone, there exist numerous distinct compression formats for sparse matrices. Even for the same compression format, there exist multiple computational algorithms, and some of them can run various computing devices. Due to the unique nature of each practical problem and the varying distribution of non-zeros within sparse matrices, there currently is not one unique format, algorithm, implementation, or configuration parameter that universally outperforms others across all matrices on a given computing device. Given that the running efficiency of SpMV is influenced by various factors like compression formats, data distribution, and hardware specifications, leveraging machine learning for optimal path selection emerges as a feasible technological approach.

In addition, in order to accelerate computations of sparse matrices across multiple computing devices, it is essential to anticipate the computational time of the segmented SpMV in advance for the purpose of load balancing. Therefore, the prediction of SpMV execution time based on machine learning stands as a significant research avenue.

\begin{figure*}[!htbp]
	\centering
	\includegraphics[width=0.6\textwidth]{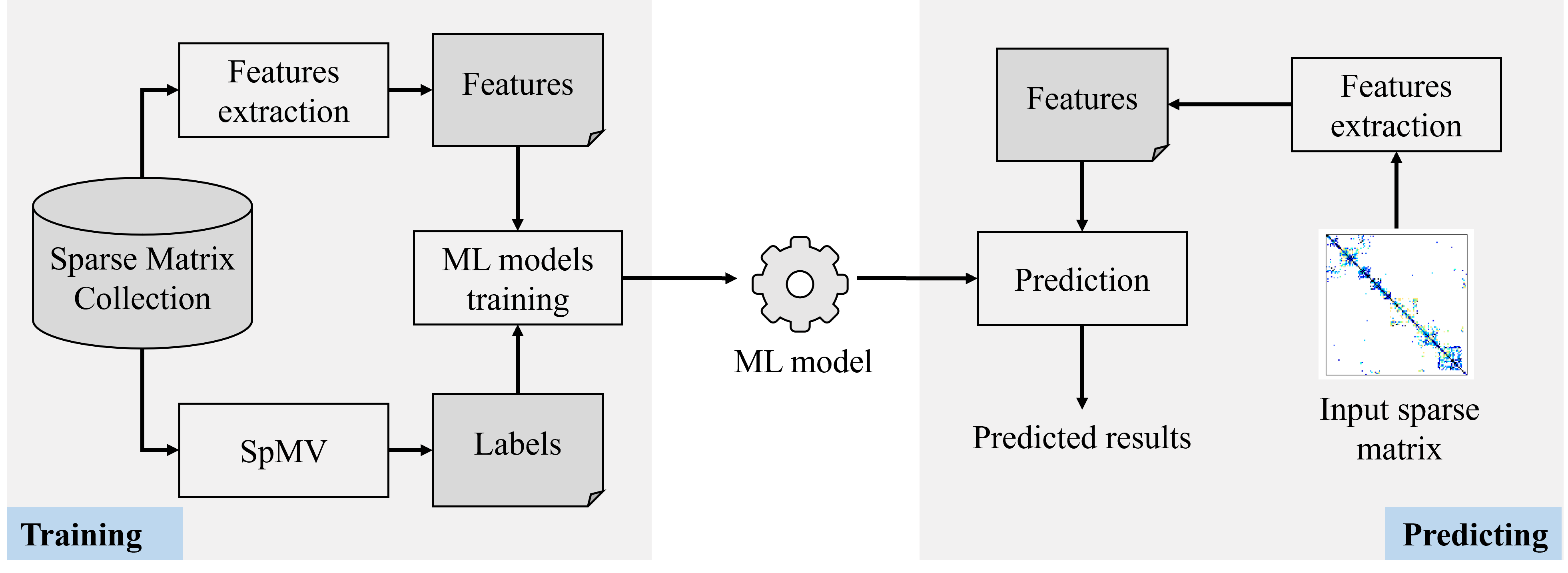} 
	\caption{Workflow of ML-based optimization \cite{GAO2024104799}.}
	\label{fig:ML-optimization}
\end{figure*}

The study of machine learning (ML) based optimization for SpMV started in approximately 2013, as illustrated in Table \ref{tab:ML-optimization}. Figure \ref{fig:ML-optimization} provides an overview of the general workflow for ML-based optimization. The labels in the figure can refer to the optimal sparse compression format, SpMV algorithms, parameter setting, and the shortest execution time of SpMV. Regression problems arise when directly predicting SpMV execution time for a given matrix, while classification problems arise for the other cases. The predicted results in Figure \ref{fig:ML-optimization} can be compression format, SpMV algorithm, parameter settings, or predicted SpMV execution time. Table \ref{tab:matrix-features} presents some commonly used matrix features that are extensively employed in this context.

\begin{table}[!htbp]
    \centering
    \caption{Classification of machine-learning based optimization.}\label{tab:ML-optimization}
        \begin{tabular}{m{2cm}m{6cm}}
        \toprule
            \textbf{Classification} & \multicolumn{1}{c}{\textbf{Contribution}} \\ \midrule
            Format/Algorithm & \cite{Armstrong2008_RL}
            \cite{SMAT2013}
            \cite{DBLP:conf/ics/SedaghatiMPPS15}
            \cite{Chen2018}
            \cite{Akrem2016ICPP_SVM}
            \cite{Mehrez2018HPCS_SVM}
            \cite{2017-IPDPSW-autoTuning}
            \cite{Akrem-BestSF}
            \cite{DBLP:journals/ppl/Hamdi-LarbiMD21}
            \cite{Nisa2018IPDPSW}
            \cite{DBLP:journals/ieicet/CuiHKT18}
            \cite{ShenXipeng2018PPoPP_bridgingGap}
            \cite{Shenxipeng2020TPDS_overheadConscious}
            \cite{10.1145/3293883.3301490}
            \cite{ijhpca/DufrechouEQ21}
            \cite{DBLP:conf/hpcc/XiaoZCHL22}
            \\ \midrule
            Parameter &  \cite{2017-IPDPSW-autoTuning}
            \cite{usman2019zaki}\cite{DBLP:journals/access/UsmanMKA19}
            \cite{DBLP:journals/access/UsmanMKA19}
            \cite{ahmed2022aaqal}
            \cite{GAO2024104799}
            \\ \midrule    
            Performance & \cite{Akrem_ICPADS2016}
            \cite{Akrem2020}
            \cite{Nisa2018IPDPSW}
            \cite{barreda2020performance}
            \cite{ barreda2021convolutional}
            \\             
            \bottomrule
        \end{tabular}
\end{table}

\begin{table*}[!ht]
    \centering
    \caption{Intensively used sparse matrix features in ML-based optimization.}\label{tab:matrix-features}
        \begin{tabular}{cll}
        \toprule
            \textbf{Feature} & \textbf{Meaning} & \textbf{Calculation Formula} \\
            \midrule 
            $m$ & Number of rows & $m$\\
            $n$ & Number of columns & $n$ \\
            $nnz$ & Number of non-zeros & $nnz$ \\
            $d$ & The density of non-zeros & $nnz/(m\times n)$ \\
            $mu$ & The average NNZ per row & $nnz/m$ \\
            $max$ & The maximum NNZ per row & $\max_{i=1}^m nnz_i$\\
            $std\_nnz$ & The standard deviation on NNZ per row & $\sqrt{\frac{1}{m}\sum_{i=1}^m (nnz_i-mu)^2}$\\
            $max-mu$ & The difference between $max$ and $mu$ & $max-mu$ \\ 
            $dia$ & The number of diagonals containing at least one non-zero element & \\
            $dis$ & The average distance between each pair of adjacent non-zero elements & $dis=\frac{1}{m}\sum_{i=1}^m dis_i$ \\
            $ave\_bw$ & The average column distances between the first and the last non-zero element & $\frac{1}{m}\sum_{i=1}^m bw_i$ \\
            $std\_bw$ & Standard deviation of column distances between the first and the last non-zero element per row & $\sqrt{\frac{1}{m}\sum_{i=1}^m(bw_i-ave\_bw)^2}$ \\
            \bottomrule
        \end{tabular}
\end{table*}

\subsection{Format or Algorithm Selection}
Armstrong et al. \cite{Armstrong2008_RL} use reinforcement learning (RL) to select the optimal compression format, focusing on three formats: CSR, COO, and BCSR. Li et al. \cite{SMAT2013}, Sedaghati et al. \cite{DBLP:conf/ics/SedaghatiMPPS15}, and Chen et al. \cite{Chen2018} employ decision tree (DT)-based models to predict the optimal format, and multiclass support vector machine (SVM)-based models are used by Benatia et al. \cite{Akrem2016ICPP_SVM} and Mehrez et al. \cite{Mehrez2018HPCS_SVM}. Hou et al. \cite{2017-IPDPSW-autoTuning} use a DT-based model to select the optimal SpMV kernel among three CSR-based kernels. BestSF, proposed by Benatia et al. \cite{Akrem-BestSF}, uses a pairwise weighted classification model to identify the best sparse format. Nisa et al. \cite{Nisa2018IPDPSW} evaluate the performance of several basic and ensemble ML algorithms for format selection, including DT, SVM, multi-layer perceptrons (MLP), and extreme gradient boosting (XGBoost). Instead of manually selecting features, He et al. \cite{DBLP:journals/ieicet/CuiHKT18} employ a feature image map and an ML tool to establish relationships between sparse matrices and SpMV implementations. Then, Zhao et al. \cite{ShenXipeng2018PPoPP_bridgingGap} develop a deep learning model to determine the optimal format. Differently, Olfa et al. \cite{DBLP:journals/ppl/Hamdi-LarbiMD21} build the ML model to predict optimal sparse format on distributed platforms.

However, in certain application scenarios, the overhead caused by extracting matrix features, model prediction, and format conversion may outweigh the potential benefits. To address this, Shen et al. \cite{Shenxipeng2020TPDS_overheadConscious} introduce a two-stage lazy-and-light prediction scheme. They employ a two-layer long short-term memory (LSTM) network to predict the total number of SpMV iterations in the first stage. If the predicted value exceeds a given threshold, the second stage is triggered. In this stage, they utilize an XGBoost model to predict the optimal compression format. Elafrou et al. \cite{10.1145/3293883.3301490} use a DT-based classifier to identify the performance bottleneck for a given sparse matrix, distinguishing between memory bandwidth bound, memory latency bound, and work imbalance bound. They propose corresponding coping strategies for different performance bottlenecks. Based on a given matrix's feature set, Dufrechou et al. \cite{ijhpca/DufrechouEQ21} employ a bagged trees-based classifier to predict the optimal SpMV method, considering two criteria: execution time and energy consumption. Different from other approaches, Xiao et al. \cite{DBLP:conf/hpcc/XiaoZCHL22} introduce DTSpMV, a DT-based selection strategy for SpMV algorithms. DTSpMV considers the sparsity features of the multiplied vector $\vect{x}$ during the iterative process in graph analytics.

\subsection{Parameter Prediction}
Hou et al. \cite{2017-IPDPSW-autoTuning} develop a DT-based model to select the optimal binning strategy from ``intra-bin" and ``inter-bin". In the ``intra-bin" strategy, each bin contains several contiguous rows, and the total number of non-zeros in different bins approximates one another, while the ``inter-bin" strategy groups rows with similar NNZ into the same bin, even if they are not adjacent. Usman et al. \cite{usman2019zaki}\cite{DBLP:journals/access/UsmanMKA19} introduce a data-driven ML tool called ZAKI, which predicts the optimal number of processes and the optimal mapping of data on the underlying hardware for SpMV on a distributed platform. They explore various ML models such as DT, random forest, and gradient boosting. Furthermore, they extend the tool and propose ZAKI+, which supports the prediction of the optimal process mapping strategy \cite{DBLP:journals/access/UsmanMKA19}. Ahmed et al. \cite{ahmed2022aaqal} present an ML-based tool to predict the fastest block size setting for BCSR. They use a set of features consisting of five basic and easy-to-compute features, along with nine more complex features. Moreover, various machine learning methods, including decision trees, random forest, gradient boosting, ridge regressor, and AdaBoost, are considered and discussed. Gao et al. \cite{GAO2024104799} revisit the thread configuration in CSR-Vector and introduce an ML-based approach to predict the optimal thread configuration. To handle irregularly distributed sparse matrices, they propose a non-zeros distribution-aware matrix partitioning method. By leveraging the ML-based method, they determine the optimal thread configuration for each sub-matrix. In the study, they further simplify the ML-based model into a lightweight decision tree, which effectively reduces the overhead associated with the online decision.

\subsection{Performance Prediction}
To enable collaborative computing across multiple GPUs in large-scale SpMV calculations and design efficient task-scheduling algorithms, determining the execution time of each matrix partition is crucial. Therefore, researchers have proposed various ML-based approaches to forecast SpMV performance. Benatia et al. \cite{Akrem_ICPADS2016}\cite{Akrem2020} utilize SVM and MLP to predict the performance of SpMV kernels with different compression formats. Building upon this work, Nisa et al. \cite{Nisa2018IPDPSW} expanded the approach by incorporating additional ML models such as XGBoost and ensembles of MLP. Subsequently, Barreda et al. \cite{barreda2020performance, barreda2021convolutional} develop offline estimators using CNN for predicting the execution time and energy consumption of SpMV. These efforts contribute to the advancement of collaborative computing and efficient task scheduling in large-scale SpMV computations.

\section{Mixed Precision-Based Optimization}\label{sec:mixed-precision}

As artificial intelligence technologies advance, hardware manufacturers are increasingly inclined to embrace low-precision floating-point number formats in their latest products. NVIDIA's H100, for instance, supports TF32, BF16, FP16, and FP8 (E4M3 and E5M2) formats, fostering a keen interest among researchers to leverage mixed-precision computing. This interest aims to optimize SpMV operations and iterative algorithms.
However, due to the non-commutative and non-distributive nature of floating-point computations, inferring the resulting calculation errors after executing a segment of code is challenging. Currently, mixed-precision approaches primarily encompass two avenues: formalized analysis-based mixed-precision computation and empirical algorithm design with practical validation. Given that SpMV is commonly employed in iterative algorithms, serving as a pivotal core within such iterative processes, the subsequent analysis focuses initially on mixed-precision iterative solving algorithms, followed by an introduction to mixed-precision SpMV computations. These studies are summarized in Table \ref{tab:mixed-precision}.

\begin{table}[!htbp]
    \centering
    \caption{Classification of mixed precision-based optimization.}\label{tab:Classification of Auto-Tuning Based Optimization}
        \begin{tabular}{cl}
        \toprule
            \textbf{Classification} & \multicolumn{1}{c}{\textbf{Contribution}} \\ \midrule
            Mixed-precision iterative algorithms & \cite{DBLP:journals/siamsc/CarsonH18}
            \cite{DBLP:journals/corr/abs-1907-10550}
            \cite{DBLP:journals/ijhpca/AliagaAGQT23}
            \cite{loe2021study}
            \cite{le2018mixed}\cite{DBLP:conf/smc2/LindquistLD20}
            \\ 
            \midrule
            Mixed-precision SpMV & \cite{DBLP:journals/taco/AhmadSH20}
            \cite{DBLP:journals/jcam/MukunokiO20}
            \cite{DBLP:conf/sbac-pad/TezcanTKKU22}
            \cite{graillat:hal-03561193}
            \cite{DBLP:journals/jocs/Isupov22}
            \cite{DBLP:journals/corr/Kouya14a}
             \\             
            \bottomrule
        \end{tabular}
        \label{tab:mixed-precision}
\end{table}

\subsection{Mixed-Precision Iterative Solving Algorithms}
\begin{figure}
    \centering
    \includegraphics[width=0.99\linewidth]{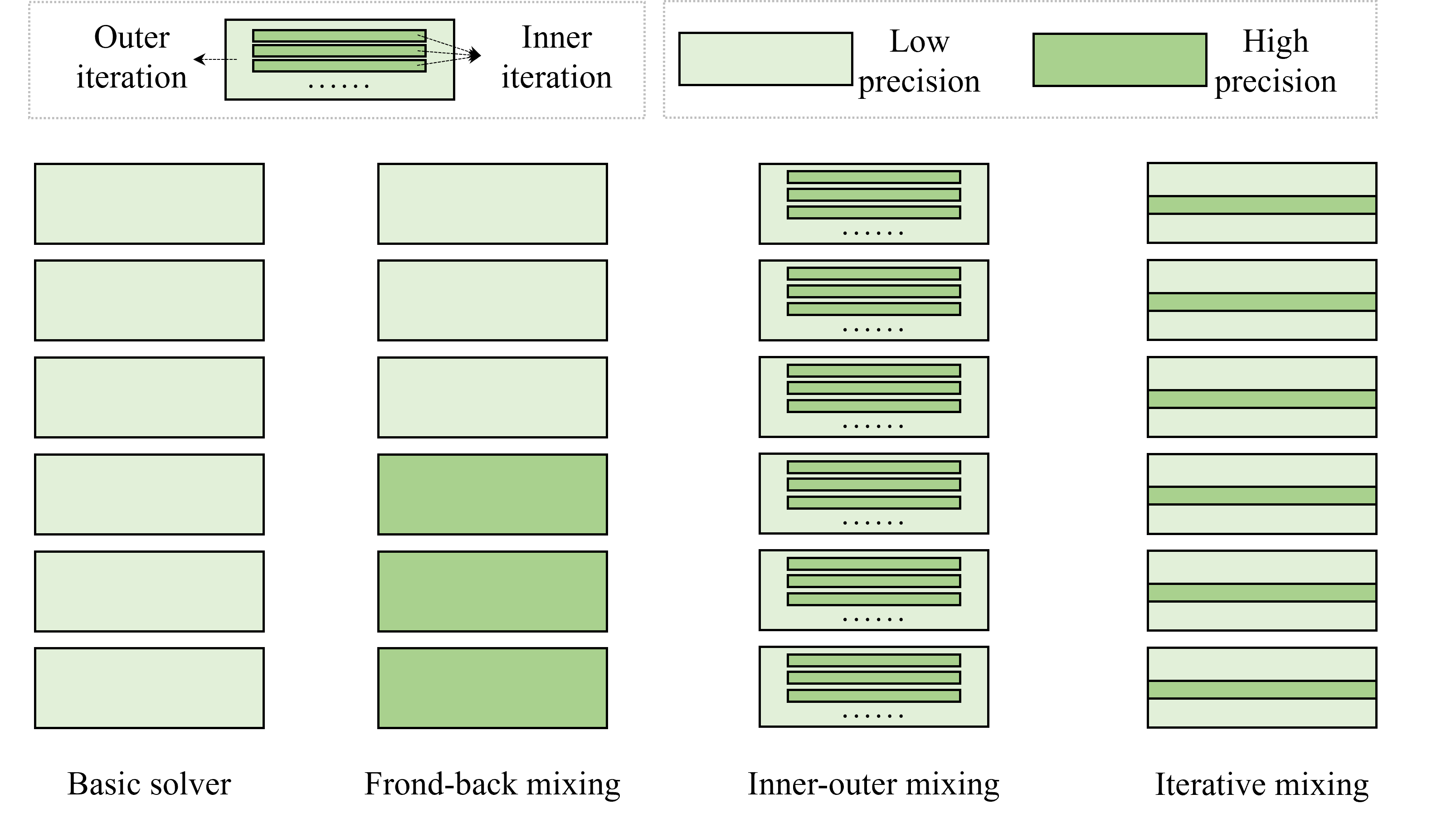}
    \caption{Typical mixed-precision iterative solving algorithms}
    \label{fig:mixed-loop}
\end{figure}

As shown in Figure \ref{fig:mixed-loop}, the typical mixed-precision iterative algorithms can be primarily summarized as three modes: front-back mixing, inner-outer mixing, and iterative mixing. The front-back mixing uses a low precision during the initial stage of an iterative algorithm, transitioning to high-precision calculations after some iterations. Inner-outer mixing, used in the inner-outer GEMRES algorithm, employs low precision for inner iteration and high precision for outer iteration. Iterative mixing, on the other hand, refers to employing different precision within a single iteration. Comparatively, front-back mixing and inner-outer mixing are more straightforward to implement, while iterative mixing demands careful consideration of how varying precision impacts iterative residuals. Moreover, it introduces the overhead of precision conversion within the same iteration.

Gratton et al. \cite{DBLP:journals/corr/abs-1907-10550} present a variable precision GMRES, which uses low precision for inner products and SpMV operations and high precision for other computations. Aliaga et al. \cite{DBLP:journals/ijhpca/AliagaAGQT23} propose to store vectors of the Krylov basis using compact format (FP16, INT16, FP32, INT32) while using double precision for computations. Lindquist et al. \cite{DBLP:conf/smc2/LindquistLD20} use double precision in the computation of residual and the update of approximate solution, while single precision is used in other computations. These methods employ the optimization principle of iterative mixing.

GMRES-FD (Float to Double) algorithm \cite{loe2021study} is a typical representative of front-back mixing. It starts in low precision (float) and switches to high precision after some iterations. Carson et al. \cite{DBLP:journals/siamsc/CarsonH18} merge the GMRES algorithm with iterative refinement, introducing GMRES-IR, which can be regarded as a hybrid of inner-outer mixing and iterative mixing. This algorithm uses up to three different floating-point precision, such as FP16, FP32, and FP64. It utilizes FP16 for LU decomposition of linear systems (preconditioning stage), employs FP64 for residual computation and approximate solution refinement (steps in outer iteration), and reserves FP32 for intermediate residual and approximate solution retention (steps in inner iteration). This approach was investigated by Gallo et al. \cite{le2018mixed} within the context of PIM architecture.

\subsection{Mixed-Precision SpMV}
The focus of mixed-precision SpMV computation lies in determining how to represent the data used in the computation with different precision. Representing all elements in the input matrix with low precision can result in significant errors in the computation result. Therefore, striking a balance between precision in representation and computational error poses a challenging problem.

\begin{table*}[!ht]
    \centering
    \caption{Classification of architecture oriented optimization.}
        \begin{tabular}{m{3cm}m{13cm}}
        \toprule
            \textbf{Architecture} & \multicolumn{1}{c}{\textbf{Contribution}} \\ \midrule
            CPU & \cite{DBLP:conf/ispdc/PabstBK12}\cite{CSR5}\cite{DBLP:conf/ccgrid/BianHDLW20}\cite{DBLP:journals/jpdc/ZhangYLTL21}\cite{DBLP:journals/ijhpca/ImYV04}\cite{DBLP:journals/aaecc/NishtalaVDY07}\cite{DBLP:conf/sc/WilliamsOVSYD07}\cite{DBLP:journals/pc/WilliamsOVSYD09}\cite{kislal2013optimizing}\cite{DBLP:conf/hpcc/YuanZS08}\cite{lee2003performance}\cite{DBLP:conf/hpcc/LiuZSQ09}\cite{DBLP:conf/icpp/ElafrouGK17}\cite{DBLP:conf/sc/TrotterELTDIU23}\cite{DBLP:conf/npc/YuMQF020}
            \\ \midrule
            \multirow{2}{*}{GPU} & \cite{Nickolls_Buck_Garland_Skadron_2008}\cite{DBLP:conf/sc/AshariSEPS14}\cite{greathouse2014efficient}\cite{CSR_adaptive_improved}\cite{CSR5}\cite{MergedSpMV_SC16}\cite{DBLP:conf/sc/FlegarA17}\cite{AnztCYDFNTTW20}\cite{DBLP:conf/icpads/GaoJLSWS21}\cite{CMRS_Koza_2014}\cite{baskaran2009optimizing}\cite{bell2008efficient}\cite{BellG09}\cite{giles2012efficient}\cite{YoshizawaT12}\cite{GAO2024104799}\cite{DBLP:conf/iccad/DengWM09}\cite{DBLP:journals/integration/HeTZLWS16}\cite{Nickolls_Buck_Garland_Skadron_2008}\cite{Garland_2008} \cite{CMRS_Koza_2014}\cite{SteinbergerZS17}\cite{LiuS15}\cite{liu2018lightspmv}\cite{DBLP:journals/tpds/KarimiADK22} \\ 
            ~ & \cite{DBLP:conf/iccsa/MukunokiT13}\cite{DBLP:conf/sc/Lu023}\cite{DBLP:conf/ppopp/ChoiSV10}\cite{NiuLDJ0T21_TileSpMV}\cite{guo2010auto}\cite{DBLP:conf/asplos/GreweL11}\cite{DBLP:conf/iccS/CevahirNM09}\cite{DBLP:conf/sbac-pad/GuoZ16}\cite{karwacki2012multi}\cite{DBLP:journals/pc/VerschoorJ12}\cite{DBLP:journals/cma/YangLC16}\cite{siamsc/KarypisK98}\cite{DBLP:journals/topc/GaoWWL16}\cite{DBLP:journals/pc/GaoZHX17}\cite{DBLP:journals/concurrency/GaoWW17}\cite{DBLP:journals/tog/Li0TCZM20}\cite{DBLP:journals/corr/abs-2209-07552}\cite{DBLP:conf/ipps/SchaaK09}\cite{DBLP:conf/europar/AbdelfattahLK15}\\ \midrule
            
            FPGA & \cite{DBLP:conf/sasp/ShanWWWWXY10}\cite{DBLP:conf/reconfig/Jain-MendonS12}\cite{DBLP:conf/fccm/KesturDC12}\cite{DBLP:conf/iccd/UmurogluJ14}\cite{DBLP:conf/fccm/FowersOSCS14}\cite{DBLP:conf/ccwc/NaherGDJ20}\cite{DBLP:conf/arc/UmurogluJ15}\cite{DBLP:journals/mam/UmurogluJ16}\cite{DBLP:conf/micro/SadiSLHPF19}\cite{DBLP:journals/tcad/HosseinabadyN20}\cite{DBLP:journals/corr/abs-2107-12371}\cite{DBLP:conf/dac/ParraviciniCSS21}\cite{DBLP:conf/aspdac/LiuL23}\cite{DBLP:conf/ccwc/NaherGDJ20}\cite{DBLP:conf/vdat/MahadurkarSK21}\cite{DBLP:journals/concurrency/NguyenMSDWW22}\cite{mi14112030}
            \\ \midrule
            Processing-in-Memory & 
            \cite{DBLP:conf/hpca/XieLGB0L0021}\cite{DBLP:conf/isca/SunLYWL21}\cite{DBLP:journals/pomacs/GiannoulaFGKGM22}
            \\
            \midrule
            Heterogeneous & 
            \cite{DBLP:conf/ics/LiuSCD13}\cite{DBLP:journals/concurrency/ChenXCLG18}\cite{DBLP:conf/cgo/XieZLGJHZ18}\cite{DBLP:journals/tjs/LiXCLYLGGX20}\cite{DBLP:conf/ics/LiuXLXYL18}\cite{DBLP:journals/isci/ChenXWTL20}\cite{DBLP:journals/iotj/XiaoCLZ20}\cite{DBLP:conf/cgo/TangZLLHLG15}\cite{DBLP:conf/hpcc/ChenXXY19}
            \\ \midrule
            Distributed & 
            \cite{DBLP:conf/3pgcic/MehrezH13}\cite{DBLP:conf/ccgrid/MiYYWL23}\cite{page2021scalability}\cite{DBLP:conf/bigdataconf/MayerMBER18}\cite{LinX17}\cite{DBLP:journals/tjs/KhodjaCGB14}\cite{DBLP:conf/fedcsis/BylinaBSS14}\cite{DBLP:conf/ics/LeeE08}\cite{usman2019zaki}\cite{ma2021developing}
            \\ 
            \bottomrule
        \end{tabular}
    \label{tab:architecture}
\end{table*}

Currently, a subset of existing work chooses the precision according to the value range of non-zeros in sparse matrices. For instance, Ahmad et al. \cite{DBLP:journals/taco/AhmadSH20} introduce a data-driven mixed-precision SpMV algorithm MpSpMV, which opts to represent all non-zeros falling within the interval $(-r, r)$ using single precision, while employing double precision for other non-zeros. MpSpMV partitions the original input matrix into two sparse sub-matrices, consequently requiring modifications to the SpMV algorithm. Mukunoki et al. \cite{DBLP:journals/jcam/MukunokiO20} adopt a similar mix-precision strategy. To enhance this approach, Tezcan et al. \cite{DBLP:conf/sbac-pad/TezcanTKKU22} propose a row-wise partitioned mixed-precision SpMV algorithm. Specifically, this method employs single-precision storage for an entire row if a certain number of non-zeros in that row fall within the range $(-r, r)$; otherwise, double-precision storage is used. This refinement aims to mitigate issues arising from precision conversion.

Another approach in existing literature assesses the necessity of employing low precision in computations based on the impact on the resulting calculation errors. For instance, in \cite{graillat:hal-03561193}, the non-zeros within each row of a sparse matrix are segmented into several groups. Subsequently, they evaluate the weight of each group in the final computation result. If the weight is relatively small, low precision representation may be employed; otherwise, high precision representation is maintained. Isupov et al. \cite{DBLP:journals/jocs/Isupov22} conduct software simulations testing various precision of bit-width ranging from 106 to 848 in different SpMV algorithms. The sparse matrix is consistently represented in double precision, while the multiplication and result vectors are represented in different high-precision. Kouya et al. \cite{DBLP:journals/corr/Kouya14a} design various precision SpMV computations based on the MPFR (a multi-precision floating-point arithmetic library used for correct rounding) and apply them to Krylov subspace-based iterative algorithms.

Compared with mixed-precision SpMV computation based on numerical range, methods assessing the impact of computational errors in mixed-precision SpMV computations yield better results. However, these approaches require more complex preprocessing, introducing substantial additional overhead. A more pressing concern arises in practical applications where the multiplied vector undergoes continual changes. Assessing it before each computation round amplifies the incurred expenses. On the other hand, while mixed-precision SpMV computation based on numerical intervals is intuitive in principle and relatively easy to implement, this method fails to address the mixed-precision representation of the multiplication vector. When the non-zeros of a sparse matrix are represented in low precision and the multiplication vector is directly represented in low precision, it introduces considerable computational errors.

\section{Architecture Oriented Optimization}\label{sec:hardware}
\subsection{Overview}
Various hardware architecture designs have attracted extensive research efforts aimed at optimizing SpMV across diverse platforms. Each hardware platform exhibits unique characteristics that significantly influence the selection and implementation of optimization strategies. In the subsequent sections, we delve into a number of research findings, as depicted in Table \ref{tab:architecture}, concentrating on the optimization of SpMV tailored for these diverse hardware architectures.

\subsection{CPU}
The CPU, being the central component of a computer system, usually has a limited yet highly powerful set of cores and is augmented with a powerful memory and cache system. Moreover, CPUs are known to accommodate a diverse range of SIMD (Single Instruction Multiple Data) instructions, enabling the execution of numerous parallel operations simultaneously. Therefore, research on SpMV in CPU systems primarily emphasizes the optimization of SIMD instructions and cache utilization.

\textbf{SIMD optimization}: To leverage SIMD units in modern multicore CPUs, Pabst et al. \cite{DBLP:conf/ispdc/PabstBK12} propose a structural-detecting algorithm for CSR-based SpMV calculation. It eliminates indirect memory access to multiplication vector when the current row is in a continuous sequence. Other techniques such as CSR5 \cite{CSR5} and CSR2 \cite{DBLP:conf/ccgrid/BianHDLW20} are designed to exploit SIMD units on CPU platforms. Zhang et al. \cite{DBLP:journals/jpdc/ZhangYLTL21} introduce aligned CSR and ELL formats for SpMV, exploiting the capabilities of NEON SIMD engines to achieve enhanced performance.

\textbf{Cache optimization}: 
Cache-blocking \cite{DBLP:journals/ijhpca/ImYV04}\cite{DBLP:journals/aaecc/NishtalaVDY07}\cite{DBLP:conf/sc/WilliamsOVSYD07}\cite{DBLP:journals/pc/WilliamsOVSYD09} partitions the sparse matrix into smaller blocks that can fit in the cache and performs computations on these blocks, which increases temporal locality by reordering memory accesses. Besides, Kislal et al. \cite{kislal2013optimizing} propose an optimization strategy that assigns computation blocks with a significant NNZ in the same column to the same CPU core or cores sharing a high-level cache for data reuse. Other computation blocks are mapped to cores sharing a low-level cache. The scheduling phase finds the order in which these blocks are scheduled to maximize data reuse. It also reorders the multiplied vector $\vect{x}$ to maximize cache utility. 

\textbf{Others}: 
Zhang et al. \cite{DBLP:conf/hpcc/YuanZS08} present a model for the CPU platform to analyze the memory access cost for CSR-based and BCSR-based SpMV. Lee et al. \cite{lee2003performance} design a CSR-based SpMV tailored for symmetric matrices by utilizing register blocking and diagonal block alignment. Liu et al. \cite{DBLP:conf/hpcc/LiuZSQ09} implement row-wise parallelization of CSR-based SpMV and evaluate four scheduling methods in OpenMP (static, dynamic, guided, and non-zeros-based scheduling). Elafrou et al. \cite{DBLP:conf/icpp/ElafrouGK17} utilize a feature-based classifier to identify performance bottlenecks on multi-cores and apply suitable optimization methods to address them. Matrix reordering is a commonly employed technique in sparse matrix operations. Trotter et al \cite{DBLP:conf/sc/TrotterELTDIU23} conduct a thorough and in-depth evaluation of matrix reordering algorithms on various multi-core platforms. Their study compares six reordering algorithms, and four matrix partitioning strategies, including regular 1D, 2D, graph, and hypergraph partitioning. They conclude that graph and hypergraph-based reordering demonstrate superior performance, with the Gray ordering algorithm exhibiting the fastest overall speed. Yu et al. \cite{DBLP:conf/npc/YuMQF020} develop a NUMA-aware SpMV algorithm specifically designed for ARM processors (Phytium 2000+). They reorder sparse matrices using hypergraph partitioning and allocate sub-matrices to different NUMA nodes with as little NUMA node communication volume as possible.

Due to the architectural features of CPUs, current SpMV optimization methods on CPUs primarily focus on leveraging SIMD and cache resources. Regarding SIMD optimization, it is important to note that different hardware vendors, or even CPUs from the same vendor, may provide different support for the width of SIMD lanes. For instance, most ARM processors support a width of 128 bits, while modern Intel processors support up to 256 bits or even 512 bits. Therefore, on different CPU platforms, the parameter configurations of these SpMV algorithms need to be changed accordingly. Cache optimization aims at maximizing the reuse of the multiplication vector $\vect{x}$. In existing work, matrix partitioning and reordering are commonly used. The performance improvements in SpMV resulting from these techniques exhibit a complex relationship with both the CPU hardware architecture and the intrinsic characteristics of the sparse matrix itself.

\subsection{GPU}
Unlike CPUs, GPUs are equipped with a substantial number of smaller cores, allowing for the simultaneous execution of a vast number of computational threads. This makes GPUs more susceptible to load imbalance due to the irregular distribution of non-zeros in sparse matrices. Consequently, load balancing for SpMV on GPUs has garnered significant attention. Moreover, the powerful parallel computing capabilities of GPUs and the inherent discrete memory access patterns in SpMV have made the improvement of memory access efficiency a crucial aspect of optimizing SpMV on GPUs. Therefore, numerous studies focused on techniques that enable coalesced memory access and improve memory locality. Besides, the development of software applications has driven continuous changes in GPU architectures, leading to research efforts tailored to specific architectures. Because of constrained memory footprints and limited computational capabilities in individual GPUs, as well as the increasing computation demands in applications, there is also a body of research focusing on optimizing SpMV on multi-GPU systems.

\subsubsection{Single GPU}

\textbf{Balanced workload}:
The irregular distribution of non-zeros results in a significant imbalance in the computational load. To deal with this, Garland et al. \cite{Nickolls_Buck_Garland_Skadron_2008} re-orchestrate irregular SpMV calculations into regular \textit{map}, \textit{scan}, and \textit{reduce} operations to mitigate load imbalance. Ashari et al. \cite{DBLP:conf/sc/AshariSEPS14} propose ACSR, which combines rows including a similar NNZ into bins and allocates bin-specific kernel to balance loads. Greathouse et al. \cite{greathouse2014efficient}\cite{CSR_adaptive_improved} propose CSR-Stream and CSR-VectorL, to tackle load imbalance issues in SpMV computations. CSR-Stream addresses load imbalances by fixing the NNZ processed by each warp. On the other hand, CSR-VectorL assigns multiple workgroups to handle extremely long rows. Subsequently, Liu et al. \cite{CSR5} present a SpMV algorithm based on the CSR5 format, aiming to achieve strict load balance. Additionally, Merrill et al. \cite{MergedSpMV_SC16} propose a merge-based SpMV algorithm to ensure an optimal balance of computational load. Flegar et al. \cite{DBLP:conf/sc/FlegarA17,AnztCYDFNTTW20} propose a balanced COO-based SpMV for GPU. They partition a sparse matrix into chunks with similar NNZ and assign one warp to process each chunk. Within a warp, if any thread's row index in the current round differs from its row index in the next round, all threads in the warp perform a warp-level segmented reduction. Gao et al. \cite{DBLP:conf/icpads/GaoJLSWS21} propose adaptive multi-row folding and non-zeros-based blocking to relieve the load imbalance in CMRS \cite{CMRS_Koza_2014}. Hola-SpMV \cite{SteinbergerZS17} consists of two kernels: the first one considers global load balancing, and the second one considers local load balancing. 

\textbf{Memory access optimization}: Coalescing global memory accesses on GPU can yield a substantial reduction in memory access overhead. CSR-Scalar is a basic parallel implementation of CSR-based SpMV on GPUs, where each matrix row is assigned to a single thread. However, its performance degrades as the NNZ per row increases due to uncoalesced memory access to column indices and non-zeros. To address the limitations, CSR-Vector \cite{baskaran2009optimizing,bell2008efficient,BellG09} assigns a block of threads ($TpR$) per row to enable coalesced memory access, and the computational results of each thread are stored in shared memory. The setting of $TpR$ is discussed in many works \cite{baskaran2009optimizing}\cite{bell2008efficient}\cite{BellG09}\cite{giles2012efficient}\cite{YoshizawaT12}\cite{GAO2024104799}. At the expense of more memory footprints, the extended vector algorithm \cite{DBLP:conf/iccad/DengWM09}\cite{DBLP:journals/integration/HeTZLWS16} allocates a new vector of the same size as the value array in the CSR format and stores the elements in the multiplied vector $\vect{x}$ into the corresponding extended vector elements, to maximize the coalesced memory access on GPU. Besides, there are some studies that specifically concentrate on improving memory access locality. Garland et al. \cite{Nickolls_Buck_Garland_Skadron_2008}\cite{Garland_2008} employ GPU's shared memory to cache the multiplied vector $\vect{x}$, enhancing memory access efficiency.  Koza et al. \cite{CMRS_Koza_2014} propose to fold a fixed number of adjacent sparse rows into one relatively dense row, which reduces the thread idleness and improves the memory access locality for multiplied vector $\vect{x}$. Karimi et al. \cite{DBLP:journals/tpds/KarimiADK22} present a series of vertical CSR formats to leverage the GPU memory model.

\textbf{Tailored for specific architecture}: Considering the changes from the NVIDIA Fermi architecture to the Kepler architecture, Mukunoki et al. \cite{DBLP:conf/iccsa/MukunokiT13} optimize the performance of CSR-based SpMV by leveraging the new architecture features. This optimization includes loading the multiplication vector using the read-only data cache, utilizing more thread blocks to avoid the use of the outer loop, and employing new shuffle instructions instead of shared memory during the reduction phase. With the emergence of the Volta architecture, NVIDIA GPUs have incorporated high-throughput tensor computing units (TCUs) specifically tailored for accelerating artificial intelligence (AI) applications. This advancement not only enhances AI performance but also opens up new avenues for optimizing scientific computing tasks. DASP, proposed by Lu et al. \cite{DBLP:conf/sc/Lu023}, is a representative work utilizing TCUs. They reorganize sparse matrices into dense structures, allowing for the full utilization of TCUs.

\begin{figure*}[!h]
	\centering
	\includegraphics[width=0.8\textwidth]{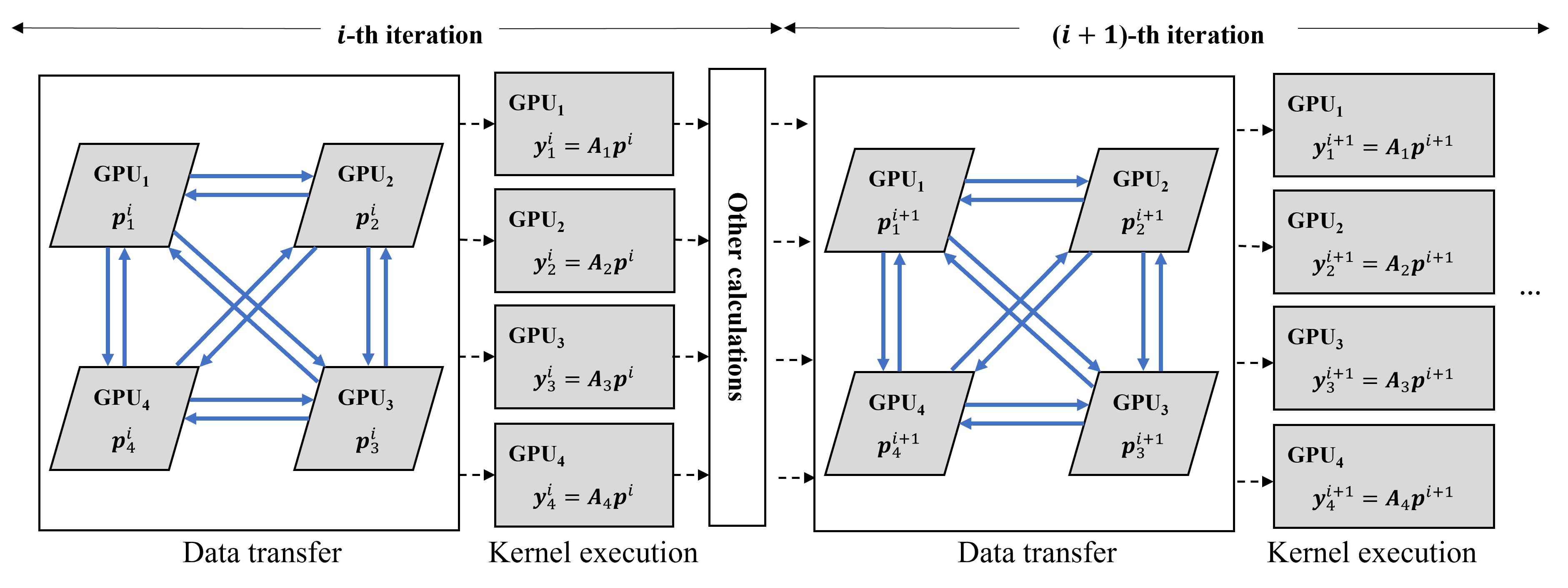} 
	\caption{Workflow of iterative algorithms on multi-GPU systems.}
	\label{fig:multi-GPU}
\end{figure*}

\textbf{Others}: Choi et al. \cite{DBLP:conf/ppopp/ChoiSV10} propose two optimized SpMV implementations, namely BCSR and BELL-based SpMV, tailored for GPU architectures. These implementations leverage CUDA's specialized data types, row alignment techniques, and interleaved memory access to achieve enhanced performance. In \cite{LiuS15}\cite{liu2018lightspmv}, vector-level and warp-level dynamic row distribution based on atomic operations and warp shuffle instructions are presented. Niu et al. \cite{NiuLDJ0T21_TileSpMV} design seven warp-level SpMV algorithms for different sparse formats on GPU. To select the optimal thread configuration for different matrices and GPU architectures, Guo et al. \cite{guo2010auto} propose an auto-tuning framework to set the number of thread blocks, the number of active warps, and the number of threads per block. In addition, Grewe et al. develop a compiler that generates and optimizes SpMV codes on GPUs using an abstract representation of sparse compression formats \cite{DBLP:conf/asplos/GreweL11}. Bell et al. 
\cite{bell2008efficient}\cite{BellG09} conduct a comprehensive evaluation of SpMV algorithms based on various sparse compression formats. Their findings indicate that structural sparse matrices generally exhibit higher SpMV performance compared with nonstructural matrices.

\subsubsection{Multiple GPU}

In contrast to SpMV computation on a single GPU, the primary obstacle encountered in performing SpMV computation on multi-GPU systems pertains to the data synchronization across GPUs. Utilizing the frequently employed iterative algorithm for SpMV invocation as an illustrative example, its workflow on a system comprising four GPUs is depicted in Figure \ref{fig:multi-GPU}. Assuming the original sparse matrix is partitioned into four sub-matrices along the rows, each sub-matrix is transmitted to one of the four GPUs before starting the iteration. During the $i$-th iteration, the vector $\vect{p}^i$, which is essential for SpMV computation, is dispersed across the memory of the four GPUs. Consequently, an all-to-all transfer is imperative to each GPU to hold the complete $\vect{p}^i$, thereby enabling the execution of the SpMV operation. Subsequently, GPU $k$ proceeds with the subsequent other operations based on its sub-vector $\vect{y}^i_k$. Before starting the $(i+1)$-th iteration, GPU $k$ possesses a sub-vector $\vect{p}^{i+1}_k$, thus necessitating all-to-all transfer. In short, the data transfer overhead introduced by multi-GPU computation significantly impacts the overall performance.

Cevahir et al. \cite{DBLP:conf/iccS/CevahirNM09} introduce a multi-GPU optimization method for CG solver. To accelerate the most time-consuming SpMV, they equally assign non-zeros of an input sparse matrix to each GPU. Guo et al. \cite{DBLP:conf/sbac-pad/GuoZ16} use OpenMP to allocate two threads, assigning the ELL part and COO part of HYB storage to two GPUs for parallel processing. The final result reduction is performed on the host side. Additionally, they utilize CUDA streaming for each GPU to achieve overlap between host-to-device transmission, kernel execution, and device-to-host transmission. Karwacki et al. \cite{karwacki2012multi} propose a multi-GPU implementation of a homogenization method for solving Markov models, where SpMV serves as the core calculation. They evenly distribute the ELL and COO parts to two GPUs. Verschoor et al. \cite{DBLP:journals/pc/VerschoorJ12} partition a similar number of 2D blocks to each GPU and sort the sub-matrix assigned to each GPU based on the number of 2D blocks included in each matrix row. Yang et al. \cite{DBLP:journals/cma/YangLC16} direct their attention towards optimizing the GMRES algorithm on a multi-GPU system. They leverage the quasi-optimal partition method \cite{siamsc/KarypisK98} provided by METIS to arrange the non-zeros of a sparse matrix near its main diagonal, aiming to minimize the transmission overhead between GPUs. Gao et al. \cite{DBLP:journals/topc/GaoWWL16}\cite{DBLP:journals/pc/GaoZHX17}\cite{DBLP:journals/concurrency/GaoWW17} establish a relationship between kernel execution time and sparse matrix features based on a manually constructed dataset for the task division of the SpMV on multi-GPU systems. Li et al. \cite{DBLP:journals/tog/Li0TCZM20} partition a matrix into multiple row blocks and ensure that the number of row blocks equals the number of GPU devices. Furthermore, they divide the corresponding row block by column into multiple 2D blocks on each GPU, enabling the overlapping of SpMV calculations for the current block and the transmission of the multiplied vector required by the next block. Chen et al. \cite{DBLP:journals/corr/abs-2209-07552} propose a multi-GPU sparse matrix representation framework called MSREP. This framework employs a fine-grained task division approach to ensure that each GPU deals with a similar NNZ, maintaining a balanced workload across multiple devices. Schaa et al. \cite{DBLP:conf/ipps/SchaaK09} propose performance prediction models for various types of multi-GPU systems. These models enable GPU developers to estimate the performance improvement of their applications on any number of GPUs without the need to invest in hardware or implement a parallel algorithm for simulation. Abdelfattah et al. \cite{DBLP:conf/europar/AbdelfattahLK15} design three kernels for different BCSR block sizes on a multi-GPU system. 

As one of the most popular accelerators in the past decade, SpMV optimization on GPUs has garnered significant attention and research. The rapid evolution of GPU architecture poses challenges to the applicability of optimization algorithms but also provides new optimization opportunities. For instance, recent generations of GPUs have introduced high-throughput TCUs, which have not been fully exploited by most current SpMV algorithms. There exists a natural conflict between the sparsity of sparse matrices and the density-friendly TCUs, necessitating further research and exploration to achieve a friendly match of the two.

\subsection{FPGA}
Field Programmable Gate Array (FPGA) is an electronic device that offers users the flexibility to program and configure specific functionalities according to their requirements. With its inherent programmable nature, FPGA stands out as a prominent technology, renowned not only for its versatility but also for its notable advantages in terms of energy efficiency and low latency. This unique combination of programmability, energy efficiency, and low latency establishes FPGA as a compelling choice for a wide range of applications across various domains.

Shan et al. \cite{DBLP:conf/sasp/ShanWWWWXY10} accelerate SpMV on FPGA by designing a pipelined processing unit and mapping tasks to the memory hierarchy based on the matrix size. Shweta et al. \cite{DBLP:conf/reconfig/Jain-MendonS12} develop an FPGA-based SpMV implementation tailored to their proposed VDCB format. The FPGA design comprises three key components: a customized memory interface, a row-column generator, and a block processing unit. Kestur et al. \cite{DBLP:conf/fccm/KesturDC12} leverage the bit-level manipulation capabilities of FPGAs to compress matrices. Yaman et al. \cite{DBLP:conf/iccd/UmurogluJ14} propose a scalable backend architecture that utilizes column-major traversal and interleaving to achieve substantial bandwidth utilization. Subsequent researchers have contributed to SpMV optimization on FPGA by focusing on memory bandwidth \cite{DBLP:conf/fccm/FowersOSCS14}\cite{DBLP:conf/ccwc/NaherGDJ20}, vector prefetching \cite{DBLP:conf/arc/UmurogluJ15}, memory access \cite{DBLP:journals/mam/UmurogluJ16}, and scalability \cite{DBLP:conf/micro/SadiSLHPF19}. Hosseinabady et al. \cite{DBLP:journals/tcad/HosseinabadyN20} propose a high-performance dataflow engine for SpMV based on FPGA, which combines the concepts of loop pipelining, dataflow graph, and dataflow to harness the majority of available memory bandwidth on FPGA. Oyarzun et al. \cite{DBLP:journals/corr/abs-2107-12371} introduce an FPGA architecture that maximizes vector reuse by introducing a cache-like architecture to address challenges in computational fluid dynamics. Parravicini et al. \cite{DBLP:conf/dac/ParraviciniCSS21} maximize bandwidth efficiency while maintaining the accuracy of Top-K SpMV, widely applied in information retrieval and recommendation systems, by utilizing streaming sparse matrix format, leveraging HBM access, and using low-precision numerical values. Liu et al. \cite{DBLP:conf/aspdac/LiuL23} present a high-bandwidth utilization SpMV accelerator using partial vector replication on FPGA, with detailed descriptions of conflict-free vector buffer for reads, conflict-free addition tree for writes, and ping-pong-like accumulation registers. Du et al. develop \cite{DBLP:journals/concurrency/NguyenMSDWW22} a high performance SpMV accelerator, named HiSparse, on HBM-equipped FPGAs. Favaro et al. \cite{mi14112030} use the roofline model to guide SpMV optimization on FPGA.

To investigate the performance impact of different parameter settings, logic synthesis is necessary after each parameter adjustment to evaluate their effects on SpMV performance. However, logic synthesis can be time-consuming, taking several hours for each synthesis operation, which hinders the process of finding the optimal parameter settings. Therefore, Naher et al. \cite{DBLP:conf/ccwc/NaherGDJ20} employ the random forest method to estimate FPGA utilization and throughput under given design parameters, significantly reducing the time overhead associated with parameter tuning. 

Deep learning algorithms require significant computational resources, making it challenging to implement them on memory and power-constrained devices. Therefore, to improve the efficiency of deep learning inference, Mahadurkar et al. \cite{DBLP:conf/vdat/MahadurkarSK21} introduce an FPGA-specific SpMV multiplier architecture. This architecture utilizes multiple MAC channels to leverage the maximum available memory bandwidth of the computing device, thereby reducing the memory bandwidth bottleneck in hardware implementation. 

The energy efficiency of computing systems is a critical consideration, particularly in resource-constrained environments where power constraints are paramount. CPUs and GPUs, with their general-purpose designs and high-performance capabilities, often exhibit higher power consumption and may not be well-suited for applications where energy efficiency is a primary concern. In contrast, FPGAs offer distinct advantages due to their inherent parallelism, reconfigurability, and ability to tailor hardware architectures to match specific computation requirements.

\subsection{Processing in Memory}
Given the memory-bound characteristics of SpMV in processor-centric systems, researchers have focused on optimizing SpMV specifically for processing-in-memory (PIM) systems. PIM architectures aim to alleviate the performance bottleneck caused by data movement between the processor and memory by integrating processing units directly into the memory subsystem. This integration allows for efficient and parallel execution of computations on the data stored in memory, thereby reducing the data transfer overhead.

Xie et al. \cite{DBLP:conf/hpca/XieLGB0L0021} propose an accelerator, referred to as SpaceA. From a hardware perspective, SpaceA leverages exceptional memory request capabilities to reduce memory access latency for data residing in non-local memory banks. It also integrates Content Addressable Memory (CAM) at the bank level to enable data reuse of the input vector. From a software perspective, they design a mapping scheme for sparse matrices to achieve workload balance and exploit the locality of the multiplied vector. A key optimization idea in the mapping scheme is the assignment of rows with similar column index patterns to the same processing element (PE). Considering that memory bus-based communication poses a significant challenge for data-intensive applications in DIMM-based near-memory Processing (NMP) systems, Sun et al. \cite{DBLP:conf/isca/SunLYWL21} propose ABC-DIMM, a software-hardware co-optimization approach. ABC-DIMM incorporates integral broadcast mechanisms for both intra- and inter-channel settings, along with a broadcast-process programming framework. Their performance evaluation in SpMV demonstrates significant performance improvement. In the realm of software optimization, Giannoula et al. \cite{DBLP:journals/pomacs/GiannoulaFGKGM22} introduce SparseP, a SpMV library designed for real PIM systems. SparseP has 25 SpMV kernels and supports popular compression formats, partitioning methods across various parallel granularities, data types, load-balancing techniques, and synchronization methods.

In summary, recent research has focused on optimizing SpMV in PIM systems to overcome the memory-bound nature of this operation. These advancements in both hardware and software optimization techniques contribute to enhancing the performance and efficiency of SpMV computations in PIM systems.

\subsection{Heterogeneous Platform}
Heterogeneous platforms typically consist of processors featuring multiple architectures, enabling the utilization of diverse processors to maximize the performance of computing systems. In accordance with heterogeneous platforms, the existing research in this field can be categorized into three main classes: CPU+GPU, Intel Xeon Phi, and Sunway architecture.

\textbf{CPU+GPU}: Indarapu et al. \cite{DBLP:conf/compute/IndarapuMK14} propose a static partition algorithm for SpMV computation, dividing the task into two parts and assigning them to CPU and GPU for parallel processing. Valeria et al. \cite{DBLP:conf/parco/CardelliniFF13} propose a static load-balancing algorithm. They first test the benchmark data of SpMV performance and GPU bandwidth and then establish a regression model based on the results to allocate computing tasks between CPU and GPU. Building upon this work, Yang et al. propose two partition methods to guide the task division between CPU and GPU. The first method is based on a probability model \cite{DBLP:journals/tc/YangLML15}, while the second method utilizes a distribution function \cite{DBLP:journals/jpdc/YangLL17}. Benatia et al. \cite{Akrem2020} utilize one-dimensional partitioning for the sparse matrix, then employ an ML model to select the optimal compression format for each partition, and finally determine the mapping from tasks to computing units (CPU and GPU) using a min-min strategy \cite{DBLP:journals/jpdc/BraunSBBMRRTYHF01}. Liu et al. \cite{DBLP:journals/pc/0002V15} employ speculative segmented sum execution on the GPU, which may yield incorrect results initially. Subsequently, the CPU rearranges the calculated results to obtain the accurate resulting vector.

\textbf{Intel Xeon Phi}:
The Xeon Phi series, developed by Intel Corporation, are many-core processors with over 50 cores, each supporting four threads. Among the prominent features of Intel Xeon Phi processors is the vector processing unit (VPU). The VPU module supports a 512-bit SIMD instruction set, officially referred to as Intel Initial Many-Core Instructions (Intel IMCI), which facilitates efficient handling of irregular memory accesses. Liu et al. \cite{DBLP:conf/ics/LiuSCD13} use the gather instruction to load elements from discrete memory locations into dense vector registers and employ the scatter instruction to achieve the reverse operation. They introduce the ESB (ELLPACK Sparse Block) format, a sparse compression scheme that utilizes a limited window sorting technique to enhance the vectorization efficiency of the SpMV kernel. Additionally, they employ a bit array to track the positions of non-zeros and use column-blocking technology to improve memory access locality. To harness the vectorization capabilities of Xeon Phi, Chen et al. \cite{DBLP:journals/concurrency/ChenXCLG18} compress non-zeros into variable-length data segments based on CSR and distribute them in memory with a contiguous address space. Differently, Tang et al. \cite{DBLP:conf/cgo/TangZLLHLG15} combine the advantages of COO and CSR and propose a new vectorized format. Subsequently, Xie et al. propose the CVR (Compressed Vectorization-oriented Sparse Row) format \cite{DBLP:conf/cgo/XieZLGJHZ18}, followed by the VBSF (Variable Blocked-$\sigma$-SIMD Format) \cite{DBLP:journals/tjs/LiXCLYLGGX20}, both of which exploit vectorization optimization techniques.

\textbf{Sunway Architecture}: 
The Sunway TaihuLight supercomputer achieved consecutive top rankings on the TOP500 List\footnote{https://www.top500.org/lists/top500} in 2016-2017. It is powered by the SW26010 many-core processors, which have four management processing elements (MPE) and 64 computing processor elements (CPE). Xue et al. \cite{DBLP:conf/ics/LiuXLXYL18} propose a dual-side partition method for sparse matrices and hardware resources. They divide the input sparse matrix into blocks, tiles, and slices at three levels. The 64 CPEs on the Sunway processor are partitioned into two levels. In the first level, they are divided horizontally into 8 groups, each containing 8 CPEs, while the second level assigns 7 CPEs within each group for computation and 1 CPE for I/O. In the first level, each group independently fetches blocks, and in the second level, each computing core fetches tiles. In the third level, each computing core processes slices independently to facilitate vectorization. Based on the HYB format, hpSpMV algorithm \cite{DBLP:conf/hpcc/ChenXXY19} assigns the ELL part to CPEs and the COO part to MPEs. Chen et al. \cite{DBLP:journals/isci/ChenXWTL20} propose a two-stage large-scale SpMV framework called tpSpMV, designed for many-core architectures and evaluated on the SW26010 CPUs. TpSpMV divides the CSR-based SpMV into two parallel execution stages: parallel partial SpMV computing and parallel accumulation. Adaptive partitioning and local caching techniques are employed in both stages to leverage the Sunway platform's architecture and reduce the delay caused by high memory access. Furthermore, techniques such as data reduction, aligned memory access, and pipelining are utilized to enhance bandwidth utilization. Xiao et al. \cite{DBLP:journals/iotj/XiaoCLZ20} present a heterogeneous work partitioning based on hybrid COO and ELL formats. A sparse matrix is first partitioned into multiple sub-matrices, and COO and ELL parts of each sub-matrix are assigned to MPE and CPE in each CG respectively. Xiao et al. \cite{DBLP:journals/tpds/XiaoLCHZL21} introduce a customized and accelerated SpMV (CASpMV) framework specifically designed for Sunway processors. Their framework uses a four-way partition method that takes into account the multi-level parallel computing architecture. Notably, it partitions long rows into shorter rows to ensure they fit within the local data memory (LDM) of each CPE core, also achieving workload balance across CPE cores. Additionally, they present a statistical model that serves as the foundation for their proposed partition auto-tuner.

\textbf{Others}: Li et al. \cite{li2023haspmv} present a heterogeneity-aware SpMV (HASpMV) algorithm designed specifically for asymmetric multicore processors (AMPs). Their approach to task assignment is guided by micro-benchmarking of bandwidth and multi-/single-core SpMV performance.

To match the requirements of software applications, diverse hardware architectures have been designed, leading to a variety of heterogeneous platforms. The design of SpMV algorithms on heterogeneous platforms must consider the characteristics of the underlying hardware and the cost associated. For instance, on a CPU+GPU heterogeneous platform, algorithm design must account for the communication overhead between the CPU and GPU, as well as the disparity in computational capabilities between the two.

\subsection{Distributed Platform}
Distributed platforms enable larger-scale SpMV computations, but careful consideration must be given to inter-node communication. Current research efforts primarily focus on reducing overall communication costs through specific matrix partitioning or reordering techniques, as well as overlapping computation and communication processes. Moreover, due to the irregular non-zeros distribution of sparse matrices, some studies have explored dynamic adaptive techniques to optimize SpMV on distributed systems.

\begin{table*}[!htbp]
    \centering
    \caption{Hardware and software of two GPU platforms.}
    \begin{tabular}{ccll}
    \toprule
        & & \textbf{Platform 1} & \textbf{Platform 2} \\ \midrule
        \multirow{4}{*}{\centering \textbf{Hardware}} & 
        \multirow{2}{*}{\centering Host} & 
            \textbf{CPU}: Intel Xeon E5-2680 v4, 2.5 GHz, 14 cores &
            \textbf{CPU}: Intel Xeon Platinum 8336C, 2.3 GHz, 32 cores \\
        & & \textbf{Memory}: 128 GB  & 
            \textbf{Memory}: 512 GB \\ 
        \cmidrule{2-4}
        & \multirow{2}{*}{Device} &
            \textbf{GPU}: Tesla V100-SXM2, 1.53 GHz, 5,120 cores  &
            \textbf{GPU}: Tesla A100-PCIE, 1.41 GHz, 6,912 cores \\
        & & \textbf{Memory}: 16 GB &
            \textbf{Memory}: 40 GB \\ 
        \midrule
        \multirow{3}{*}{\centering \textbf{Software}} & 
        OS & 64-bit Ubuntu 18.04 & CentOS 7.9.2009 \\ 
        & Compiler &  
        \textbf{nvcc}: 12.0; \textbf{gcc/g++}: 8.4.0 &
        \textbf{nvcc}: 12.0; \textbf{gcc/g++}: 9.3.1 \\ 
        & Library & \multicolumn{2}{l}{cuSPARSE: 12.0; CUSP: 0.5.1; MAGMA: 2.7.2; Ginkgo: 1.7.0} \\ \bottomrule
    \end{tabular}
     \label{tab:environment-setup}
\end{table*}
\begin{table*}[!htbp]
    \centering
    \caption{Detailed information of existing libraries.}
    \begin{tabular}{ccccc}
    \toprule
        \textbf{Library} & \textbf{Sparse formats/SpMV algorithm} & \textbf{Architecture} & \textbf{Open source} & \textbf{Last update} \\ \midrule
        cuSPARSE & COO,CSR,CSC,SELL & GPU & No & Nov-23 \\
        Ginkgo & COO,CSR,ELL,SELL,HYB & CPU/GPU & Yes & Nov-23 \\ 
        ViennaCL & COO,CSR,ELL,HYB,SELL & CPU/GPU & Yes & Jul-19 \\
        MAGMA & CSR,ELL,SELLP & CPU/GPU & Yes & Aug-23 \\
        Merge-SpMV & Merge-based SpMV & CPU/GPU & Yes & Feb-17 \\ 
        Intel MKL & COO,CSR,BSR,DIA & CPU & No & Jul-23 \\ 
        CUSP & COO,CSR,DIA,ELL,HYB & CPU/GPU & Yes & Apr-15 \\ 
        yaSpMV & BCCOO/BCCOO+ & GPU & Yes & Nov-14 \\ 
        CSR5 & CSR5 & CPU/GPU & Yes & Dec-20 \\
        hipSparse & CSR,BSR,HYB & GPU & No & Oct-23 \\ \bottomrule
    \end{tabular}
    \label{tab:spmv-library}
\end{table*}

\textbf{Matrix partitioning and reordering}: Matrix partitioning is closely related to the inter-node communication overhead. There are some studies utilizing graph and hypergraph partitioning to improve locality and minimize the amount of data communication. For instance, Mehrez et al. \cite{DBLP:conf/3pgcic/MehrezH13} propose a combination of the 1D/2D hypergraph model and the S-GBNZ (Sorted Generalized Fragmentation with Balanced Number of Non-zeros) algorithm to enhance the performance of distributed SpMV. Their findings suggest employing 1D hypergraph partitioning for inter-node decomposition and the S-GBNZ algorithm for intra-node decomposition. Mi et al. \cite{DBLP:conf/ccgrid/MiYYWL23} employ graph partitioning to reorder the sparse matrix and then resize sub-matrices stored locally to balance computation and communication among computation nodes. Page et al. \cite{page2021scalability} leverage hypergraph partitioners from the HYPRE package \cite{DBLP:conf/bigdataconf/MayerMBER18} to mitigate communication overhead. However, due to limitations in current graph partitioners when handling graphs with highly imbalanced edge distribution, they employ an additional optimization method called vertex delegation to eliminate vertices with exceptionally high degrees. Besides, Lin et al. \cite{LinX17} propose an efficient reordering technique to reduce inter-node communication of the Jacobi preconditioned CG method (Jacobi\_PCG) on a multi-GPU cluster. Khodja et al. \cite{DBLP:journals/tjs/KhodjaCGB14} propose two methods to reduce communication overhead: the first method uses compressed storage to store the exchanged vector, while the second method employs hypergraph partitioning. Different from these studies, Bilina et al. \cite{DBLP:conf/fedcsis/BylinaBSS14} partition SpMV into a 2D task grid of $p\times p$ on a distributed system comprising $p^2$ CPUs. To ensure that each process has the necessary multiplication vector, they initially send the data in the first column of the grid to the processes situated on the main diagonal of the 2D grid. Subsequently, these processes perform a column-wise broadcast of the data. 

\textbf{Runtime adaptive techniques}: Instead of static partitioning, Lee et al. \cite{DBLP:conf/ics/LeeE08} utilize a dynamic and adaptive runtime tuning approach on distributed systems. This approach has two steps: the first step dynamically reassigns matrix rows to processes to achieve a balanced computational load, while the second step determines the optimal communication method based on the profiling generated in the first step. Usman et al. \cite{usman2019zaki}\cite{DBLP:journals/access/UsmanMKA19} introduce ZAKI/ZAKI+, a data-driven and machine learning-based tool, to predict the optimal number of processes and optimal mapping of the processes and data on distributed-memory architectures.

\textbf{Overlap of communication and calculation}: Ma et al. \cite{ma2021developing} develop a distributed block SpMV algorithm that employs consecutive block row partitioning for a block sparse matrix. Additionally, an MPI process is utilized to assign one partition to one GPU. Each partition is stored in two distinct parts: the main diagonal part (MDP) and the off-diagonal part (ODP). Notably, MDP does not require communication, whereas ODP does. This separate storage scheme enables the overlap of the calculation process for MDP and the communication process for ODP.

In summary, efforts to optimize distributed SpMV computations are aimed at addressing inter-node communication challenges. Hypergraph partitioning, two-dimensional processor grid-based partitioning, and computation and communication overlapping are the most commonly used optimization techniques. By reducing communication costs and improving the overall efficiency of computation and communication, these strategies contribute to the optimization of SpMV on distributed systems.

\section{Performance Evaluation}\label{sec:performance}
\subsection{Experimental Setup}
In this section, we evaluate the performance of several widely used SpMV implementations on two GPU platforms. The hardware and software specifications of these platforms can be found in Table \ref{tab:environment-setup}. We use the well-known dataset SuiteSparse Matrix Collection that has more than 2,800 sparse matrices \cite{Davis2011_SuiteSparse}. The storage and computation of floating-point values are performed in double precision.

As presented in Table \ref{tab:spmv-library}, we summarize detailed information of existing libraries that support SpMV operations, including the supported sparse formats, algorithms, architectures,  accessibility, and the date of its last update. 

We conducted an evaluation of various algorithms available in mainstream sparse libraries, including cuSPARSE \cite{cusparse}, CUSP \cite{Cusp}, MAGMA \cite{Magma-sparse1,Magma-sparse2}, and Ginkgo \cite{anzt2020ginkgo}, as well as two popular SpMV algorithms: CSR5-based SpMV \cite{CSR5} and Merge-based SpMV \cite{merge-spmv,MergedSpMV_SC16}. We test the performance of cuSPARSE by calling \textit{cusparseSpMV}, with the sparse matrices stored in COO or CSR formats. In the case of COO storage, we test \textit{CUSPARSE\_SPMV\_COO\_ALG1} and \textit{CUSPARSE\_SPMV\_COO\_ARG2}, referred to as COO1 and COO2, respectively. For CSR encoding, we evaluated  \textit{CUSPARSE\_SPMV\_CSR\_ALG1} and \textit{CUSPARSE\_SPMV\_CSR2\_ARG2}, denoted as CSR1 and CSR2, respectively. Five SpMV algorithms based on COO, CSR, ELL, DIA, and HYB formats in CUSP library were tested. Additionally, we evaluated three SpMV algorithms in the library MAGMA, employing the CSR, ELL, and SELLP formats, and four SpMV algorithms in the library Ginkgo, employing the COO, CSR, ELL, and HYB formats. 

\begin{figure*}[htbp]
	\centering
	\subfloat[COO1 on V100] 
	{	\includegraphics[width=0.24\linewidth]{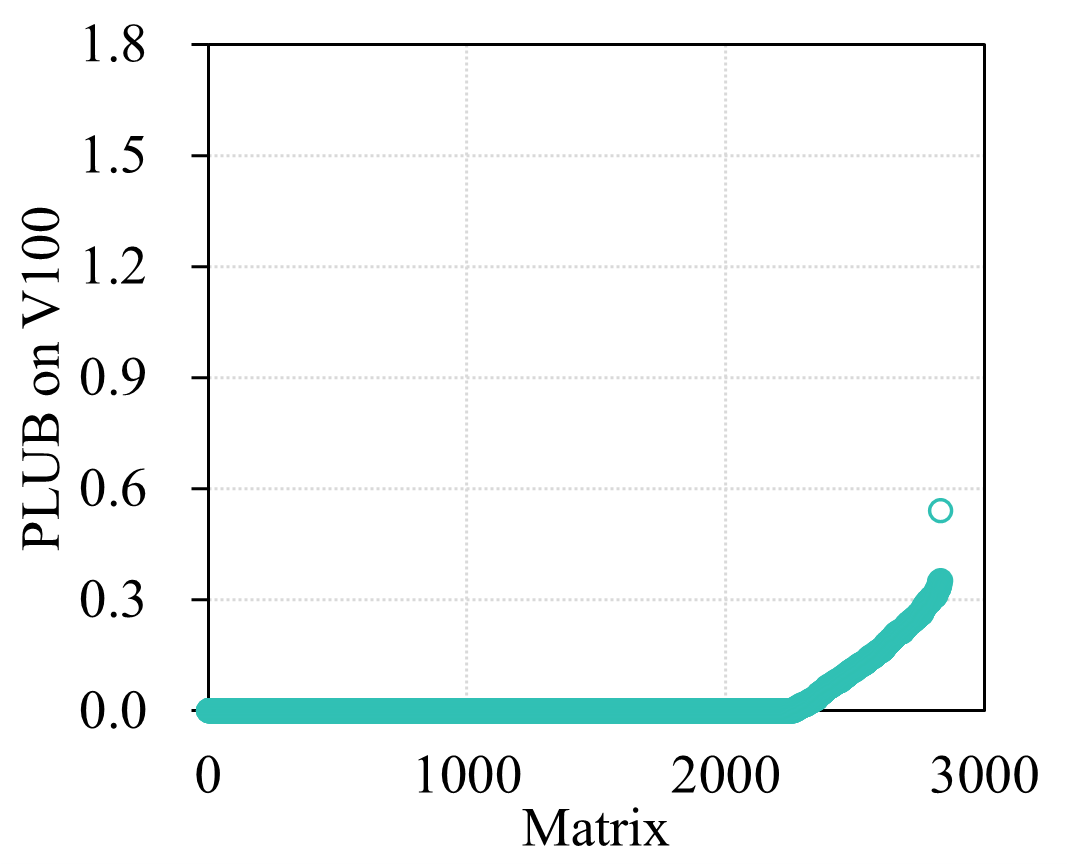}
	}
	\subfloat[COO2 on V100] {
		\includegraphics[width=0.24\linewidth]{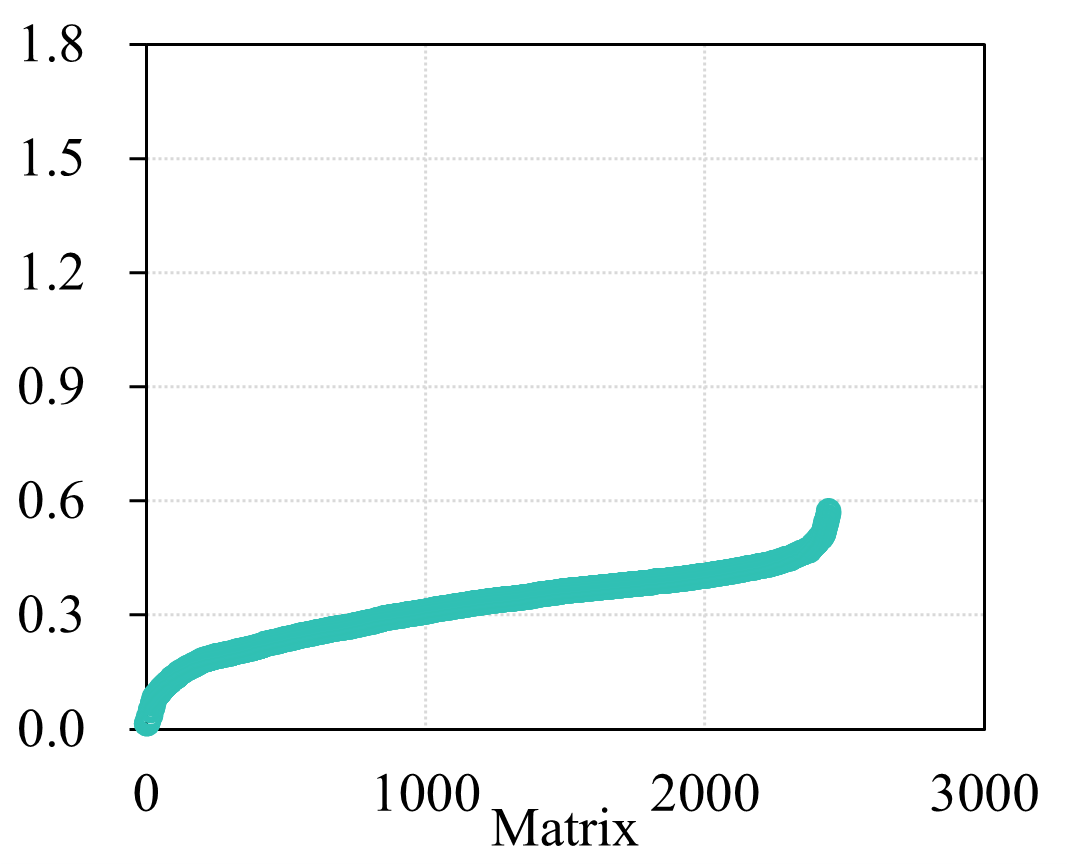}
	}
        \subfloat[CSR1 on V100] {
		\includegraphics[width=0.24\linewidth]{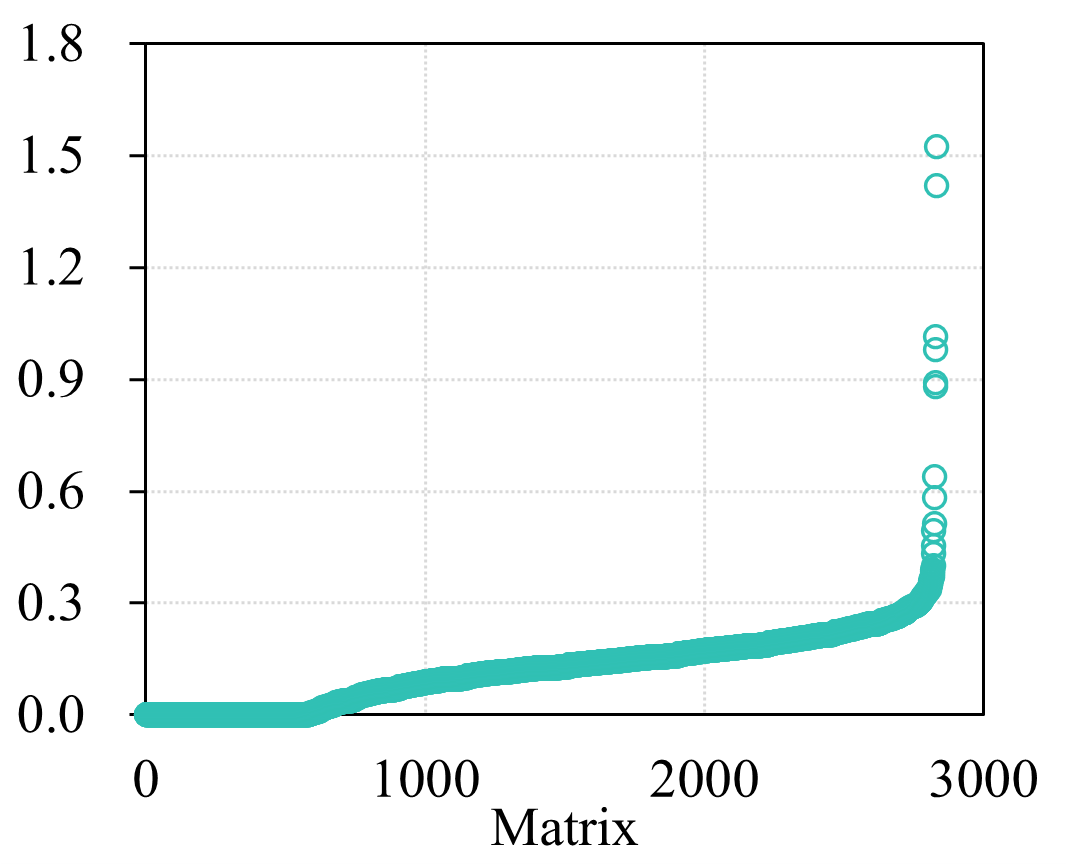}
	}
        \subfloat[CSR2 on V100] {
		\includegraphics[width=0.24\linewidth]{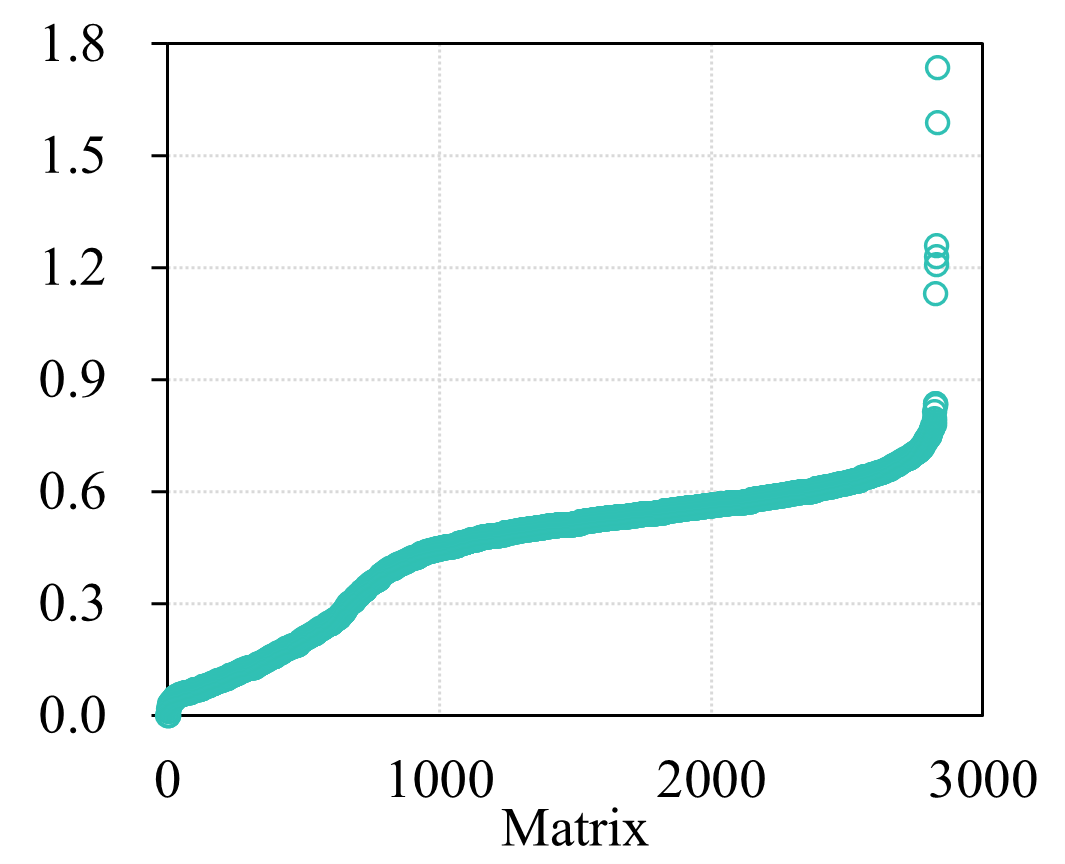}
	}
        \hfill
        \subfloat[COO1 on A100] 
	{	\includegraphics[width=0.24\linewidth]               {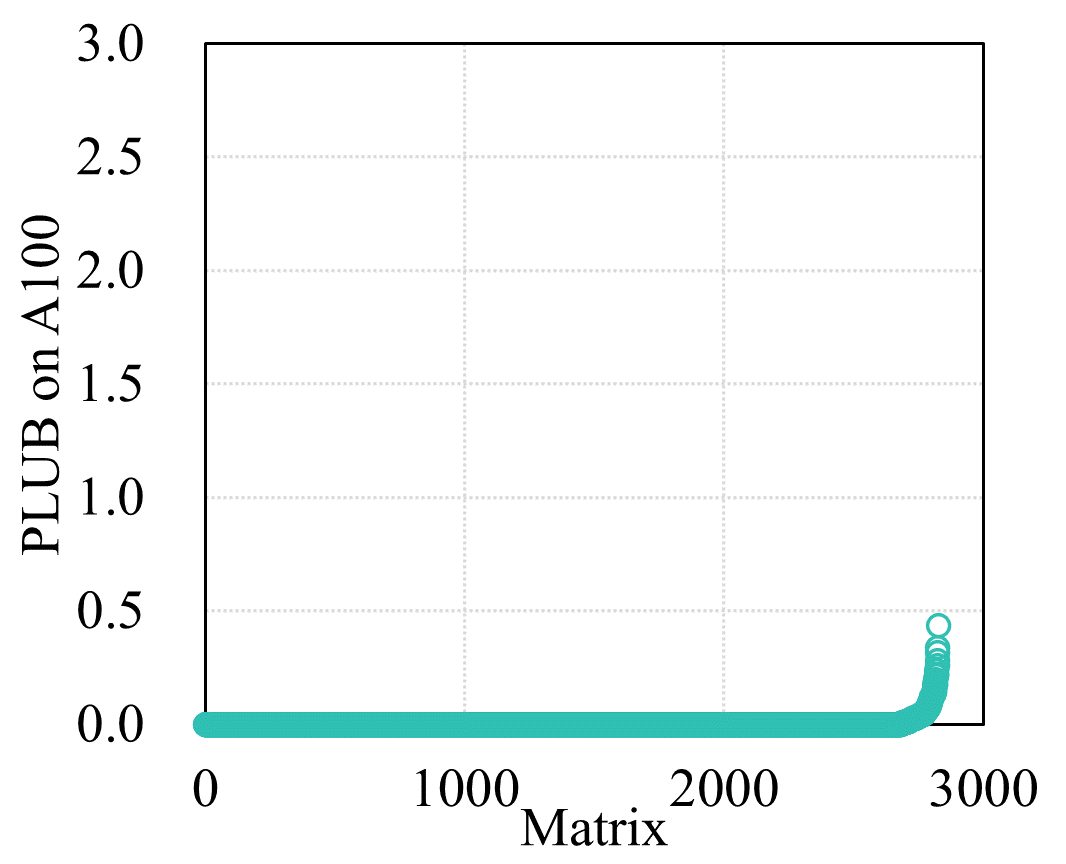}
	}
	\subfloat[COO2 on A100] {
		\includegraphics[width=0.24\linewidth]{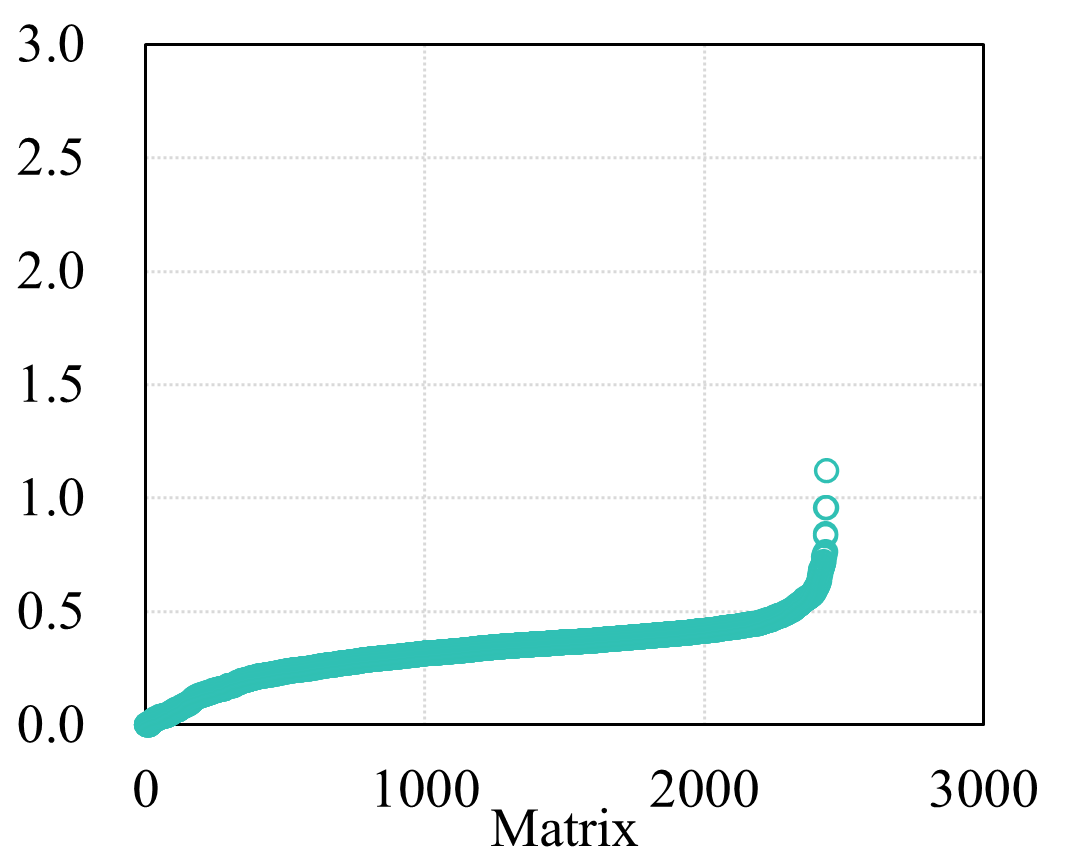}
	}
        \subfloat[CSR1 on A100] {
		\includegraphics[width=0.24\linewidth]{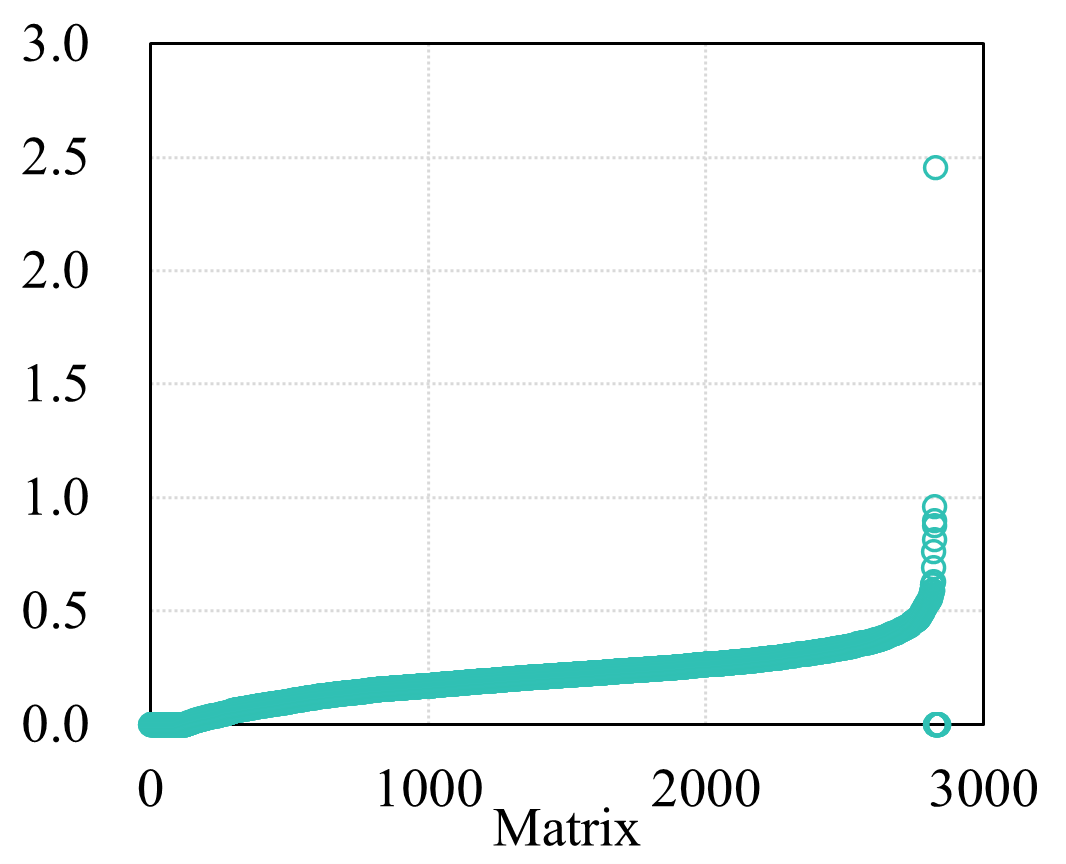}
	}
        \subfloat[CSR2 on A100] {
		\includegraphics[width=0.24\linewidth]{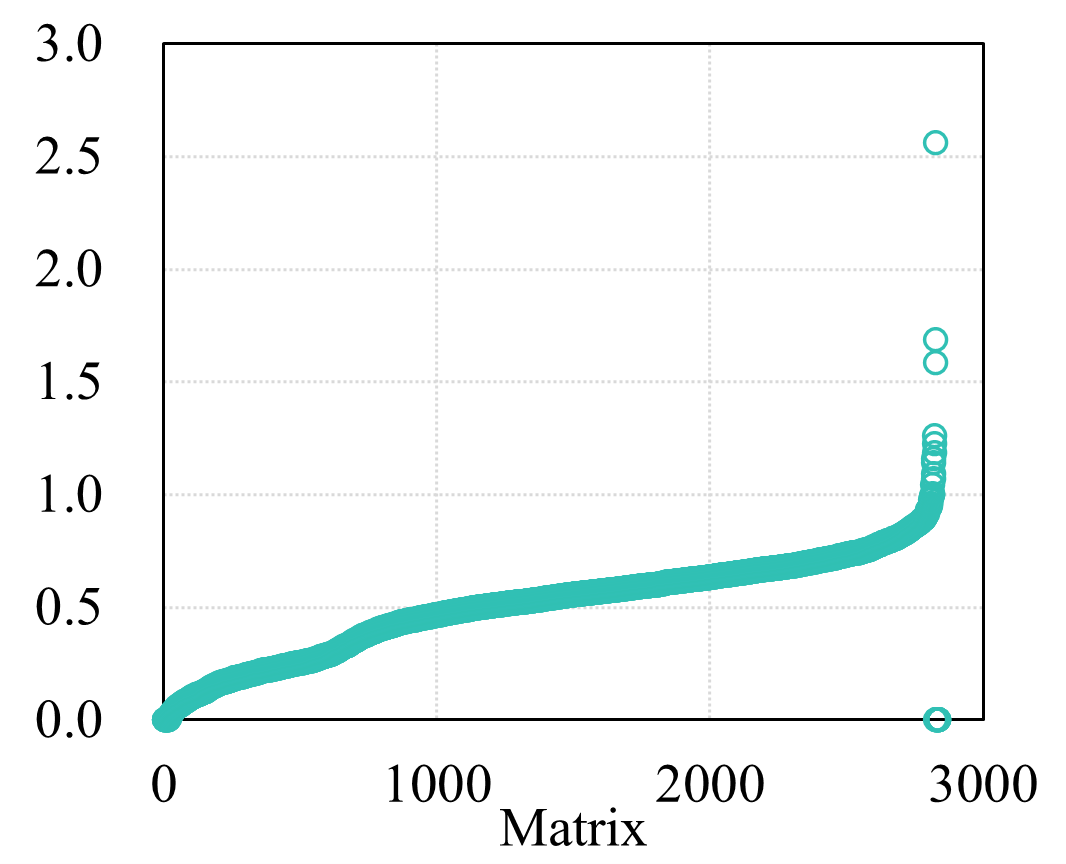}
	}
	\caption{PLUB of each SpMV algorithm with respect to the best-performing SpMV algorithm in cuSPARSE for each sparse matrix. A smaller PLUB indicates better performance.}
	\label{fig:cusparse-PLUB}
\end{figure*}

In this section, we use several important metrics for comparing different algorithms:

The first metric is the performance loss of an algorithm under the best-performing algorithm, referred to as PLUB. Equation \ref{equ:PLUB} illustrates its calculation formula, where $t_k^i$ represents the running time of algorithm $k$ for sparse matrix $i$, and $t_s^i$ denotes the shortest running time among several algorithms for sparse matrix $i$. For instance, in the case of cuSPARSE, the calculation of $t_s^i$ is performed according to Equation \ref{equ:t_s}. PLUB serves as a measure to assess the disparity between a given algorithm and the optimal algorithm. In the subsequent performance evaluation, we primarily use this metric to select the SpMV algorithm with the best overall performance in each sparse library. 
\begin{equation}\label{equ:PLUB}
    PLUB_k^i = \frac{t_{k}^i-t_s^i}{t_s^i}.
\end{equation}
\begin{equation}\label{equ:t_s}
    t_s^i = min\{t_{COO1}^i, t_{COO2}^i, t_{CSR1}^i, t_{CSR2}^i\}
\end{equation}

In addition to PLUB, we also compare the number of sparse matrices for which an algorithm takes the shortest running time, i.e. achieves the highest floating-point operations per second (FLOPS), referred to as $N_h$. This metric provides insights into the algorithm's proficiency in handling various matrix types.

Thirdly, we compare the number of sparse matrices that an algorithm successfully processes, denoted as $N_s$, indicating its overall availability in various scenarios. 

Another important metric is the GFLOPS (Giga Floating-point Operations Per Second) achieved by each algorithm. GFLOPS of algorithm $k$ for sparse matrix $i$, denoted as ${GFLOPS}_k^i$, is calculated as Equation \ref{equ:gflops}. It provides insights into the computational efficiency of an algorithm.

\begin{equation}\label{equ:gflops}
    GFLOPS_k^i = \frac{nnz\times 2}{10^9\times t_k^i}
\end{equation}

The final commonly used evaluation metric is the speedup. Let algorithm $b$ be considered as the baseline, then the speedup of algorithm $k$ over algorithm $b$ for sparse matrix $i$ is calculated as the ratio of $GFLOPS_k^i$ to $GFLOPS_b^i$.

In the subsequent evaluation, we compare the PLUB of various SpMV algorithms within each sparse library. We identify the algorithm with the lowest average PLUB as the representative of that library. Subsequently, we compare the GFLOPS achieved by these representatives with the other two SpMV algorithms (CSR5-based SpMV and Merged-based SpMV).

\begin{table}[!htbp]
    \centering
    \caption{Overall performance comparison of different SpMV algorithms in cuSPARSE.}
    \begin{tabular}{ccrrrr}
    \toprule
        \multicolumn{2}{c}{\textbf{Algorithm}} & \textbf{COO1} & \textbf{COO2} & \textbf{CSR1} & \textbf{CSR2} \\ \midrule
        \multirow{2}{*}{PLUB(\%)} & V100 & \textbf{2.94} & 31.99 & 12.27 & 44.64 \\ 
        ~ & A100 & \textbf{0.33} & 32.41 & 21.22 & 50.66 \\ \midrule
        \multirow{2}{*}{$N_h$} & V100 & \textbf{2,258} & 0 & 574 & 3 \\
        ~ & A100 & \textbf{2,679} & 11 & 125 & 12 \\ \midrule
         \multirow{2}{*}{$N_s$} & V100 & 2,831 & 2,443 & \textbf{2,832} & \textbf{2,832} \\
        ~ & A100 & \textbf{2,826} & 2,438 & \textbf{2,826} & \textbf{2,826} \\ 
        \bottomrule
    \end{tabular}
    \label{tab:cusparse-PLUB}
\end{table}

\begin{table}[!ht]
    \centering
    \caption{Overall performance comparison of different SpMV algorithms in CUSP.}
    \begin{tabular}{ccrrrrr}
    \toprule
        \multicolumn{2}{c}{\textbf{Algorithm}} & \textbf{COO} & \textbf{CSR} & \textbf{ELL} & \textbf{DIA} &\textbf{ }HYB \\ \midrule
        \multirow{2}{*}{PLUB(\%)} & V100 & 416 & \textbf{58} & 61 & 252 & 406  \\ 
        ~ & A100 & 365 & 207 & \textbf{60} & 212 & 348 \\ \midrule
        \multirow{2}{*}{$N_h$} & V100 & 70 & \textbf{2,159} & 479 & 66 & 79 \\
        ~ & A100 & 78 & \textbf{1,942} & 665 & 77 & 86 \\ \midrule
         \multirow{2}{*}{$N_s$} & V100 & \textbf{2,829} & \textbf{2,829} & 1,971 & 953 & 2,828 \\
        ~ & A100 & \textbf{2,823} & \textbf{2,823} & 1,966 & 953 & 2,821 \\ 
        \bottomrule
    \end{tabular}
    \label{tab:cusp-PLUB}
\end{table}

\begin{figure*}[!htbp]
	\centering
	\subfloat[COO] 
	{	\includegraphics[width=0.19\linewidth]{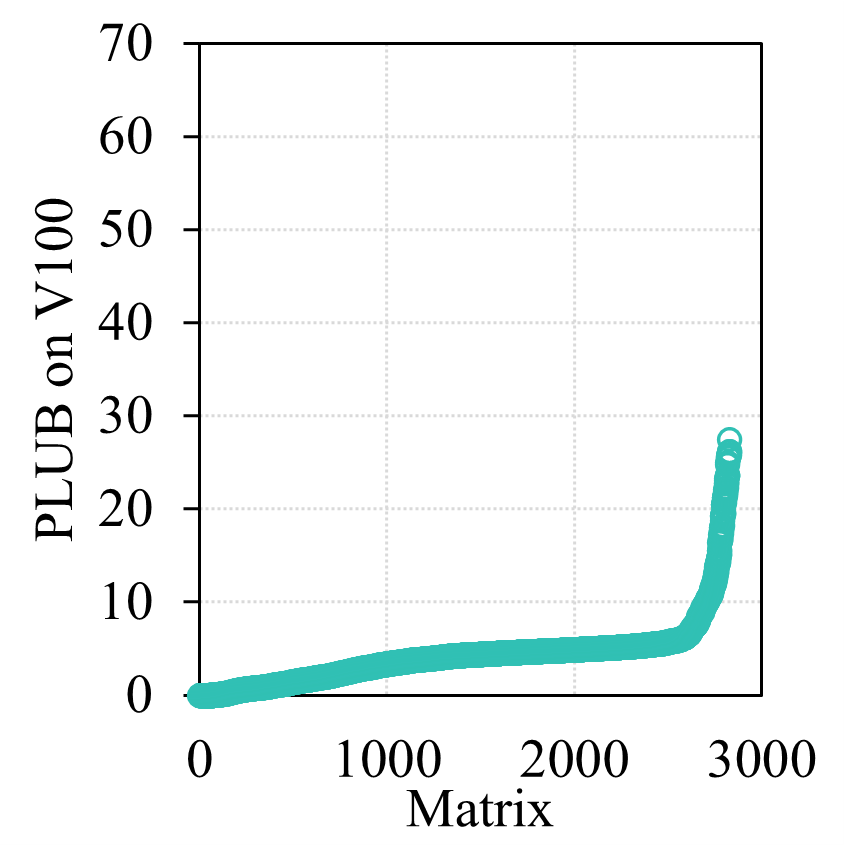}
	}
	\subfloat[CSR] {
		\includegraphics[width=0.19\linewidth]{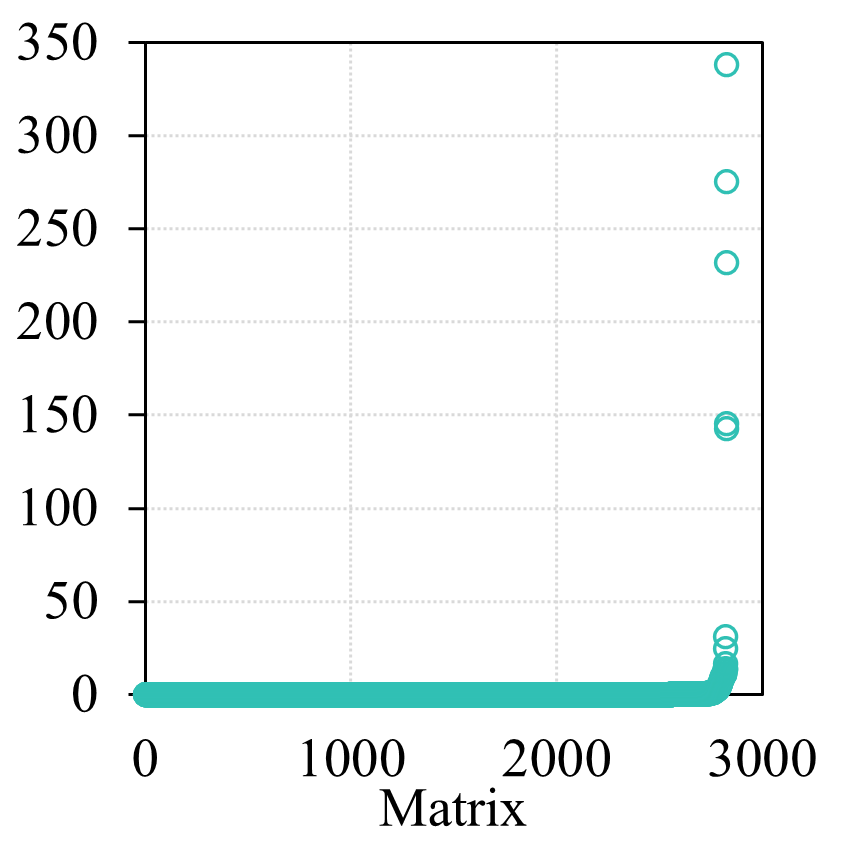}
	}
        \subfloat[ELL] {
		\includegraphics[width=0.19\linewidth]{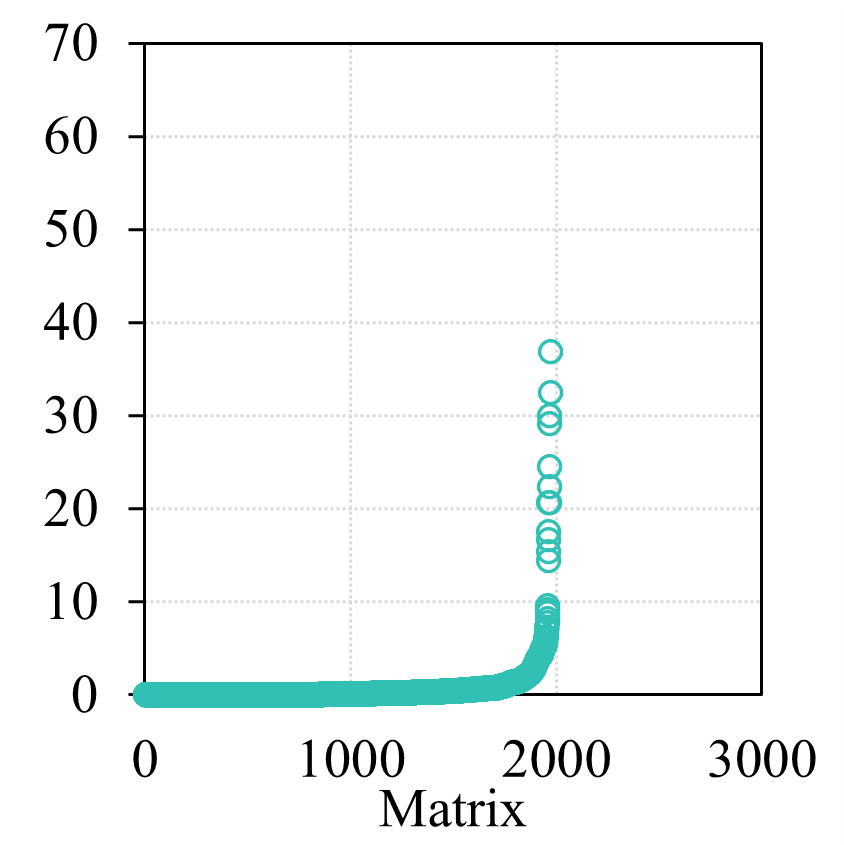}
	}
        \subfloat[DIA] {
		\includegraphics[width=0.19\linewidth]{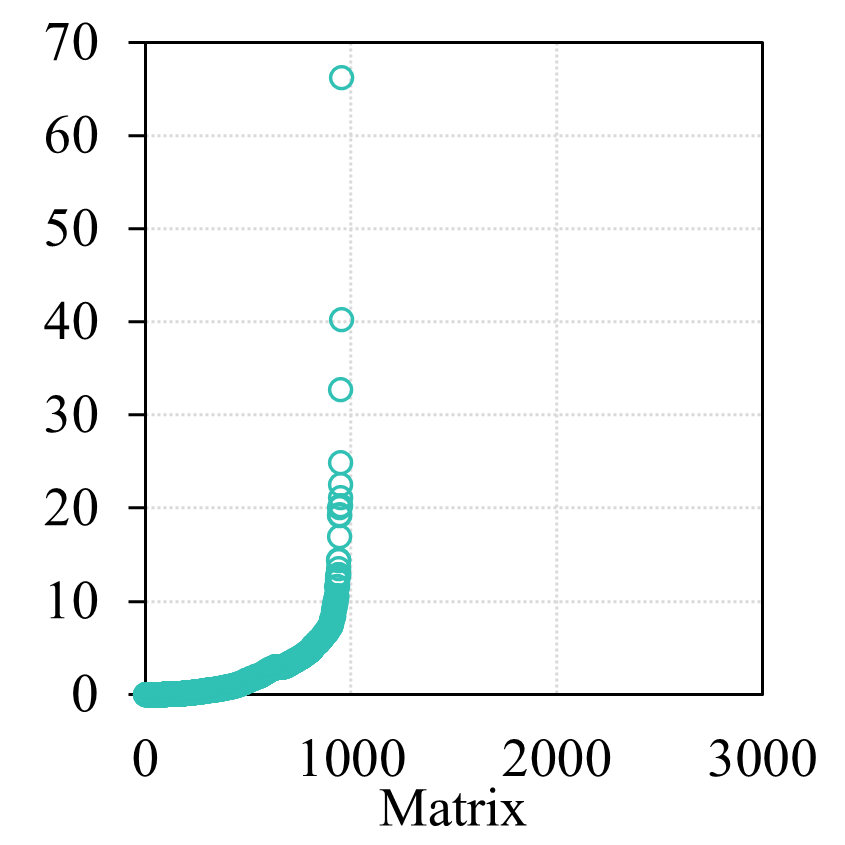}
	}
        \subfloat[HYB] {
		\includegraphics[width=0.19\linewidth]{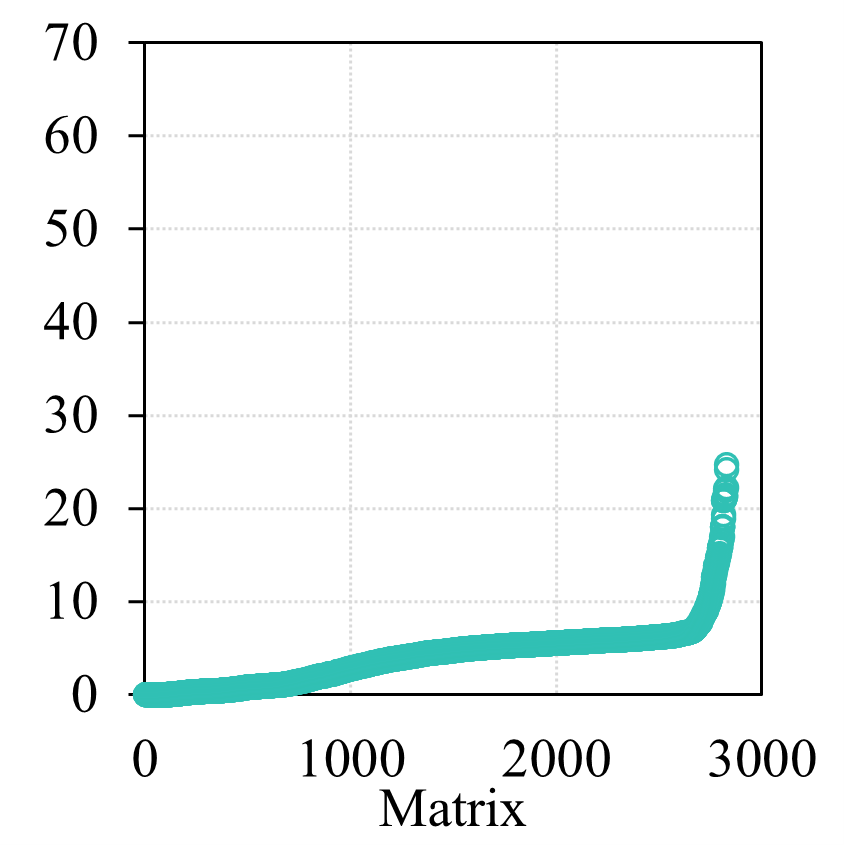}
	}
        \hfill
        \subfloat[COO] 
	{	\includegraphics[width=0.19\linewidth]{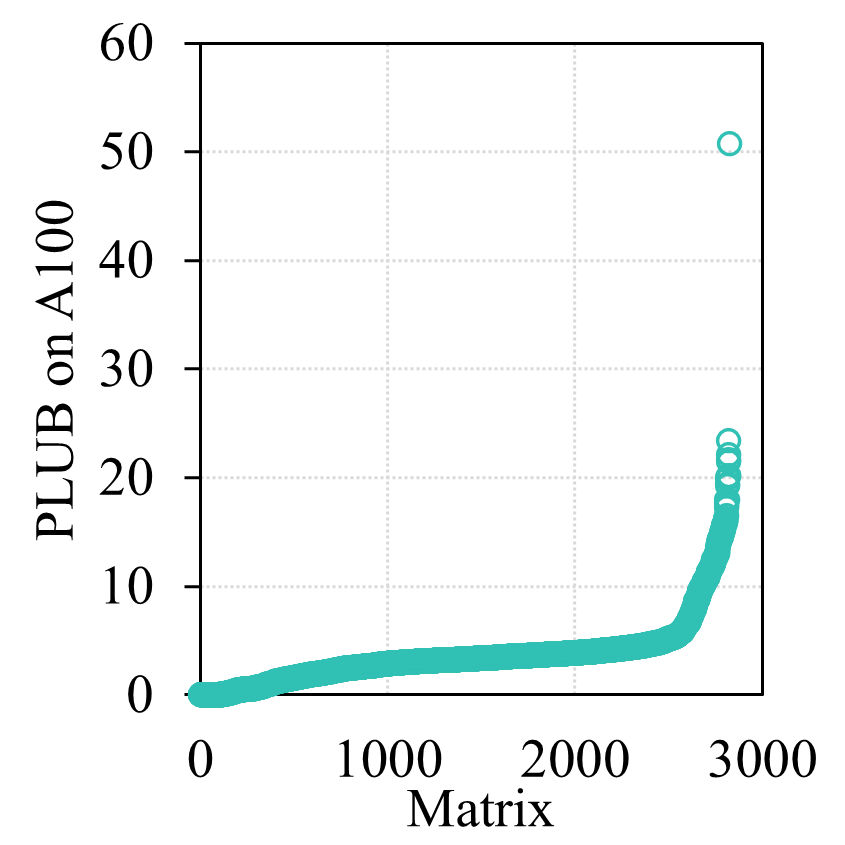}
	}
	\subfloat[CSR] {
		\includegraphics[width=0.19\linewidth]{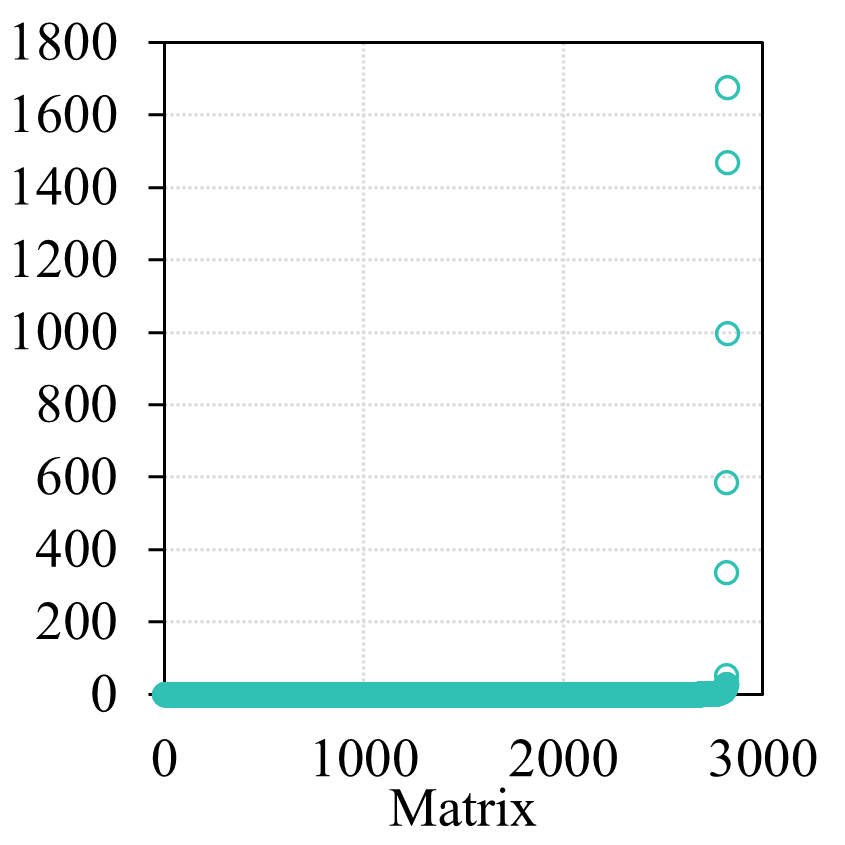}
	}
        \subfloat[ELL] {
		\includegraphics[width=0.19\linewidth]{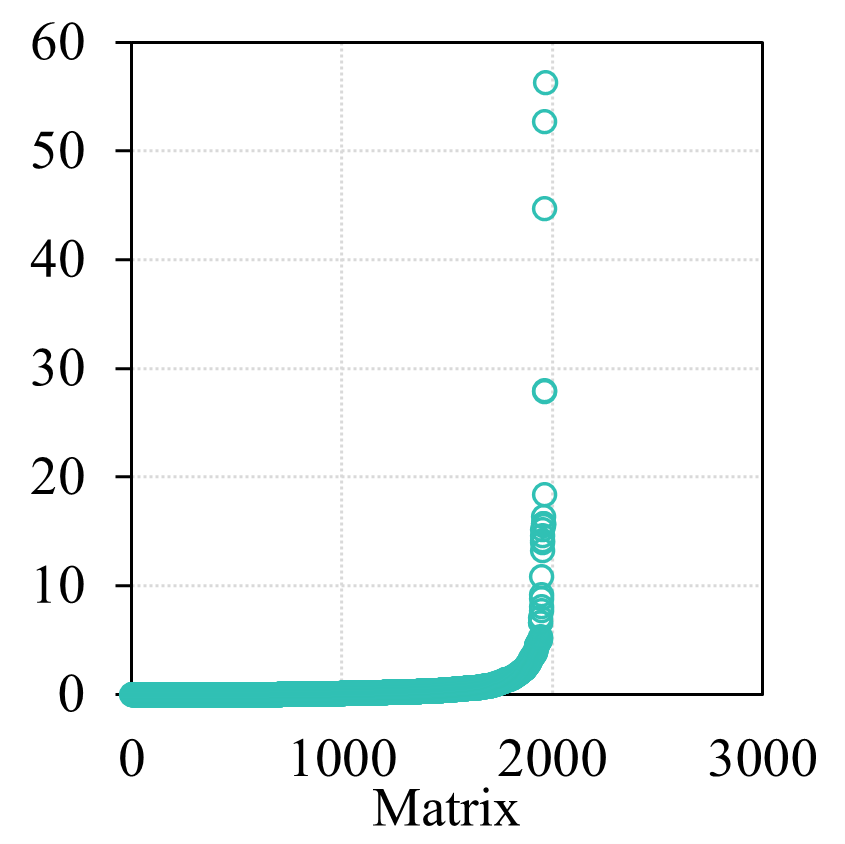}
	}
        \subfloat[DIA] {
		\includegraphics[width=0.19\linewidth]{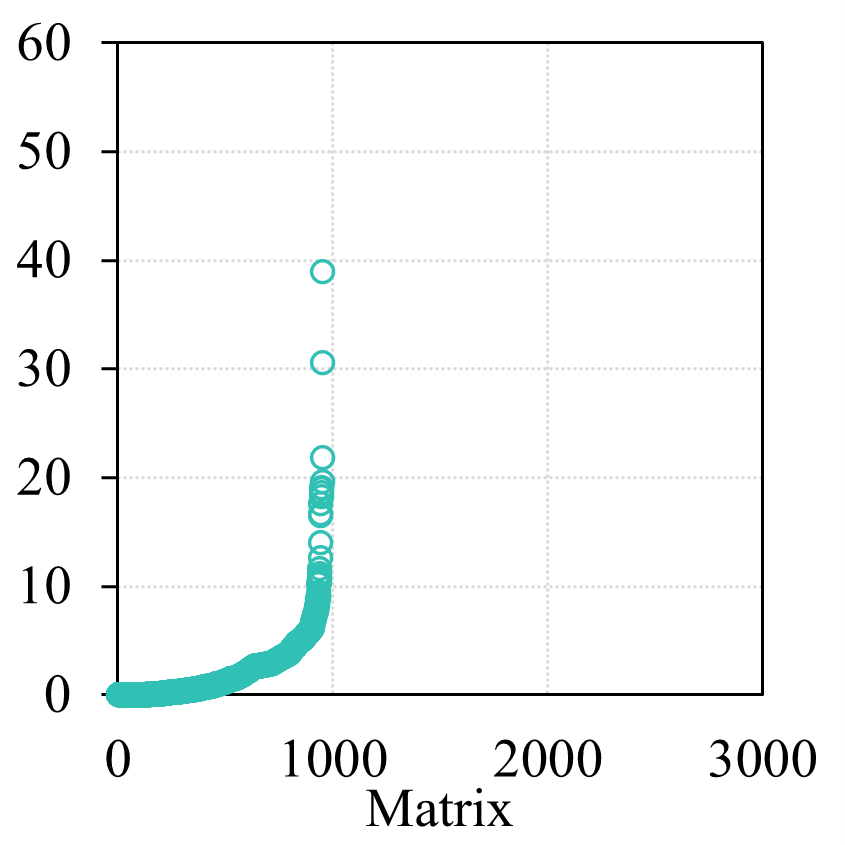}
	}
        \subfloat[HYB] {
		\includegraphics[width=0.19\linewidth]{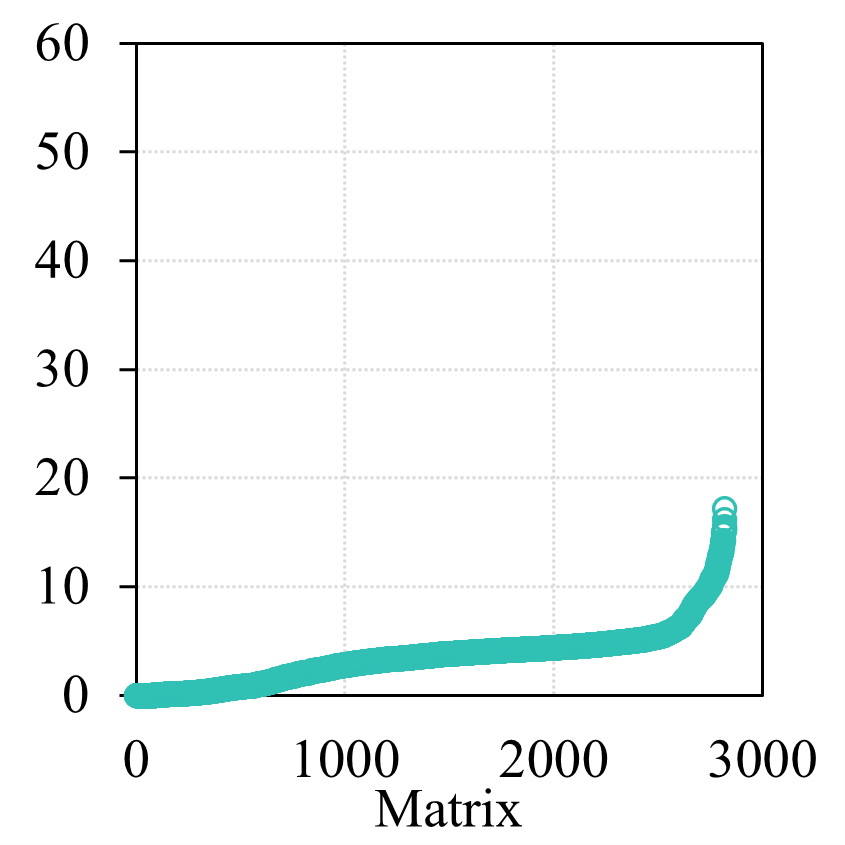}
	}
	\caption{PLUB of each SpMV algorithm with respect to the best-performing SpMV algorithm in CUSP for each sparse matrix. A smaller PLUB indicates better performance.}
	\label{fig:cusp-PLUB}
\end{figure*}

\subsection{Experimental Result}
\begin{figure*}[!htbp]
	\centering
	\subfloat[CSR] 
	{	\includegraphics[width=0.3\linewidth]{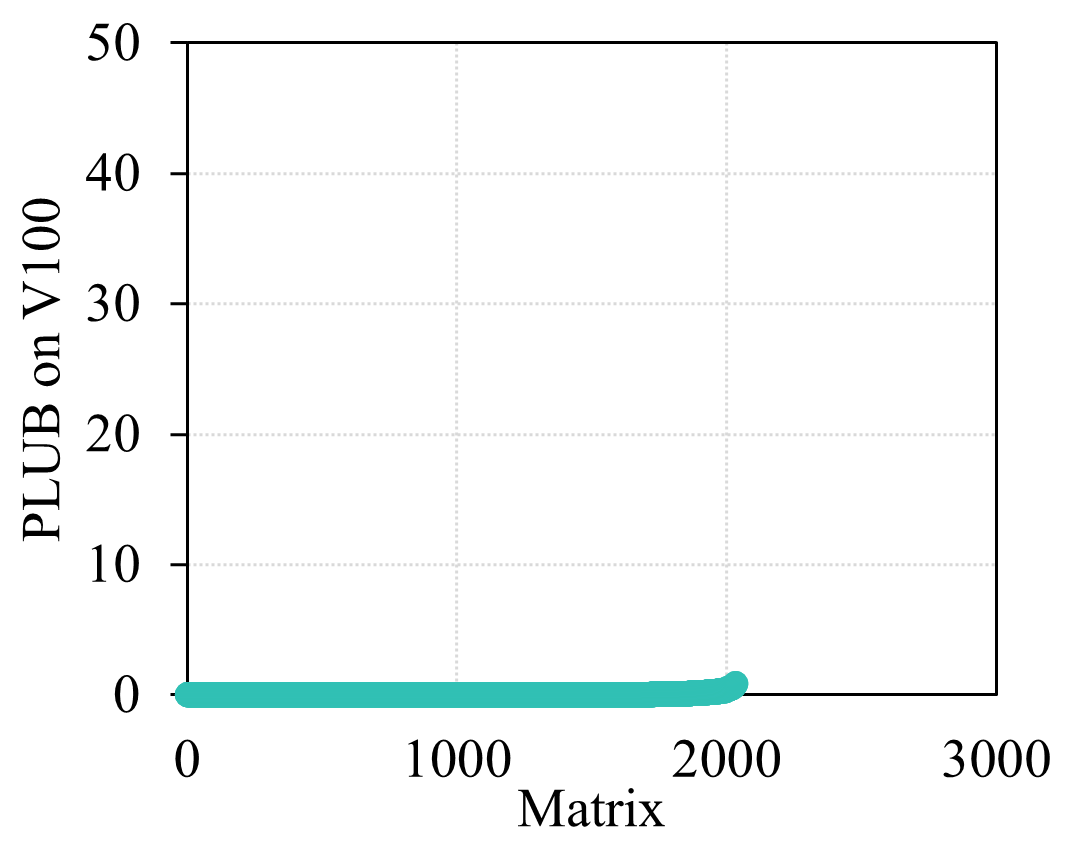}
	}
	\subfloat[ELL] {
		\includegraphics[width=0.3\linewidth]{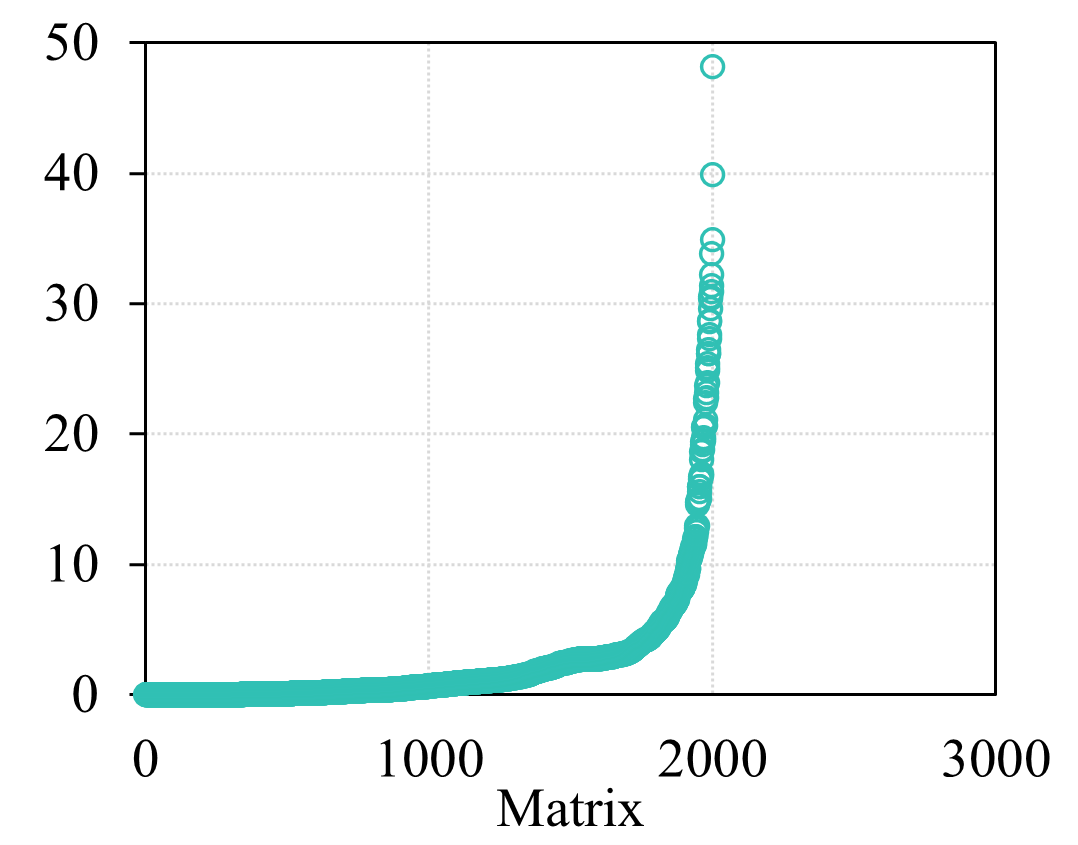}
	}
        \subfloat[SELLP] {
		\includegraphics[width=0.3\linewidth]{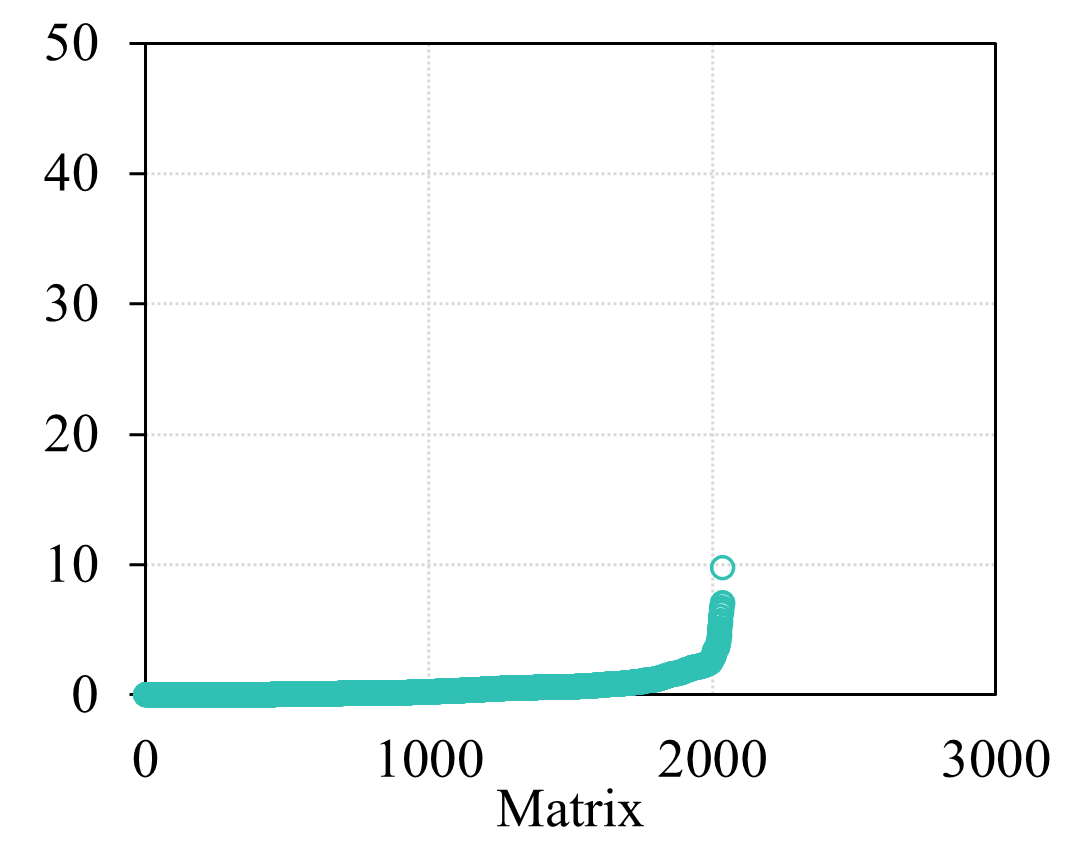}
	}
        \hfill
        \subfloat[CSR] 
	{	\includegraphics[width=0.3\linewidth]{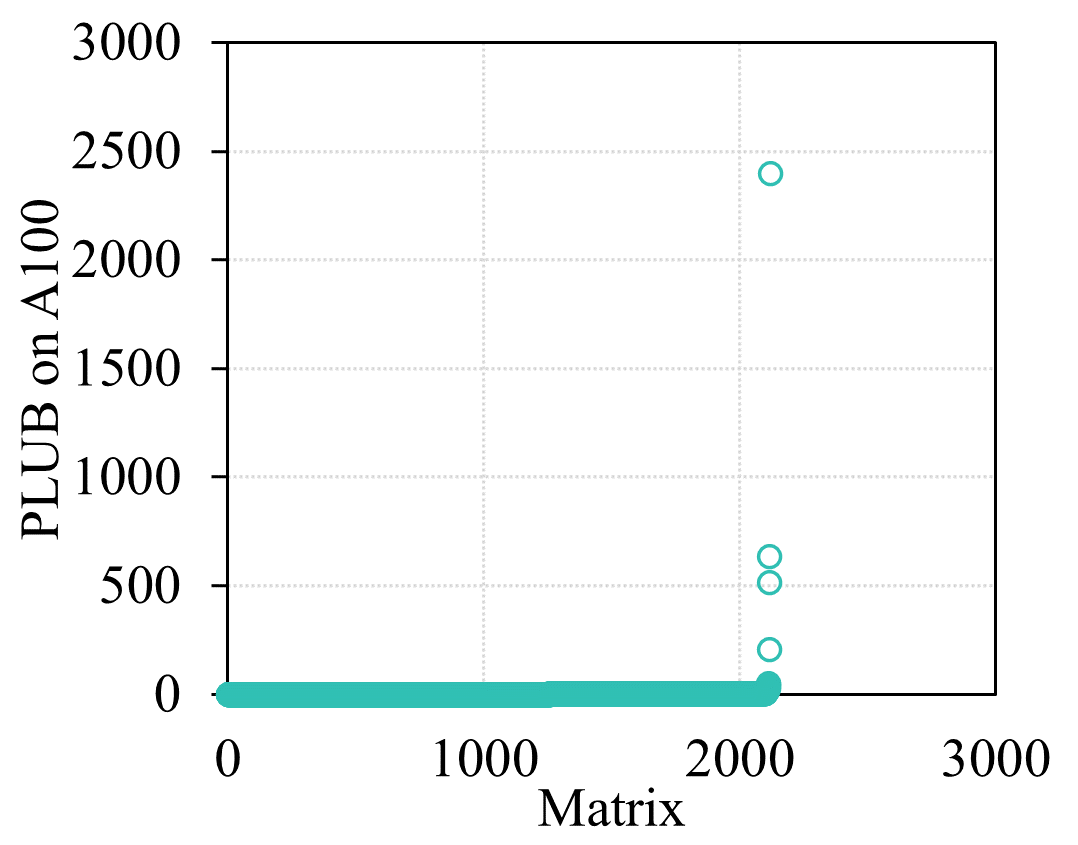}
	}
	\subfloat[ELL] {
		\includegraphics[width=0.3\linewidth]{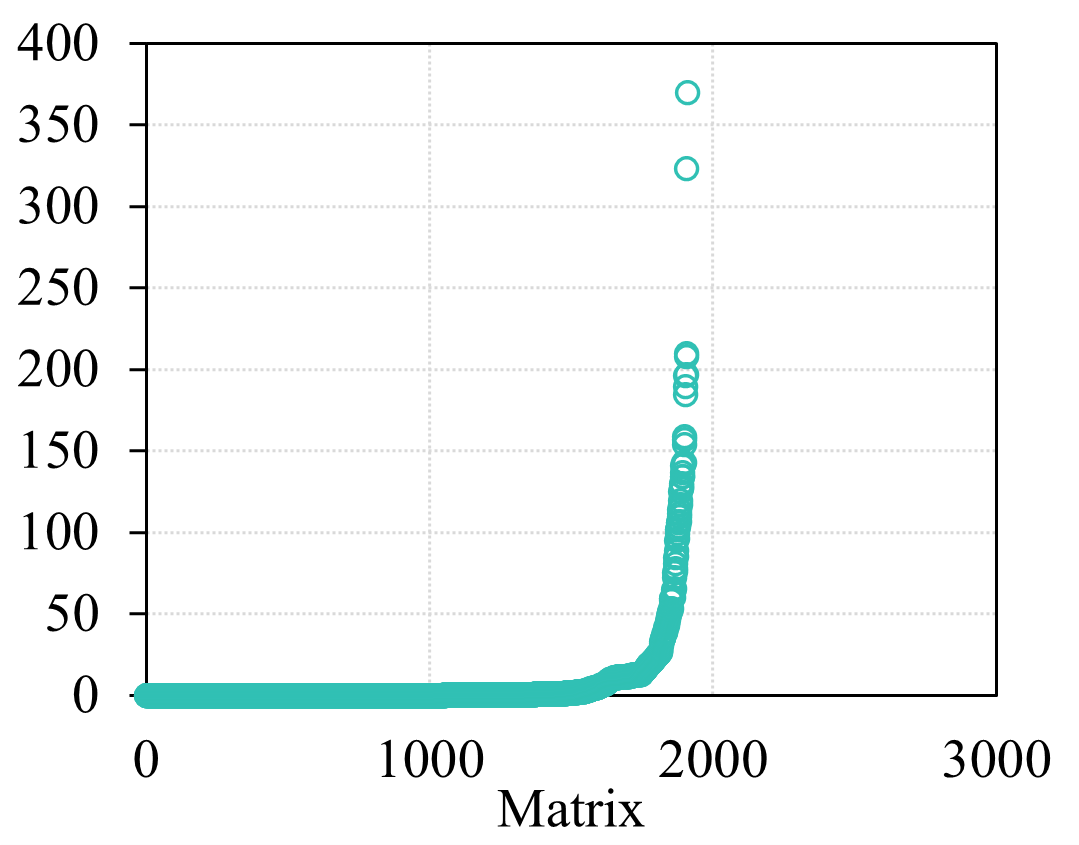}
	}
        \subfloat[SELLP] {
		\includegraphics[width=0.3\linewidth]{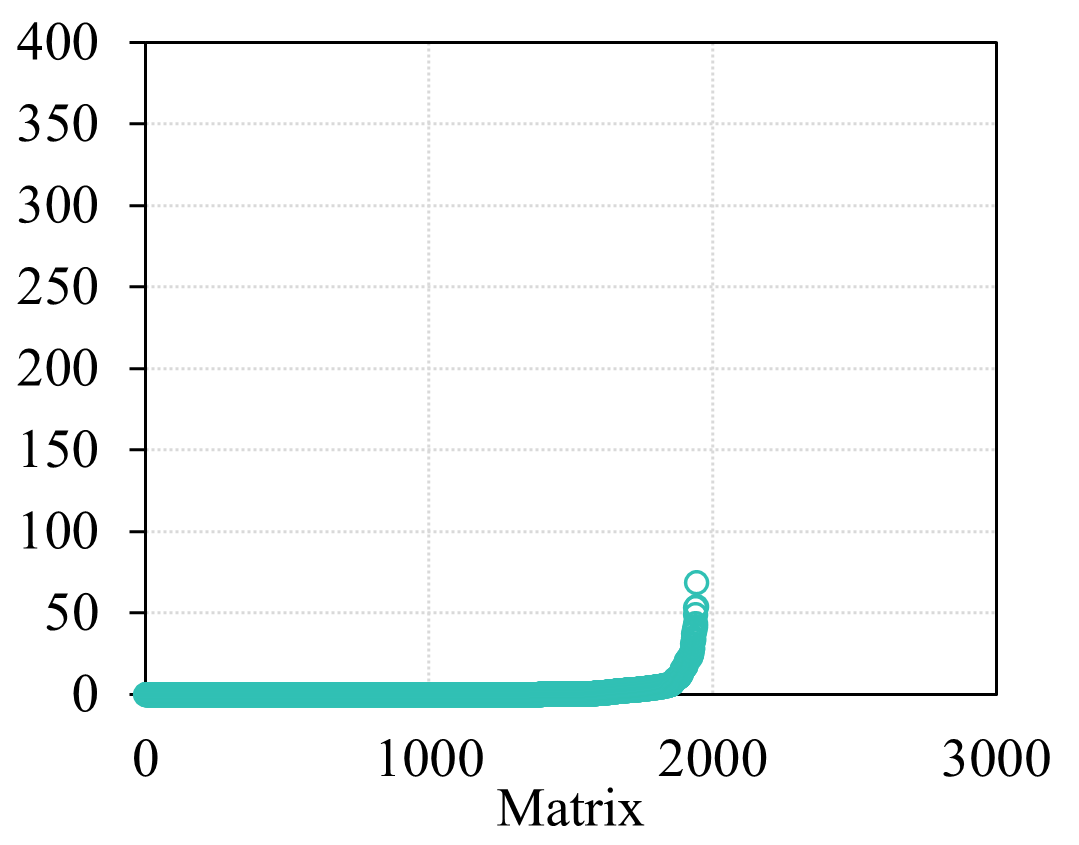}
	}
	\caption{PLUB of each SpMV algorithm with respect to the best-performing SpMV algorithm in MAGMA for each sparse matrix. A smaller PLUB indicates better performance.}
	\label{fig:magma-PLUB}
\end{figure*}

\begin{figure*}[!htbp]
	\centering
	\subfloat[COO] 
	{	\includegraphics[width=0.24\linewidth]{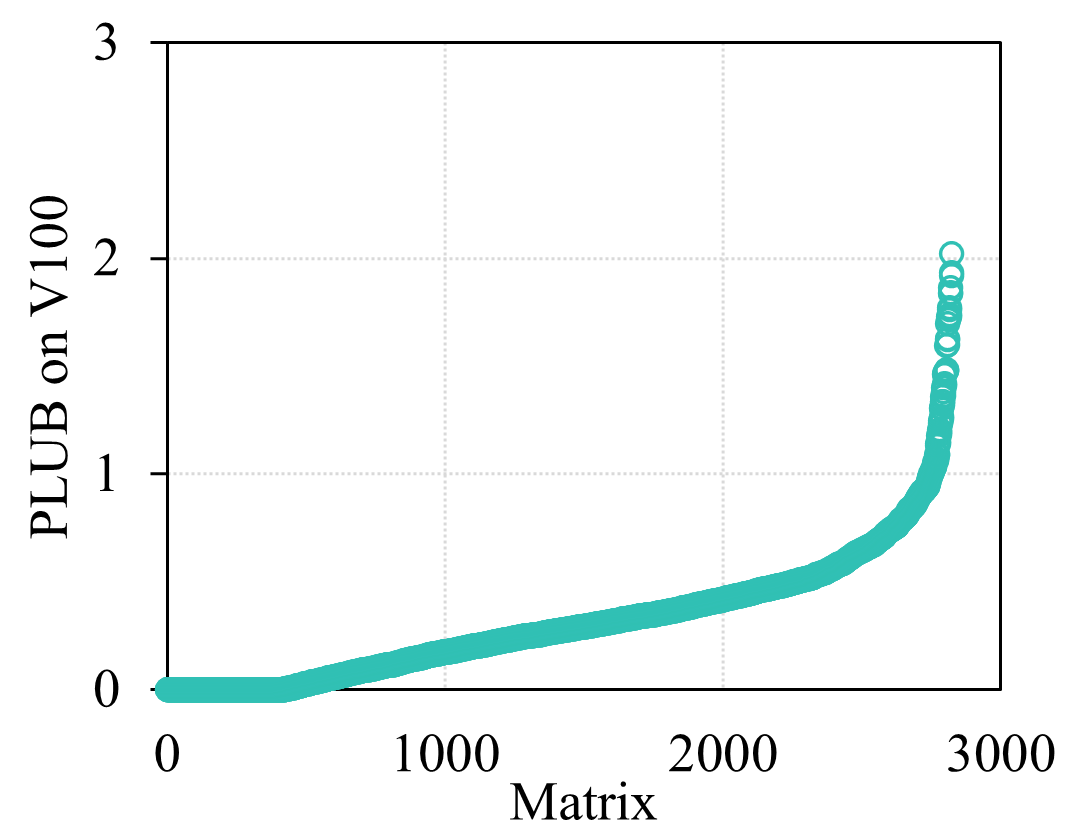}
	}
	\subfloat[CSR] {
		\includegraphics[width=0.24\linewidth]{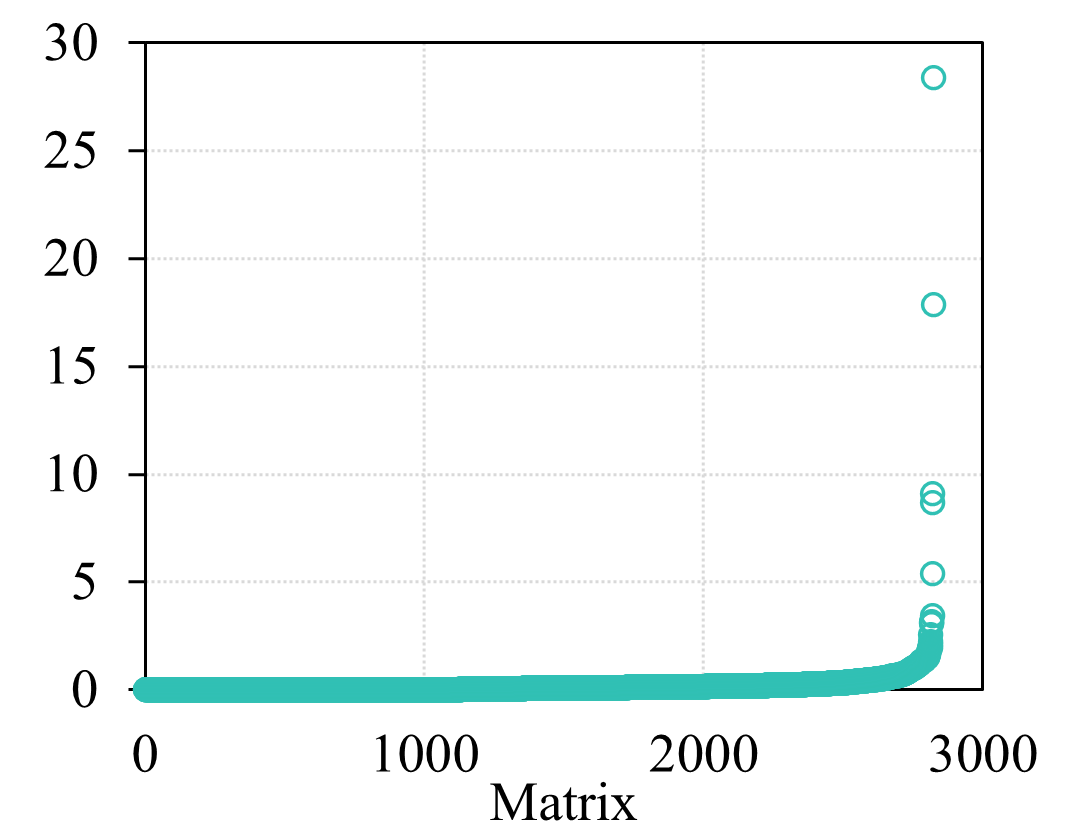}
	}
        \subfloat[ELL] {
		\includegraphics[width=0.24\linewidth]{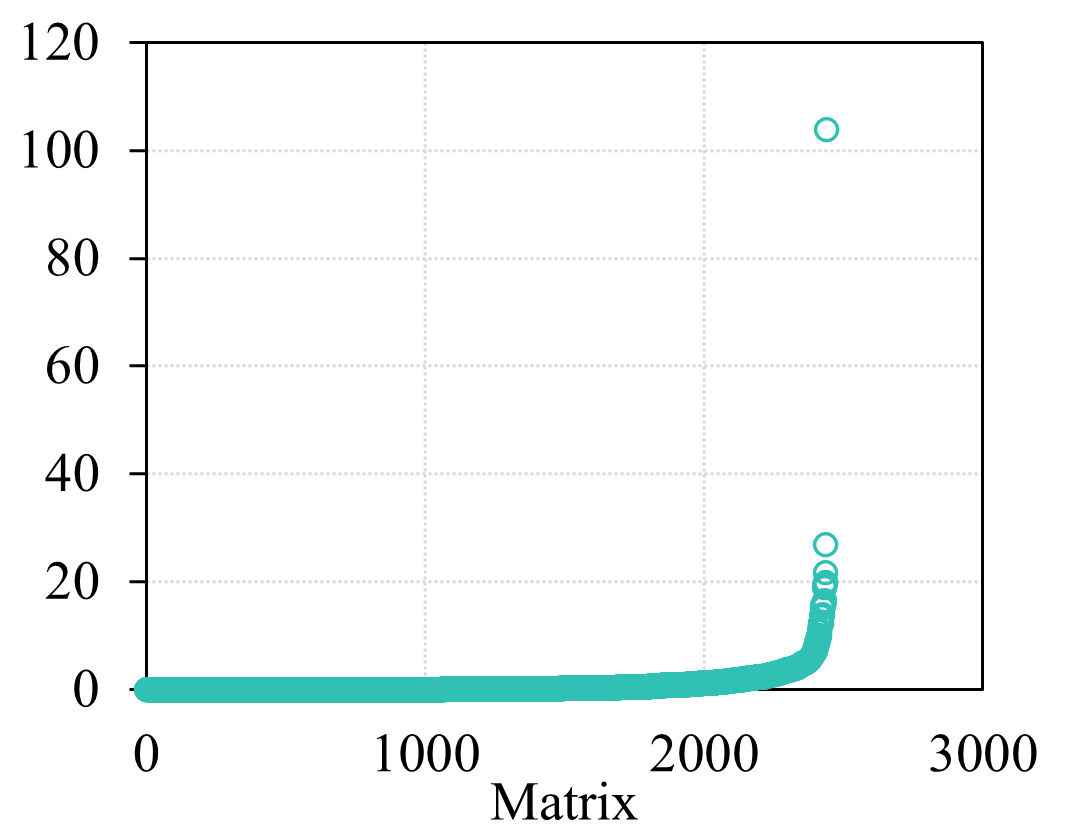}
	}
        \subfloat[HYB] {
		\includegraphics[width=0.24\linewidth]{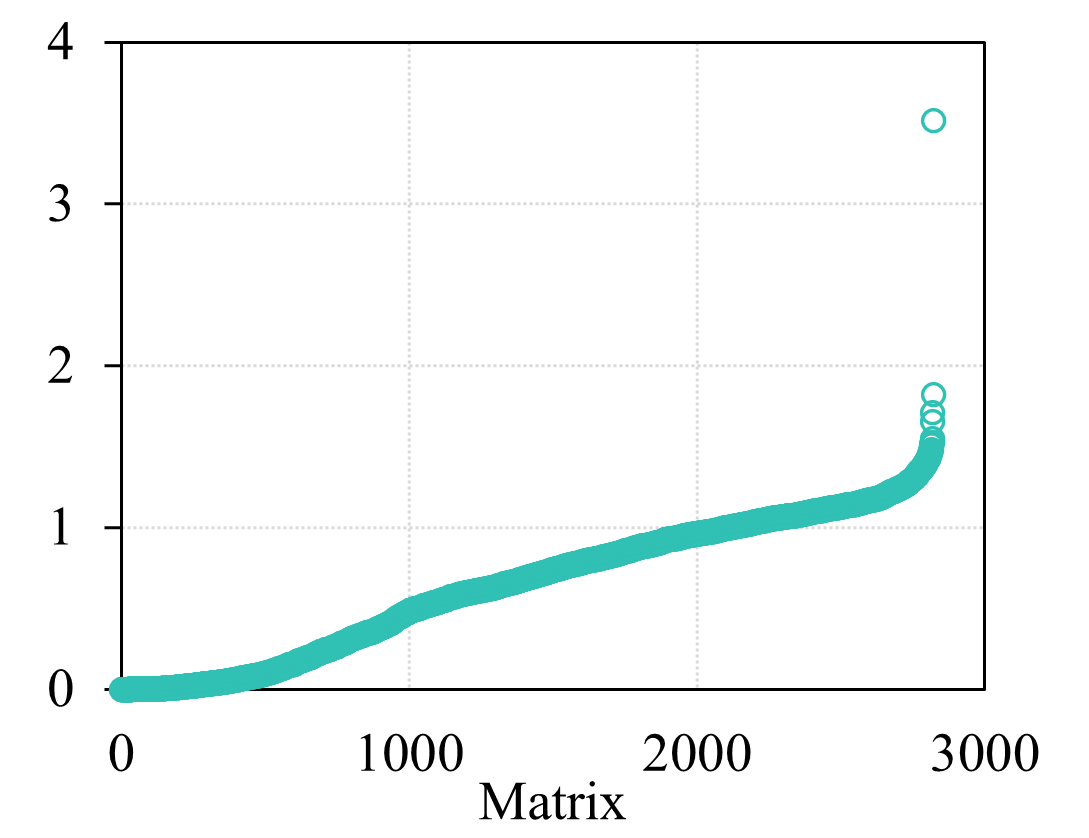}
	}
        \hfill
        \subfloat[COO] 
	{	\includegraphics[width=0.24\linewidth]               {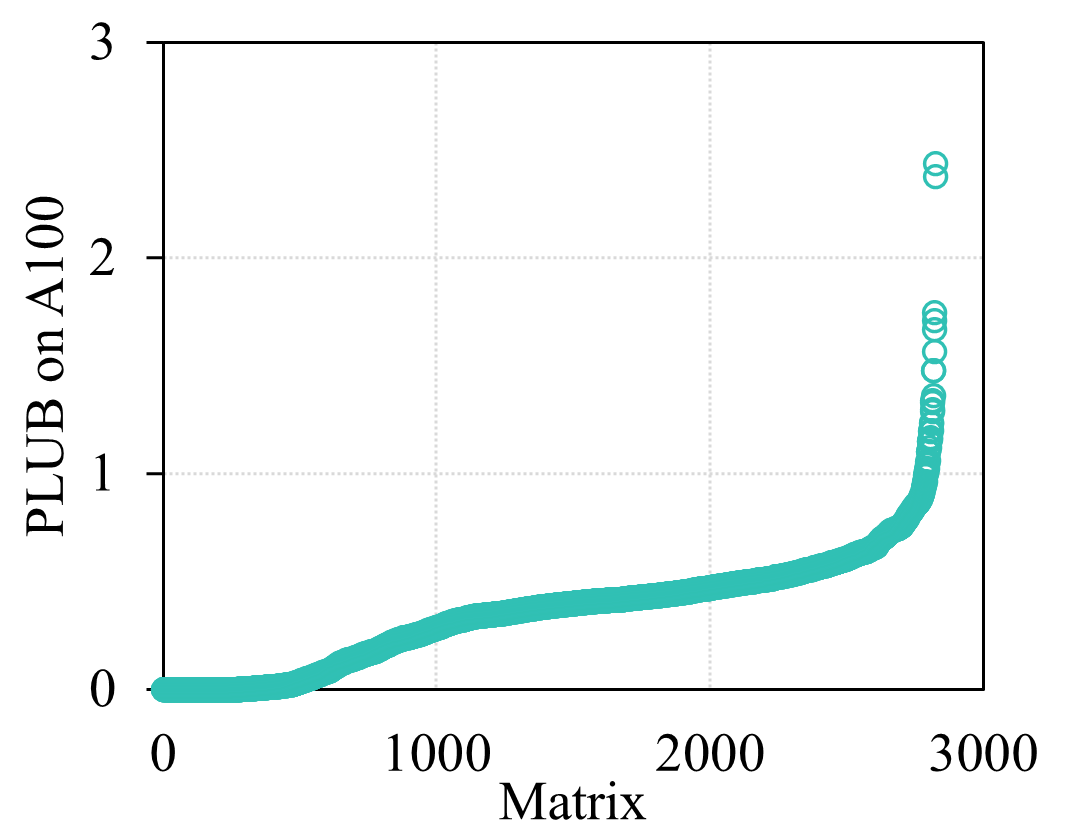}
	}
	\subfloat[CSR] {
		\includegraphics[width=0.24\linewidth]{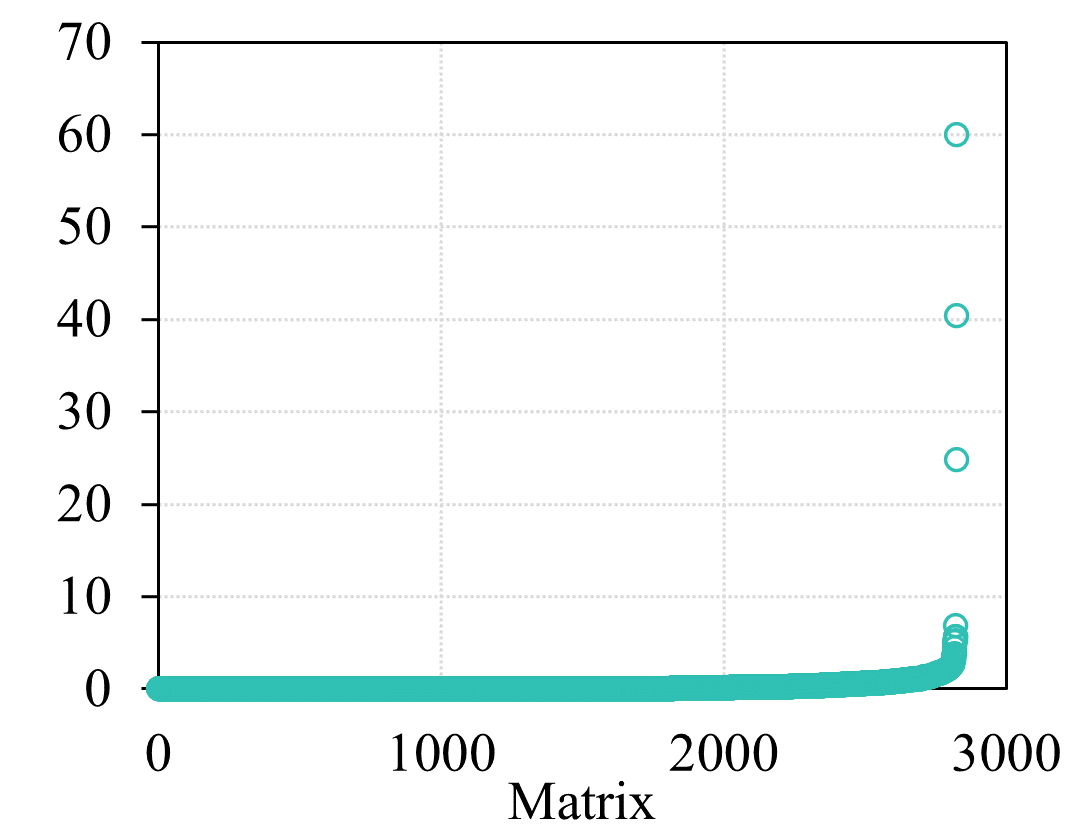}
	}
        \subfloat[ELL] {
		\includegraphics[width=0.24\linewidth]{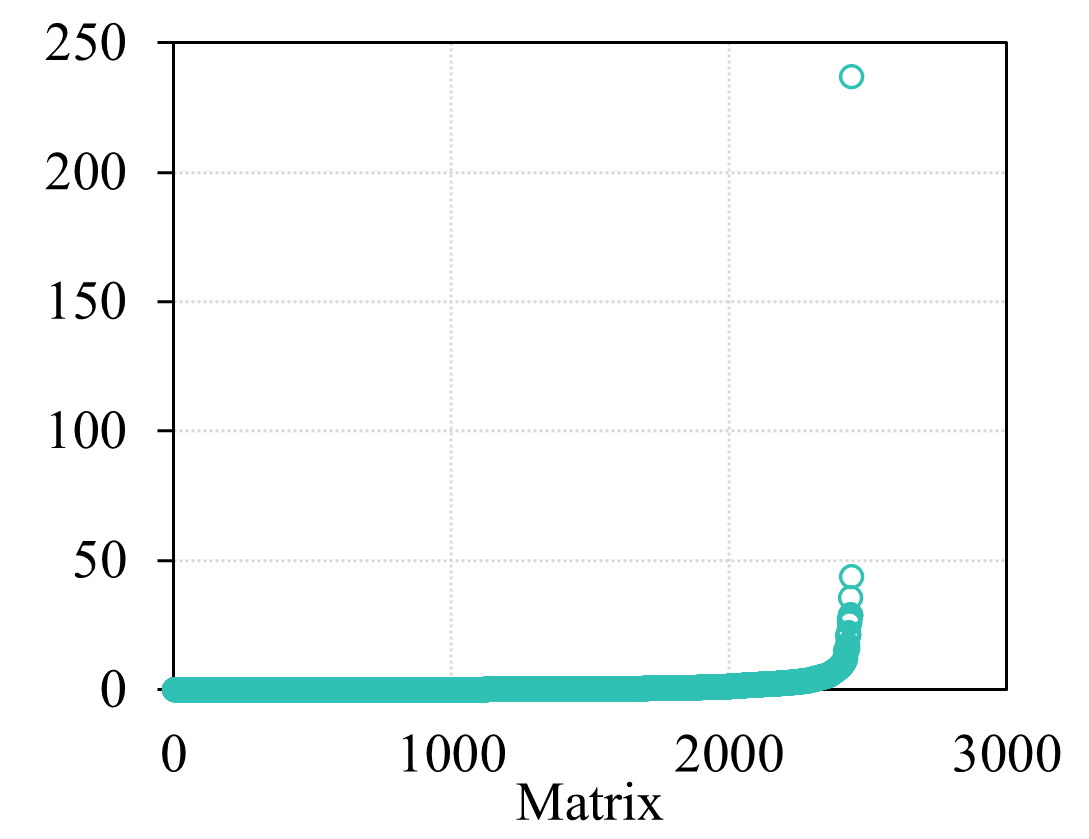}
	}
        \subfloat[HYB] {
		\includegraphics[width=0.24\linewidth]{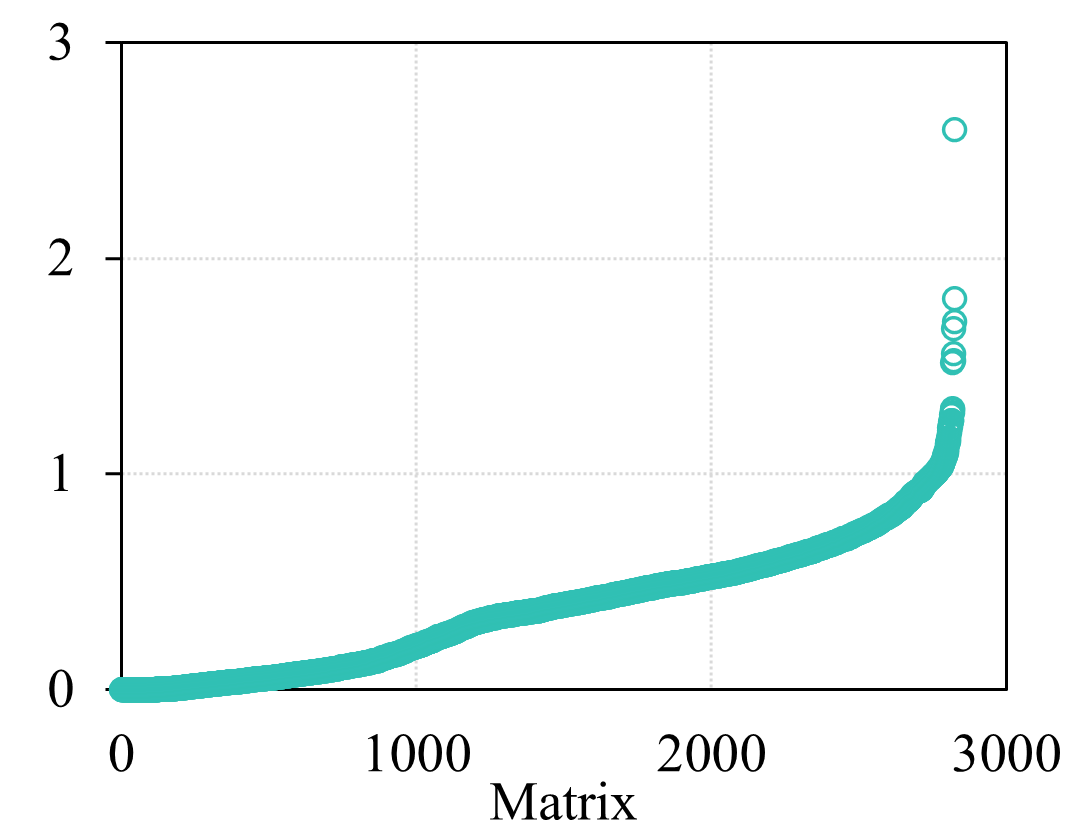}
	}
	\caption{PLUB of each SpMV algorithm with respect to the best-performing SpMV algorithm in Ginkgo for each sparse matrix. A smaller PLUB indicates better performance.}
	\label{fig:ginkgo-PLUB}
\end{figure*}

Figures \ref{fig:cusparse-PLUB}, \ref{fig:cusp-PLUB}, \ref{fig:magma-PLUB}, and \ref{fig:ginkgo-PLUB} illustrate the PLUB of various SpMV algorithms in the library cuSPARSE, CUSP, MAGMA, and Ginkgo, respectively, in comparison with the best-performing algorithm in each library. Tables \ref{tab:cusparse-PLUB}, \ref{tab:cusp-PLUB}, \ref{tab:magma-PLUB}, and \ref{tab:ginkgo-PLUB} present the average PLUB, the number of matrices $N_h$ for which each algorithm achieves the highest GFLOPS, and the number of matrices $N_s$ for which each algorithm successfully executes. We can make the following observations:

(1) In cuSPARSE, the COO1 algorithm achieves optimal performance for more than 2,000 matrices. On V100, the CSR1 algorithm outperforms others across over 500 matrices. On A100, CSR1 achieves optimal performance for over 100 matrices, while COO2 and CSR2 only perform best for a few matrices on both two GPUs. In the subsequent GFLOPS comparison, we choose COO1 as the representative algorithm from the cuSPARSE library, owing to its exceptional performance relative to other algorithms in the same library.

(2) In CUSP, the CSR-based SpMV algorithm achieves optimal performance for nearly 2,000 matrices, demonstrating superior performance to other algorithms in the library. However, it is worth noting that this algorithm exhibits significantly higher performance loss on a few sparse matrices, especially on A100, where its highest PLUB exceeds 1,600. The ELL-based SpMV algorithm achieves optimal performance for over 400 matrices, but due to the inherent characteristics of the ELL format, it fails to execute for approximately 900 matrices. The DIA format poses even more significant challenges, with the DIA-based SpMV algorithm successfully running for fewer than 1,000 matrices. The COO-based and HYB-based SpMV algorithms exhibit similar PLUB. In conclusion, we select the CSR-based SpMV algorithm with lower overall PLUB and higher $N_h$ and $N_s$ as the representative algorithm from the CUSP library.

(3) In MAGMA, all three SpMV algorithms execute successfully for nearly 2,000 sparse matrices. The CSR-based SpMV algorithm achieves the lowest average PLUB (3\%) on V100, while on A100, the SELLP-based SpMV algorithm achieves the lowest average PLUB (137\%). Therefore, we select the CSR-based SpMV algorithm as the representative algorithm from the MAGMA library on V100, and the SELLP-based SpMV algorithm as the representative on A100.

(4) In Ginkgo, the CSR-based SpMV algorithm achieves the lowest average PLUB, followed by the COO-based SpMV, then the HYB-based SpMV. The ELL-based SpMV exhibits the highest average PLUB. The CSR-based SpMV algorithm achieves optimal performance for over 900 matrices on V100 and over 1,400 matrices on A100. The COO-based and ELL-based SpMV algorithms demonstrate optimal performance for approximately 200 to 600 matrices, while the HYB-based SpMV only performs best for a few matrices. In contrast, the ELL-based SpMV algorithm fails to execute for nearly 400 matrices. Overall, we select the CSR-based SpMV, which demonstrates the best overall performance, as the representative from the Ginkgo library.

(5) In libraries other than cuSPARSE, the PLUB of different SpMV algorithms ranges from tens to hundreds, and even thousands, for some matrices. However, in cuSPARSE, the PLUB of all four algorithms is below 3, with values on V100 even below 1.8. This indicates that these algorithms in the cuSPARSE library exhibit relatively stable performance, with minimal disparities from the optimal performance.

\begin{table}[!ht]
    \centering
    \caption{Overall performance comparison of different SpMV algorithms in MAGMA.}
    \begin{tabular}{ccrrr}
    \toprule
        \multicolumn{2}{c}{\textbf{Algorithm}} & \textbf{CSR} & \textbf{ELL} & \textbf{SELLP} \\ \midrule
        \multirow{2}{*}{PLUB(\%)} & V100 & \textbf{3} & 212 & 57 \\ 
        ~ & A100 & 325 & 620 & \textbf{137} \\ \midrule
        \multirow{2}{*}{$N_h$} & V100 & \textbf{1,612} & 218 & 231 \\
        ~ & A100 & 686 & \textbf{776} & 658 \\ \midrule
         \multirow{2}{*}{$N_s$} & V100 & 2,033 & 2,001 & \textbf{2,037}\\
        ~ & A100 & \textbf{2,118} & 1,909 & 1,943 \\ 
        \bottomrule
    \end{tabular}
    \label{tab:magma-PLUB}
\end{table}

\begin{figure*}[!htbp]
	\centering
	\subfloat[cuSPARSE] 
	{	\includegraphics[width=0.33\linewidth]{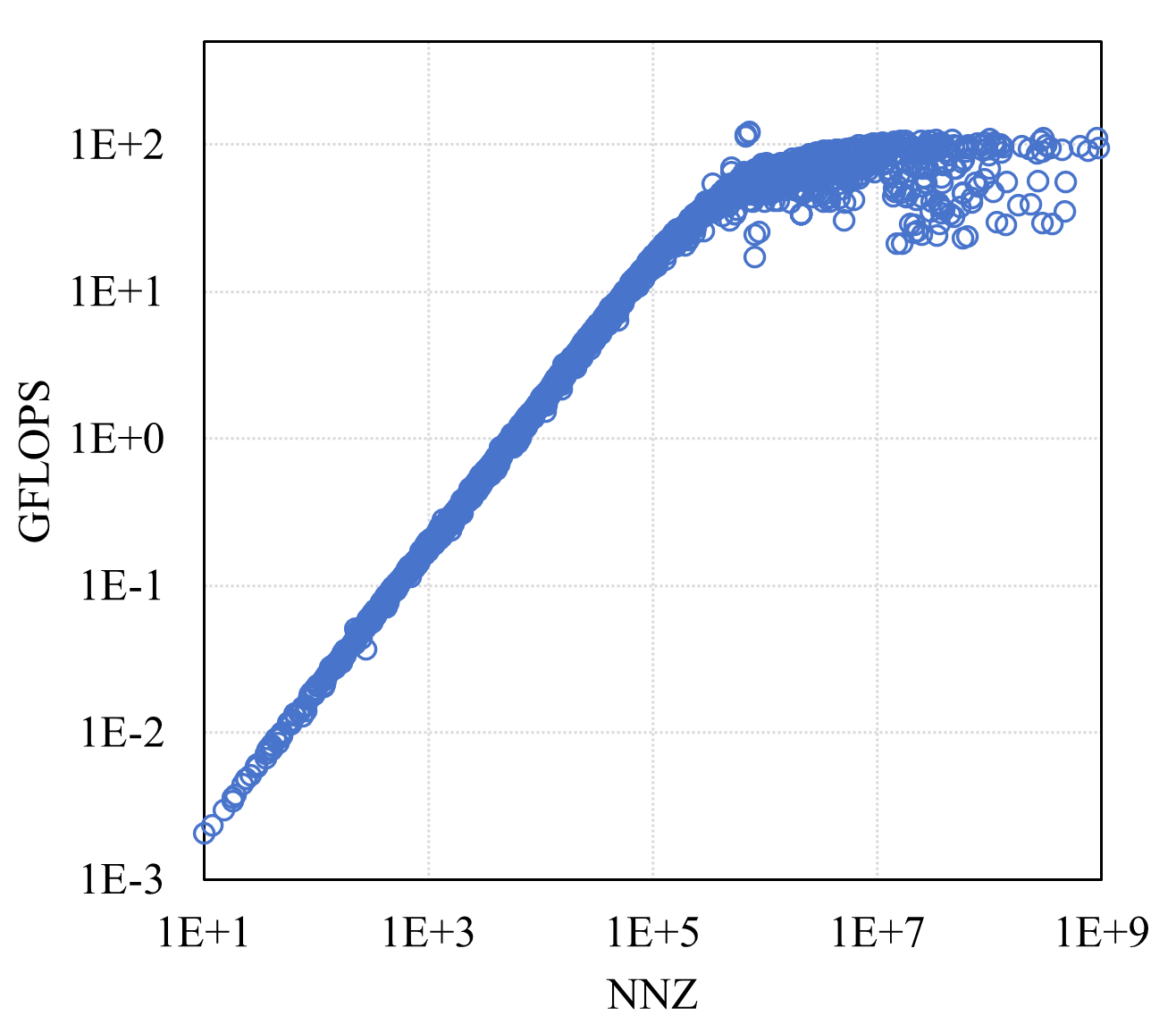}}
	\subfloat[CUSP] {
		\includegraphics[width=0.33\linewidth]{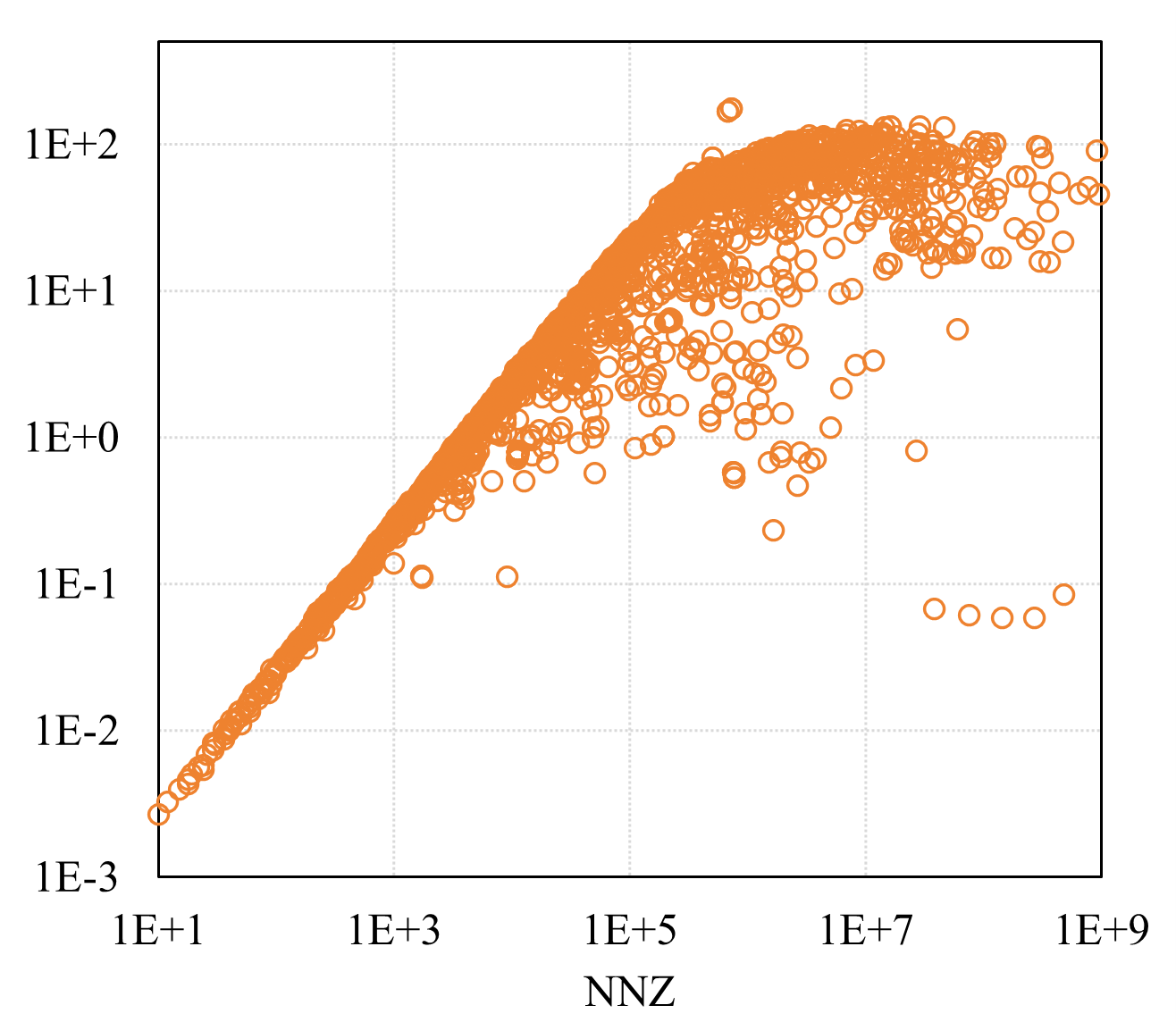}
	}
        \subfloat[Merge] {
		\includegraphics[width=0.33\linewidth]{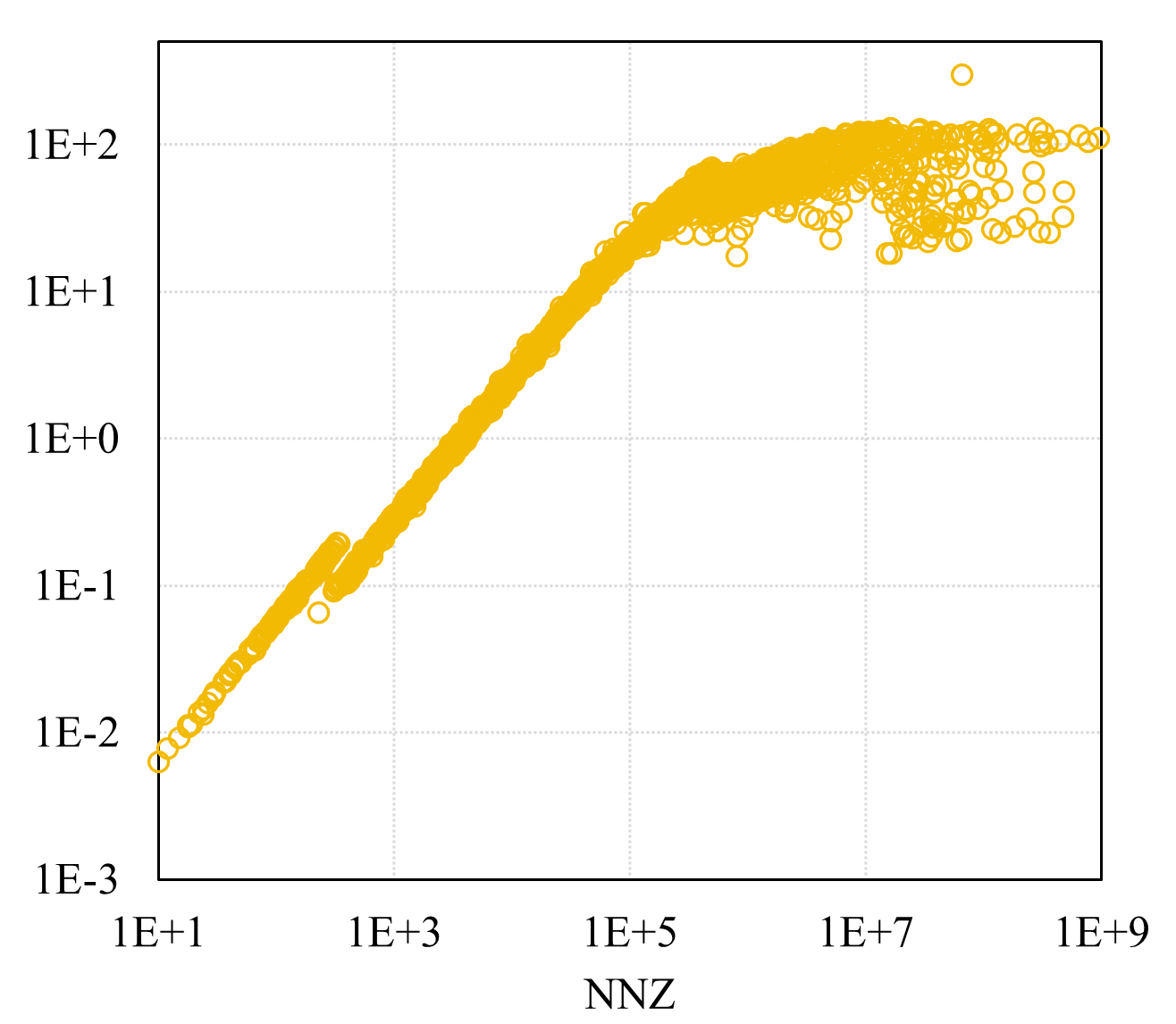}
	}
        \hfill
        \subfloat[MAGMA] {
		\includegraphics[width=0.33\linewidth]{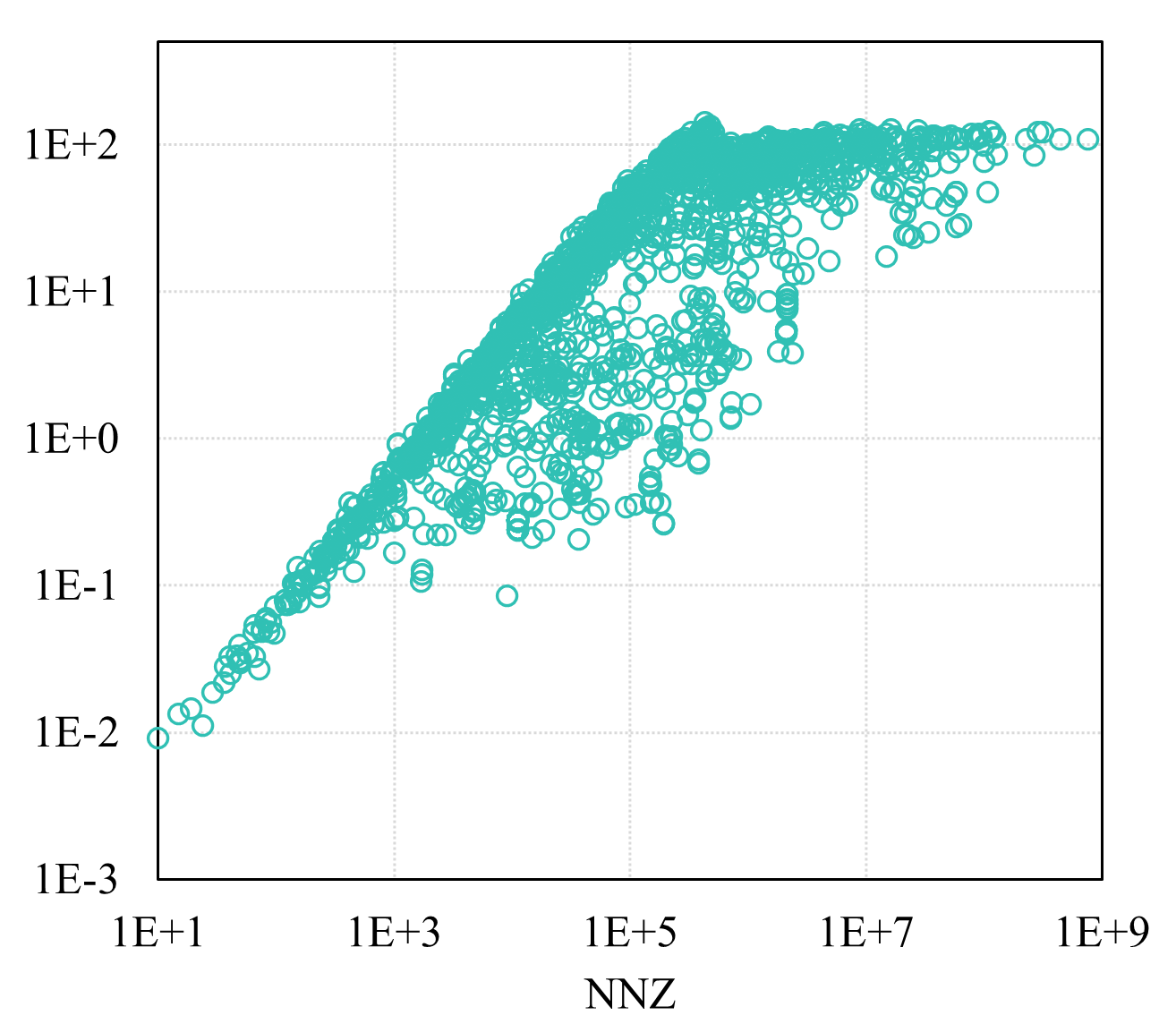}
	}
        \subfloat[CSR5] {
		\includegraphics[width=0.33\linewidth]{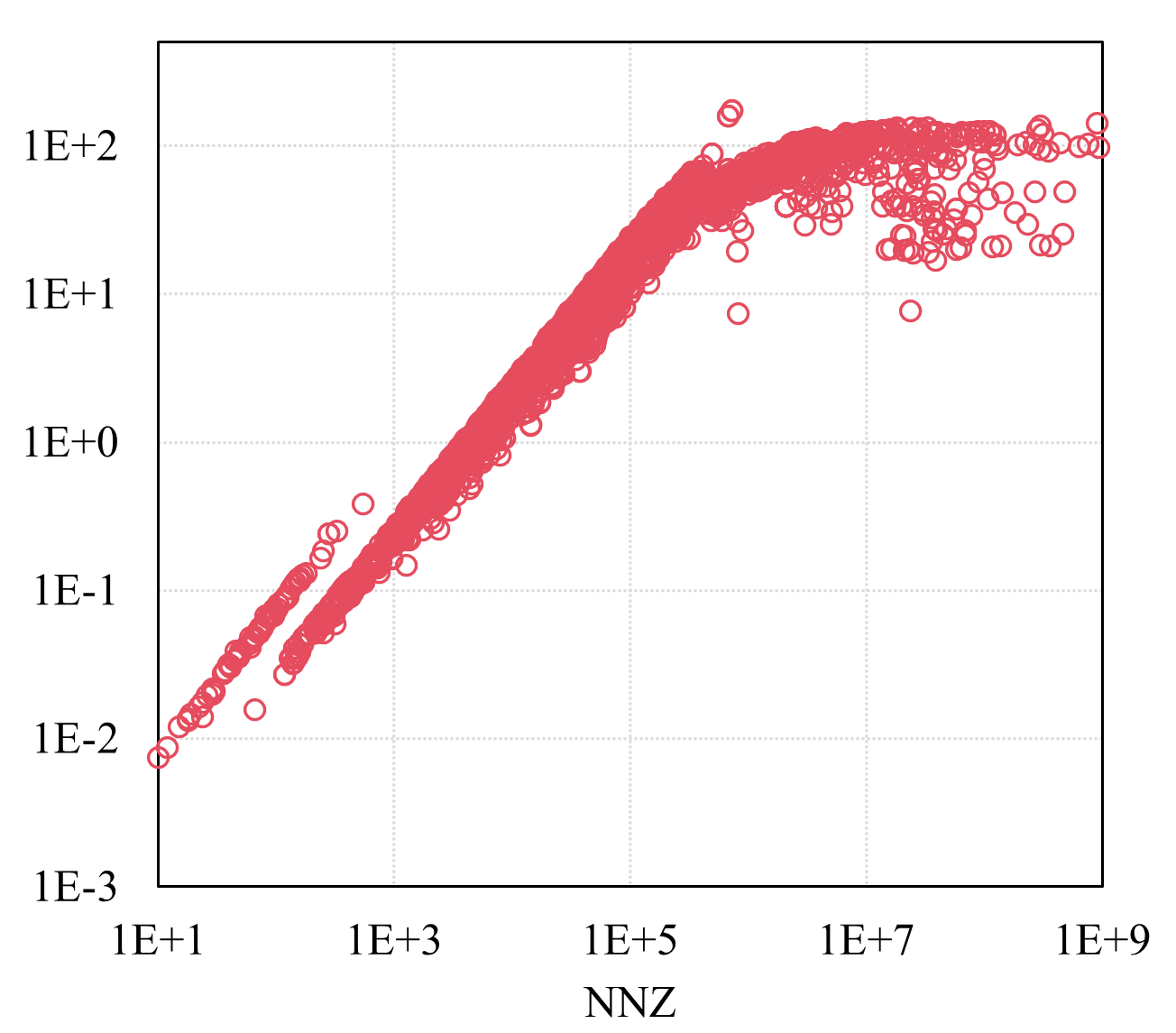}
	}
        \subfloat[Ginkgo] {
		\includegraphics[width=0.33\linewidth]{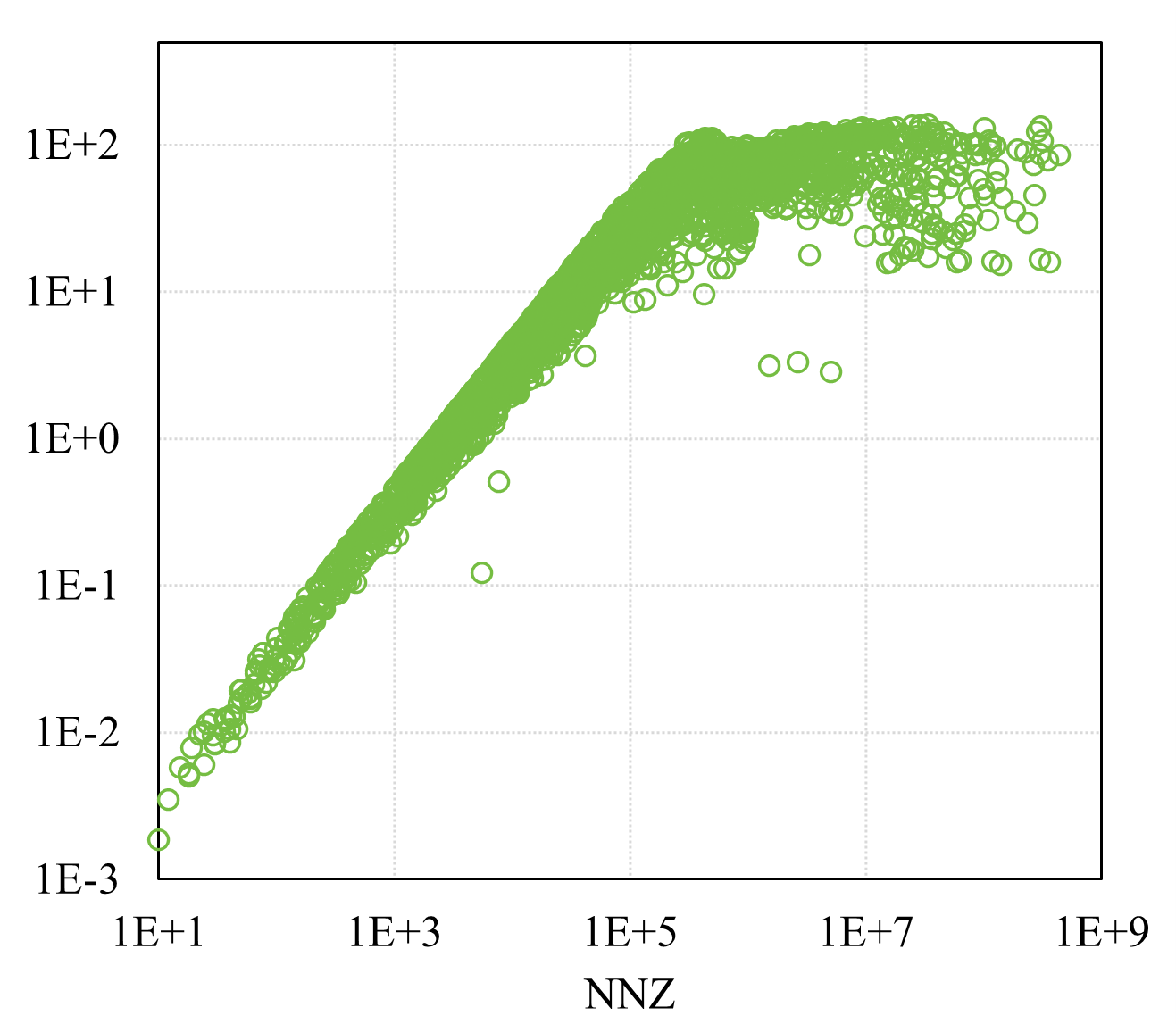}
	}
	\caption{GFLOPS of different SpMV algorithms on Tesla V100.}
	\label{fig:GFLOPS_V100}
\end{figure*}

\begin{table}[!htbp]
    \centering
    \caption{Overall performance comparison of different SpMV algorithms in Ginkgo.}
    \begin{tabular}{ccrrrr}
    \toprule
        \multicolumn{2}{c}{\textbf{Algorithm}} & \textbf{COO} & \textbf{CSR} & \textbf{ELL} & \textbf{HYB} \\ \midrule
        \multirow{2}{*}{PLUB(\%)} & V100 & 32 & \textbf{19} & 86 & 64 \\ 
        ~ & A100 & 35 & \textbf{27} & 117 & 37 \\ \midrule
        \multirow{2}{*}{$N_h$} & V100 & 416 & \textbf{977} & 620 & 24 \\
        ~ & A100 & 277 & \textbf{1,487} & 209 & 4 \\ \midrule
         \multirow{2}{*}{$N_s$} & V100 & \textbf{2,823} & \textbf{2,823} & 2,438 & 2,822\\
        ~ & A100 & \textbf{2,825} & \textbf{2,825} & 2,441 & \textbf{2,825} \\     
        \bottomrule
    \end{tabular}
    \label{tab:ginkgo-PLUB}
\end{table}

\begin{figure*}[!htbp]
	\centering
	\subfloat[cuSPARSE] 
	{	\includegraphics[width=0.33\linewidth]{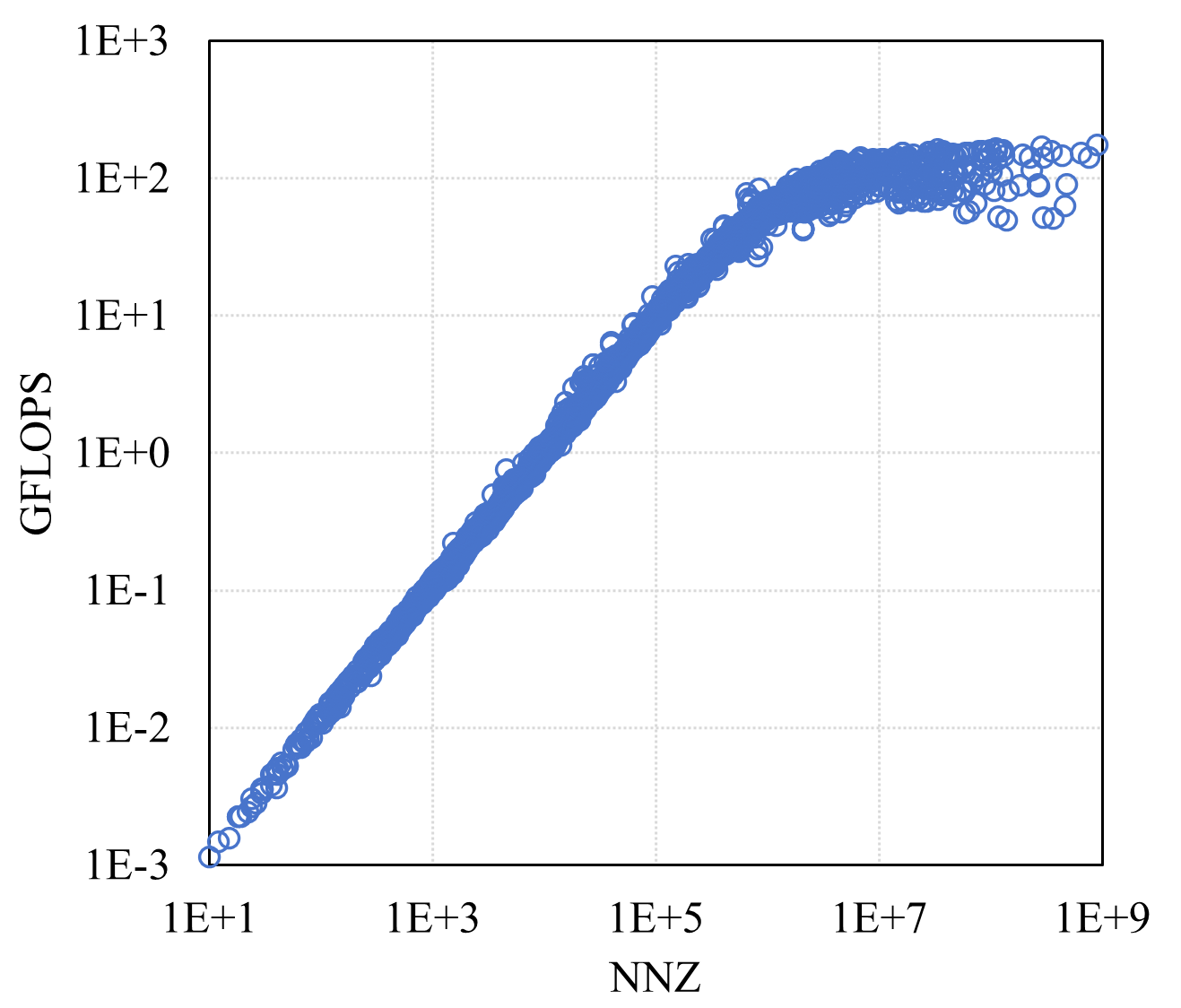}}
	\subfloat[CUSP] {
		\includegraphics[width=0.33\linewidth]{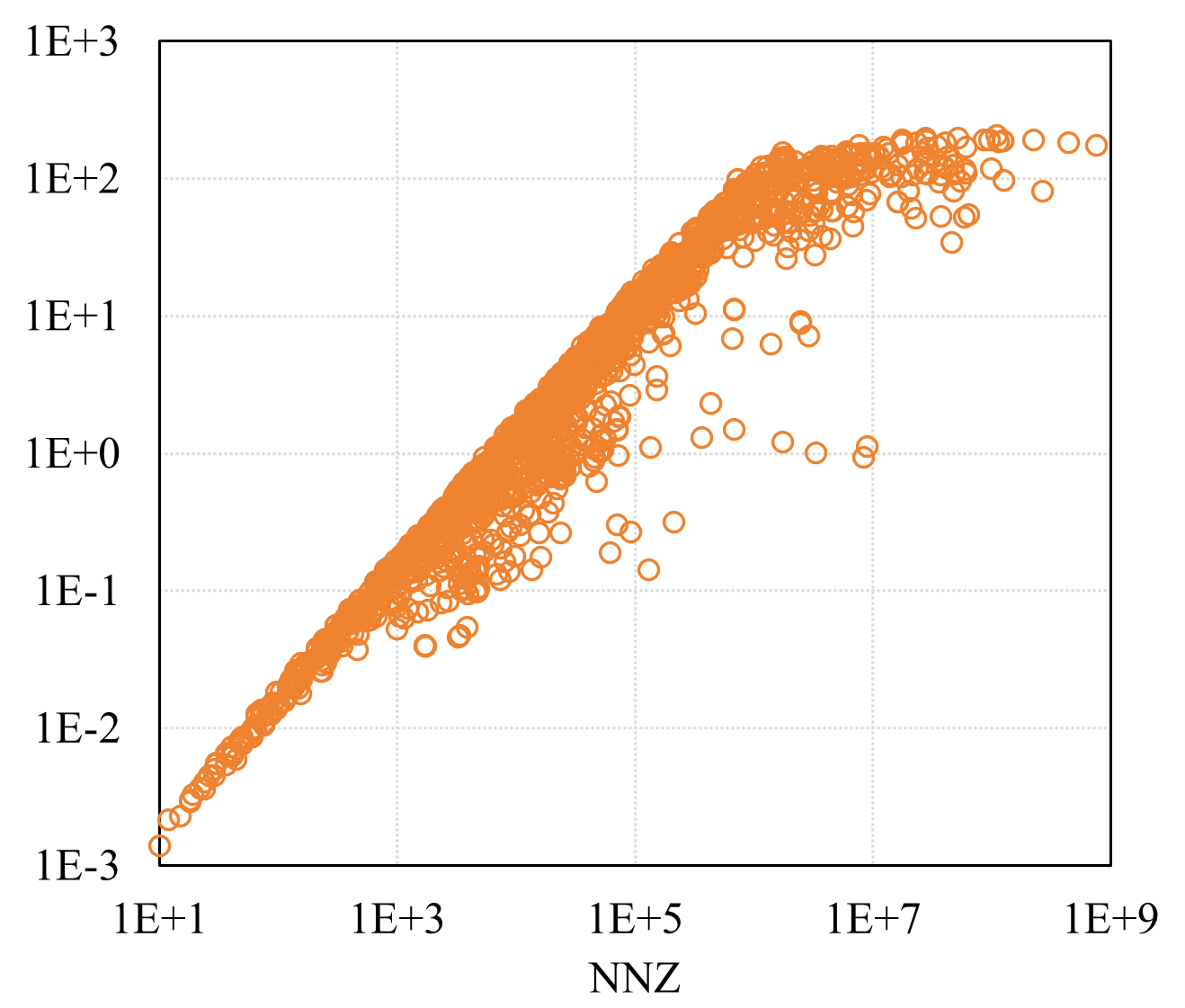}
	}
        \subfloat[Merge] {
		\includegraphics[width=0.33\linewidth]{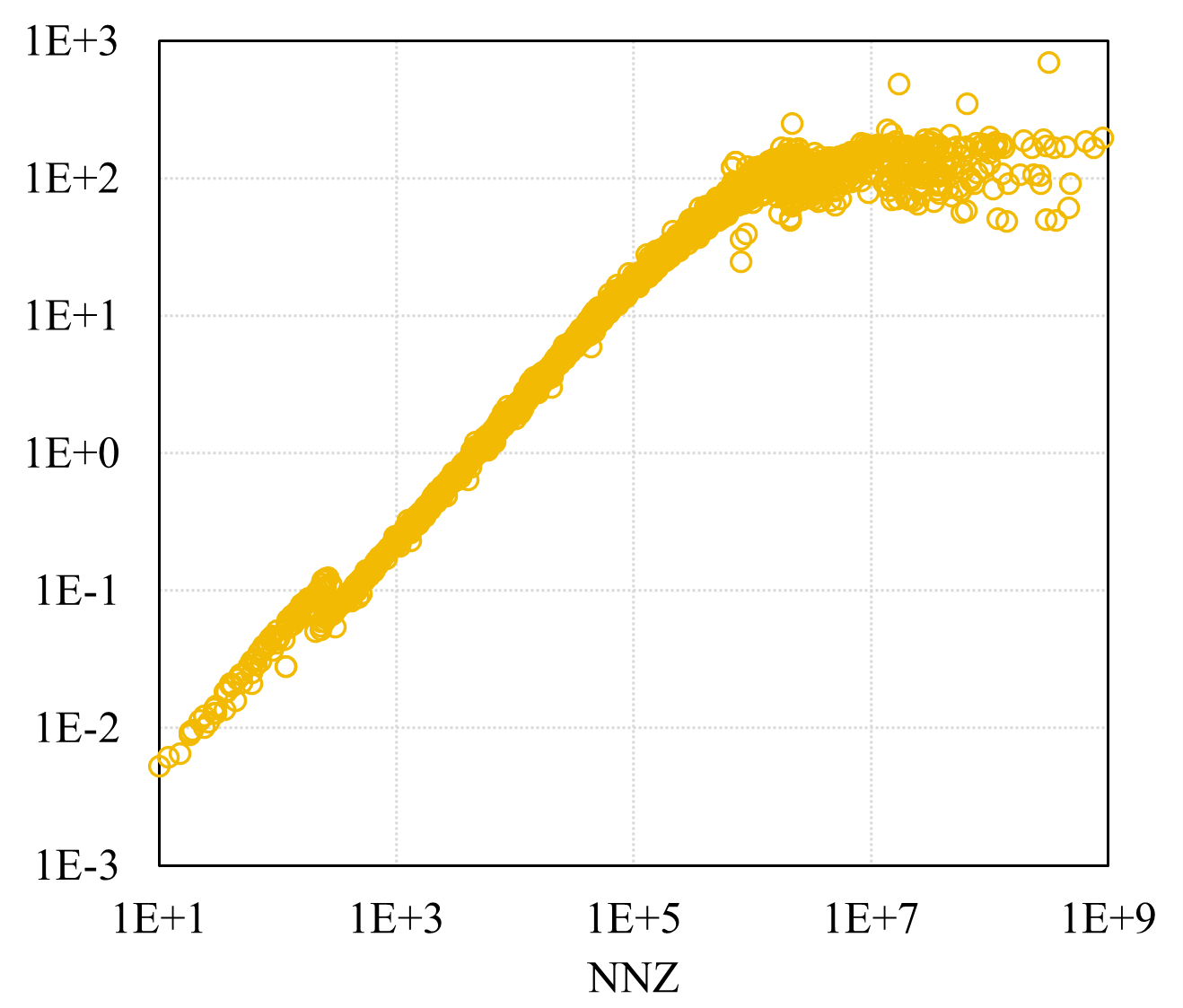}
	}
        \hfill
        \subfloat[MAGMA] {
		\includegraphics[width=0.33\linewidth]{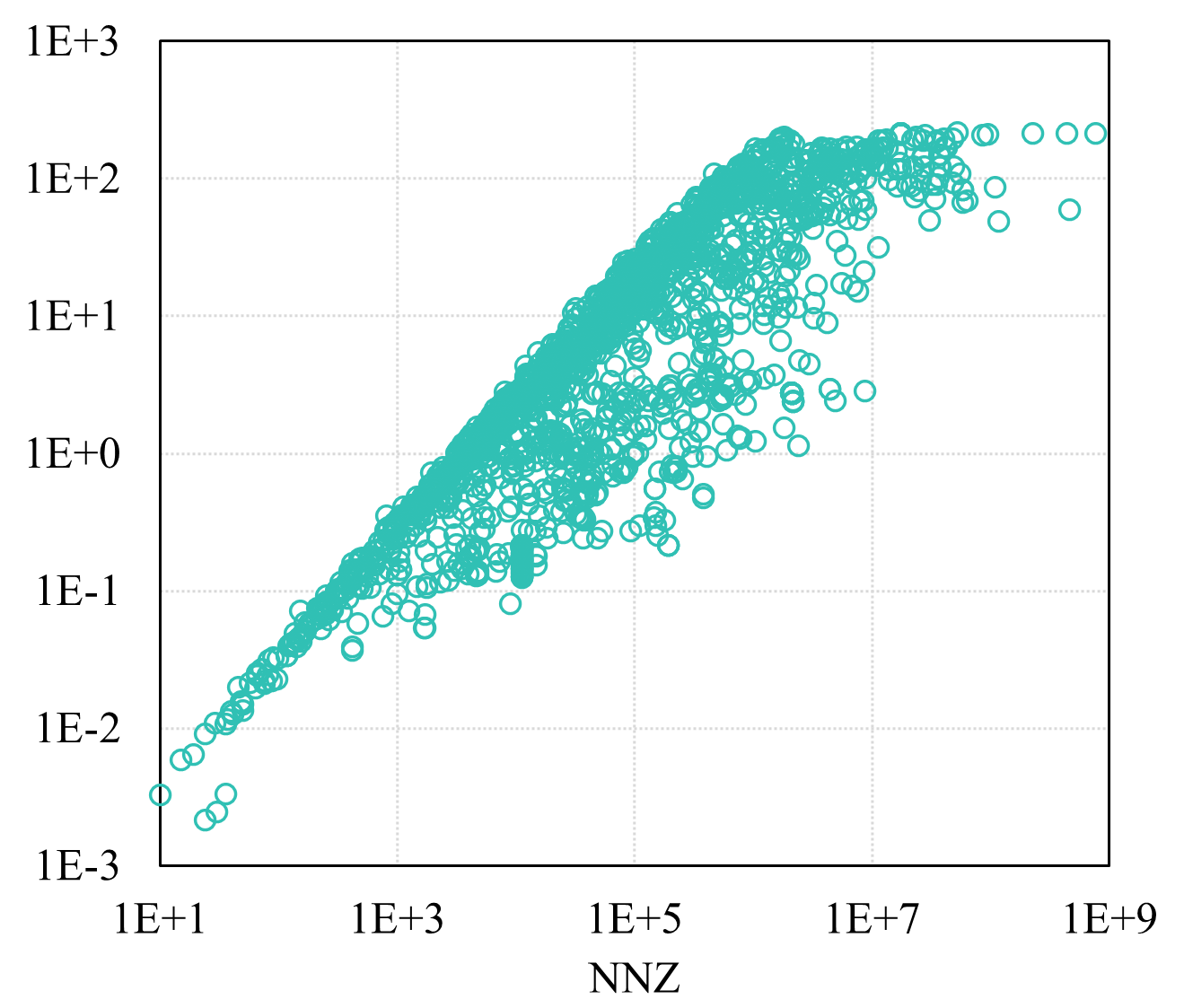}
	}
        \subfloat[CSR5] {
		\includegraphics[width=0.33\linewidth]{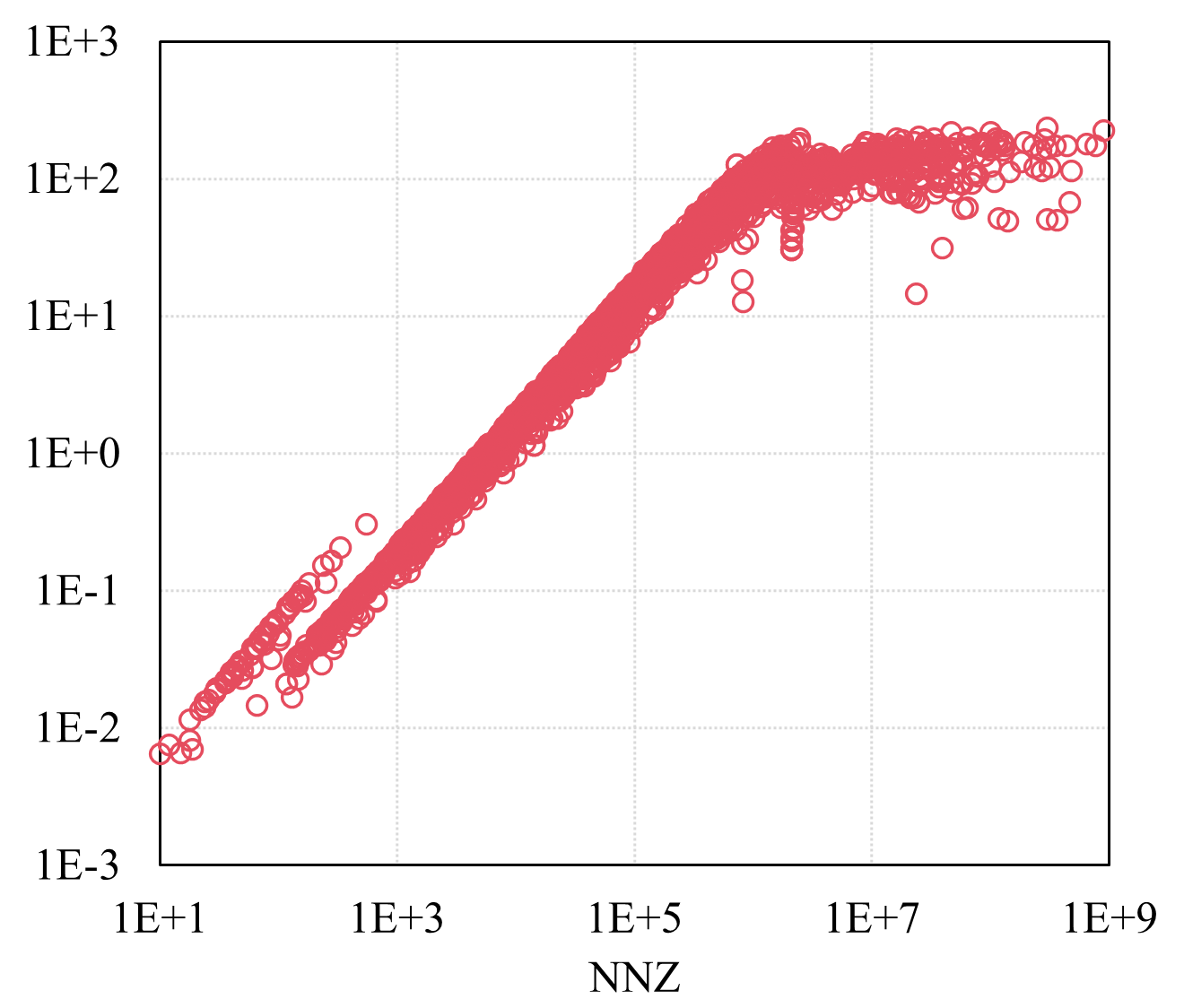}
	}
        \subfloat[Ginkgo] {
		\includegraphics[width=0.33\linewidth]{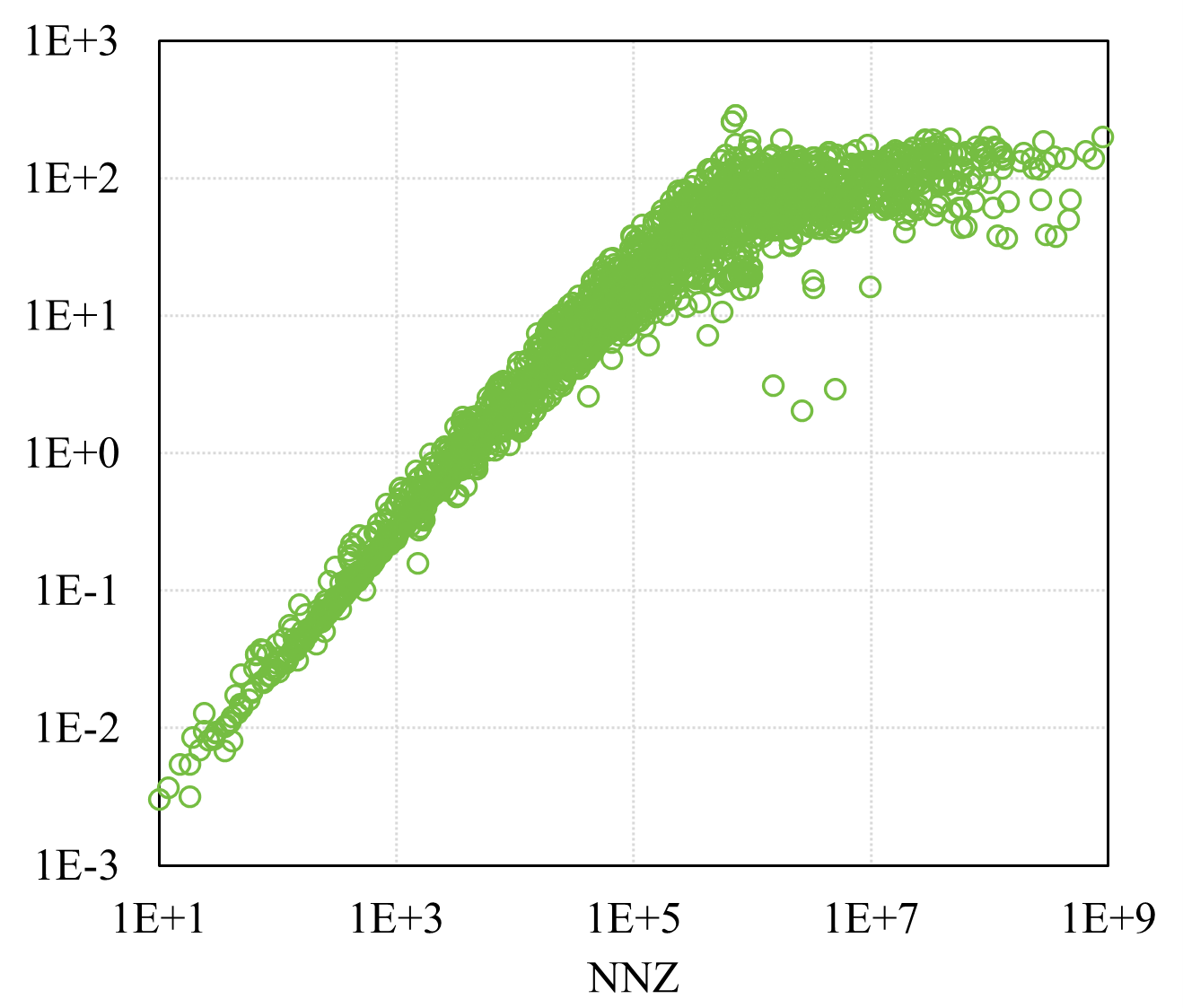}
	}
	\caption{GFLOPS of different SpMV algorithms on Tesla A100.}
	\label{fig:GFLOPS_A100}
\end{figure*}

Figures \ref{fig:GFLOPS_V100} and \ref{fig:GFLOPS_A100} illustrate the GFLOPS of the representative algorithms from these sparse libraries and two other mainstream algorithms on V100 and A100, respectively. The results are sorted in ascending order based on the number of non-zeros. It can be observed that all algorithms exhibit an increasing trend of GFLOPS with an increasing number of non-zeros. The performance of CUSP and MAGMA appears to be relatively unstable (as indicated by the scattered points towards the lower area of the figure), with lower GFLOPS compared with other algorithms for several matrices of similar NNZ, while the performance of cuSPARSE and Merge algorithms demonstrates more stability. Additionally, it can be observed that Merge, MAGMA, and CSR5 achieve higher GFLOPS than the other three algorithms for some small-sized sparse matrices. Table \ref{tab:speedup} presents the average and maximum speedups of the other SpMV algorithms with respect to cuSPARSE. Table \ref{tab:all-PLUB} presents the average PLUB and the number of matrices for which each algorithm achieves the highest GFLOPS. We can observe that MAGMA achieves the highest average speedup on V100, while on A100, Ginkgo emerges as the winner. Ginkgo exhibits the lowest average PLUB on V100, while on A100, the Merge-based SpMV algorithm achieves the lowest average PLUB.

\begin{table}[!htbp]
    \centering
    \caption{Average and maximum speedups of CUSP, Merge, MAGMA, CSR5, and Ginkgo over cuSPARSE.}
    \scalebox{0.95}{
    \begin{tabular}{ccrrrrr}
    \toprule
        \multicolumn{2}{c}{\textbf{Algorithm}} & \textbf{CUSP} & \textbf{Merge} & \textbf{MAGMA} & \textbf{CSR5} & \textbf{Ginkgo} \\ \midrule
        \multirow{2}{*}{Average} & V100 & 1.07 & 1.38 & \textbf{1.80} & 1.21 & 1.58  \\ 
        ~ & A100 &  1.13 &  1.89 &  1.55 &  1.59 &  \textbf{1.97} \\ \midrule
        \multirow{2}{*}{Maximum} & V100 & 2.04 & \textbf{31.44} & 5.12 & 6.51 & 3.34 \\
        ~ & A100 & 2.01 & 5.79 & 4.04 & \textbf{6.96} & 5.49 \\
        \bottomrule
    \end{tabular}}
    \label{tab:speedup}
\end{table}

\begin{table}[!htbp]
    \centering
    \caption{Average PLUB and the number of sparse matrices ($N_h$) for which each SpMV algorithm achieves the highest GFLOPS.}
    \scalebox{0.8}{
    \begin{tabular}{ccrrrrrr}
    \toprule
        \multicolumn{2}{c}{\textbf{Algorithm}} & \textbf{cuSPARSE} & \textbf{CUSP} & \textbf{Merge} & \textbf{MAGMA} & \textbf{CSR5} & \textbf{Ginkgo} \\ \midrule
        \multirow{2}{*}{PLUB(\%)} & V100 & 102 & 343 & 47 & 233 & 79 & \textbf{33} \\ 
        ~ & A100 & 131 & 243 & \textbf{22} & 389 & 52 & 36 \\ \midrule
        \multirow{2}{*}{$N_h$} & V100 & 72 & 33 & 358 & \textbf{1,071} & 406 & 891 \\
        ~ & A100 & 11 & 31 & 673 & 509 & 396 & \textbf{1,218} \\
        \bottomrule
    \end{tabular}}
    \label{tab:all-PLUB}
\end{table}

\section{Challenges and Future Work}\label{sec:challenge}

\subsection{Preprocessing Overhead Awareness}
The performance of SpMV is heavily impacted by the choice of compression format for sparse matrices. Consequently, researchers propose a number of new compression formats or introduce preprocessing operations like reordering to enhance memory access and computational efficiency in SpMV computations. Although these preprocessing operations can lead to significant performance improvements in executing the SpMV kernel, the gains are somewhat offset by the high overhead associated with preprocessing or format conversion. This limitation constrains the applicability of the proposed optimization algorithms, rendering them suitable primarily for applications with a substantial number of iterations to offset the overhead. Current evaluations of SpMV optimization research either neglect to assess the preprocessing overhead or provide such assessments without actual application verification. In the latter case, the preprocessing overhead is typically presented as an average number of SpMV kernels. Table \ref{tab:overhead} summaries the normalized overhead, reported in existing work, in terms of a single SpMV running time. We can observe that the preprocessing overhead is approximately equal to several, tens, or even hundreds of times the running time of a single SpMV operation. The number of iterations in SpMV-dominant iterative algorithms depends on the specific problem, and it is possible to converge in just tens or hundreds of iterations, while in some cases, it may require thousands of iterations without convergence. Therefore, application verification that considers the preprocessing overhead for each tested sparse matrix is more persuasive than just presenting a rough average number of SpMV kernels.

\begin{table}[!ht]
    \centering
    \caption{Preprocessing overhead (the ratio between the preprocessing time and a single SpMV running time)}\label{tab:overhead}
    \begin{tabular}{cl|cl}
        \toprule
        \textbf{Overhead} & \textbf{Contribution} & \textbf{Overhead} & \textbf{Contribution}\\ \midrule    
        3 & ACSR \cite{DBLP:conf/sc/AshariSEPS14} & 40 & OSKI \cite{vuduc2005oski} \\ 
        4 & VBSF \cite{DBLP:journals/tjs/LiXCLYLGGX20} & 68 & MMSparse \cite{TAN2020521} \\ 
        5 & SURAA \cite{SURAA} & 75 & BestSF \cite{Akrem-BestSF} \\
        7 & SMAT \cite{SMAT2013} & 215 & Cray reordering \cite{DBLP:conf/sc/TrotterELTDIU23} \\
        12 & CSR2 \cite{DBLP:conf/ccgrid/BianHDLW20} & 484 & pOSKI \cite{byun2012autotuning} \\
        10$\sim$20 & CSR5 \cite{CSR5} & 304 & AMB \cite{2016_AMB} \\
        5$\sim$20 & SMARTER \cite{DBLP:journals/toms/TanLL18} & $\le$10 & TileSpMV \cite{DBLP:journals/isci/ChenXWTL20} \\
        25 & clSpMV \cite{cocktail2012} & 2$\sim$280 & Ginkgo \cite{AnztCYDFNTTW20} \\
        38 & GOSpMV \cite{zhang2009automatic} & 500$\sim$2,000 & \cite{chen2022explicit} \\
        \bottomrule
    \end{tabular}
\end{table}

\begin{figure*}[!h]
	\centering
	\includegraphics[width=0.6\textwidth]{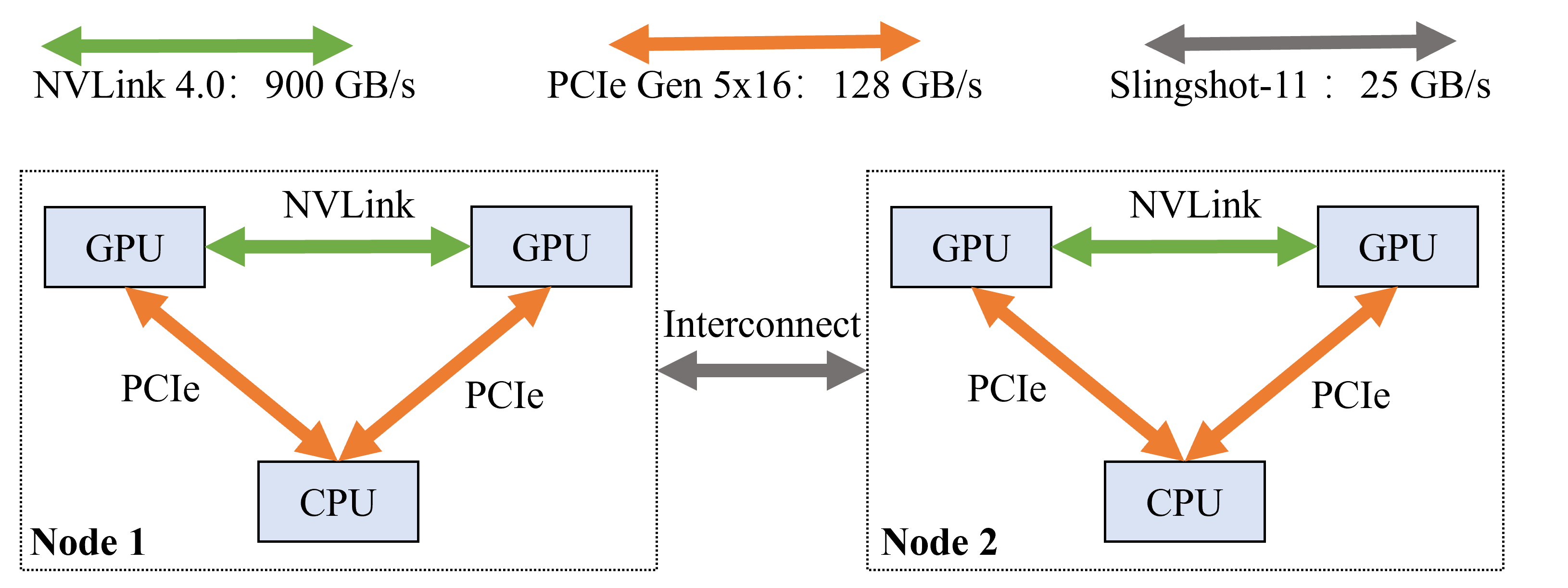} 
	\caption{Data transmission in distributed computing.}
	\label{fig:transfer}
\end{figure*}

\subsection{Transmission Overhead Redunction}
The demand for large-scale computing in scientific and engineering fields has led to increased attention on GPU-oriented SpMV optimization research. Compared with modern multi-core platforms, GPUs with many-core processing units excel at processing large-scale, highly parallel applications, resulting in significant performance improvements for SpMV. However, a notable drawback is the requirement to transfer data involved in the calculations. Figure \ref{fig:transfer} presents an overview of data transmission in GPU-centered high-performance computing platform. In addition to inter-node communication, which also exists in CPU-centered cluster computing, GPU computing introduces communication between CPU and GPU via PCIe, as well as communication between GPUs via NVLink. The gap between limited communication bandwidth and powerful computational capabilities restricts the performance of applications. For example, investigations by Gao et al. \cite{DBLP:journals/tpds/GaoJTWS22} have revealed that although the transmission of sparse matrices between CPU and GPU only occurs once in iterative algorithms, its overhead is comparable to dozens or even hundreds of kernel execution for some sparse matrices. Besides, in iterative solvers dominated by the SpMV operator, communication between multiple GPUs and across multiple nodes is required at each iteration, significantly limiting the performance of the iterative algorithm. Presently, most of the research focuses on optimizing the SpMV kernel execution, there is a need for more research focused on optimizing the communication process, such as data compression, computation and communication overlap, and communication-minimized matrix partitioning and task scheduling.

\subsection{Generative Algorithm Design}

The performance of the SpMV algorithm is greatly impacted by the distribution of non-zeros in sparse matrices. Presently, most SpMV algorithms are developed based on well-known sparse matrix distributions. However, there may exist undiscovered or underexplored sparse matrix distributions that could necessitate the redesign of compression formats or SpMV algorithms. This task can be laborious and time-consuming, even for experts. Furthermore, there is no guarantee that the existing compression formats or SpMV algorithms for certain sparse matrix distributions are optimal, as the current performance of SpMV still falls short of the theoretical peak of hardware platforms.

Consequently, there is a strong appeal and potential for designing a framework that can automatically generate sparse compression formats or SpMV algorithms. Du et al. \cite{DBLP:conf/sc/DuLWLTS22} have made a promising initial attempt in this direction. However, this field is still in its early stages, and there are numerous intricate details that require careful consideration and further research.

\subsection{Mixed-Precision Optimization}

The rapid advancement of artificial intelligence (AI) has resulted in increased hardware support for low-precision calculations, driving the development of mixed-precision optimization technologies that combine low-precision and high-precision representation and computation. Mixed-precision optimization has achieved remarkable success in the AI domain, where mainstream intelligent models are trained using a combination of 16-bit floating-point format (FP16) and single-precision floating-point format (FP32), rather than the traditional unified FP32 precision.

However, the application of mixed-precision optimization in SpMV calculations is still in its early stages. In contrast to AI, high-performance computing applications are more sensitive to and have higher precision requirements for floating-point operations. This presents significant challenges for mixed-precision-based SpMV optimization. On the software side, mixed-precision optimization can introduce divergence in iterative solvers. While careful numerical analysis enables the use of low-precision representations for the unchanged sparse matrix during the iteration process, accurately determining the appropriate precision representation for the changing multiplication vector $\vect{x}$ poses a challenge. On the hardware side, current mainstream hardware has limited support for low precision. When designing mixed-precision optimization algorithms, researchers must consider the hardware's support for low-precision representation and computations or resort to software simulation to validate their ideas. Moreover, the many-core processors, such as GPUs, exhibit different computing capabilities for different precisions due to the varying number of hardware units dedicated to each precision. Therefore, it is crucial to consider the disparities in underlying hardware when evaluating the performance of mixed-precision optimization techniques.

\subsection{Application-Oriented Optimization}
The research on universal SpMV has a longstanding history, with researchers dedicating extensive efforts to propose numerous new formats and algorithms. However, the diminishing performance benefits achieved through this approach highlight the need for further exploration in optimizing SpMV for specific application fields. This is largely due to the distinct distribution of non-zeros in sparse matrices across various applications, as well as the varying shapes of matrices.

For instance, when discretizing partial differential equations using the finite difference method (FDM), the coefficient matrix typically exhibits diagonal dominance. In graph applications, where many sparse matrices have fixed non-zeros set to 1, the storage of value arrays is unnecessary, resulting in a significant reduction in memory accesses during SpMV calculations. Moving forward, it is highly desirable to witness more SpMV optimizations tailored to specific applications, leveraging the unique characteristics and requirements of each domain.

\section{Conclusion}\label{sec:conclusion}
SpMV is widely applicable and significantly impacts the performance of various applications, which has garnered considerable attention from researchers over the past two decades. In this literature survey, we first highlight two typical applications of SpMV. Subsequently, we detailedly analyze compressed formats for sparse matrices, as they are the foundation of numerous optimization algorithms. We then delve into different aspects of SpMV optimization, including classical optimization algorithm design, auto-tuning and machine learning-based approaches, mix-precision optimization, and architecture-oriented optimization. We also present a comprehensive experimental evaluation that compares the performance of state-of-the-art SpMV implementations. This discussion and analysis of existing work lead to several challenges and future work that we conclude with. 

\bibliographystyle{IEEEtran}
\bibliography{IEEEabrv,sample-base.bib}

\begin{thebibliography}{100}
\providecommand{\url}[1]{#1}
\csname url@samestyle\endcsname
\providecommand{\newblock}{\relax}
\providecommand{\bibinfo}[2]{#2}
\providecommand{\BIBentrySTDinterwordspacing}{\spaceskip=0pt\relax}
\providecommand{\BIBentryALTinterwordstretchfactor}{4}
\providecommand{\BIBentryALTinterwordspacing}{\spaceskip=\fontdimen2\font plus
\BIBentryALTinterwordstretchfactor\fontdimen3\font minus
  \fontdimen4\font\relax}
\providecommand{\BIBforeignlanguage}[2]{{%
\expandafter\ifx\csname l@#1\endcsname\relax
\typeout{** WARNING: IEEEtran.bst: No hyphenation pattern has been}%
\typeout{** loaded for the language `#1'. Using the pattern for}%
\typeout{** the default language instead.}%
\else
\language=\csname l@#1\endcsname
\fi
#2}}
\providecommand{\BIBdecl}{\relax}
\BIBdecl

\bibitem{kulkarni2014survey}
M.~A.~V. Kulkarni and P.~Barde, ``A survey on performance modelling and
  optimization techniques for spmv on gpus,'' \emph{Int. J. Comput. Sci. Inf.
  Technol}, vol.~5, pp. 7577--7582, 2014.

\bibitem{grossman2016survey}
M.~Grossman, C.~Thiele, M.~Araya-Polo, F.~Frank, F.~O. Alpak, and V.~Sarkar,
  ``A survey of sparse matrix-vector multiplication performance on large
  matrices,'' \emph{arXiv preprint arXiv:1608.00636}, 2016.

\bibitem{DBLP:journals/toms/FilipponeCBF17}
\BIBentryALTinterwordspacing
S.~Filippone, V.~Cardellini, D.~Barbieri, and A.~Fanfarillo, ``Sparse
  matrix-vector multiplication on gpgpus,'' \emph{{ACM} Trans. Math. Softw.},
  vol.~43, no.~4, pp. 30:1--30:49, 2017. [Online]. Available:
  \url{https://doi.org/10.1145/3017994}
\BIBentrySTDinterwordspacing

\bibitem{wang2020research}
Q.~Wang, M.~Li, J.~Pang, and D.~Zhu, ``Research on performance optimization for
  sparse matrix-vector multiplication in multi/many-core architecture,'' in
  \emph{2020 2nd International Conference on Information Technology and
  Computer Application (ITCA)}.\hskip 1em plus 0.5em minus 0.4em\relax IEEE,
  2020, pp. 350--362.

\bibitem{10.1145/3604606}
\BIBentryALTinterwordspacing
G.~Xiao, C.~Yin, T.~Zhou, X.~Li, Y.~Chen, and K.~Li, ``A survey of accelerating
  parallel sparse linear algebra,'' \emph{ACM Comput. Surv.}, vol.~56, no.~1,
  aug 2023. [Online]. Available: \url{https://doi.org/10.1145/3604606}
\BIBentrySTDinterwordspacing

\bibitem{DBLP:conf/sc/FuZWLLYZGLZ0023}
\BIBentryALTinterwordspacing
X.~Fu, B.~Zhang, T.~Wang, W.~Li, Y.~Lu, E.~Yi, J.~Zhao, X.~Geng, F.~Li,
  J.~Zhang, Z.~Jin, and W.~Liu, ``Pangulu: {A} scalable regular two-dimensional
  block-cyclic sparse direct solver on distributed heterogeneous systems,'' in
  \emph{Proceedings of the International Conference for High Performance
  Computing, Networking, Storage and Analysis, {SC} 2023, Denver, CO, USA,
  November 12-17, 2023}, D.~Arnold, R.~M. Badia, and K.~M. Mohror, Eds.\hskip
  1em plus 0.5em minus 0.4em\relax {ACM}, 2023, pp. 51:1--51:14. [Online].
  Available: \url{https://doi.org/10.1145/3581784.3607050}
\BIBentrySTDinterwordspacing

\bibitem{saad1986gmres}
Y.~Saad and M.~H. Schultz, ``Gmres: A generalized minimal residual algorithm
  for solving nonsymmetric linear systems,'' \emph{SIAM Journal on scientific
  and statistical computing}, vol.~7, no.~3, pp. 856--869, 1986.

\bibitem{DBLP:journals/jcam/SuzukiFI23}
\BIBentryALTinterwordspacing
K.~Suzuki, T.~Fukaya, and T.~Iwashita, ``A novel {ILU} preconditioning method
  with a block structure suitable for {SIMD} vectorization,'' \emph{J. Comput.
  Appl. Math.}, vol. 419, p. 114687, 2023. [Online]. Available:
  \url{https://doi.org/10.1016/j.cam.2022.114687}
\BIBentrySTDinterwordspacing

\bibitem{DBLP:journals/csur/ShiZZJHLH18}
\BIBentryALTinterwordspacing
X.~Shi, Z.~Zheng, Y.~Zhou, H.~Jin, L.~He, B.~Liu, and Q.~Hua, ``Graph
  processing on gpus: {A} survey,'' \emph{{ACM} Comput. Surv.}, vol.~50, no.~6,
  pp. 81:1--81:35, 2018. [Online]. Available:
  \url{https://doi.org/10.1145/3128571}
\BIBentrySTDinterwordspacing

\bibitem{DBLP:journals/kais/TongFP08}
\BIBentryALTinterwordspacing
H.~Tong, C.~Faloutsos, and J.~Pan, ``Random walk with restart: fast solutions
  and applications,'' \emph{Knowl. Inf. Syst.}, vol.~14, no.~3, pp. 327--346,
  2008. [Online]. Available: \url{https://doi.org/10.1007/s10115-007-0094-2}
\BIBentrySTDinterwordspacing

\bibitem{DBLP:conf/sc/AshariSEPS14}
\BIBentryALTinterwordspacing
A.~Ashari, N.~Sedaghati, J.~Eisenlohr, S.~Parthasarathy, and P.~Sadayappan,
  ``Fast sparse matrix-vector multiplication on gpus for graph applications,''
  in \emph{International Conference for High Performance Computing, Networking,
  Storage and Analysis, {SC} 2014, New Orleans, LA, USA, November 16-21, 2014},
  T.~Damkroger and J.~J. Dongarra, Eds.\hskip 1em plus 0.5em minus 0.4em\relax
  {IEEE} Computer Society, 2014, pp. 781--792. [Online]. Available:
  \url{https://doi.org/10.1109/SC.2014.69}
\BIBentrySTDinterwordspacing

\bibitem{DBLP:conf/icpp/WuWSYWX10}
\BIBentryALTinterwordspacing
T.~Wu, B.~Wang, Y.~Shan, F.~Yan, Y.~Wang, and N.~Xu, ``Efficient pagerank and
  spmv computation on {AMD} gpus,'' in \emph{39th International Conference on
  Parallel Processing, {ICPP} 2010, San Diego, California, USA, 13-16 September
  2010}.\hskip 1em plus 0.5em minus 0.4em\relax {IEEE} Computer Society, 2010,
  pp. 81--89. [Online]. Available: \url{https://doi.org/10.1109/ICPP.2010.17}
\BIBentrySTDinterwordspacing

\bibitem{DBLP:journals/jacm/Kleinberg99}
\BIBentryALTinterwordspacing
J.~M. Kleinberg, ``Authoritative sources in a hyperlinked environment,''
  \emph{J. {ACM}}, vol.~46, no.~5, pp. 604--632, 1999. [Online]. Available:
  \url{https://doi.org/10.1145/324133.324140}
\BIBentrySTDinterwordspacing

\bibitem{DBLP:journals/cn/BrinP12}
\BIBentryALTinterwordspacing
S.~Brin and L.~Page, ``Reprint of: The anatomy of a large-scale hypertextual
  web search engine,'' \emph{Comput. Networks}, vol.~56, no.~18, pp.
  3825--3833, 2012. [Online]. Available:
  \url{https://doi.org/10.1016/j.comnet.2012.10.007}
\BIBentrySTDinterwordspacing

\bibitem{paolini1989data}
G.~V. Paolini and G.~R.~D. Brozolo, ``Data structures to vectorize cg
  algorithms for general sparsity patterns,'' \emph{BIT Numerical Mathematics},
  vol.~29, no.~4, pp. 703--718, 1989.

\bibitem{DBLP:journals/pc/Peters91}
\BIBentryALTinterwordspacing
A.~Peters, ``Sparse matrix vector multiplication techniques on the {IBM} 3090
  {VF},'' \emph{Parallel Comput.}, vol.~17, no.~12, pp. 1409--1424, 1991.
  [Online]. Available: \url{https://doi.org/10.1016/S0167-8191(05)80007-9}
\BIBentrySTDinterwordspacing

\bibitem{saad1994sparskit}
Y.~Saad, ``{SPARSKIT: A Basic Took Kit for Sparse Matrix Computations, Version
  2},'' \emph{http://www. cs. umn. edu/saad/software/SPARSKIT/sparskit. html},
  1994.

\bibitem{BellG09}
\BIBentryALTinterwordspacing
N.~Bell and M.~Garland, ``{Implementing Sparse Matrix-Vector Multiplication on
  Throughput-Oriented Processors},'' in \emph{Proceedings of the {ACM/IEEE}
  Conference on High Performance Computing, {SC}, November 14-20, 2009,
  Portland, Oregon, {USA}}.\hskip 1em plus 0.5em minus 0.4em\relax {ACM}, 2009,
  pp. 1--11. [Online]. Available: \url{https://doi.org/10.1145/1654059.1654078}
\BIBentrySTDinterwordspacing

\bibitem{Sengupta_Harris_Zhang_Owens_2007}
S.~Sengupta, M.~Harris, Y.~Zhang, and J.~D. Owens, ``{Scan Primitives for GPU
  Computing},'' in \emph{Proceedings of the 22nd ACM SIGGRAPH/EUROGRAPHICS
  Symposium on Graphics Hardware}, ser. GH'07.\hskip 1em plus 0.5em minus
  0.4em\relax Goslar, DEU: Eurographics Association, 2007, p. 97–106.

\bibitem{Garland_2008}
\BIBentryALTinterwordspacing
M.~Garland, ``{Sparse Matrix Computations on Manycore GPU’s},'' in
  \emph{Proceedings of the 45th annual Design Automation Conference}, ser. DAC
  ’08.\hskip 1em plus 0.5em minus 0.4em\relax New York, NY, USA: Association
  for Computing Machinery, Jun 2008, p. 2–6. [Online]. Available:
  \url{https://doi.org/10.1145/1391469.1391473}
\BIBentrySTDinterwordspacing

\bibitem{DBLP:journals/procedia/DangS12}
\BIBentryALTinterwordspacing
H.~Dang and B.~Schmidt, ``{The Sliced COO Format for Sparse Matrix-Vector
  Multiplication on CUDA-Enabled GPUs},'' in \emph{Proceedings of the
  International Conference on Computational Science, {ICCS}, Omaha, Nebraska,
  USA, 4-6 June, 2012}, ser. Procedia Computer Science, H.~H. Ali, Y.~Shi,
  D.~Khazanchi, M.~Lees, G.~D. van Albada, J.~J. Dongarra, and P.~M.~A. Sloot,
  Eds., vol.~9.\hskip 1em plus 0.5em minus 0.4em\relax Elsevier, 2012, pp.
  57--66. [Online]. Available:
  \url{https://doi.org/10.1016/j.procs.2012.04.007}
\BIBentrySTDinterwordspacing

\bibitem{DBLP:conf/iccS/DAzevedoFM05}
\BIBentryALTinterwordspacing
E.~F. D'Azevedo, M.~R. Fahey, and R.~T. Mills, ``Vectorized sparse matrix
  multiply for compressed row storage format,'' in \emph{Computational Science
  - {ICCS} 2005, 5th International Conference, Atlanta, GA, USA, May 22-25,
  2005, Proceedings, Part {I}}, ser. Lecture Notes in Computer Science, V.~S.
  Sunderam, G.~D. van Albada, P.~M.~A. Sloot, and J.~J. Dongarra, Eds., vol.
  3514.\hskip 1em plus 0.5em minus 0.4em\relax Springer, 2005, pp. 99--106.
  [Online]. Available: \url{https://doi.org/10.1007/11428831\_13}
\BIBentrySTDinterwordspacing

\bibitem{DBLP:conf/hipeac/MonakovLA10}
\BIBentryALTinterwordspacing
A.~Monakov, A.~Lokhmotov, and A.~Avetisyan, ``{Automatically Tuning Sparse
  Matrix-Vector Multiplication for GPU Architectures},'' in \emph{High
  Performance Embedded Architectures and Compilers, 5th International
  Conference, HiPEAC 2010, Pisa, Italy, January 25-27, 2010. Proceedings}, ser.
  Lecture Notes in Computer Science, Y.~N. Patt, P.~Foglia, E.~Duesterwald,
  P.~Faraboschi, and X.~Martorell, Eds., vol. 5952.\hskip 1em plus 0.5em minus
  0.4em\relax Springer, 2010, pp. 111--125. [Online]. Available:
  \url{https://doi.org/10.1007/978-3-642-11515-8\_10}
\BIBentrySTDinterwordspacing

\bibitem{barbieri2015three}
D.~Barbieri, V.~Cardellini, A.~Fanfarillo, and S.~Filippone, ``{Three Storage
  Formats for Sparse Matrices on GPGPUs},'' 2015.

\bibitem{pJDS2012}
M.~Kreutzer, G.~Hager, G.~Wellein, H.~Fehske, A.~Basermann, and A.~R. Bishop,
  ``{Sparse Matrix-Vector Multiplication on GPGPU Clusters: A New Storage
  Format and a Scalable Implementation},'' in \emph{2012 IEEE 26th
  International Parallel and Distributed Processing Symposium Workshops PhD
  Forum}, 2012, pp. 1696--1702.

\bibitem{kreutzer2014unified}
M.~Kreutzer, G.~Hager, G.~Wellein, H.~Fehske, and A.~R. Bishop, ``{A Unified
  Sparse Matrix Data Format for Efficient General Sparse Matrix-Vector
  Multiplication on Modern Processors With Wide SIMD Units},'' \emph{SIAM
  Journal on Scientific Computing}, vol.~36, no.~5, pp. C401--C423, 2014.

\bibitem{DBLP:conf/hpcc/YuanZSW10}
\BIBentryALTinterwordspacing
L.~Yuan, Y.~Zhang, X.~Sun, and T.~Wang, ``{Optimizing Sparse Matrix Vector
  Multiplication Using Diagonal Storage Matrix Format},'' in \emph{12th {IEEE}
  International Conference on High Performance Computing and Communications,
  {HPCC}, 1-3 September 2010, Melbourne, Australia}.\hskip 1em plus 0.5em minus
  0.4em\relax {IEEE}, 2010, pp. 585--590. [Online]. Available:
  \url{https://doi.org/10.1109/HPCC.2010.67}
\BIBentrySTDinterwordspacing

\bibitem{DBLP:conf/europar/SunZWLZL11}
\BIBentryALTinterwordspacing
X.~Sun, Y.~Zhang, T.~Wang, G.~Long, X.~Zhang, and Y.~Li, ``{CRSD: Application
  Specific Auto-tuning of SpMV for Diagonal Sparse Matrices},'' in
  \emph{Euro-Par Parallel Processing - 17th International Conference, Euro-Par,
  Bordeaux, France, August 29 - September 2, 2011, Proceedings, Part {II}},
  ser. Lecture Notes in Computer Science, vol. 6853.\hskip 1em plus 0.5em minus
  0.4em\relax Springer, 2011, pp. 316--327. [Online]. Available:
  \url{https://doi.org/10.1007/978-3-642-23397-5\_32}
\BIBentrySTDinterwordspacing

\bibitem{DBLP:conf/icpp/SunZWZYR11}
\BIBentryALTinterwordspacing
X.~Sun, Y.~Zhang, T.~Wang, X.~Zhang, L.~Yuan, and L.~Rao, ``{Optimizing SpMV
  for Diagonal Sparse Matrices on GPU},'' in \emph{International Conference on
  Parallel Processing, {ICPP}, Taipei, Taiwan, September 13-16, 2011}.\hskip
  1em plus 0.5em minus 0.4em\relax {IEEE} Computer Society, 2011, pp. 492--501.
  [Online]. Available: \url{https://doi.org/10.1109/ICPP.2011.53}
\BIBentrySTDinterwordspacing

\bibitem{DBLP:conf/sc/PinarH99}
\BIBentryALTinterwordspacing
A.~Pinar and M.~T. Heath, ``Improving performance of sparse matrix-vector
  multiplication,'' in \emph{Proceedings of the {ACM/IEEE} Conference on
  Supercomputing, {SC} 1999, November 13-19, 1999, Portland, Oregon,
  {USA}}.\hskip 1em plus 0.5em minus 0.4em\relax {ACM}, 1999, p.~30. [Online].
  Available: \url{https://doi.org/10.1145/331532.331562}
\BIBentrySTDinterwordspacing

\bibitem{DBLP:journals/tjs/LiXCLYLGGX20}
\BIBentryALTinterwordspacing
Y.~Li, P.~Xie, X.~Chen, J.~Liu, B.~Yang, S.~Li, C.~Gong, X.~Gan, and H.~Xu,
  ``{VBSF:} a new storage format for {SIMD} sparse matrix-vector multiplication
  on modern processors,'' \emph{The Journal of Supercomputing}, vol.~76, no.~3,
  pp. 2063--2081, 2020. [Online]. Available:
  \url{https://doi.org/10.1007/s11227-019-02835-4}
\BIBentrySTDinterwordspacing

\bibitem{vuduc2003automatic}
R.~W. Vuduc, \emph{{Automatic Performance Tuning of Sparse Matrix
  Kernels}}.\hskip 1em plus 0.5em minus 0.4em\relax University of California,
  Berkeley, 2003.

\bibitem{AshariSES14_BRC}
\BIBentryALTinterwordspacing
A.~Ashari, N.~Sedaghati, J.~Eisenlohr, and P.~Sadayappan, ``{An Efficient
  Two-Dimensional Blocking Strategy for Sparse Matrix-Vector Multiplication on
  GPUs},'' in \emph{International Conference on Supercomputing, ICS'14,
  Muenchen, Germany, June 10-13, 2014}, A.~Bode, M.~Gerndt, P.~Stenstr{\"{o}}m,
  L.~Rauchwerger, B.~P. Miller, and M.~Schulz, Eds.\hskip 1em plus 0.5em minus
  0.4em\relax {ACM}, 2014, pp. 273--282. [Online]. Available:
  \url{https://doi.org/10.1145/2597652.2597678}
\BIBentrySTDinterwordspacing

\bibitem{DBLP:conf/reconfig/Jain-MendonS12}
\BIBentryALTinterwordspacing
S.~Jain{-}Mendon and R.~Sass, ``A case study of streaming storage format for
  sparse matrices,'' in \emph{2012 International Conference on Reconfigurable
  Computing and FPGAs, ReConFig 2012, Cancun, Mexico, December 5-7,
  2012}.\hskip 1em plus 0.5em minus 0.4em\relax {IEEE}, 2012, pp. 1--6.
  [Online]. Available: \url{https://doi.org/10.1109/ReConFig.2012.6416788}
\BIBentrySTDinterwordspacing

\bibitem{Yan2014}
\BIBentryALTinterwordspacing
S.~Yan, C.~Li, Y.~Zhang, and H.~Zhou, ``{yaSpMV: Yet Another SpMV Framework on
  GPUs},'' in \emph{Proceedings of the 19th ACM SIGPLAN Symposium on Principles
  and Practice of Parallel Programming}, ser. PPoPP '14.\hskip 1em plus 0.5em
  minus 0.4em\relax New York, NY, USA: ACM, 2014, pp. 107--118. [Online].
  Available: \url{http://doi.acm.org/10.1145/2555243.2555255}
\BIBentrySTDinterwordspacing

\bibitem{DBLP:conf/asplos/GodwinHS12}
\BIBentryALTinterwordspacing
J.~Godwin, J.~Holewinski, and P.~Sadayappan, ``{High-Performance Sparse
  Matrix-Vector Multiplication on GPUs for Structured Grid Computations},'' in
  \emph{The 5th Annual Workshop on General Purpose Processing with Graphics
  Processing Units, GPGPU-5, London, United Kingdom, March 3}, D.~R. Kaeli,
  J.~Cavazos, and E.~Sun, Eds.\hskip 1em plus 0.5em minus 0.4em\relax {ACM},
  2012, pp. 47--56. [Online]. Available:
  \url{https://doi.org/10.1145/2159430.2159436}
\BIBentrySTDinterwordspacing

\bibitem{DBLP:conf/ppopp/ChoiSV10}
\BIBentryALTinterwordspacing
J.~Choi, A.~Singh, and R.~W. Vuduc, ``{Model-Driven Autotuning of Sparse
  Matrix-Vector Multiply on GPUs},'' in \emph{Proceedings of the 15th {ACM}
  {SIGPLAN} Symposium on Principles and Practice of Parallel Programming,
  {PPOPP}, Bangalore, India, January 9-14, 2010}.\hskip 1em plus 0.5em minus
  0.4em\relax {ACM}, 2010, pp. 115--126. [Online]. Available:
  \url{https://doi.org/10.1145/1693453.1693471}
\BIBentrySTDinterwordspacing

\bibitem{CSR5}
\BIBentryALTinterwordspacing
W.~Liu and B.~Vinter, ``{CSR5: An Efficient Storage Format for Cross-Platform
  Sparse Matrix-Vector Multiplication},'' in \emph{Proceedings of the 29th
  {ACM} on International Conference on Supercomputing, ICS'15, Newport
  Beach/Irvine, CA, USA, June 08 - 11, 2015}, L.~N. Bhuyan, F.~Chong, and
  V.~Sarkar, Eds.\hskip 1em plus 0.5em minus 0.4em\relax {ACM}, 2015, pp.
  339--350. [Online]. Available: \url{https://doi.org/10.1145/2751205.2751209}
\BIBentrySTDinterwordspacing

\bibitem{DBLP:conf/ccgrid/BianHDLW20}
\BIBentryALTinterwordspacing
H.~Bian, J.~Huang, R.~Dong, L.~Liu, and X.~Wang, ``{CSR2: A New Format for
  SIMD-Accelerated SpMV},'' in \emph{20th {IEEE/ACM} International Symposium on
  Cluster, Cloud and Internet Computing, {CCGRID} 2020, Melbourne, Australia,
  May 11-14, 2020}.\hskip 1em plus 0.5em minus 0.4em\relax {IEEE}, 2020, pp.
  350--359. [Online]. Available:
  \url{https://doi.org/10.1109/CCGrid49817.2020.00-58}
\BIBentrySTDinterwordspacing

\bibitem{DBLP:conf/ppopp/KourtisKGK11}
\BIBentryALTinterwordspacing
K.~Kourtis, V.~Karakasis, G.~I. Goumas, and N.~Koziris, ``{CSX:} an extended
  compression format for spmv on shared memory systems,'' in \emph{Proceedings
  of the 16th {ACM} {SIGPLAN} Symposium on Principles and Practice of Parallel
  Programming, {PPOPP} 2011, San Antonio, TX, USA, February 12-16, 2011},
  C.~Cascaval and P.~Yew, Eds.\hskip 1em plus 0.5em minus 0.4em\relax {ACM},
  2011, pp. 247--256. [Online]. Available:
  \url{https://doi.org/10.1145/1941553.1941587}
\BIBentrySTDinterwordspacing

\bibitem{cocktail2012}
\BIBentryALTinterwordspacing
B.-Y. Su and K.~Keutzer, ``{ClSpMV: A Cross-Platform OpenCL SpMV Framework on
  GPUs},'' in \emph{Proceedings of the 26th ACM International Conference on
  Supercomputing}, ser. ICS'12.\hskip 1em plus 0.5em minus 0.4em\relax New
  York, NY, USA: Association for Computing Machinery, 2012, p. 353–364.
  [Online]. Available: \url{https://doi.org/10.1145/2304576.2304624}
\BIBentrySTDinterwordspacing

\bibitem{bell2008efficient}
N.~Bell and M.~Garland, ``{Efficient Sparse Matrix-Vector Multiplication on
  CUDA},'' NVIDIA Corporation, NVIDIA Technical Report NVR-2008-004, Dec. 2008.

\bibitem{AnztCYDFNTTW20}
\BIBentryALTinterwordspacing
H.~Anzt, T.~Cojean, C.~Yen{-}Chen, J.~J. Dongarra, G.~Flegar, P.~Nayak,
  S.~Tomov, Y.~M. Tsai, and W.~Wang, ``{Load-Balancing Sparse Matrix Vector
  Product Kernels on GPUs},'' \emph{ACM Transactions on Parallel Computing},
  vol.~7, no.~1, pp. 2:1--2:26, 2020. [Online]. Available:
  \url{https://doi.org/10.1145/3380930}
\BIBentrySTDinterwordspacing

\bibitem{DBLP:conf/hipc/MaringantiAP09}
\BIBentryALTinterwordspacing
A.~Maringanti, V.~Athavale, and S.~B. Patkar, ``{Acceleration of Conjugate
  Gradient Method for Circuit Simulation Using CUDA},'' in \emph{16th
  International Conference on High Performance Computing, December 16-19, 2009,
  Kochi, India, Proceedings}, Y.~Yang, M.~Parashar, R.~Muralidhar, and V.~K.
  Prasanna, Eds.\hskip 1em plus 0.5em minus 0.4em\relax {IEEE} Computer
  Society, 2009, pp. 438--444. [Online]. Available:
  \url{https://doi.org/10.1109/HIPC.2009.5433184}
\BIBentrySTDinterwordspacing

\bibitem{DBLP:journals/jcss/YangLL18}
\BIBentryALTinterwordspacing
W.~Yang, K.~Li, and K.~Li, ``A parallel computing method using blocked format
  with optimal partitioning for spmv on {GPU},'' \emph{J. Comput. Syst. Sci.},
  vol.~92, pp. 152--170, 2018. [Online]. Available:
  \url{https://doi.org/10.1016/j.jcss.2017.09.010}
\BIBentrySTDinterwordspacing

\bibitem{2016_AMB}
\BIBentryALTinterwordspacing
Y.~Nagasaka, A.~Nukada, and S.~Matsuoka, ``{Adaptive Multi-Level Blocking
  Optimization for Sparse Matrix Vector Multiplication on GPU},''
  \emph{Procedia Computer Science}, vol.~80, pp. 131--142, 2016, international
  Conference on Computational Science 2016, ICCS 2016, 6-8 June 2016, San
  Diego, California, USA. [Online]. Available:
  \url{https://www.sciencedirect.com/science/article/pii/S187705091630655X}
\BIBentrySTDinterwordspacing

\bibitem{bolz2003sparse}
J.~Bolz, I.~Farmer, E.~Grinspun, and P.~Schr{\"o}der, ``{Sparse Matrix Solvers
  on the GPU: Conjugate Gradients and Multigrid},'' \emph{ACM Transactions on
  Graphics (TOG)}, vol.~22, no.~3, pp. 917--924, 2003.

\bibitem{DBLP:journals/ijhpca/YangLLSW14}
\BIBentryALTinterwordspacing
W.~Yang, K.~Li, Y.~Liu, L.~Shi, and L.~Wan, ``{Optimization of Quasi-Diagonal
  Matrix-Vector Multiplication on GPU},'' \emph{The International Journal of
  High Performance Computing Applications}, vol.~28, no.~2, pp. 183--195, 2014.
  [Online]. Available: \url{https://doi.org/10.1177/1094342013501126}
\BIBentrySTDinterwordspacing

\bibitem{DBLP:journals/tpds/GaoJTWS22}
\BIBentryALTinterwordspacing
J.~Gao, W.~Ji, Z.~Tan, Y.~Wang, and F.~Shi, ``Taichi: {A} hybrid compression
  format for binary sparse matrix-vector multiplication on {GPU},''
  \emph{{IEEE} Trans. Parallel Distributed Syst.}, vol.~33, no.~12, pp.
  3732--3745, 2022. [Online]. Available:
  \url{https://doi.org/10.1109/TPDS.2022.3170501}
\BIBentrySTDinterwordspacing

\bibitem{DBLP:conf/cgo/TangZLLHLG15}
\BIBentryALTinterwordspacing
W.~T. Tang, R.~Zhao, M.~Lu, Y.~Liang, H.~P. Huyng, X.~Li, and R.~S.~M. Goh,
  ``Optimizing and auto-tuning scale-free sparse matrix-vector multiplication
  on intel xeon phi,'' in \emph{Proceedings of the 13th Annual {IEEE/ACM}
  International Symposium on Code Generation and Optimization, {CGO} 2015, San
  Francisco, CA, USA, February 07 - 11, 2015}, K.~Olukotun, A.~Smith, R.~Hundt,
  and J.~Mars, Eds.\hskip 1em plus 0.5em minus 0.4em\relax {IEEE} Computer
  Society, 2015, pp. 136--145. [Online]. Available:
  \url{https://doi.org/10.1109/CGO.2015.7054194}
\BIBentrySTDinterwordspacing

\bibitem{DBLP:conf/samos/MonakovA09}
\BIBentryALTinterwordspacing
A.~Monakov and A.~Avetisyan, ``Implementing blocked sparse matrix-vector
  multiplication on {NVIDIA} gpus,'' in \emph{Embedded Computer Systems:
  Architectures, Modeling, and Simulation, 9th International Workshop, {SAMOS}
  2009, Samos, Greece, July 20-23, 2009. Proceedings}, ser. Lecture Notes in
  Computer Science, K.~Bertels, N.~J. Dimopoulos, C.~Silvano, and S.~Wong,
  Eds., vol. 5657.\hskip 1em plus 0.5em minus 0.4em\relax Springer, 2009, pp.
  289--297. [Online]. Available:
  \url{https://doi.org/10.1007/978-3-642-03138-0\_32}
\BIBentrySTDinterwordspacing

\bibitem{TAN2020521}
\BIBentryALTinterwordspacing
Z.~Tan, W.~Ji, J.~Gao, Y.~Zhao, A.~Benatia, Y.~Wang, and F.~Shi, ``{MMSparse:
  2D Partitioning of Sparse Matrix Based on Mathematical Morphology},''
  \emph{Future Generation Computer Systems}, vol. 108, pp. 521 -- 532, 2020.
  [Online]. Available:
  \url{http://www.sciencedirect.com/science/article/pii/S0167739X19327967}
\BIBentrySTDinterwordspacing

\bibitem{NiuLDJ0T21_TileSpMV}
\BIBentryALTinterwordspacing
Y.~Niu, Z.~Lu, M.~Dong, Z.~Jin, W.~Liu, and G.~Tan, ``{TileSpMV: A Tiled
  Algorithm for Sparse Matrix-Vector Multiplication on GPUs},'' in \emph{35th
  {IEEE} International Parallel and Distributed Processing Symposium, {IPDPS}
  2021, Portland, OR, USA, May 17-21, 2021}.\hskip 1em plus 0.5em minus
  0.4em\relax {IEEE}, 2021, pp. 68--78. [Online]. Available:
  \url{https://doi.org/10.1109/IPDPS49936.2021.00016}
\BIBentrySTDinterwordspacing

\bibitem{DBLP:conf/ics/WillcockL06}
\BIBentryALTinterwordspacing
J.~Willcock and A.~Lumsdaine, ``Accelerating sparse matrix computations via
  data compression,'' in \emph{Proceedings of the 20th Annual International
  Conference on Supercomputing, {ICS} 2006, Cairns, Queensland, Australia, June
  28 - July 01, 2006}, G.~K. Egan and Y.~Muraoka, Eds.\hskip 1em plus 0.5em
  minus 0.4em\relax {ACM}, 2006, pp. 307--316. [Online]. Available:
  \url{https://doi.org/10.1145/1183401.1183444}
\BIBentrySTDinterwordspacing

\bibitem{DBLP:conf/cf/KourtisGK08}
\BIBentryALTinterwordspacing
K.~Kourtis, G.~I. Goumas, and N.~Koziris, ``Optimizing sparse matrix-vector
  multiplication using index and value compression,'' in \emph{Proceedings of
  the 5th Conference on Computing Frontiers, 2008, Ischia, Italy, May 5-7,
  2008}, A.~Ram{\'{\i}}rez, G.~Bilardi, and M.~Gschwind, Eds.\hskip 1em plus
  0.5em minus 0.4em\relax {ACM}, 2008, pp. 87--96. [Online]. Available:
  \url{https://doi.org/10.1145/1366230.1366244}
\BIBentrySTDinterwordspacing

\bibitem{TangTRWCKGTW13}
\BIBentryALTinterwordspacing
W.~T. Tang, W.~J. Tan, R.~Ray, Y.~W. Wong, W.~Chen, S.~Kuo, R.~S.~M. Goh, S.~J.
  Turner, and W.~Wong, ``{Accelerating Sparse Matrix-Vector Multiplication on
  GPUs Using Bit-Representation-Optimized Schemes},'' in \emph{International
  Conference for High Performance Computing, Networking, Storage and Analysis,
  SC'13, Denver, CO, {USA} - November 17 - 21, 2013}, W.~Gropp and S.~Matsuoka,
  Eds.\hskip 1em plus 0.5em minus 0.4em\relax {ACM}, 2013, pp. 26:1--26:12.
  [Online]. Available: \url{https://doi.org/10.1145/2503210.2503234}
\BIBentrySTDinterwordspacing

\bibitem{Tang_TPDS_2015}
W.~T. Tang, W.~J. Tan, R.~S.~M. Goh, S.~J. Turner, and W.-F. Wong, ``A family
  of bit-representation-optimized formats for fast sparse matrix-vector
  multiplication on the gpu,'' \emph{IEEE Transactions on Parallel and
  Distributed Systems}, vol.~26, no.~9, pp. 2373--2385, 2015.

\bibitem{DBLP:conf/fccm/KesturDC12}
\BIBentryALTinterwordspacing
S.~Kestur, J.~D. Davis, and E.~S. Chung, ``Towards a universal {FPGA}
  matrix-vector multiplication architecture,'' in \emph{2012 {IEEE} 20th Annual
  International Symposium on Field-Programmable Custom Computing Machines,
  {FCCM} 2012, 29 April - 1 May 2012, Toronto, Ontario, Canada}.\hskip 1em plus
  0.5em minus 0.4em\relax {IEEE} Computer Society, 2012, pp. 9--16. [Online].
  Available: \url{https://doi.org/10.1109/FCCM.2012.12}
\BIBentrySTDinterwordspacing

\bibitem{DBLP:journals/ijhpca/Mellor-CrummeyG04}
\BIBentryALTinterwordspacing
J.~M. Mellor{-}Crummey and J.~Garvin, ``Optimizing sparse matrix - vector
  product computations using unroll and jam,'' \emph{Int. J. High Perform.
  Comput. Appl.}, vol.~18, no.~2, pp. 225--236, 2004. [Online]. Available:
  \url{https://doi.org/10.1177/1094342004038951}
\BIBentrySTDinterwordspacing

\bibitem{DBLP:journals/concurrency/VazquezFG11}
\BIBentryALTinterwordspacing
F.~V{\'{a}}zquez, J.~Fern{\'{a}}ndez, and E.~M. Garz{\'{o}}n, ``A new approach
  for sparse matrix vector product on {NVIDIA} gpus,'' \emph{Concurr. Comput.
  Pract. Exp.}, vol.~23, no.~8, pp. 815--826, 2011. [Online]. Available:
  \url{https://doi.org/10.1002/cpe.1658}
\BIBentrySTDinterwordspacing

\bibitem{DBLP:journals/ijhsc/AndersonS89}
\BIBentryALTinterwordspacing
E.~J. Anderson and Y.~Saad, ``Solving sparse triangular linear systems on
  parallel computers,'' \emph{Int. J. High Speed Comput.}, vol.~1, no.~1, pp.
  73--95, 1989. [Online]. Available:
  \url{https://doi.org/10.1142/S0129053389000056}
\BIBentrySTDinterwordspacing

\bibitem{DBLP:journals/pc/Melhem88}
\BIBentryALTinterwordspacing
R.~G. Melhem, ``Parallel solution of linear systems with striped sparse
  matrices,'' \emph{Parallel Comput.}, vol.~6, no.~2, pp. 165--184, 1988.
  [Online]. Available: \url{https://doi.org/10.1016/0167-8191(88)90082-8}
\BIBentrySTDinterwordspacing

\bibitem{DBLP:journals/ipl/MontagneE04}
\BIBentryALTinterwordspacing
E.~Montagne and A.~Ekambaram, ``An optimal storage format for sparse
  matrices,'' \emph{Inf. Process. Lett.}, vol.~90, no.~2, pp. 87--92, 2004.
  [Online]. Available: \url{https://doi.org/10.1016/j.ipl.2004.01.014}
\BIBentrySTDinterwordspacing

\bibitem{baskaran2008optimizing}
M.~M. Baskaran and R.~Bordawekar, ``{Optimizing Sparse Matrix-Vector
  Multiplication on GPUs Using Compile-Time and Run-Time Strategies},''
  \emph{IBM Reserach Report, RC24704 (W0812-047)}, 2008.

\bibitem{DBLP:conf/ccgrid/ZeinR10}
\BIBentryALTinterwordspacing
A.~H.~E. Zein and A.~P. Rendell, ``{From Sparse Matrix to Optimal GPU CUDA
  Sparse Matrix Vector Product Implementation},'' in \emph{10th {IEEE/ACM}
  International Conference on Cluster, Cloud and Grid Computing, CCGrid, 17-20
  May 2010, Melbourne, Victoria, Australia}.\hskip 1em plus 0.5em minus
  0.4em\relax {IEEE} Computer Society, 2010, pp. 808--813. [Online]. Available:
  \url{https://doi.org/10.1109/CCGRID.2010.81}
\BIBentrySTDinterwordspacing

\bibitem{DBLP:journals/concurrency/ZeinR12}
\BIBentryALTinterwordspacing
------, ``{Generating Optimal CUDA Sparse Matrix-Vector Product Implementations
  for Evolving GPU Hardware},'' \emph{Concurrency Computation: Practice and
  Experience}, vol.~24, no.~1, pp. 3--13, 2012. [Online]. Available:
  \url{https://doi.org/10.1002/cpe.1732}
\BIBentrySTDinterwordspacing

\bibitem{Cusp}
\BIBentryALTinterwordspacing
S.~Dalton, N.~Bell, L.~Olson, and M.~Garland, ``{Cusp: Generic Parallel
  Algorithms for Sparse Matrix and Graph Computations},'' 2014, version 0.5.0.
  [Online]. Available: \url{http://cusplibrary.github.io/}
\BIBentrySTDinterwordspacing

\bibitem{DBLP:conf/iccsa/MukunokiT13}
\BIBentryALTinterwordspacing
D.~Mukunoki and D.~Takahashi, ``{Optimization of Sparse Matrix-Vector
  Multiplication for CRS Format on NVIDIA Kepler Architecture GPUs},'' in
  \emph{Computational Science and Its Applications - {ICCSA} 13th International
  Conference, Ho Chi Minh City, Vietnam, June 24-27, 2013, Proceedings, Part
  {V}}, ser. Lecture Notes in Computer Science, vol. 7975.\hskip 1em plus 0.5em
  minus 0.4em\relax Springer, 2013, pp. 211--223. [Online]. Available:
  \url{https://doi.org/10.1007/978-3-642-39640-3\_15}
\BIBentrySTDinterwordspacing

\bibitem{giles2012efficient}
I.~Reguly and M.~Giles, ``{Efficient Sparse Matrix-Vector Multiplication on
  Cache-Based GPUs},'' in \emph{2012 Innovative Parallel Computing (InPar)},
  May 2012, pp. 1--12.

\bibitem{YoshizawaT12}
\BIBentryALTinterwordspacing
H.~Yoshizawa and D.~Takahashi, ``{Automatic Tuning of Sparse Matrix-Vector
  Multiplication for CRS Format on GPUs},'' in \emph{15th {IEEE} International
  Conference on Computational Science and Engineering, {CSE} 2012, Paphos,
  Cyprus, December 5-7, 2012}.\hskip 1em plus 0.5em minus 0.4em\relax {IEEE}
  Computer Society, 2012, pp. 130--136. [Online]. Available:
  \url{https://doi.org/10.1109/ICCSE.2012.28}
\BIBentrySTDinterwordspacing

\bibitem{CMRS_Koza_2014}
\BIBentryALTinterwordspacing
Z.~Koza, M.~Matyka, S.~Szkoda, and {\L}.~Miros{\l}aw, ``{Compressed Multirow
  Storage Format for Sparse Matrices on Graphics Processing Units},''
  \emph{SIAM Journal on Scientific Computing}, vol.~36, no.~2, pp. C219--C239,
  2014. [Online]. Available: \url{https://doi.org/10.1137/120900216}
\BIBentrySTDinterwordspacing

\bibitem{anzt2020ginkgo}
\BIBentryALTinterwordspacing
H.~Anzt, T.~Cojean, G.~Flegar, F.~G{\"{o}}bel, T.~Gr{\"{u}}tzmacher, P.~Nayak,
  T.~Ribizel, Y.~M. Tsai, and E.~S. Quintana{-}Ort{\'{\i}}, ``{Ginkgo: A Modern
  Linear Operator Algebra Framework for High Performance Computing},''
  \emph{ACM Transactions on Mathematical Software}, vol.~48, no.~1, pp.
  2:1--2:33, 2022. [Online]. Available: \url{https://doi.org/10.1145/3480935}
\BIBentrySTDinterwordspacing

\bibitem{DBLP:conf/sc/FlegarA17}
\BIBentryALTinterwordspacing
G.~Flegar and H.~Anzt, ``Overcoming load imbalance for irregular sparse
  matrices,'' in \emph{Proceedings of the Seventh Workshop on Irregular
  Applications: Architectures and Algorithms, IA3@SC 2017, Denver, CO, USA,
  November 12 - 17, 2017}.\hskip 1em plus 0.5em minus 0.4em\relax {ACM}, 2017,
  pp. 2:1--2:8. [Online]. Available:
  \url{https://doi.org/10.1145/3149704.3149767}
\BIBentrySTDinterwordspacing

\bibitem{DBLP:conf/hpcc/LiuZSQ09}
\BIBentryALTinterwordspacing
S.~Liu, Y.~Zhang, X.~Sun, and R.~Qiu, ``{Performance Evaluation of
  Multithreaded Sparse Matrix-Vector Multiplication Using OpenMP},'' in
  \emph{11th {IEEE} International Conference on High Performance Computing and
  Communications, {HPCC}, 25-27 June 2009, Seoul, Korea}.\hskip 1em plus 0.5em
  minus 0.4em\relax {IEEE}, 2009, pp. 659--665. [Online]. Available:
  \url{https://doi.org/10.1109/HPCC.2009.75}
\BIBentrySTDinterwordspacing

\bibitem{Feng_Jin_Zheng_Hu_Zeng_Shao_2011}
\BIBentryALTinterwordspacing
X.~Feng, H.~Jin, R.~Zheng, K.~Hu, J.~Zeng, and Z.~Shao, ``Optimization of
  sparse matrix-vector multiplication with variant {CSR} on gpus,'' in
  \emph{17th {IEEE} International Conference on Parallel and Distributed
  Systems, {ICPADS}, Tainan, Taiwan, December 7-9, 2011}.\hskip 1em plus 0.5em
  minus 0.4em\relax {IEEE} Computer Society, 2011, pp. 165--172. [Online].
  Available: \url{https://doi.org/10.1109/ICPADS.2011.91}
\BIBentrySTDinterwordspacing

\bibitem{greathouse2014efficient}
J.~L. Greathouse and M.~Daga, ``{Efficient Sparse Matrix-Vector Multiplication
  on GPUs Using the CSR Storage Format},'' in \emph{SC'14: Proceedings of the
  International Conference for High Performance Computing, Networking, Storage
  and Analysis}.\hskip 1em plus 0.5em minus 0.4em\relax IEEE, 2014, pp.
  769--780.

\bibitem{CSR_adaptive_improved}
M.~{Daga} and J.~L. {Greathouse}, ``{Structural Agnostic SpMV: Adapting
  CSR-Adaptive for Irregular Matrices},'' in \emph{2015 IEEE 22nd International
  Conference on High Performance Computing (HiPC)}, 2015, pp. 64--74.

\bibitem{DBLP:conf/icpads/GaoJLSWS21}
\BIBentryALTinterwordspacing
J.~Gao, W.~Ji, J.~Liu, S.~Shao, Y.~Wang, and F.~Shi, ``{AMF-CSR:} adaptive
  multi-row folding of {CSR} for spmv on {GPU},'' in \emph{27th {IEEE}
  International Conference on Parallel and Distributed Systems, {ICPADS} 2021,
  Beijing, China, December 14-16, 2021}.\hskip 1em plus 0.5em minus 0.4em\relax
  {IEEE}, 2021, pp. 418--425. [Online]. Available:
  \url{https://doi.org/10.1109/ICPADS53394.2021.00058}
\BIBentrySTDinterwordspacing

\bibitem{LiuS15}
\BIBentryALTinterwordspacing
Y.~Liu and B.~Schmidt, ``{LightSpMV: Faster CSR-Based Sparse Matrix-Vector
  Multiplication on CUDA-Enabled GPUs},'' in \emph{26th {IEEE} International
  Conference on Application-specific Systems, Architectures and Processors,
  {ASAP} 2015, Toronto, ON, Canada, July 27-29, 2015}.\hskip 1em plus 0.5em
  minus 0.4em\relax {IEEE} Computer Society, 2015, pp. 82--89. [Online].
  Available: \url{https://doi.org/10.1109/ASAP.2015.7245713}
\BIBentrySTDinterwordspacing

\bibitem{liu2018lightspmv}
------, ``{LightSpMV: Faster CUDA-Compatible Sparse Matrix-Vector
  Multiplication Using Compressed Sparse Rows},'' \emph{Journal of Signal
  Processing Systems}, vol.~90, no.~1, pp. 69--86, 2018.

\bibitem{MergedSpMV_SC16}
\BIBentryALTinterwordspacing
D.~Merrill and M.~Garland, ``{Merge-Based Parallel Sparse Matrix-Vector
  Multiplication},'' in \emph{Proceedings of the International Conference for
  High Performance Computing, Networking, Storage and Analysis, {SC} 2016, Salt
  Lake City, UT, USA, November 13-18, 2016}, J.~West and C.~M. Pancake,
  Eds.\hskip 1em plus 0.5em minus 0.4em\relax {IEEE} Computer Society, 2016,
  pp. 678--689. [Online]. Available: \url{https://doi.org/10.1109/SC.2016.57}
\BIBentrySTDinterwordspacing

\bibitem{DBLP:conf/europar/FlegarQ17}
\BIBentryALTinterwordspacing
G.~Flegar and E.~S. Quintana{-}Ort{\'{\i}}, ``{Balanced CSR Sparse
  Matrix-Vector Product on Graphics Processors},'' in \emph{Euro-Par 2017:
  Parallel Processing - 23rd International Conference on Parallel and
  Distributed Computing, Santiago de Compostela, Spain, August 28 - September
  1, 2017, Proceedings}, ser. Lecture Notes in Computer Science, F.~F. Rivera,
  T.~F. Pena, and J.~C. Cabaleiro, Eds., vol. 10417.\hskip 1em plus 0.5em minus
  0.4em\relax Springer, 2017, pp. 697--709. [Online]. Available:
  \url{https://doi.org/10.1007/978-3-319-64203-1\_50}
\BIBentrySTDinterwordspacing

\bibitem{SteinbergerZS17}
\BIBentryALTinterwordspacing
M.~Steinberger, R.~Zayer, and H.~Seidel, ``{Globally Homogeneous, Locally
  Adaptive Sparse Matrix-Vector Multiplication on the GPU},'' in
  \emph{Proceedings of the International Conference on Supercomputing, {ICS}
  2017, Chicago, IL, USA, June 14-16, 2017}, W.~D. Gropp, P.~Beckman, Z.~Li,
  and F.~J. Cazorla, Eds.\hskip 1em plus 0.5em minus 0.4em\relax {ACM}, 2017,
  pp. 13:1--13:11. [Online]. Available:
  \url{https://doi.org/10.1145/3079079.3079086}
\BIBentrySTDinterwordspacing

\bibitem{DBLP:conf/iccS/ImY01}
\BIBentryALTinterwordspacing
E.~Im and K.~A. Yelick, ``{Optimizing Sparse Matrix Computations for Register
  Reuse in SPARSITY},'' in \emph{Computational Science - {ICCS} 2001,
  International Conference, San Francisco, CA, USA, May 28-30, 2001.
  Proceedings, Part {I}}, ser. Lecture Notes in Computer Science, V.~N.
  Alexandrov, J.~J. Dongarra, B.~A. Juliano, R.~S. Renner, and C.~J.~K. Tan,
  Eds., vol. 2073.\hskip 1em plus 0.5em minus 0.4em\relax Springer, 2001, pp.
  127--136. [Online]. Available:
  \url{https://doi.org/10.1007/3-540-45545-0\_22}
\BIBentrySTDinterwordspacing

\bibitem{DBLP:journals/ijhpca/ImYV04}
\BIBentryALTinterwordspacing
E.~Im, K.~A. Yelick, and R.~W. Vuduc, ``{Sparsity: Optimization Framework for
  Sparse Matrix Kernels},'' \emph{The International Journal of High Performance
  Computing Applications}, vol.~18, no.~1, pp. 135--158, 2004. [Online].
  Available: \url{https://doi.org/10.1177/1094342004041296}
\BIBentrySTDinterwordspacing

\bibitem{DBLP:conf/sc/VuducDYKNL02}
\BIBentryALTinterwordspacing
R.~Vuduc, J.~Demmel, K.~A. Yelick, S.~Kamil, R.~Nishtala, and B.~C. Lee,
  ``{Performance Optimizations and Bounds for Sparse Matrix-Vector Multiply},''
  in \emph{Proceedings of the 2002 {ACM/IEEE} conference on Supercomputing,
  Baltimore, Maryland, USA, November 16-22, 2002, {CD-ROM}}, R.~C. Giles, D.~A.
  Reed, and K.~Kelley, Eds.\hskip 1em plus 0.5em minus 0.4em\relax {IEEE}
  Computer Society, 2002, pp. 35:1--35:35. [Online]. Available:
  \url{https://doi.org/10.1109/SC.2002.10025}
\BIBentrySTDinterwordspacing

\bibitem{vuduc2005oski}
R.~Vuduc, J.~W. Demmel, and K.~A. Yelick, ``{OSKI: A Library of Automatically
  Tuned Sparse Matrix Kernels},'' in \emph{Journal of Physics: Conference
  Series}, vol.~16, no.~1.\hskip 1em plus 0.5em minus 0.4em\relax IOP
  Publishing, 2005, p. 521.

\bibitem{DBLP:journals/aaecc/NishtalaVDY07}
\BIBentryALTinterwordspacing
R.~Nishtala, R.~W. Vuduc, J.~Demmel, and K.~A. Yelick, ``When cache blocking of
  sparse matrix vector multiply works and why,'' \emph{Appl. Algebra Eng.
  Commun. Comput.}, vol.~18, no.~3, pp. 297--311, 2007. [Online]. Available:
  \url{https://doi.org/10.1007/s00200-007-0038-9}
\BIBentrySTDinterwordspacing

\bibitem{zhang2009automatic}
X.~Zhang, Y.~Zhang, X.~Sun, F.~Liu, S.~Liu, Y.~Tang, and Y.~Li, ``Automatic
  performance tuning of spmv on gpgpu,'' \emph{HPC Asia, Kaohsiung, Taiwan,
  China}, pp. 173--179, 2009.

\bibitem{DBLP:journals/tpds/GuoWC14}
\BIBentryALTinterwordspacing
P.~Guo, L.~Wang, and P.~Chen, ``{A Performance Modeling and Optimization
  Analysis Tool for Sparse Matrix-Vector Multiplication on GPUs},'' \emph{IEEE
  Transactions on Parallel and Distributed Systems}, vol.~25, no.~5, pp.
  1112--1123, 2014. [Online]. Available:
  \url{https://doi.org/10.1109/TPDS.2013.123}
\BIBentrySTDinterwordspacing

\bibitem{LiKenli-TPDS2016}
K.~{Li}, W.~{Yang}, and K.~{Li}, ``{Performance Analysis and Optimization for
  SpMV on GPU Using Probabilistic Modeling},'' \emph{IEEE Transactions on
  Parallel and Distributed Systems}, vol.~26, no.~1, pp. 196--205, 2015.

\bibitem{VazquezFG12}
\BIBentryALTinterwordspacing
F.~V{\'{a}}zquez, J.~Fern{\'{a}}ndez, and E.~M. Garz{\'{o}}n, ``{Automatic
  Tuning of the Sparse Matrix Vector Product on GPUs Based on the ELLR-T
  Approach},'' \emph{Parallel Computing}, vol.~38, no.~8, pp. 408--420, 2012.
  [Online]. Available: \url{https://doi.org/10.1016/j.parco.2011.08.003}
\BIBentrySTDinterwordspacing

\bibitem{DBLP:conf/IEEEcit/VazquezOFG10}
\BIBentryALTinterwordspacing
F.~V{\'{a}}zquez, G.~O. L{\'{o}}pez, J.~Fern{\'{a}}ndez, and E.~M.
  Garz{\'{o}}n, ``Improving the performance of the sparse matrix vector product
  with gpus,'' in \emph{10th {IEEE} International Conference on Computer and
  Information Technology, {CIT} 2010, Bradford, West Yorkshire, UK, June
  29-July 1, 2010}.\hskip 1em plus 0.5em minus 0.4em\relax {IEEE} Computer
  Society, 2010, pp. 1146--1151. [Online]. Available:
  \url{https://doi.org/10.1109/CIT.2010.208}
\BIBentrySTDinterwordspacing

\bibitem{DBLP:journals/chinaf/LiHZZ15}
\BIBentryALTinterwordspacing
S.~Li, C.~Hu, J.~Zhang, and Y.~Zhang, ``Automatic tuning of sparse
  matrix-vector multiplication on multicore clusters,'' \emph{Sci. China Inf.
  Sci.}, vol.~58, no.~9, pp. 1--14, 2015. [Online]. Available:
  \url{https://doi.org/10.1007/s11432-014-5254-x}
\BIBentrySTDinterwordspacing

\bibitem{DBLP:conf/hpcc/ChenXXY19}
\BIBentryALTinterwordspacing
Y.~Chen, G.~Xiao, Z.~Xiao, and W.~Yang, ``hpspmv: {A} heterogeneous parallel
  computing scheme for spmv on the sunway taihulight supercomputer,'' in
  \emph{21st {IEEE} International Conference on High Performance Computing and
  Communications; 17th {IEEE} International Conference on Smart City; 5th
  {IEEE} International Conference on Data Science and Systems,
  HPCC/SmartCity/DSS 2019, Zhangjiajie, China, August 10-12, 2019}, Z.~Xiao,
  L.~T. Yang, P.~Balaji, T.~Li, K.~Li, and A.~Y. Zomaya, Eds.\hskip 1em plus
  0.5em minus 0.4em\relax {IEEE}, 2019, pp. 989--995. [Online]. Available:
  \url{https://doi.org/10.1109/HPCC/SmartCity/DSS.2019.00142}
\BIBentrySTDinterwordspacing

\bibitem{guo2010auto}
P.~Guo and L.~Wang, ``{Auto-Tuning CUDA Parameters for Sparse Matrix-Vector
  Multiplication on GPUs},'' in \emph{International Conference on Computational
  and Information Sciences}.\hskip 1em plus 0.5em minus 0.4em\relax IEEE, 2010,
  pp. 1154--1157.

\bibitem{DBLP:conf/supercomputer/Abu-SufahK13}
\BIBentryALTinterwordspacing
W.~A. Abu{-}Sufah and A.~A. Karim, ``Auto-tuning of sparse matrix-vector
  multiplication on graphics processors,'' in \emph{Supercomputing - 28th
  International Supercomputing Conference, {ISC} 2013, Leipzig, Germany, June
  16-20, 2013. Proceedings}, ser. Lecture Notes in Computer Science, J.~M.
  Kunkel, T.~Ludwig, and H.~W. Meuer, Eds., vol. 7905.\hskip 1em plus 0.5em
  minus 0.4em\relax Springer, 2013, pp. 151--164. [Online]. Available:
  \url{https://doi.org/10.1007/978-3-642-38750-0\_12}
\BIBentrySTDinterwordspacing

\bibitem{DBLP:conf/hpcc/Abu-SufahK12}
\BIBentryALTinterwordspacing
------, ``An effective approach for implementing sparse matrix-vector
  multiplication on graphics processing units,'' in \emph{14th {IEEE}
  International Conference on High Performance Computing and Communication {\&}
  9th {IEEE} International Conference on Embedded Software and Systems,
  {HPCC-ICESS} 2012, Liverpool, United Kingdom, June 25-27, 2012}, G.~Min,
  J.~Hu, L.~C. Liu, L.~T. Yang, S.~Seelam, and L.~Lef{\`{e}}vre, Eds.\hskip 1em
  plus 0.5em minus 0.4em\relax {IEEE} Computer Society, 2012, pp. 453--460.
  [Online]. Available: \url{https://doi.org/10.1109/HPCC.2012.68}
\BIBentrySTDinterwordspacing

\bibitem{Armstrong2008_RL}
W.~Armstrong and A.~P. Rendell, ``{Reinforcement Learning for Automated
  Performance Tuning: Initial Evaluation for Sparse Matrix Format Selection},''
  in \emph{2008 IEEE International Conference on Cluster Computing}, 2008, pp.
  411--420.

\bibitem{SMAT2013}
\BIBentryALTinterwordspacing
J.~Li, G.~Tan, M.~Chen, and N.~Sun, ``{SMAT: An Input Adaptive Auto-Tuner for
  Sparse Matrix-Vector Multiplication},'' \emph{ACM SIGPLAN Notices}, vol.~48,
  no.~6, p. 117–126, jun 2013. [Online]. Available:
  \url{https://doi.org/10.1145/2499370.2462181}
\BIBentrySTDinterwordspacing

\bibitem{DBLP:conf/ics/SedaghatiMPPS15}
\BIBentryALTinterwordspacing
N.~Sedaghati, T.~Mu, L.~Pouchet, S.~Parthasarathy, and P.~Sadayappan,
  ``Automatic selection of sparse matrix representation on gpus,'' in
  \emph{Proceedings of the 29th {ACM} on International Conference on
  Supercomputing, ICS'15, Newport Beach/Irvine, CA, USA, June 08 - 11, 2015},
  L.~N. Bhuyan, F.~Chong, and V.~Sarkar, Eds.\hskip 1em plus 0.5em minus
  0.4em\relax {ACM}, 2015, pp. 99--108. [Online]. Available:
  \url{https://doi.org/10.1145/2751205.2751244}
\BIBentrySTDinterwordspacing

\bibitem{Chen2018}
S.~{Chen}, J.~{Fang}, D.~{Chen}, C.~{Xu}, and Z.~{Wang}, ``{Adaptive
  Optimization of Sparse Matrix-Vector Multiplication on Emerging Many-Core
  Architectures},'' in \emph{2018 IEEE 20th International Conference on High
  Performance Computing and Communications; IEEE 16th International Conference
  on Smart City; IEEE 4th International Conference on Data Science and Systems
  (HPCC/SmartCity/DSS)}, 2018, pp. 649--658.

\bibitem{Akrem2016ICPP_SVM}
A.~Benatia, W.~Ji, Y.~Wang, and F.~Shi, ``{Sparse Matrix Format Selection with
  Multiclass SVM for SpMV on GPU},'' in \emph{2016 45th International
  Conference on Parallel Processing (ICPP)}, 2016, pp. 496--505.

\bibitem{Mehrez2018HPCS_SVM}
I.~Mehrez, O.~Hamdi-Larbi, T.~Dufaud, and N.~Emad, ``{Machine Learning for
  Optimal Compression Format Prediction on Multiprocessor Platform},'' in
  \emph{2018 International Conference on High Performance Computing Simulation
  (HPCS)}, 2018, pp. 213--220.

\bibitem{2017-IPDPSW-autoTuning}
K.~{Hou}, W.~{Feng}, and S.~{Che}, ``{Auto-Tuning Strategies for Parallelizing
  Sparse Matrix-Vector (SpMV) Multiplication on Multi- and Many-Core
  Processors},'' in \emph{2017 IEEE International Parallel and Distributed
  Processing Symposium Workshops (IPDPSW)}, 2017, pp. 713--722.

\bibitem{Akrem-BestSF}
\BIBentryALTinterwordspacing
A.~Benatia, W.~Ji, Y.~Wang, and F.~Shi, ``{BestSF: A Sparse Meta-Format for
  Optimizing SpMV on GPU},'' \emph{ACM Transactions on Architecture and Code
  Optimization}, vol.~15, no.~3, Sep. 2018. [Online]. Available:
  \url{https://doi.org/10.1145/3226228}
\BIBentrySTDinterwordspacing

\bibitem{DBLP:journals/ppl/Hamdi-LarbiMD21}
\BIBentryALTinterwordspacing
O.~Hamdi{-}Larbi, I.~Mehrez, and T.~Dufaud, ``Machine learning to design an
  auto-tuning system for the best compressed format detection for parallel
  sparse computations,'' \emph{Parallel Process. Lett.}, vol.~31, no.~4, pp.
  2\,150\,019:1--2\,150\,019:37, 2021. [Online]. Available:
  \url{https://doi.org/10.1142/S0129626421500195}
\BIBentrySTDinterwordspacing

\bibitem{Nisa2018IPDPSW}
I.~Nisa, C.~Siegel, A.~S. Rajam, A.~Vishnu, and P.~Sadayappan, ``{Effective
  Machine Learning Based Format Selection and Performance Modeling for SpMV on
  GPUs},'' in \emph{2018 IEEE International Parallel and Distributed Processing
  Symposium Workshops (IPDPSW)}, 2018, pp. 1056--1065.

\bibitem{DBLP:journals/ieicet/CuiHKT18}
\BIBentryALTinterwordspacing
H.~Cui, S.~Hirasawa, H.~Kobayashi, and H.~Takizawa, ``A machine learning-based
  approach for selecting spmv kernels and matrix storage formats,''
  \emph{{IEICE} Trans. Inf. Syst.}, vol. 101-D, no.~9, pp. 2307--2314, 2018.
  [Online]. Available: \url{https://doi.org/10.1587/transinf.2017EDP7176}
\BIBentrySTDinterwordspacing

\bibitem{ShenXipeng2018PPoPP_bridgingGap}
\BIBentryALTinterwordspacing
Y.~Zhao, J.~Li, C.~Liao, and X.~Shen, ``{Bridging the Gap Between Deep Learning
  and Sparse Matrix Format Selection},'' in \emph{Proceedings of the 23rd ACM
  SIGPLAN Symposium on Principles and Practice of Parallel Programming}, ser.
  PPoPP '18.\hskip 1em plus 0.5em minus 0.4em\relax New York, NY, USA:
  Association for Computing Machinery, 2018, p. 94–108. [Online]. Available:
  \url{https://doi.org/10.1145/3178487.3178495}
\BIBentrySTDinterwordspacing

\bibitem{Shenxipeng2020TPDS_overheadConscious}
W.~Zhou, Y.~Zhao, X.~Shen, and W.~Chen, ``{Enabling Runtime SpMV Format
  Selection through an Overhead Conscious Method},'' \emph{IEEE Transactions on
  Parallel and Distributed Systems}, vol.~31, no.~1, pp. 80--93, 2020.

\bibitem{10.1145/3293883.3301490}
\BIBentryALTinterwordspacing
A.~Elafrou, G.~Goumas, and N.~Koziris, ``{BASMAT: Bottleneck-Aware Sparse
  Matrix-Vector Multiplication Auto-Tuning on GPGPUs},'' in \emph{Proceedings
  of the 24th Symposium on Principles and Practice of Parallel Programming
  (PPoPP), Washington, District of Columbia}.\hskip 1em plus 0.5em minus
  0.4em\relax New York, NY, USA: Association for Computing Machinery, 2019, p.
  423–424. [Online]. Available: \url{https://doi.org/10.1145/3293883.3301490}
\BIBentrySTDinterwordspacing

\bibitem{ijhpca/DufrechouEQ21}
\BIBentryALTinterwordspacing
E.~Dufrechou, P.~Ezzatti, and E.~S. Quintana{-}Ort{\'{\i}}, ``{Selecting
  Optimal SpMV Realizations for GPUs via Machine Learning},'' \emph{The
  International Journal of High Performance Computing Applications}, vol.~35,
  no.~3, 2021. [Online]. Available:
  \url{https://doi.org/10.1177/1094342021990738}
\BIBentrySTDinterwordspacing

\bibitem{DBLP:conf/hpcc/XiaoZCHL22}
\BIBentryALTinterwordspacing
G.~Xiao, T.~Zhou, Y.~Chen, Y.~Hu, and K.~Li, ``Dtspmv: An adaptive spmv
  framework for graph analysis on gpus,'' in \emph{24th {IEEE} Int Conf on High
  Performance Computing {\&} Communications; 8th Int Conf on Data Science {\&}
  Systems; 20th Int Conf on Smart City; 8th Int Conf on Dependability in
  Sensor, Cloud {\&} Big Data Systems {\&} Application,
  HPCC/DSS/SmartCity/DependSys 2022, Hainan, China, December 18-20,
  2022}.\hskip 1em plus 0.5em minus 0.4em\relax {IEEE}, 2022, pp. 35--42.
  [Online]. Available:
  \url{https://doi.org/10.1109/HPCC-DSS-SmartCity-DependSys57074.2022.00039}
\BIBentrySTDinterwordspacing

\bibitem{usman2019zaki}
S.~Usman, R.~Mehmood, I.~Katib, A.~Albeshri, and S.~M. Altowaijri, ``Zaki: A
  smart method and tool for automatic performance optimization of parallel spmv
  computations on distributed memory machines,'' \emph{Mobile Networks and
  Applications}, pp. 1--20, 2019.

\bibitem{DBLP:journals/access/UsmanMKA19}
\BIBentryALTinterwordspacing
S.~Usman, R.~Mehmood, I.~A. Katib, and A.~Albeshri, ``{ZAKI+:} {A} machine
  learning based process mapping tool for spmv computations on distributed
  memory architectures,'' \emph{{IEEE} Access}, vol.~7, pp. 81\,279--81\,296,
  2019. [Online]. Available: \url{https://doi.org/10.1109/ACCESS.2019.2923565}
\BIBentrySTDinterwordspacing

\bibitem{ahmed2022aaqal}
M.~Ahmed, S.~Usman, N.~A. Shah, M.~U. Ashraf, A.~M. Alghamdi, A.~A. Bahadded,
  and K.~A. Almarhabi, ``{AAQAL: A Machine Learning-Based Tool for Performance
  Optimization of Parallel SpMV Computations Using Block CSR},'' \emph{Applied
  Sciences}, vol.~12, no.~14, p. 7073, 2022.

\bibitem{GAO2024104799}
\BIBentryALTinterwordspacing
J.~Gao, W.~Ji, J.~Liu, Y.~Wang, and F.~Shi, ``Revisiting thread configuration
  of spmv kernels on gpu: A machine learning based approach,'' \emph{Journal of
  Parallel and Distributed Computing}, vol. 185, p. 104799, 2024. [Online].
  Available:
  \url{https://www.sciencedirect.com/science/article/pii/S0743731523001697}
\BIBentrySTDinterwordspacing

\bibitem{Akrem_ICPADS2016}
A.~Benatia, W.~Ji, Y.~Wang, and F.~Shi, ``{Machine Learning Approach for the
  Predicting Performance of SpMV on GPU},'' in \emph{2016 IEEE 22nd
  International Conference on Parallel and Distributed Systems (ICPADS)}, 2016,
  pp. 894--901.

\bibitem{Akrem2020}
\BIBentryALTinterwordspacing
------, ``{Sparse Matrix Partitioning for Optimizing SpMV on CPU-GPU
  Heterogeneous Platforms},'' \emph{The International Journal of High
  Performance Computing Applications}, vol.~34, no.~1, pp. 66--80, 2020.
  [Online]. Available: \url{https://doi.org/10.1177/1094342019886628}
\BIBentrySTDinterwordspacing

\bibitem{barreda2020performance}
M.~Barreda, M.~F. Dolz, M.~A. Casta{\~n}o, P.~Alonso-Jord{\'a}, and E.~S.
  Quintana-Orti, ``{Performance Modeling of the Sparse Matrix-Vector Product
  via Convolutional Neural Networks},'' \emph{The Journal of Supercomputing},
  vol.~76, no.~11, pp. 8883--8900, 2020.

\bibitem{barreda2021convolutional}
M.~Barreda, M.~F. Dolz, and M.~A. Castano, ``{Convolutional Neural Nets for
  Estimating the Run Time and Energy Consumption of the Sparse Matrix-Vector
  Product},'' \emph{The International Journal of High Performance Computing
  Applications}, vol.~35, no.~3, pp. 268--281, 2021.

\bibitem{DBLP:journals/siamsc/CarsonH18}
\BIBentryALTinterwordspacing
E.~C. Carson and N.~J. Higham, ``Accelerating the solution of linear systems by
  iterative refinement in three precisions,'' \emph{{SIAM} J. Sci. Comput.},
  vol.~40, no.~2, 2018. [Online]. Available:
  \url{https://doi.org/10.1137/17M1140819}
\BIBentrySTDinterwordspacing

\bibitem{DBLP:journals/corr/abs-1907-10550}
\BIBentryALTinterwordspacing
S.~Gratton, E.~Simon, D.~Titley{-}P{\'{e}}loquin, and P.~L. Toint, ``Exploiting
  variable precision in {GMRES},'' \emph{CoRR}, vol. abs/1907.10550, 2019.
  [Online]. Available: \url{http://arxiv.org/abs/1907.10550}
\BIBentrySTDinterwordspacing

\bibitem{DBLP:journals/ijhpca/AliagaAGQT23}
\BIBentryALTinterwordspacing
J.~I. Aliaga, H.~Anzt, T.~Gr{\"{u}}tzmacher, E.~S. Quintana{-}Ort{\'{\i}}, and
  A.~E. Tom{\'{a}}s, ``Compressed basis {GMRES} on high-performance graphics
  processing units,'' \emph{Int. J. High Perform. Comput. Appl.}, vol.~37,
  no.~2, pp. 82--100, 2023. [Online]. Available:
  \url{https://doi.org/10.1177/10943420221115140}
\BIBentrySTDinterwordspacing

\bibitem{loe2021study}
J.~A. Loe, C.~A. Glusa, I.~Yamazaki, E.~G. Boman, and S.~Rajamanickam, ``A
  study of mixed precision strategies for gmres on gpus,'' \emph{arXiv preprint
  arXiv:2109.01232}, 2021.

\bibitem{le2018mixed}
M.~Le~Gallo, A.~Sebastian, R.~Mathis, M.~Manica, H.~Giefers, T.~Tuma, C.~Bekas,
  A.~Curioni, and E.~Eleftheriou, ``Mixed-precision in-memory computing,''
  \emph{Nature Electronics}, vol.~1, no.~4, pp. 246--253, 2018.

\bibitem{DBLP:conf/smc2/LindquistLD20}
\BIBentryALTinterwordspacing
N.~Lindquist, P.~Luszczek, and J.~J. Dongarra, ``Improving the performance of
  the {GMRES} method using mixed-precision techniques,'' in \emph{Driving
  Scientific and Engineering Discoveries Through the Convergence of HPC, Big
  Data and {AI} - 17th Smoky Mountains Computational Sciences and Engineering
  Conference, {SMC} 2020, Oak Ridge, TN, USA, August 26-28, 2020, Revised
  Selected Papers}, ser. Communications in Computer and Information Science,
  J.~Nichols, B.~Verastegui, A.~B. Maccabe, O.~R. Hernandez, S.~Parete{-}Koon,
  and T.~Ahearn, Eds., vol. 1315.\hskip 1em plus 0.5em minus 0.4em\relax
  Springer, 2020, pp. 51--66. [Online]. Available:
  \url{https://doi.org/10.1007/978-3-030-63393-6\_4}
\BIBentrySTDinterwordspacing

\bibitem{DBLP:journals/taco/AhmadSH20}
\BIBentryALTinterwordspacing
K.~Ahmad, H.~Sundar, and M.~W. Hall, ``Data-driven mixed precision sparse
  matrix vector multiplication for gpus,'' \emph{{ACM} Trans. Archit. Code
  Optim.}, vol.~16, no.~4, pp. 51:1--51:24, 2020. [Online]. Available:
  \url{https://doi.org/10.1145/3371275}
\BIBentrySTDinterwordspacing

\bibitem{DBLP:journals/jcam/MukunokiO20}
\BIBentryALTinterwordspacing
D.~Mukunoki and T.~Ogita, ``Performance and energy consumption of accurate and
  mixed-precision linear algebra kernels on gpus,'' \emph{J. Comput. Appl.
  Math.}, vol. 372, p. 112701, 2020. [Online]. Available:
  \url{https://doi.org/10.1016/j.cam.2019.112701}
\BIBentrySTDinterwordspacing

\bibitem{DBLP:conf/sbac-pad/TezcanTKKU22}
\BIBentryALTinterwordspacing
E.~Tezcan, T.~Torun, F.~Kosar, K.~Kaya, and D.~Unat, ``Mixed and
  multi-precision spmv for gpus with row-wise precision selection,'' in
  \emph{2022 {IEEE} 34th International Symposium on Computer Architecture and
  High Performance Computing (SBAC-PAD), Bordeaux, France, November 2-5,
  2022}.\hskip 1em plus 0.5em minus 0.4em\relax {IEEE}, 2022, pp. 31--40.
  [Online]. Available: \url{https://doi.org/10.1109/SBAC-PAD55451.2022.00014}
\BIBentrySTDinterwordspacing

\bibitem{graillat:hal-03561193}
\BIBentryALTinterwordspacing
S.~Graillat, F.~J{\'e}z{\'e}quel, T.~Mary, and R.~Molina, ``{Adaptive precision
  matrix-vector product},'' Feb. 2022, working paper or preprint. [Online].
  Available: \url{https://hal.science/hal-03561193}
\BIBentrySTDinterwordspacing

\bibitem{DBLP:journals/jocs/Isupov22}
\BIBentryALTinterwordspacing
K.~Isupov, ``Multiple-precision sparse matrix-vector multiplication on gpus,''
  \emph{J. Comput. Sci.}, vol.~61, p. 101609, 2022. [Online]. Available:
  \url{https://doi.org/10.1016/j.jocs.2022.101609}
\BIBentrySTDinterwordspacing

\bibitem{DBLP:journals/corr/Kouya14a}
\BIBentryALTinterwordspacing
T.~Kouya, ``A highly efficient implementation of multiple precision sparse
  matrix-vector multiplication and its application to product-type krylov
  subspace methods,'' \emph{CoRR}, vol. abs/1411.2377, 2014. [Online].
  Available: \url{http://arxiv.org/abs/1411.2377}
\BIBentrySTDinterwordspacing

\bibitem{DBLP:conf/ispdc/PabstBK12}
\BIBentryALTinterwordspacing
H.~Pabst, B.~Bachmayer, and M.~Klemm, ``Performance of a structure-detecting
  spmv using the {CSR} matrix representation,'' in \emph{11th International
  Symposium on Parallel and Distributed Computing, {ISPDC} 2012, Munich,
  Germany, June 25-29, 2012}, M.~Bader, H.~Bungartz, D.~Grigoras, M.~Mehl,
  R.~Mundani, and R.~Potolea, Eds.\hskip 1em plus 0.5em minus 0.4em\relax
  {IEEE} Computer Society, 2012, pp. 3--10. [Online]. Available:
  \url{https://doi.org/10.1109/ISPDC.2012.9}
\BIBentrySTDinterwordspacing

\bibitem{DBLP:journals/jpdc/ZhangYLTL21}
\BIBentryALTinterwordspacing
Y.~Zhang, W.~Yang, K.~Li, D.~Tang, and K.~Li, ``Performance analysis and
  optimization for spmv based on aligned storage formats on an {ARM}
  processor,'' \emph{J. Parallel Distributed Comput.}, vol. 158, pp. 126--137,
  2021. [Online]. Available: \url{https://doi.org/10.1016/j.jpdc.2021.08.002}
\BIBentrySTDinterwordspacing

\bibitem{DBLP:conf/sc/WilliamsOVSYD07}
\BIBentryALTinterwordspacing
S.~Williams, L.~Oliker, R.~W. Vuduc, J.~Shalf, K.~A. Yelick, and J.~Demmel,
  ``Optimization of sparse matrix-vector multiplication on emerging multicore
  platforms,'' in \emph{Proceedings of the {ACM/IEEE} Conference on High
  Performance Networking and Computing, {SC} 2007, November 10-16, 2007, Reno,
  Nevada, {USA}}, B.~Verastegui, Ed.\hskip 1em plus 0.5em minus 0.4em\relax
  {ACM} Press, 2007, p.~38. [Online]. Available:
  \url{https://doi.org/10.1145/1362622.1362674}
\BIBentrySTDinterwordspacing

\bibitem{DBLP:journals/pc/WilliamsOVSYD09}
\BIBentryALTinterwordspacing
------, ``Optimization of sparse matrix-vector multiplication on emerging
  multicore platforms,'' \emph{Parallel Comput.}, vol.~35, no.~3, pp. 178--194,
  2009. [Online]. Available: \url{https://doi.org/10.1016/j.parco.2008.12.006}
\BIBentrySTDinterwordspacing

\bibitem{kislal2013optimizing}
O.~Kislal, W.~Ding, M.~Kandemir, and I.~Demirkiran, ``Optimizing sparse matrix
  vector multiplication on emerging multicores,'' in \emph{2013 IEEE 6th
  International Workshop on Multi-/Many-core Computing Systems
  (MuCoCoS)}.\hskip 1em plus 0.5em minus 0.4em\relax IEEE, 2013, pp. 1--10.

\bibitem{DBLP:conf/hpcc/YuanZS08}
\BIBentryALTinterwordspacing
E.~Yuan, Y.~Zhang, and X.~Sun, ``Memory access complexity analysis of spmv in
  {RAM} (h) model,'' in \emph{10th {IEEE} International Conference on High
  Performance Computing and Communications, {HPCC} 2008, 25-27 Sept. 2008,
  Dalian, China}.\hskip 1em plus 0.5em minus 0.4em\relax {IEEE} Computer
  Society, 2008, pp. 913--920. [Online]. Available:
  \url{https://doi.org/10.1109/HPCC.2008.130}
\BIBentrySTDinterwordspacing

\bibitem{lee2003performance}
B.~C. Lee, R.~Vuduc, J.~W. Demmel, K.~A. Yelick, M.~deLorimier, and L.~Zhong,
  ``Performance optimizations and bounds for sparse symmetric matrix-multiple
  vector multiply,'' \emph{University of California, Berkeley, Berkeley, CA,
  USA, Tech. Rep. UCB/CSD-03-1297}, 2003.

\bibitem{DBLP:conf/icpp/ElafrouGK17}
\BIBentryALTinterwordspacing
A.~Elafrou, G.~I. Goumas, and N.~Koziris, ``Performance analysis and
  optimization of sparse matrix-vector multiplication on modern multi- and
  many-core processors,'' in \emph{46th International Conference on Parallel
  Processing, {ICPP} 2017, Bristol, United Kingdom, August 14-17, 2017}.\hskip
  1em plus 0.5em minus 0.4em\relax {IEEE} Computer Society, 2017, pp. 292--301.
  [Online]. Available: \url{https://doi.org/10.1109/ICPP.2017.38}
\BIBentrySTDinterwordspacing

\bibitem{DBLP:conf/sc/TrotterELTDIU23}
\BIBentryALTinterwordspacing
J.~D. Trotter, S.~Ekmek{\c{c}}ibasi, J.~Langguth, T.~Torun, E.~D{\"{u}}zakin,
  A.~Ilic, and D.~Unat, ``Bringing order to sparsity: {A} sparse matrix
  reordering study on multicore cpus,'' in \emph{Proceedings of the
  International Conference for High Performance Computing, Networking, Storage
  and Analysis, {SC} 2023, Denver, CO, USA, November 12-17, 2023}, D.~Arnold,
  R.~M. Badia, and K.~M. Mohror, Eds.\hskip 1em plus 0.5em minus 0.4em\relax
  {ACM}, 2023, pp. 31:1--31:13. [Online]. Available:
  \url{https://doi.org/10.1145/3581784.3607046}
\BIBentrySTDinterwordspacing

\bibitem{DBLP:conf/npc/YuMQF020}
\BIBentryALTinterwordspacing
X.~Yu, H.~Ma, Z.~Qu, J.~Fang, and W.~Liu, ``Numa-aware optimization of sparse
  matrix-vector multiplication on armv8-based many-core architectures,'' in
  \emph{Network and Parallel Computing - 17th {IFIP} {WG} 10.3 International
  Conference, {NPC} 2020, Zhengzhou, China, September 28-30, 2020, Revised
  Selected Papers}, ser. Lecture Notes in Computer Science, X.~He, E.~Shao, and
  G.~Tan, Eds., vol. 12639.\hskip 1em plus 0.5em minus 0.4em\relax Springer,
  2020, pp. 231--242. [Online]. Available:
  \url{https://doi.org/10.1007/978-3-030-79478-1\_20}
\BIBentrySTDinterwordspacing

\bibitem{Nickolls_Buck_Garland_Skadron_2008}
J.~Nickolls, I.~Buck, M.~Garland, and K.~Skadron, ``{Scalable Parallel
  Programming With CUDA: Is CUDA the Parallel Programming Model That
  Application Developers Have Been Waiting For?}'' \emph{Queue}, vol.~6, no.~2,
  p. 40–53, Mar 2008.

\bibitem{baskaran2009optimizing}
M.~M. Baskaran and R.~Bordawekar, ``{Optimizing Sparse Matrix-Vector
  Multiplication on GPUs},'' \emph{IBM Research Report RC24704}, no.
  W0812--047, 2009.

\bibitem{DBLP:conf/iccad/DengWM09}
\BIBentryALTinterwordspacing
Y.~Deng, B.~D. Wang, and S.~Mu, ``Taming irregular {EDA} applications on
  gpus,'' in \emph{2009 International Conference on Computer-Aided Design,
  {ICCAD} 2009, San Jose, CA, USA, November 2-5, 2009}, J.~S. Roychowdhury,
  Ed.\hskip 1em plus 0.5em minus 0.4em\relax {ACM}, 2009, pp. 539--546.
  [Online]. Available: \url{https://doi.org/10.1145/1687399.1687501}
\BIBentrySTDinterwordspacing

\bibitem{DBLP:journals/integration/HeTZLWS16}
\BIBentryALTinterwordspacing
K.~He, S.~X. Tan, H.~Zhao, X.~Liu, H.~Wang, and G.~Shi, ``Parallel {GMRES}
  solver for fast analysis of large linear dynamic systems on {GPU}
  platforms,'' \emph{Integr.}, vol.~52, pp. 10--22, 2016. [Online]. Available:
  \url{https://doi.org/10.1016/j.vlsi.2015.07.005}
\BIBentrySTDinterwordspacing

\bibitem{DBLP:journals/tpds/KarimiADK22}
\BIBentryALTinterwordspacing
E.~Karimi, N.~B. Agostini, S.~Dong, and D.~R. Kaeli, ``{VCSR:} an efficient
  {GPU} memory-aware sparse format,'' \emph{{IEEE} Trans. Parallel Distributed
  Syst.}, vol.~33, no.~10, pp. 3977--3989, 2022. [Online]. Available:
  \url{https://doi.org/10.1109/TPDS.2022.3177291}
\BIBentrySTDinterwordspacing

\bibitem{DBLP:conf/sc/Lu023}
\BIBentryALTinterwordspacing
Y.~Lu and W.~Liu, ``{DASP:} specific dense matrix multiply-accumulate units
  accelerated general sparse matrix-vector multiplication,'' in
  \emph{Proceedings of the International Conference for High Performance
  Computing, Networking, Storage and Analysis, {SC} 2023, Denver, CO, USA,
  November 12-17, 2023}, D.~Arnold, R.~M. Badia, and K.~M. Mohror, Eds.\hskip
  1em plus 0.5em minus 0.4em\relax {ACM}, 2023, pp. 73:1--73:14. [Online].
  Available: \url{https://doi.org/10.1145/3581784.3607051}
\BIBentrySTDinterwordspacing

\bibitem{DBLP:conf/asplos/GreweL11}
\BIBentryALTinterwordspacing
D.~Grewe and A.~Lokhmotov, ``Automatically generating and tuning {GPU} code for
  sparse matrix-vector multiplication from a high-level representation,'' in
  \emph{Proceedings of 4th Workshop on General Purpose Processing on Graphics
  Processing Units, {GPGPU} 2011, Newport Beach, CA, USA, March 5, 2011}.\hskip
  1em plus 0.5em minus 0.4em\relax {ACM}, 2011, p.~12. [Online]. Available:
  \url{https://doi.org/10.1145/1964179.1964196}
\BIBentrySTDinterwordspacing

\bibitem{DBLP:conf/iccS/CevahirNM09}
\BIBentryALTinterwordspacing
A.~Cevahir, A.~Nukada, and S.~Matsuoka, ``{Fast Conjugate Gradients with
  Multiple GPUs},'' in \emph{Computational Science - {ICCS}, 9th International
  Conference, Baton Rouge, LA, USA, May 25-27, 2009, Proceedings, Part {I}},
  ser. Lecture Notes in Computer Science, G.~Allen, J.~Nabrzyski, E.~Seidel,
  G.~D. van Albada, J.~J. Dongarra, and P.~M.~A. Sloot, Eds., vol. 5544.\hskip
  1em plus 0.5em minus 0.4em\relax Springer, 2009, pp. 893--903. [Online].
  Available: \url{https://doi.org/10.1007/978-3-642-01970-8\_90}
\BIBentrySTDinterwordspacing

\bibitem{DBLP:conf/sbac-pad/GuoZ16}
\BIBentryALTinterwordspacing
P.~Guo and C.~Zhang, ``{Performance Optimization for SpMV on Multi-GPU Systems
  Using Threads and Multiple Streams},'' in \emph{International Symposium on
  Computer Architecture and High Performance Computing Workshops, {SBAC-PAD}
  Workshops , Los Angeles, CA, USA, October 26-28, 2016}.\hskip 1em plus 0.5em
  minus 0.4em\relax {IEEE} Computer Society, 2016, pp. 67--72. [Online].
  Available: \url{https://doi.org/10.1109/SBAC-PADW.2016.20}
\BIBentrySTDinterwordspacing

\bibitem{karwacki2012multi}
M.~Karwacki, B.~Bylina, and J.~Bylina, ``{Multi-GPU Implementation of the
  Uniformization Method for Solving Markov Models},'' in \emph{Federated
  conference on computer science and information systems (FedCSIS)}.\hskip 1em
  plus 0.5em minus 0.4em\relax IEEE, 2012, pp. 533--537.

\bibitem{DBLP:journals/pc/VerschoorJ12}
\BIBentryALTinterwordspacing
M.~Verschoor and A.~C. Jalba, ``{Analysis and Performance Estimation of the
  Conjugate Gradient Method on Multiple GPUs},'' \emph{Parallel Computing},
  vol.~38, no. 10-11, pp. 552--575, 2012. [Online]. Available:
  \url{https://doi.org/10.1016/j.parco.2012.07.002}
\BIBentrySTDinterwordspacing

\bibitem{DBLP:journals/cma/YangLC16}
\BIBentryALTinterwordspacing
B.~Yang, H.~Liu, and Z.~Chen, ``{Preconditioned GMRES Solver on Multiple-GPU
  Architecture},'' \emph{Computers and Mathematics with Applications}, vol.~72,
  no.~4, pp. 1076--1095, 2016. [Online]. Available:
  \url{https://doi.org/10.1016/j.camwa.2016.06.027}
\BIBentrySTDinterwordspacing

\bibitem{siamsc/KarypisK98}
\BIBentryALTinterwordspacing
G.~Karypis and V.~Kumar, ``{A Fast and High Quality Multilevel Scheme for
  Partitioning Irregular Graphs},'' \emph{SIAM Journal on Scientific
  Computing}, vol.~20, no.~1, pp. 359--392, 1998. [Online]. Available:
  \url{https://doi.org/10.1137/S1064827595287997}
\BIBentrySTDinterwordspacing

\bibitem{DBLP:journals/topc/GaoWWL16}
\BIBentryALTinterwordspacing
J.~Gao, Y.~Wang, J.~Wang, and R.~Liang, ``{Adaptive Optimization Modeling of
  Preconditioned Conjugate Gradient on Multi-GPUs},'' \emph{ACM Transactions on
  Parallel Computing}, vol.~3, no.~3, pp. 16:1--16:33, 2016. [Online].
  Available: \url{https://doi.org/10.1145/2990849}
\BIBentrySTDinterwordspacing

\bibitem{DBLP:journals/pc/GaoZHX17}
\BIBentryALTinterwordspacing
J.~Gao, Y.~Zhou, G.~He, and Y.~Xia, ``{A Multi-GPU Parallel Optimization Model
  for the Preconditioned Conjugate Gradient Algorithm},'' \emph{Parallel
  Computing}, vol.~63, pp. 1--16, 2017. [Online]. Available:
  \url{https://doi.org/10.1016/j.parco.2017.04.003}
\BIBentrySTDinterwordspacing

\bibitem{DBLP:journals/concurrency/GaoWW17}
\BIBentryALTinterwordspacing
J.~Gao, Y.~Wang, and J.~Wang, ``{A Novel Multi-Graphics Processing Unit
  Parallel Optimization Framework for the Sparse Matrix-Vector
  Multiplication},'' \emph{Concurrency Computation Practice and Experience},
  vol.~29, no.~5, 2017. [Online]. Available:
  \url{https://doi.org/10.1002/cpe.3936}
\BIBentrySTDinterwordspacing

\bibitem{DBLP:journals/tog/Li0TCZM20}
\BIBentryALTinterwordspacing
C.~Li, M.~Tang, R.~Tong, M.~Cai, J.~Zhao, and D.~Manocha, ``P-cloth:
  interactive complex cloth simulation on multi-gpu systems using dynamic
  matrix assembly and pipelined implicit integrators,'' \emph{{ACM} Trans.
  Graph.}, vol.~39, no.~6, pp. 180:1--180:15, 2020. [Online]. Available:
  \url{https://doi.org/10.1145/3414685.3417763}
\BIBentrySTDinterwordspacing

\bibitem{DBLP:journals/corr/abs-2209-07552}
\BIBentryALTinterwordspacing
J.~Chen, C.~Xie, J.~S. Firoz, J.~Li, S.~L. Song, K.~J. Barker, M.~Raugas, and
  A.~Li, ``{MSREP:} {A} fast yet light sparse matrix framework for multi-gpu
  systems,'' \emph{CoRR}, vol. abs/2209.07552, 2022. [Online]. Available:
  \url{https://doi.org/10.48550/arXiv.2209.07552}
\BIBentrySTDinterwordspacing

\bibitem{DBLP:conf/ipps/SchaaK09}
\BIBentryALTinterwordspacing
D.~Schaa and D.~R. Kaeli, ``{Exploring the Multiple-GPU Design Space},'' in
  \emph{23rd {IEEE} International Symposium on Parallel and Distributed
  Processing, {IPDPS} 2009, Rome, Italy, May 23-29, 2009}.\hskip 1em plus 0.5em
  minus 0.4em\relax {IEEE}, 2009, pp. 1--12. [Online]. Available:
  \url{https://doi.org/10.1109/IPDPS.2009.5161068}
\BIBentrySTDinterwordspacing

\bibitem{DBLP:conf/europar/AbdelfattahLK15}
\BIBentryALTinterwordspacing
A.~Abdelfattah, H.~Ltaief, and D.~E. Keyes, ``{High Performance Multi-GPU SpMV
  for Multi-Component PDE-Based Applications},'' in \emph{Euro-Par: Parallel
  Processing - 21st International Conference on Parallel and Distributed
  Computing, Vienna, Austria, August 24-28, 2015, Proceedings}, ser. Lecture
  Notes in Computer Science, J.~L. Tr{\"{a}}ff, S.~Hunold, and F.~Versaci,
  Eds., vol. 9233.\hskip 1em plus 0.5em minus 0.4em\relax Springer, 2015, pp.
  601--612. [Online]. Available:
  \url{https://doi.org/10.1007/978-3-662-48096-0\_46}
\BIBentrySTDinterwordspacing

\bibitem{DBLP:conf/sasp/ShanWWWWXY10}
\BIBentryALTinterwordspacing
Y.~Shan, T.~Wu, Y.~Wang, B.~Wang, Z.~Wang, N.~Xu, and H.~Yang, ``{FPGA} and
  {GPU} implementation of large scale spmv,'' in \emph{{IEEE} 8th Symposium on
  Application Specific Processors, {SASP} 2010, Anaheim, CA, USA, June 13-14,
  2010}.\hskip 1em plus 0.5em minus 0.4em\relax {IEEE} Computer Society, 2010,
  pp. 64--70. [Online]. Available:
  \url{https://doi.org/10.1109/SASP.2010.5521144}
\BIBentrySTDinterwordspacing

\bibitem{DBLP:conf/iccd/UmurogluJ14}
\BIBentryALTinterwordspacing
Y.~Umuroglu and M.~Jahre, ``An energy efficient column-major backend for {FPGA}
  spmv accelerators,'' in \emph{32nd {IEEE} International Conference on
  Computer Design, {ICCD} 2014, Seoul, South Korea, October 19-22, 2014}.\hskip
  1em plus 0.5em minus 0.4em\relax {IEEE} Computer Society, 2014, pp. 432--439.
  [Online]. Available: \url{https://doi.org/10.1109/ICCD.2014.6974716}
\BIBentrySTDinterwordspacing

\bibitem{DBLP:conf/fccm/FowersOSCS14}
\BIBentryALTinterwordspacing
J.~Fowers, K.~Ovtcharov, K.~Strauss, E.~S. Chung, and G.~Stitt, ``A high memory
  bandwidth {FPGA} accelerator for sparse matrix-vector multiplication,'' in
  \emph{22nd {IEEE} Annual International Symposium on Field-Programmable Custom
  Computing Machines, {FCCM} 2014, Boston, MA, USA, May 11-13, 2014}.\hskip 1em
  plus 0.5em minus 0.4em\relax {IEEE} Computer Society, 2014, pp. 36--43.
  [Online]. Available: \url{https://doi.org/10.1109/FCCM.2014.23}
\BIBentrySTDinterwordspacing

\bibitem{DBLP:conf/ccwc/NaherGDJ20}
\BIBentryALTinterwordspacing
J.~Naher, C.~Gloster, C.~C. Doss, and S.~S. Jadhav, ``Using machine learning to
  estimate utilization and throughput for opencl-based matrix-vector
  multiplication {(MVM)},'' in \emph{10th Annual Computing and Communication
  Workshop and Conference, {CCWC} 2020, Las Vegas, NV, USA, January 6-8,
  2020}.\hskip 1em plus 0.5em minus 0.4em\relax {IEEE}, 2020, pp. 365--372.
  [Online]. Available: \url{https://doi.org/10.1109/CCWC47524.2020.9031173}
\BIBentrySTDinterwordspacing

\bibitem{DBLP:conf/arc/UmurogluJ15}
\BIBentryALTinterwordspacing
Y.~Umuroglu and M.~Jahre, ``A vector caching scheme for streaming {FPGA} spmv
  accelerators,'' in \emph{Applied Reconfigurable Computing - 11th
  International Symposium, {ARC} 2015, Bochum, Germany, April 13-17, 2015,
  Proceedings}, ser. Lecture Notes in Computer Science, K.~Sano, D.~Soudris,
  M.~H{\"{u}}bner, and P.~C. Diniz, Eds., vol. 9040.\hskip 1em plus 0.5em minus
  0.4em\relax Springer, 2015, pp. 15--26. [Online]. Available:
  \url{https://doi.org/10.1007/978-3-319-16214-0\_2}
\BIBentrySTDinterwordspacing

\bibitem{DBLP:journals/mam/UmurogluJ16}
\BIBentryALTinterwordspacing
------, ``Random access schemes for efficient {FPGA} spmv acceleration,''
  \emph{Microprocess. Microsystems}, vol.~47, pp. 321--332, 2016. [Online].
  Available: \url{https://doi.org/10.1016/j.micpro.2016.02.015}
\BIBentrySTDinterwordspacing

\bibitem{DBLP:conf/micro/SadiSLHPF19}
\BIBentryALTinterwordspacing
F.~Sadi, J.~Sweeney, T.~M. Low, J.~C. Hoe, L.~T. Pileggi, and F.~Franchetti,
  ``Efficient spmv operation for large and highly sparse matrices using
  scalable multi-way merge parallelization,'' in \emph{Proceedings of the 52nd
  Annual {IEEE/ACM} International Symposium on Microarchitecture, {MICRO} 2019,
  Columbus, OH, USA, October 12-16, 2019}.\hskip 1em plus 0.5em minus
  0.4em\relax {ACM}, 2019, pp. 347--358. [Online]. Available:
  \url{https://doi.org/10.1145/3352460.3358330}
\BIBentrySTDinterwordspacing

\bibitem{DBLP:journals/tcad/HosseinabadyN20}
\BIBentryALTinterwordspacing
M.~Hosseinabady and J.~L. N{\'{u}}{\~{n}}ez{-}Y{\'{a}}{\~{n}}ez, ``A streaming
  dataflow engine for sparse matrix-vector multiplication using high-level
  synthesis,'' \emph{{IEEE} Trans. Comput. Aided Des. Integr. Circuits Syst.},
  vol.~39, no.~6, pp. 1272--1285, 2020. [Online]. Available:
  \url{https://doi.org/10.1109/TCAD.2019.2912923}
\BIBentrySTDinterwordspacing

\bibitem{DBLP:journals/corr/abs-2107-12371}
\BIBentryALTinterwordspacing
G.~Oyarzun, D.~Peyrolon, C.~{\'{A}}lvarez, and X.~Martorell, ``An {FPGA} cached
  sparse matrix vector product (spmv) for unstructured computational fluid
  dynamics simulations,'' \emph{CoRR}, vol. abs/2107.12371, 2021. [Online].
  Available: \url{https://arxiv.org/abs/2107.12371}
\BIBentrySTDinterwordspacing

\bibitem{DBLP:conf/dac/ParraviciniCSS21}
\BIBentryALTinterwordspacing
A.~Parravicini, L.~G. Cellamare, M.~Siracusa, and M.~D. Santambrogio, ``Scaling
  up {HBM} efficiency of top-k spmv for approximate embedding similarity on
  fpgas,'' in \emph{58th {ACM/IEEE} Design Automation Conference, {DAC} 2021,
  San Francisco, CA, USA, December 5-9, 2021}.\hskip 1em plus 0.5em minus
  0.4em\relax {IEEE}, 2021, pp. 799--804. [Online]. Available:
  \url{https://doi.org/10.1109/DAC18074.2021.9586203}
\BIBentrySTDinterwordspacing

\bibitem{DBLP:conf/aspdac/LiuL23}
\BIBentryALTinterwordspacing
B.~Liu and D.~Liu, ``Towards high-bandwidth-utilization spmv on fpgas via
  partial vector duplication,'' in \emph{Proceedings of the 28th Asia and South
  Pacific Design Automation Conference, {ASPDAC} 2023, Tokyo, Japan, January
  16-19, 2023}, A.~Takahashi, Ed.\hskip 1em plus 0.5em minus 0.4em\relax {ACM},
  2023, pp. 33--38. [Online]. Available:
  \url{https://doi.org/10.1145/3566097.3567839}
\BIBentrySTDinterwordspacing

\bibitem{DBLP:conf/vdat/MahadurkarSK21}
\BIBentryALTinterwordspacing
M.~Mahadurkar, N.~Sivanandan, and S.~Kala, ``Hardware acceleration of spmv
  multiplier for deep learning,'' in \emph{25th International Symposium on
  {VLSI} Design and Test, {VDAT} 2021, Surat, India, September 16-18,
  2021}.\hskip 1em plus 0.5em minus 0.4em\relax {IEEE}, 2021, pp. 1--6.
  [Online]. Available: \url{https://doi.org/10.1109/VDAT53777.2021.9600988}
\BIBentrySTDinterwordspacing

\bibitem{DBLP:journals/concurrency/NguyenMSDWW22}
\BIBentryALTinterwordspacing
T.~Nguyen, C.~MacLean, M.~Siracusa, D.~Doerfler, N.~J. Wright, and S.~Williams,
  ``Fpga-based {HPC} accelerators: An evaluation on performance and energy
  efficiency,'' \emph{Concurr. Comput. Pract. Exp.}, vol.~34, no.~20, 2022.
  [Online]. Available: \url{https://doi.org/10.1002/cpe.6570}
\BIBentrySTDinterwordspacing

\bibitem{mi14112030}
\BIBentryALTinterwordspacing
F.~Favaro, E.~Dufrechou, J.~P. Oliver, and P.~Ezzatti, ``Optimizing the
  performance of the sparse matrix–vector multiplication kernel in fpga
  guided by the roofline model,'' \emph{Micromachines}, vol.~14, no.~11, p.
  2030, Oct. 2023. [Online]. Available:
  \url{http://dx.doi.org/10.3390/mi14112030}
\BIBentrySTDinterwordspacing

\bibitem{DBLP:conf/hpca/XieLGB0L0021}
\BIBentryALTinterwordspacing
X.~Xie, Z.~Liang, P.~Gu, A.~Basak, L.~Deng, L.~Liang, X.~Hu, and Y.~Xie,
  ``Spacea: Sparse matrix vector multiplication on processing-in-memory
  accelerator,'' in \emph{{IEEE} International Symposium on High-Performance
  Computer Architecture, {HPCA} 2021, Seoul, South Korea, February 27 - March
  3, 2021}.\hskip 1em plus 0.5em minus 0.4em\relax {IEEE}, 2021, pp. 570--583.
  [Online]. Available: \url{https://doi.org/10.1109/HPCA51647.2021.00055}
\BIBentrySTDinterwordspacing

\bibitem{DBLP:conf/isca/SunLYWL21}
\BIBentryALTinterwordspacing
W.~Sun, Z.~Li, S.~Yin, S.~Wei, and L.~Liu, ``{ABC-DIMM:} alleviating the
  bottleneck of communication in dimm-based near-memory processing with
  inter-dimm broadcast,'' in \emph{48th {ACM/IEEE} Annual International
  Symposium on Computer Architecture, {ISCA} 2021, Valencia, Spain, June 14-18,
  2021}.\hskip 1em plus 0.5em minus 0.4em\relax {IEEE}, 2021, pp. 237--250.
  [Online]. Available: \url{https://doi.org/10.1109/ISCA52012.2021.00027}
\BIBentrySTDinterwordspacing

\bibitem{DBLP:journals/pomacs/GiannoulaFGKGM22}
\BIBentryALTinterwordspacing
C.~Giannoula, I.~Fernandez, J.~G{\'{o}}mez{-}Luna, N.~Koziris, G.~I. Goumas,
  and O.~Mutlu, ``Sparsep: Towards efficient sparse matrix vector
  multiplication on real processing-in-memory architectures,'' \emph{Proc.
  {ACM} Meas. Anal. Comput. Syst.}, vol.~6, no.~1, pp. 21:1--21:49, 2022.
  [Online]. Available: \url{https://doi.org/10.1145/3508041}
\BIBentrySTDinterwordspacing

\bibitem{DBLP:conf/ics/LiuSCD13}
\BIBentryALTinterwordspacing
X.~Liu, M.~Smelyanskiy, E.~Chow, and P.~Dubey, ``Efficient sparse matrix-vector
  multiplication on x86-based many-core processors,'' in \emph{International
  Conference on Supercomputing, ICS'13, Eugene, OR, {USA} - June 10 - 14,
  2013}, A.~D. Malony, M.~Nemirovsky, and S.~P. Midkiff, Eds.\hskip 1em plus
  0.5em minus 0.4em\relax {ACM}, 2013, pp. 273--282. [Online]. Available:
  \url{https://doi.org/10.1145/2464996.2465013}
\BIBentrySTDinterwordspacing

\bibitem{DBLP:journals/concurrency/ChenXCLG18}
\BIBentryALTinterwordspacing
X.~Chen, P.~Xie, L.~Chi, J.~Liu, and C.~Gong, ``An efficient {SIMD} compression
  format for sparse matrix-vector multiplication,'' \emph{Concurr. Comput.
  Pract. Exp.}, vol.~30, no.~23, 2018. [Online]. Available:
  \url{https://doi.org/10.1002/cpe.4800}
\BIBentrySTDinterwordspacing

\bibitem{DBLP:conf/cgo/XieZLGJHZ18}
\BIBentryALTinterwordspacing
B.~Xie, J.~Zhan, X.~Liu, W.~Gao, Z.~Jia, X.~He, and L.~Zhang, ``{CVR:}
  efficient vectorization of spmv on x86 processors,'' in \emph{Proceedings of
  the 2018 International Symposium on Code Generation and Optimization, {CGO}
  2018, V{\"{o}}sendorf / Vienna, Austria, February 24-28, 2018}, J.~Knoop,
  M.~Schordan, T.~Johnson, and M.~F.~P. O'Boyle, Eds.\hskip 1em plus 0.5em
  minus 0.4em\relax {ACM}, 2018, pp. 149--162. [Online]. Available:
  \url{https://doi.org/10.1145/3168818}
\BIBentrySTDinterwordspacing

\bibitem{DBLP:conf/ics/LiuXLXYL18}
\BIBentryALTinterwordspacing
C.~Liu, B.~Xie, X.~Liu, W.~Xue, H.~Yang, and X.~Liu, ``Towards efficient spmv
  on sunway manycore architectures,'' in \emph{Proceedings of the 32nd
  International Conference on Supercomputing, {ICS} 2018, Beijing, China, June
  12-15, 2018}.\hskip 1em plus 0.5em minus 0.4em\relax {ACM}, 2018, pp.
  363--373. [Online]. Available: \url{https://doi.org/10.1145/3205289.3205313}
\BIBentrySTDinterwordspacing

\bibitem{DBLP:journals/isci/ChenXWTL20}
\BIBentryALTinterwordspacing
Y.~Chen, G.~Xiao, F.~Wu, Z.~Tang, and K.~Li, ``tpspmv: {A} two-phase
  large-scale sparse matrix-vector multiplication kernel for manycore
  architectures,'' \emph{Inf. Sci.}, vol. 523, pp. 279--295, 2020. [Online].
  Available: \url{https://doi.org/10.1016/j.ins.2020.03.020}
\BIBentrySTDinterwordspacing

\bibitem{DBLP:journals/iotj/XiaoCLZ20}
\BIBentryALTinterwordspacing
G.~Xiao, Y.~Chen, C.~Liu, and X.~Zhou, ``ahspmv: An autotuning hybrid computing
  scheme for spmv on the sunway architecture,'' \emph{{IEEE} Internet Things
  J.}, vol.~7, no.~3, pp. 1736--1744, 2020. [Online]. Available:
  \url{https://doi.org/10.1109/JIOT.2019.2947257}
\BIBentrySTDinterwordspacing

\bibitem{DBLP:conf/3pgcic/MehrezH13}
\BIBentryALTinterwordspacing
I.~Mehrez and O.~Hamdi{-}Larbi, ``{SMVP} distribution using hypergraph model
  and {S-GBNZ} algorithm,'' in \emph{Eighth International Conference on P2P,
  Parallel, Grid, Cloud and Internet Computing, 3PGCIC 2013, Compiegne, France,
  October 28-30, 2013}, F.~Xhafa, L.~Barolli, D.~Nace, S.~Venticinque, and
  A.~Bui, Eds.\hskip 1em plus 0.5em minus 0.4em\relax {IEEE}, 2013, pp.
  235--241. [Online]. Available: \url{https://doi.org/10.1109/3PGCIC.2013.41}
\BIBentrySTDinterwordspacing

\bibitem{DBLP:conf/ccgrid/MiYYWL23}
\BIBentryALTinterwordspacing
H.~Mi, X.~Yu, X.~Yu, S.~Wu, and W.~Liu, ``Balancing computation and
  communication in distributed sparse matrix-vector multiplication,'' in
  \emph{23rd {IEEE/ACM} International Symposium on Cluster, Cloud and Internet
  Computing, CCGrid 2023, Bangalore, India, May 1-4, 2023}, Y.~Simmhan,
  I.~Altintas, A.~L. Varbanescu, P.~Balaji, A.~S. Prasad, and L.~Carnevale,
  Eds.\hskip 1em plus 0.5em minus 0.4em\relax {IEEE}, 2023, pp. 535--544.
  [Online]. Available: \url{https://doi.org/10.1109/CCGrid57682.2023.00056}
\BIBentrySTDinterwordspacing

\bibitem{page2021scalability}
B.~A. Page and P.~M. Kogge, ``Scalability of hybrid spmv with hypergraph
  partitioning and vertex delegation for communication avoidance,'' in
  \emph{International Conference on High Performance Computing \& Simulation
  (HPCS 2020)}, 2021.

\bibitem{DBLP:conf/bigdataconf/MayerMBER18}
\BIBentryALTinterwordspacing
C.~Mayer, R.~Mayer, S.~Bhowmik, L.~Epple, and K.~Rothermel, ``{HYPE:} massive
  hypergraph partitioning with neighborhood expansion,'' in \emph{{IEEE}
  International Conference on Big Data {(IEEE} BigData 2018), Seattle, WA, USA,
  December 10-13, 2018}, N.~Abe, H.~Liu, C.~Pu, X.~Hu, N.~K. Ahmed, M.~Qiao,
  Y.~Song, D.~Kossmann, B.~Liu, K.~Lee, J.~Tang, J.~He, and J.~S. Saltz,
  Eds.\hskip 1em plus 0.5em minus 0.4em\relax {IEEE}, 2018, pp. 458--467.
  [Online]. Available: \url{https://doi.org/10.1109/BigData.2018.8621968}
\BIBentrySTDinterwordspacing

\bibitem{LinX17}
\BIBentryALTinterwordspacing
S.~Lin and Z.~Xie, ``{A Jacobi{\_}PCG Solver for Sparse Linear Systems on
  Multi-GPU Cluster},'' \emph{The Journal of Supercomputing}, vol.~73, no.~1,
  pp. 433--454, 2017. [Online]. Available:
  \url{https://doi.org/10.1007/s11227-016-1887-4}
\BIBentrySTDinterwordspacing

\bibitem{DBLP:journals/tjs/KhodjaCGB14}
\BIBentryALTinterwordspacing
L.~Z. Khodja, R.~Couturier, A.~Giersch, and J.~M. Bahi, ``Parallel sparse
  linear solver with {GMRES} method using minimization techniques of
  communications for {GPU} clusters,'' \emph{J. Supercomput.}, vol.~69, no.~1,
  pp. 200--224, 2014. [Online]. Available:
  \url{https://doi.org/10.1007/s11227-014-1143-8}
\BIBentrySTDinterwordspacing

\bibitem{DBLP:conf/fedcsis/BylinaBSS14}
\BIBentryALTinterwordspacing
B.~Bylina, J.~Bylina, P.~Stpiczynski, and D.~Szalkowski, ``Performance analysis
  of multicore and multinodal implementation of spmv operation,'' in
  \emph{Proceedings of the 2014 Federated Conference on Computer Science and
  Information Systems, Warsaw, Poland, September 7-10, 2014}, ser. Annals of
  Computer Science and Information Systems, M.~Ganzha, L.~A. Maciaszek, and
  M.~Paprzycki, Eds., vol.~2, 2014, pp. 569--576. [Online]. Available:
  \url{https://doi.org/10.15439/2014F313}
\BIBentrySTDinterwordspacing

\bibitem{DBLP:conf/ics/LeeE08}
\BIBentryALTinterwordspacing
S.~Lee and R.~Eigenmann, ``Adaptive runtime tuning of parallel sparse
  matrix-vector multiplication on distributed memory systems,'' in
  \emph{Proceedings of the 22nd Annual International Conference on
  Supercomputing, {ICS} 2008, Island of Kos, Greece, June 7-12, 2008}, P.~Zhou,
  Ed.\hskip 1em plus 0.5em minus 0.4em\relax {ACM}, 2008, pp. 195--204.
  [Online]. Available: \url{https://doi.org/10.1145/1375527.1375558}
\BIBentrySTDinterwordspacing

\bibitem{ma2021developing}
W.~Ma, Y.~Hu, W.~Yuan, and X.~Liu, ``Developing a multi-gpu-enabled
  preconditioned gmres with inexact triangular solves for block sparse
  matrices,'' \emph{Mathematical Problems in Engineering}, vol. 2021, pp.
  1--17, 2021.

\bibitem{DBLP:conf/compute/IndarapuMK14}
\BIBentryALTinterwordspacing
S.~R. K.~B. Indarapu, M.~K. Maramreddy, and K.~Kothapalli, ``Architecture- and
  workload- aware heterogeneous algorithms for sparse matrix vector
  multiplication,'' in \emph{Proceedings of the 7th {ACM} India Computing
  Conference, {COMPUTE} 2014, Nagpur, India, October 9-11, 2014},
  P.~Bhattacharyya, P.~J. Narayanan, and S.~Padmanabhuni, Eds.\hskip 1em plus
  0.5em minus 0.4em\relax {ACM}, 2014, pp. 3:1--3:9. [Online]. Available:
  \url{https://doi.org/10.1145/2675744.2675749}
\BIBentrySTDinterwordspacing

\bibitem{DBLP:conf/parco/CardelliniFF13}
\BIBentryALTinterwordspacing
V.~Cardellini, A.~Fanfarillo, and S.~Filippone, ``Heterogeneous sparse matrix
  computations on hybrid {GPU/CPU} platforms,'' in \emph{Parallel Computing:
  Accelerating Computational Science and Engineering (CSE), Proceedings of the
  International Conference on Parallel Computing, ParCo 2013, 10-13 September
  2013, Garching (near Munich), Germany}, ser. Advances in Parallel Computing,
  M.~Bader, A.~Bode, H.~Bungartz, M.~Gerndt, G.~R. Joubert, and F.~J. Peters,
  Eds., vol.~25.\hskip 1em plus 0.5em minus 0.4em\relax {IOS} Press, 2013, pp.
  203--212. [Online]. Available:
  \url{https://doi.org/10.3233/978-1-61499-381-0-203}
\BIBentrySTDinterwordspacing

\bibitem{DBLP:journals/tc/YangLML15}
\BIBentryALTinterwordspacing
W.~Yang, K.~Li, Z.~Mo, and K.~Li, ``Performance optimization using partitioned
  spmv on gpus and multicore cpus,'' \emph{{IEEE} Trans. Computers}, vol.~64,
  no.~9, pp. 2623--2636, 2015. [Online]. Available:
  \url{https://doi.org/10.1109/TC.2014.2366731}
\BIBentrySTDinterwordspacing

\bibitem{DBLP:journals/jpdc/YangLL17}
\BIBentryALTinterwordspacing
W.~Yang, K.~Li, and K.~Li, ``A hybrid computing method of spmv on {CPU-GPU}
  heterogeneous computing systems,'' \emph{J. Parallel Distributed Comput.},
  vol. 104, pp. 49--60, 2017. [Online]. Available:
  \url{https://doi.org/10.1016/j.jpdc.2016.12.023}
\BIBentrySTDinterwordspacing

\bibitem{DBLP:journals/jpdc/BraunSBBMRRTYHF01}
\BIBentryALTinterwordspacing
T.~D. Braun, H.~J. Siegel, N.~Beck, L.~B{\"{o}}l{\"{o}}ni, M.~Maheswaran, A.~I.
  Reuther, J.~P. Robertson, M.~D. Theys, B.~Yao, D.~A. Hensgen, and R.~F.
  Freund, ``A comparison of eleven static heuristics for mapping a class of
  independent tasks onto heterogeneous distributed computing systems,''
  \emph{J. Parallel Distributed Comput.}, vol.~61, no.~6, pp. 810--837, 2001.
  [Online]. Available: \url{https://doi.org/10.1006/jpdc.2000.1714}
\BIBentrySTDinterwordspacing

\bibitem{DBLP:journals/pc/0002V15}
\BIBentryALTinterwordspacing
W.~Liu and B.~Vinter, ``Speculative segmented sum for sparse matrix-vector
  multiplication on heterogeneous processors,'' \emph{Parallel Comput.},
  vol.~49, pp. 179--193, 2015. [Online]. Available:
  \url{https://doi.org/10.1016/j.parco.2015.04.004}
\BIBentrySTDinterwordspacing

\bibitem{DBLP:journals/tpds/XiaoLCHZL21}
\BIBentryALTinterwordspacing
G.~Xiao, K.~Li, Y.~Chen, W.~He, A.~Y. Zomaya, and T.~Li, ``Caspmv: {A}
  customized and accelerative spmv framework for the sunway taihulight,''
  \emph{{IEEE} Trans. Parallel Distributed Syst.}, vol.~32, no.~1, pp.
  131--146, 2021. [Online]. Available:
  \url{https://doi.org/10.1109/TPDS.2019.2907537}
\BIBentrySTDinterwordspacing

\bibitem{li2023haspmv}
W.~Li, H.~Cheng, Z.~Lu, Y.~Lu, and W.~Liu, ``Haspmv: Heterogeneity-aware sparse
  matrix-vector multiplication on modern asymmetric multicore processors,'' in
  \emph{2023 IEEE International Conference on Cluster Computing
  (CLUSTER)}.\hskip 1em plus 0.5em minus 0.4em\relax IEEE Computer Society,
  2023, pp. 209--220.

\bibitem{Davis2011_SuiteSparse}
\BIBentryALTinterwordspacing
T.~A. Davis and Y.~Hu, ``{The University of Florida Sparse Matrix
  Collection},'' \emph{ACM Transactions on Mathematical Software}, vol.~38,
  no.~1, pp. 1:1--1:25, 2011. [Online]. Available:
  \url{https://doi.org/10.1145/2049662.2049663}
\BIBentrySTDinterwordspacing

\bibitem{cusparse}
\BIBentryALTinterwordspacing
NVIDIA, ``{NVIDIA cuSPARSE Library},'' 2023. [Online]. Available:
  \url{https://docs.nvidia.com/cuda/archive/12.0.0/index.html}
\BIBentrySTDinterwordspacing

\bibitem{Magma-sparse1}
H.~Anzt, W.~Sawyer, S.~Tomov, P.~Luszczek, I.~Yamazaki, and J.~Dongarra,
  ``{Optimizing Krylov Subspace Solvers on Graphics Processing Units},'' in
  \emph{Fourth International Workshop on Accelerators and Hybrid Exascale
  Systems (AsHES), IPDPS 2014}, IEEE.\hskip 1em plus 0.5em minus 0.4em\relax
  Phoenix, AZ: IEEE, 05-2014 2014.

\bibitem{Magma-sparse2}
I.~Yamazaki, H.~Anzt, S.~Tomov, M.~Hoemmen, and J.~Dongarra, ``{Improving the
  Performance of CA-GMRES on Multicores with Multiple GPUs},'' in \emph{IPDPS
  2014}.\hskip 1em plus 0.5em minus 0.4em\relax Phoenix, AZ: IEEE, 05-2014
  2014.

\bibitem{merge-spmv}
D.~Merrill, ``{Merge-based Parallel Sparse Matrix-Vector Multiplication},''
  \url{https://github.com/dumerrill/merge-spmv}, (Accessed on 10/5/2023).

\bibitem{SURAA}
\BIBentryALTinterwordspacing
T.~Muhammed, R.~Mehmood, A.~Albeshri, and I.~Katib, ``{SURAA: A Novel Method
  and Tool for Loadbalanced and Coalesced SpMV Computations on GPUs},''
  \emph{Applied Sciences}, vol.~9, no.~5, p. 947, Mar 2019. [Online].
  Available: \url{http://dx.doi.org/10.3390/app9050947}
\BIBentrySTDinterwordspacing

\bibitem{byun2012autotuning}
J.-H. Byun, R.~Lin, K.~A. Yelick, and J.~Demmel, ``Autotuning sparse
  matrix-vector multiplication for multicore,'' \emph{EECS, UC Berkeley, Tech.
  Rep}, 2012.

\bibitem{DBLP:journals/toms/TanLL18}
\BIBentryALTinterwordspacing
G.~Tan, J.~Liu, and J.~Li, ``{Design and Implementation of Adaptive SpMV
  Library for Multicore and Many-Core Architecture},'' \emph{ACM Transactions
  on Mathematical Software}, vol.~44, no.~4, pp. 46:1--46:25, 2018. [Online].
  Available: \url{https://doi.org/10.1145/3218823}
\BIBentrySTDinterwordspacing

\bibitem{chen2022explicit}
C.~Chen, ``Explicit caching hyb: a new high-performance spmv framework on
  gpgpu,'' \emph{arXiv preprint arXiv:2204.06666}, 2022.

\bibitem{DBLP:conf/sc/DuLWLTS22}
\BIBentryALTinterwordspacing
Z.~Du, J.~Li, Y.~Wang, X.~Li, G.~Tan, and N.~Sun, ``{AlphaSparse: Generating
  High Performance SpMV Codes Directly from Sparse Matrices},'' in
  \emph{{SC22:} International Conference for High Performance Computing,
  Networking, Storage and Analysis, Dallas, TX, USA, November 13-18,
  2022}.\hskip 1em plus 0.5em minus 0.4em\relax {IEEE}, 2022, pp. 1--15.
  [Online]. Available: \url{https://doi.org/10.1109/SC41404.2022.00071}
\BIBentrySTDinterwordspacing

\end{thebibliography}

\end{document}